\documentstyle[psfig]{report}
\def\TL{\hfil$\displaystyle{##}$}
\def\TR{$\displaystyle{{}##}$\hfil}
\def\TC{\hfil$\displaystyle{##}$\hfil}
\def\TT{\hbox{##}}

\def\comment#1{}
\def\fixit#1{}

\def\tf#1#2{{\textstyle{#1 \over #2}}}
\def\df#1#2{{\displaystyle{#1 \over #2}}}

\def\mop#1{\mathop{\rm #1}\nolimits}

\def\coth{\mop{coth}}
\def\csch{\mop{csch}}
\def\sech{\mop{sech}}
\def\Vol{\mop{Vol}}
\def\vol{\mop{vol}}
\def\diag{\mop{diag}}
\def\tr{\mop{tr}}
\def\Disc{\mop{Disc}}

\def\overleftrightarrow#1{\vbox{\ialign{##\crcr
     $\leftrightarrow$\crcr\noalign{\kern-0pt\nointerlineskip}
     $\hfil\displaystyle{#1}\hfil$\crcr}}}

\def\SU{{\rm SU}}

\def\lsim{\mathrel{\mathstrut\smash{\ooalign{\raise2.5pt\hbox{$<$}\cr\lower2.5pt\hbox{$\sim$}}}}}
\def\gsim{\mathrel{\mathstrut\smash{\ooalign{\raise2.5pt\hbox{$>$}\cr\lower2.5pt\hbox{$\sim$}}}}}

\def\slashed#1{\ooalign{\hfil\hfil/\hfil\cr $#1$}}

\def\sqr#1#2{{\vcenter{\vbox{\hrule height.#2pt
         \hbox{\vrule width.#2pt height#1pt \kern#1pt
            \vrule width.#2pt}
         \hrule height.#2pt}}}}
\def\square{\mathop{\mathchoice\sqr56\sqr56\sqr{3.75}4\sqr34\,}\nolimits}

\def\href#1#2{#2}  

\def\lbldef#1#2{\expandafter\gdef\csname #1\endcsname {#2}}
\def\eqn#1#2{\lbldef{#1}{(\ref{#1})}%
\begin{equation} #2 \label{#1} \end{equation}}
\def\eqalign#1{\vcenter{\openup1\jot
    \halign{\strut\span\TL & \span\TR\cr #1 \cr
   }}}
\def\eno#1{(\ref{#1})}

\begin{document}

\begin{titlepage}

\begin{center}

\ 

\vskip 1.5cm

{\LARGE {\bf Dynamics of D-brane Black Holes}}

\vskip 1.2cm

{\large Steven S. Gubser\footnote{{\tt ssgubser@born.harvard.edu}}}

\vspace{1cm}

{\bf Abstract}
\end{center}

\noindent
We explore the interplay between black holes in supergravity and
quantum field theories on the world-volumes of D-branes.  Each sheds
light on the other in various ways.  The world-volume description
provides a statistical mechanical picture of black hole entropy and a
manifestly unitary account of Hawking radiation.  The black hole
description elucidates the nature of the couplings between the
world-volume theory and supergravity in the bulk.  Finally, the
computation of Green's functions in the world-volume theory can be
mapped in a large $N$ limit to a classical minimization problem in
supergravity.

A brief summary of black hole entropy calculations for D-brane black
holes is followed by a detailed study of particle absorption by black
holes whose string theory description involves D-branes intersecting
along a string.  A two-dimensional conformal field theory with
large central charge describes the low-energy excitations of this
string.  In the string theory description, Hawking radiation is simply
the time-reversal of particle absorption.  The greybody factors
(deviations in the power spectrum from Planck's law) are
characteristic of conformal field theory at finite temperature.

Particle absorption by extremal three-branes is examined next, with
particular attention to the implications for supersymmetric gauge
theory in four dimensions.  A fascinating duality between supergravity
and gauge theory emerges from the study of these processes.  Fields of
supergravity are dual to local operators in the gauge theory.  A
non-renormalization theorem of ${\cal N} = 4$ gauge theory helps
explain why certain aspects of the duality can be explored
perturbatively.  Anomalous dimensions of a large class of local
operators in the gauge theory are shown to become large at strong
't~Hooft coupling, signaling a possible simplification of ${\cal N} =
4$ gauge theory in this limit.

\vskip0.8cm

\begin{center}
A dissertation presented to the faculty of Princeton University \\
in candidacy for the degree of Doctor of Philosophy
\end{center}

\begin{center}
Recommended for acceptance by the Department of Physics
\end{center}

\vskip0.8cm
June 1998

\vskip1cm
\copyright Copyright by Steven Scott Gubser, 1998.  All rights reserved.

\end{titlepage}

\newpage
\vskip1cm
\begin{center}
{\Large Acknowledgements}
\end{center}

\vskip0.3cm

The research summarized in this thesis was supported in part by DOE
Grant No. DE-FG02-91ER40671, by the NSF Presidential Young
Investigator Award PHY-9157482, by the James S.~McDonnell Foundation
under Grant No. 91-48, and by the Hertz Foundation.

I would like to thank my collaborators: Curt Callan, Aki Hashimoto,
Michael Krasnitz, Juan Maldacena, Amanda Peet, Alexander Polyakov,
Arkady Tseytlin, and particularly my advisor, Igor Klebanov.  Without
them not a tenth of the research in this dissertation would have been
realized. 

Thanks to all my teachers, particularly E.~Lieb, P.~Meyers, H.~Osborn,
B.~van Fraassen, A.~Eisenkraft, P.~Ryan, P.~Sarnak, S.~Sondhi, and
D.~Gross.

I am grateful to Laurel Lerner for her encouragement, both on matters
academic and pianistic.

Finally, thanks to my brother Charles, for leading the way, to my
grandparents, for many years of encouragement, and to my parents.

\newpage
\tableofcontents
\newpage

\chapter{Introduction}
\label{Introduction}

The usual conception of how string theory (see for example \cite{GSW})
makes contact with the physical world is that string theory is
compactified from ten to four dimensions on a manifold of size on the
order $10^{-32}\,{\rm cm}$---roughly as small compared to a proton as
a proton is to Mount Everest.  All the ``fundamental'' particles of
the Standard Model of particle physics are supposed to be realized as
different excitation modes of strings.  The most powerful of particle
accelerators probe distances only about a sixth of the way (on a
logarithmic scale) between the proton's size and the string scale, so
it is difficult to imagine how the ``stringy'' character of elementary
particles could ever be directly observed.  There is however an
extensive literature investigating the phenomenological implications
of string theory and how contact may yet be made with experiment (for
a recent review see \cite{dineTASI}).

String theory is also a quantum theory of gravity.  In fact, it is the
only consistent quantum theory to date which contains gravity in any
dynamical sense.  (In two and three dimensions, theories of quantum
gravity can be formulated, but they are non-dynamical in the sense
that gravitons do not propagate).\fixit{Make sure: ask Lorenzo}
Another avenue, then, for connecting string theory with the physical
world is via quantum gravitational effects.  As an example one might
consider Hawking radiation from black holes.  This is an essentially
quantum mechanical effect: whereas black holes classically cannot
radiate (due to the causal structure of their geometry), quantum
mechanically they radiate as blackbodies.  For many years a
fundamental, microscopic description of this effect was almost as much
a mystery in string theory as in General Relativity.  However, recent
advances in the description of string-theoretic solitons have enabled
string theorists to give an account of Hawking radiation for a certain
limited class of black holes.  It is worth emphasizing that we are
still very far from experiment, particularly since the black holes
amenable to treatment in string theory in general carry a large
electric charge.  Astrophysically, such a charge would quickly be
neutralized by infalling matter.

The recent developments in the understanding of dualities in string
theory (for a review see \cite{schwarzTASI}) are very much at the
heart of progress in the string-theoretic description of black holes.
A particularly prominent role has been played by the so-called
D-branes \cite{JP}, which are objects derived via a detailed
understanding of T-duality in open and closed string theories.  Closed
string theories require that strings cannot have ends, whereas open
strings by definition do have ends.  However, the full consistency of
closed string theories under duality symmetries requires the existence
of D-branes, which can be thought of as special surfaces on which
strings are allowed to end.  But they are more than this mathematical
characterization suggests: they are dynamical objects in themselves,
with a definite tension which at weak coupling is large compared to
the tension of fundamental strings.  The large tension of D-branes
suggests that they may be suitable objects from which to build black
holes.  The development of this idea and its implications for the
description of dynamical effects like Hawking radiation are two of the
main themes of this thesis.

The link between extremal or near-extremal black holes and the
world-volume theories describing excitations of branes has been
explored in a now-voluminous literature.  Various aspects of the
general program of understanding this link have been initiated in
\cite{sv,cm} (entropy of effective string models), \cite{dmw,dmOne}
(Hawking radiation processes), and \cite{kleb,gukt} (absorption into
extremal three-branes).  A tangentially related subject, the probing
of sub-string-scale geometries using D-branes, grew from the initial
work of \cite{kabat} into the vast subject of matrix theory
\cite{bfss}.

The excitations of D-branes at low energies are described by
supersymmetric gauge theories \cite{Witten}.  This observation opens
up yet another avenue for connecting string theory with mainstream
physics: if on one hand D-branes describe black holes, and on the
other hand are characterized at low energies by gauge theory, then
perhaps some information about gauge theory is encoded in the
supergravity solutions of the corresponding black holes.  The
realization that there are concrete ways of developing this
correspondence is at present one of the main sources of excitement in
the string theory community.  There are even hints that supersymmetry
may no longer be an essential ingredient in the gravity--gauge theory
correspondence.  Earlier work on extracting gauge theory physics from
intersecting brane configurations \cite{WitMQCD} relies very little on
supergravity, but has yielded some remarkable insights into the
geometric character of supersymmetric gauge theory (for a review see
\cite{GivKut}).

In the past few months, with the appearance of the papers
\cite{jthroat,acft,gkPol,WitHolOne,jWilson}, our entire understanding
of the link through string theory and D-branes between supergravity
and gauge theory has evolved rapidly.  The essential ingredients in
the new approach are the following two insights: 1) information about
the infrared behavior of gauge theories is encoded in the near-horizon
geometry of black branes in supergravity; and 2) perturbations to the
gauge theory can be implemented by constraints on the supergravity
fields on a hypersurface which surrounds the brane at a distance close
to where the near-horizon geometry merges with asymptotically flat
space.  The second insight, achieved in various forms in
\cite{acft,gkPol,WitHolOne,jWilson}, has been termed ``holography'' by
Witten.  Although this term (originally proposed by 't~Hooft and
Susskind as applicable to black holes) has been used in a somewhat 
murky fashion in the string theory literature, its current application
is appropriate in the following sense.  The dynamics of the gauge
theory is mirrored in the supergravity theory subjected to boundary
conditions on the hypersurface, which is of one lower dimension than
the ambient space (but of the same dimension as the world-volume of
the gauge theory).  Thus one views the gauge theory as a hologram of
the supergravity.

All but the last two chapters of this thesis are devoted to the
pre-holography era of black holes in string theory.
Chapter~\ref{EntUnp} presents a brief investigation of the
entropy of black holes built out of D-branes.  The focus is on the two
examples which have been most intensively studied in the literature:
D3-branes in ten-dimensional spacetime, and three-charge black holes
with regular horizon in five-dimensional spacetime.  The remarkable
insight that the D-brane approach yields is that the entropy of
certain black holes can be explained in terms of the statistical
mechanics of the world-volume theory of D-branes.

The remainder of the thesis is split into two parts.  The first part,
chapters~\ref{FixedScalars}-\ref{PhotonAbsorption}, is based on the
papers \cite{cgkt,gpartial,gphoton}.  The focus is on the string
theory description of Hawking radiation and particle absorption by
near-extremal black holes in four and five dimensions.  The string
theory model for these black holes is an ``effective string'' whose
excitations are described by a conformal field theory (CFT).  The
influential paper \cite{sv} includes a computation of the central
charge of this conformal field theory via a moduli space argument.  In
chapter~\ref{EntUnp} we develop a more concise but limited version of
this argument using index theorems.  Here the effective string is
realized as a D3-brane with two of its spatial dimensions wrapped
around a four-dimensional manifold.  A more familiar description,
developed initially in \cite{cm}, is to view the effective string as
some number of D1-branes embedded in D5-branes (whose four extra
dimensions are again wrapped around a four-dimensional manifold).  A
black hole in five dimensions can be obtained by wrapping the
effective string around a circle.  Black holes in four dimensions have
also been treated via effective strings which can be constructed
within string theory \cite{jaFirst,hlm}, but which arise most
naturally intersecting five-branes in M-theory \cite{kt,jaw}.

In the extensive recent literature comparing black hole entropy with
microscopic state counts, black holes which admit an effective string
description have always proven the most reliably tractable.
Agreements between the Bekenstein-Hawking entropy and the statistical
mechanical entropy are no longer considered miraculous in the case of
supersymmetric black holes, because the two calculations are different
ways of counting states that preserve supersymmetry.  These so-called
BPS states are generally protected against mass renormalizations by
their short multiplet structure under the supersymmetry algebra.  More
surprising are successful comparisons of near-extremal black hole
entropy with statistical mechanical analyses, because in such cases
there is no supersymmetry to prevent microstates from changing their
masses as one proceeds from the region of weak coupling where the
statistical mechanical analysis is performed to the region of strong
coupling where supergravity is applicable.  In chapter~\ref{EntUnp} we
will see that such agreements are indeed rather the exception than the
rule.  However, comparisons of the entropy of near-extremal effective
string black holes have succeeded in certain limits \cite{cm}.  As it
turns out, the central charge alone determines the effective string
model's entropy in the relevant limits, usually termed the ``dilute
gas regime.''\footnote{This is only true once the proper
identification of charges and angular momenta has been made in the
effective string description.}  Thus the status of the central charge
as a non-renormalized quantity (namely a measure of the density of BPS
states) basically guarantees the success of the entropy comparisons in
the dilute gas regime.

The original realization \cite{sv} of the effective string CFT was as
a supersymmetric non-linear sigma model whose target space was a
symmetric power of a four dimensional manifold.  It is straightforward
to extract the central charge and proceed with entropy calculations,
but it has proven more difficult to determine what are the primary
fields and how they couple to the fields of supergravity.  Precisely
these questions can be explored via calculations of absorption and
Hawking emission by black holes.

In the semi-classical supergravity description of black holes, Hawking
radiation and particle absorption are radically different phenomena:
absorption is a classical effect, while Hawking radiation is strictly
quantum mechanical.  To be precise, Hawking's formula (see for example
\cite{bd}) relating the emission rate $d\Gamma$ of massless particles 
into an element of
wave-number phase space $d^3 k$ to the absorption cross-section
$\sigma$ is
  \eqn{HawkingRate}{
   d\Gamma = \hbar c {\sigma \over e^{E/k_B T} - 1} 
     {d^3 k \over (2 \pi)^3}
  }
 where the temperature is given in terms of the surface gravity
$\kappa$ at the horizon:
  \eqn{HawkingTemp}{
   T = {\hbar \over c} {\kappa \over 2 \pi k_B} \ .
  }
 Planck's constant is involved in two essential ways, and both the
Hawking rate and the Hawking temperature vanish as $\hbar \to 0$.
From here on we will adopt units in which $\hbar = c = k_B = 1$.

In string theory the relation between absorption and Hawking radiation
is simply time-reversal.  Absorption into an extremal black hole is
mediated by $S$-matrix elements which can be calculated from a
knowledge of couplings between supergravity and the effective string.
Hawking radiation is impossible for an extremal black hole, but the
effective string description extends easily to the near-extremal case,
and then the same $S$-matrix element that describes particle
absorption also mediates Hawking radiation.  Time-reversal symmetry is
broken only by the kinematics of the thermal field theory on the
effective string.  Absorption and emission processes are related by
detailed balance, so that an effective string black hole can be in
equilibrium with a thermal sea of particles in the ambient space (it
should be remarked at this point that unlike Schwarzschild black
holes, near-extremal black holes generically have positive specific
heat, so thermal equilibrium is possible).  Indeed, in the statistical
mechanical string theory picture, \HawkingRate\ is nothing more than
the statement of detailed balance.  A concrete illustration of this
point will be developed in chapter~\ref{FixedScalars} in the context
of fixed scalars.

\begin{figure}
    \centerline{\psfig{figure=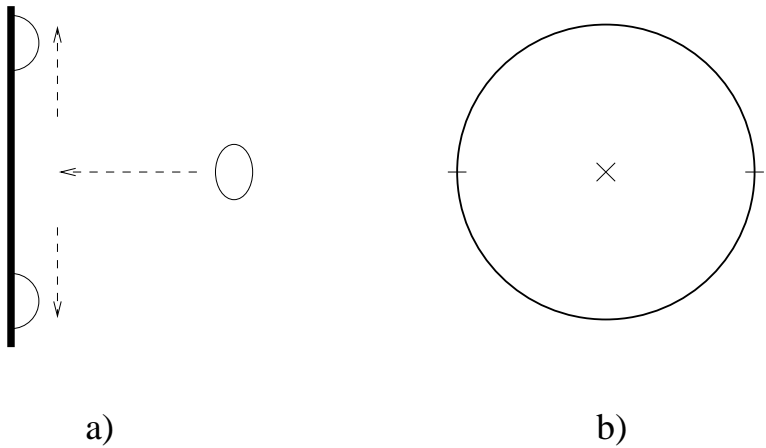,width=4in}}
   \caption{a) Particle absorption by the effective string;
            b) The string disk diagram for absorption.}
   \label{figF}
  \end{figure}

Supergravities in dimensions less than ten with extended supersymmetry
emerge from the low-energy limit of string theory, and in general have
a large number of scalar fields \cite{cremmer}.  For the purpose of
studying effective string black holes, we will primarily be interested
in maximally supersymmetric theories in four or five dimensions, which
admit extremal black holes with regular horizons (\cite{cvetic} and
references therein).  Some of the scalar fields can be regarded as
moduli of the compactified geometry that takes the theory down from
ten dimensions to four or five.  In general, small fluctuations of
these scalars obey the curved space wave equation (the massless
Klein-Gordon equation), and are termed ``minimal scalars.''  Their
absorption into black holes can be described in the effective string
language as a massless closed string state hitting a set of
intersecting D-branes and turning into a pair of open strings that run
in opposite directions along the $1+1$-dimensional intersection
manifold, as depicted in figure~\ref{figF}a).  In principle these
absorption amplitudes can be computed via a string disk diagram,
figure~\ref{figF}b), with one insertion in the bulk (the closed string
vertex) and two on the boundary (the open string vertices).
Calculations along these lines have been explored \cite{aki}, but
difficulties arise due to the complicated nature of the D-brane bound
state which constitutes the effective string.  It proves more
insightful to propose as a low-energy effective action the
Dirac-Born-Infeld (DBI) action \cite{dbiOne}.  Because this action
includes among its fields the induced metric on the effective string,
the dilaton, and other fields of supergravity (in a manner recently
systematized in \cite{cederOne,cederTwo,jhs,bt} for certain simpler
brane configurations than the effective string), there is an implicit
prescription for calculating the coupling between supergravity fields
and effective string excitations.

In view of the extensive literature on minimal scalars, our primary
focus will be on another class, the so-called ``fixed scalars''
(\cite{cgkt} and references therein) whose absorption cross-section
at low energies (and consequently Hawking emission) is suppressed by
powers of the energy relative to that of the minimal scalars.  In the
effective string description, fixed scalars couple to no less than
four excitations on the effective string.  In figure~\ref{figF}b), one
would have four rather than two boundary insertions.  In the DBI
effective action, the coupling turns out to be quartic rather than
quadratic in the elementary fields.  Morally speaking, one can view
the suppression of the absorption cross-section as due to the amount
of phase space available at low energies.  From the point of view of
supergravity, the suppression is due to a potential barrier, which
might also be thought of as a variable mass term in the Klein-Gordon
equation.

The standard WKB method for barrier penetration does not apply to the
calculation of the low-energy cross-section because the wavelength is
longer than the width of the barrier.  Instead, when the linearized
equations of motion are not exactly soluble, one can use approximation
techniques first developed in \cite{unruh}.  A substantial part of
chapters~\ref{FixedScalars}-\ref{ThreeBraneAbs} is devoted to the
application of these techniques to various extremal and near-extremal
black holes.  In particular, chapter~\ref{PartialWaves} is devoted to
an analysis of partial waves of a minimal scalar falling into
near-extremal black holes in five dimensions, while
chapter~\ref{ThreeBraneAbs} focuses on partial waves of minimal
scalars falling into D3-branes in ten dimensions.  Aside from a
derivation of a version of the Optical Theorem in arbitrary
dimensions, the semi-classical calculations in these chapters are
fairly nondescript.  The real interest lies in the relevance to a
microscopic string theory description.

The description of partial wave absorption in the effective string
model is somewhat problematic because of our limited understanding of
the underlying conformal field theory.  It is clear \cite{brekEx} that
angular momentum is carried by the fermionic excitations of the
effective string.  The Pauli Exclusion principle would then seem to
imply that only a finite number of partial waves are capable of
coupling to the effective string.  The analysis of
chapter~\ref{PartialWaves} suggests that this number is large and
possibly related to the principle of cosmic censorship.

In chapter~\ref{PhotonAbsorption}, an attempt is made to move a few
steps closer to astrophysical reality by considering particles with
spin (photons or chiral fermions) falling into a black hole in four
dimensions.  The effective string description is again found to be in
remarkable agreement with the functional form of the cross-sections.
Supersymmetry still plays an essential role in the discussion, since
the black hole under consideration is realized in pure ${\cal N} = 4$
supergravity.  This is not a theory which descends in any simple way
from string theory, so the effective string model lacks a microscopic
justification (in contrast to the case of the D1-D5 bound state in
five dimensions or the four-charge black hole in ${\cal N} = 8$
four-dimensional supergravity).  However, taking it as a useful
effective description in terms of conformal field theory we explore
the properties of conformal Green's functions at finite temperature.
The powerful constraints of conformal invariance essentially dictate
the energy dependence of the cross-section, also termed the greybody
factor.

The second part of the thesis,
chapters~\ref{ThreeBraneAbs}-\ref{GreensFunctions}, is based on the
papers \cite{gukt,gkThree,gkPol}, and is concerned primarily with the
relation via string theory of ten-dimensional type~IIB supergravity
with four-dimensional ${\cal N} = 4$ supersymmetric Yang-Mills (SYM)
theory.  The supergravity theory is the low-energy limit of type~IIB
string theory, which is one of the two maximally supersymmetric string
theories in ten dimensions.  One of the charged branes in this theory
is the D3-brane, which (following \cite{JP}) one views as a
$3+1$-dimensional hypersurface (usually flat) in $9+1$-dimensional
spacetime, on which strings are allowed to end.  One of the early
discoveries \cite{Witten} was that the low-energy effective theory
describing excitations of $N$ coincident D3-branes is ${\cal N} = 4$
SYM theory with gauge group $U(N)$.  It has long been hoped that this
theory, on account of its supersymmetry, conformal symmetry, and
finiteness, would provide one of the simplest examples of an
interacting quantum field theory in $3+1$ dimensions.  However, the
theory has proven notoriously difficult to solve in any sense similar
to the exactly solvable quantum field theories in $1+1$
dimensions---for example rational conformal field theories
\cite{Ginsparg,Moore,fqs}.\fixit{Need to correct the bbl file title
for Moore: RCFT not rcft}\footnote{See however \cite{hw} for recent
progress on extracting correlation functions of ${\cal N} = 4$ SYM
using superspace techniques.}  For this reason, any handle one can get
on ${\cal N} = 4$ SYM via D3-branes is of tremendous interest,
independent of black hole physics.

In fact, the same supergravity methods used to explore low-energy
absorption processes in the case of four- and five-dimensional black
holes apply equally to the case of the extremal D3-brane.  In
chapter~\ref{ThreeBraneAbs} these methods are compared with tree-level
calculations in the world-volume theory.  It is shown that the energy
dependence of absorption cross-sections is predicted reliably by the
gauge theory;\footnote{In light of recent developments
\cite{gkPol,WitHolOne} relating Kaluza-Klein masses in the
near-horizon geometry to the conformal dimensions of gauge theory
operators, this agreement illustrates how the dimensions of operators
coupling to arbitrary partial waves can be computed successfully at
weak coupling.} however it is only for low partial waves (the $s$-wave
or the $p$-wave) for the infalling particle that we have been able to
obtain the properly normalized cross-section through tree-level
calculations.  In a sense even this level of agreement between gauge
theory and supergravity is surprising, because the gauge theory
calculations are valid only when the 't~Hooft coupling $g_{YM}
\sqrt{N}$ is small, whereas the supergravity calculations are reliable
only when the spacetime curvatures are small.  These turn out to be
exactly opposite limits.  In chapter~\ref{SchwingerTerms} we advance
an explanation for why the tree-level gauge theory computation for the
$s$-wave was in fact reliable.  The absorption cross-sections are
related to the cuts in two-point functions in the gauge theory: to be
precise,
  \eqn{TwoCut}{
   \lim_{\kappa \to 0} {i \omega \over \kappa^2} \sigma = 
     \Disc \Pi_2^{\rm SYM}(p^2) \bigg|_{p^\mu = (\omega,0,0,0)} \ .
  }
 In \TwoCut, $\kappa$ is the gravitational constant ($2\kappa^2 = 16
\pi G_N$), $\Pi_2^{\rm SYM}$ is a Green's function in the gauge
theory, and $p^\mu = (\omega,0,0,0)$ is imposed to reflect the
four-dimensional components (parallel to the D3-brane) of the
ten-dimensional momentum $p^M$ of the infalling particle.  The
``decoupling'' limit $\kappa \to 0$ is taken in order to extract from
the cross-section the part which is relevant to the renormalizable
gauge theory.  This corresponds to the leading dependence of $\sigma$
on the energy $\omega$.  In principle, the full cross-section contains
information about the full (non-renormalizable) theory describing
fluctuations of the D3-branes; this is thought to be a non-abelian
extension of the Dirac-Born-Infeld (DBI) action \cite{Ark}.  While the
dynamics of the DBI theory may be reflected in a very complicated
two-point function $\Pi_2^{\rm DBI}$, certain two-point functions in
the renormalizable gauge theory are quite simple and well-understood
on account of its conformal symmetry.  The example considered in
chapter~\ref{SchwingerTerms} is the two-point function of the
stress-energy tensor, which pertains to the absorption of gravitons
polarized parallel to the brane.  This two-point function is in fact
related to a conformal anomaly, and as such can be computed reliably
at one loop.  The corresponding statement regarding the graviton
cross-section is that it may be computed reliably at tree level: all
radiative corrections cancel.

\begin{figure}
    \centerline{\psfig{figure=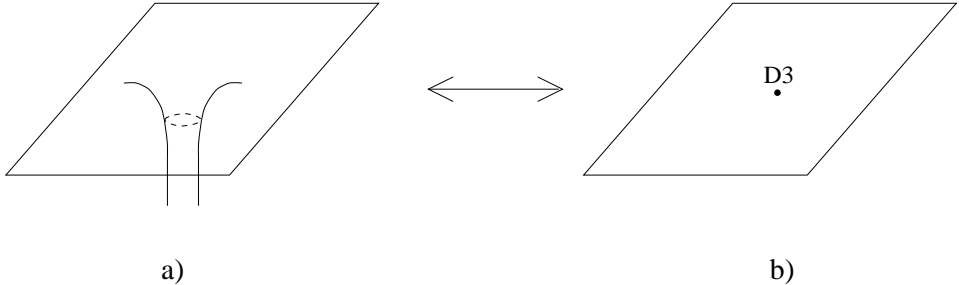,width=5in}}
   \caption{The throat-brane correspondence.}
   \label{figE}
  \end{figure}

In hindsight the formula \TwoCut\ should have immediately suggested
the far-reaching extension of the gauge theory--supergravity
correspondence which we formulate in chapter~\ref{GreensFunctions}.
As a preliminary to the statement of this extension, it is perhaps
useful to review a heuristic picture of the throat-brane
correspondence developed in conversations with Prof.~C.~Callan.
String theory offers two different descriptions of three-branes, each
valid in a different regime of couplings.  The first is the low-energy
supergravity solution, pictured schematically in figure~\ref{figE}a).
 The near-horizon region (pictured as a long throat in the figure) has
the geometry of anti-de~Sitter space, $AdS_5$, times a sphere $S^5$ of
constant radius.  It is joined in the ``mouth region'' (the vicinity
of the dotted line) to the asymptotically flat region.  The second
description, figure~\ref{figE}b) is simply a cluster of $N$ coincident
D3-branes in flat space.  Because these two pictures describe the same
physical system, the response to perturbations (i.e.\ infalling
particles from asymptotically flat infinity) should be the same.
Experience with the linearized equations of motion in supergravity
teaches us that perturbations which are the same far from the brane
are also the same to leading order in small energies when propagated
in the curved geometry down to the dotted line as when propagated to
the location of the D-branes in flat space.  Now imagine solving
supergravity in the throat subject to boundary conditions imposed on
the dotted line.  In view of the physical equivalence of a) and b),
that solution should encode the same physics as the D-brane
world-volume gauge theory, when subjected to the same external
perturbations.

The claim can be made precise in the following fashion.  Consider a
field $\phi$ in supergravity which couples to a local operator ${\cal
O}$ in the world-volume.  For the sake of having a mathematically
well-defined problem, consider Wick-rotating both the gauge theory and
the supergravity to a Euclidean signature (this is a well-defined
operation on the supergravity side because the near-horizon metric is
conformal to flat space).  Let us define $K[\phi_0]$ as the minimum of
the supergravity action $S[\phi]$ in the throat, subject to Dirichlet
boundary conditions $\phi=\phi_0$ imposed at the mouth.  In the
low-energy limit, this is equivalent to imposing asymptotic boundary
conditions on supergravity fields in anti-de~Sitter spacetime.  Then
the extension of \TwoCut\ can be formulated as an identification of
the free energy of the gauge theory with the supergravity action:
 \def\maximum{\mathop{\rm maximum}}
  \eqn{Favorite}{\eqalign{
   Z_{\rm YM}[\phi_0] &= e^{-\beta F_{YM}[\phi_0]} = 
    \left\langle \exp\left(-\int d^4 x \, \phi_0 {\cal O}
    \right) \right\rangle  \cr
    &= e^{-K[\phi_0]} = 
       \maximum_{\phi=\phi_0\ {\rm on}\ \partial AdS} e^{-S[\phi]} \ .
  }}
 The main caveat to this identification is that the validity of
classical supergravity requires that the number of D3-branes be large
so that curvatures are small.  More precisely the 't~Hooft coupling
$g_{YM} \sqrt{N}$ must be large.  It is possible that quantum loop
corrections to the supergravity action (obtained by replacing the
maximum in the last expression by a formal path integral, evaluated
via the saddle point approximation) can provide at least the leading
$1/N^2$ corrections to the large $N$ behavior of $F_{YM}$.  A truly
satisfactory treatment of such effects and the (possibly more
important) $\alpha'$ corrections must be carried out via the full type
IIB string theory in the near-horizon background geometry.
Unfortunately, this geometry is supported by a Ramond-Ramond five-form
flux, and it is not known how to incorporate such a field into the
usual non-linear sigma model approach to string theory.

From the point of view of quantum field theory, however, an exciting
consequence of \Favorite\ is that in the strong coupling limit, ${\cal
N} = 4$ gauge theory seems to simplify to something which can be
captured by a classical minimization problem.\footnote{The following
ideas arose in a discussion with Prof.~T.~Banks.}  In particular, as
will be suggested in chapter~\ref{GreensFunctions}, all operators in
the gauge theory except chiral primaries and their descendants (the
operators corresponding to fields in the supergravity theory) may be
expected to acquire a large mass dimension in the strong coupling
limit.  It is conceivable that they even drop out of the operator
algebra in this limit, leaving us with a greatly reduced set of finite
gauge-invariant Green's functions.  It would be the realization of the
dreams of decades to obtain simple expressions for these Green's
functions, or at least simple equations for them.

\chapter{Entropy considerations}
\label{EntUnp}

A large segment of the recent string theory literature, starting with
\cite{sv}, is devoted to comparisons of the Bekenstein-Hawking entropy
for a black hole or brane with entropies derived by counting states in
some microscopic theory proposed to describe the black hole degrees of
freedom.  Finding the Bekenstein-Hawking entropy is straightforward,
since it is one quarter of the horizon area in Planck units.  As a
rule, the microscopic calculations rely on either conformal invariance
or the assumption of weak coupling to aid in the counting.

In section~\ref{Coincide} we calculate the Bekenstein-Hawking entropy
of clusters of coincident D$p$-branes and compare with the free field
theory result derived from the matter content of the world-volume
theory.  These results represent an extension of the work of
\cite{gkp} and have also appeared in various forms in the literature
\cite{ktPrime,hp,skBH,imsy}.

In section~\ref{Curve} we exhibit a derivation of the effective string
model of five-di\-men\-sion\-al black holes starting from a D3-brane
wrapped around a two-cycle of a four-di\-men\-sion\-al compact
manifold.  The methods are similar to those developed in \cite{jaw} to
study effective string descriptions of four-di\-men\-sion\-al black
holes.  Index theorems relating the dimensions of cohomology groups to
integrals of characteristic classes are used to calculate the central
charge of the CFT characterizing fluctuations of the effective string.

\section{Thermodynamics of coincident branes}
\label{Coincide}

Coincident near-extremal black $p$-branes are described by the
solution \cite{jpTASI,hp}
  \eqn{SMetric}{\eqalign{
   ds^2 &= -A(r) dt^2 + B(r) dr^2 + C(r) r^2 d\Omega_{n+1}^2 + 
            D(r) dy^i dy^i  \cr
        &= -{h \over \sqrt{f}} dt^2 + {\sqrt{f} \over h} dr^2 + 
           \sqrt{f} r^2 d\Omega_{n+1}^2 + 
           {1 \over \sqrt{f}} dy^i dy^i \cr
   e^{2 \phi} &= f^{(n-4)/2} \ ,
  }}
 where
  \eqn{HandF}{
   h = 1 - {r_0^n \over r^n} \qquad\qquad 
   f = 1 + {r_0^n \sinh^2 \alpha \over r^n} \ .
  }
 In \eno{SMetric}, $ds^2 = G_{\mu\nu} dx^\mu dx^\nu$ is the string
metric, related to the Einstein metric $g_{\mu\nu}$ by $g_{\mu\nu} =
e^{-\phi/2} G_{\mu\nu}$.  The $y^i$ are the $p$ spatial coordinates of
the $p$-brane world-volume.  For convenience we set $n=7-p$.

Newton's constant in uncompactified type~II string theory is
\cite{hms}
  \eqn{NewtonG}{
   16 \pi G = 2 \kappa^2 = (2\pi)^7 g^2 (\alpha')^4 \ .
  }
 The right hand side is the value at infinity, where the dilaton
$\phi$ vanishes.  With varying $\phi$, the string coupling is really
$g e^\phi$.  The Planck length $\ell_{\rm Pl}$ therefore acquires a
factor $e^{\phi/4}$.  We have to include this factor when we calculate
the Bekenstein-Hawking entropy: that entropy is roughly one ``string
bit'' per unit Planck area $\ell_{\rm Pl}^8$.  More precisely,
  \eqn{BekHawk}{
   S_{\rm cl} = {A_{\rm str} \over 4 G e^{2 \phi}}
         = {4 \pi \Omega_{n+1} \over 2 \kappa^2} V r_0^{n+1}
            \cosh \alpha \ ,
  }
 where $A_{\rm str}$ is measured using the string metric and 
$\Omega_{n+1}$ is the volume of the sphere $S^{n+1}$:
  \eqn{SphereVol}{
   \Omega_m = \Vol S^m = 2 {\pi^{m+1 \over 2} \over 
    \Gamma\left( {m+1 \over 2} \right)} \ .
  }

The mass of black $p$-branes was computed in general by Lu in
\cite{lu}.  To apply his results it is necessary to use the Einstein
metric.  The result is
  \eqn{ADMMass}{
   M_{\rm ADM} = {\Omega_{n+1} V \over 2 \kappa^2} r_0^n
    \left( {n+2 \over 2} + {n \over 2} \cosh 2\alpha \right) \ .
  }
 The non-extremal energy is defined as the difference between this
mass and its extremal limit:
  \eqn{Ecl}{
   E_{\rm cl} = M_{\rm ADM} - M_0 = 
    M_0 \left( {n+2 \over n} \csch 2\alpha + \coth 2\alpha - 1 \right)
  }
 where
  \eqn{Mzero}{
   M_0 = {Q_p V \over (2 \pi)^p g \alpha'^{p+1 \over 2}} \ .
  }
 Lastly, one must consider the charge of the $p$-brane, which is
quantized in a manner consistent with the Dirac quantization condition
\cite{jpTASI}.  For BPS states (extremal branes), the charge $\mu_p$
is related to the tension $\tau_p$:
  \eqn{DTension}{
   2 \kappa^2 \tau_p^2 = \mu_p^2 = (2 \pi)^{7-2p} (\alpha')^{3-p} \ .
  }
 For non-extremal branes, one has the relation \cite{hp}
  \eqn{QuantRA}{
   {\textstyle{1 \over 2}} r_0^n \sinh 2\alpha = 
    {2 \kappa^2 \over n \Omega_{n+1}} Q_p \tau_p
    = {Q_p \sqrt{ 2\kappa^2 (2 \pi)^{7-2p} (\alpha')^{3-p}} \over
       n \Omega_{n+1}}
    = {(2 \pi)^n \over n \Omega_{n+1}} \sqrt{\alpha'}^n g Q_p \ .
  }
 Given $S$ and $E$, we can compute $T$ from thermodynamics:
  \eqn{TempThermo}{
   T_{\rm cl} = \left( \partial E_{\rm cl} \over \partial S_{\rm cl} \right)_{Q,V}
         = {n \over 4 \pi r_0 \cosh \alpha} \ .
  }
 This value precisely matches with Hawking's formula 
  $T = {\kappa \over 2 \pi}$
 where $\kappa$ is the surface gravity, computed according to the
formula 
  \eqn{SurfKappa}{
   \kappa = \left. {\partial A / \partial r \over 2 \sqrt{AB}}
    \right|_{r=r_0} \ ,
  }
 where $A$ and $B$ are the metric components appearing in \SMetric.
Note that \SurfKappa\ gives the same answer for the string metric as
for the Einstein metric.  This will always happen unless the dilaton
does something singular on the horizon, which would seem unphysical
because the horizon is supposed to be locally undetectable to an
infalling observer.

By straightforward state counting, a free gas of $8 Q_p^2$ species of
bosons and an equal number of fermions at temperature $T$ occupying a
volume $V$ can be shown to have free energy
  \eqn{FreeF}{
   F_{\rm free} = -8 Q_p^2 {\Omega_{p-1} \over (2\pi)^p} 
        \Gamma(p) \zeta(p+1) (2-2^{-p})
        V T^{p+1} \ .
  }
 It follows that the energy and entropy are given by
  \eqn{EandS}{
   E_{\rm free} = -pF_{\rm free} \qquad\qquad 
   S_{\rm free} = -{p+1 \over T} F_{\rm free} \ .
  }
 The $p=3$ case agrees with the results of \cite{gkp}.  Note however
that here we are taking $V$ and $T$ as the independent variables which
are identified between GR and D-branes.

The quantity which can most directly be compared with field theory is
the free energy:
  \eqn{Fcl}{
   F_{\rm cl} = E_{\rm cl} - T_{\rm cl} S_{\rm cl} 
     = -M_0 \left( {n-2 \over n} \csch 2\alpha - \coth 2\alpha + 
        1 \right) \ .
  }
 If we define $\chi$ by $F_{\rm cl} = F_{\rm free} \chi$, then 
  \eqn{chiDef}{
   \chi = c \left[ {g Q_p \over (\sqrt{\alpha'} T)^{n-4}}
    \right]^{-{n-4 \over n-2}} 
  }
 where $c$ is a pure number:
  \eqn{cDef}{
   c = {4^{4n-3 \over n-2} (n-2) n^{-3n \over n-2} 
    \pi^{n^2-3n+10 \over 2 (n-2)} \over 2^8 - 2^n}
    {\Gamma\left( {7-n \over 2} \right) 
     \Gamma\left( {n+2 \over 2} \right)^{2 \over n-2} \over
     \Gamma(7-n) \zeta(8-n)} \ .
  }
 When $n=4$ (i.e.{} for the three-brane), $\chi = c = 3/4$, a result first
obtained in \cite{gkp,sEntUnp}.  Results have been obtained in
\cite{ktPrime,imsy}.  For $n \neq 4$, the ratio in square brackets in
\chiDef\ is precisely the 't~Hooft coupling of the world-volume
gauge theory multiplied by the power of the temperature needed to make
it dimensionless.  This has led to the conjecture \cite{skBH} that the
semi-classical result $F_{\rm cl}$ constitutes a prediction of gravity
about the entropy of maximally supersymmetric gauge theory at strong
coupling.  More generally, one would expect that the exact free energy
interpolates between $F_{\rm free}$ and $F_{\rm cl}$ as one proceeds
from weak to strong coupling.  It remains a mystery how one might make
any concrete check on this interesting speculation.

\section{Embedding of three-branes and the effective string}
\label{Curve}
%
\def\SU{{\rm SU}}
\def\eps{\epsilon}
\def\ch{\mop{ch}}
\def\td{\mop{td}}

In \cite{jaw}, methods of algebraic geometry were used to make an
exact count of the degrees of freedom on the effective string that
underlies microscopic models of four-dimensional black holes
\cite{hlm,kt}.  These powerful methods can be adapted to the analogous
computation for five-dimensional black holes, where the effective
string \cite{sv,cm} is a left-right symmetric conformal field theory
with $(4,4)$ supersymmetry.  When the effective string is realized as
$Q_1$ D1-branes bound to $Q_5$ D5-branes wound around $T^5$, it was
argued in \cite{cm,hs} that the number of left-moving bosons (real
scalars) was $N_L^B = 4 Q_1 Q_5$.  The central charge consequently was
found to be $c = 6 Q_1 Q_5$, and the microscopic entropy of the
conformal field theory (CFT) at level $(N_L,N_R)$,
  \eqn{MicEn}{
   S = 2 \pi \sqrt{Q_1 Q_5} (\sqrt{N_L} + \sqrt{N_R}) \ ,
  }
 agreed with the Bekenstein-Hawking entropy of the geometry with the
same charges.

The purpose of this section is to obtain the formula $N_L^B = 4 Q_1
Q_5$ from a T-dual picture where the D1-D5 system is replaced by two
sets of intersecting D3-branes.  We will restrict ourselves to BPS
brane configurations, with the understanding that once the central
charge of the effective string CFT is established the general
dilute-gas formula \eno{MicEn} follows immediately for near-extremal
black holes as well as extremal ones.  The extremal entropy calculation can
be regarded as a subcase of \cite{sv} where the analysis is particularly
simple.

The D3-brane must span the entire $S^1$ which the effective string
wraps, so its embedding in the transverse $T^4$ is specified by a
two-cycle.  For a more general manifold $M$ of complex dimension $2$,
the embedding of the D3-brane is a complex curve $P$.  In order to
preserve some supersymmetry, $M$ must be either $T^4$ or K3, and $P$
must be a holomorphic cycle.  Using the same letter $P$ to denote the
dual of $P$ in integral cohomology, the requirement of holomorphicity
is $\bar\partial P = 0$.  Carrying over the assumptions of \cite{jaw}
that $M$ is much larger than the string scale, that $S^1$ is much
larger yet, and that $P$ is a very ample divisor of $M$, one is left
with the fairly simple problem of determining the moduli of the
effective string from Kaluza-Klein (KK) reduction of the degrees of
freedom on a D3-brane wrapping a given homology cycle.

The real scalar bosons on the effective string come from two sources:
the moduli of the D3-brane embedding and the D3-brane gauge field,
KK-reduced to $1+1$ dimensions.  Explicitly,
  \eqn{NLB}{
   N_L^B = d_P + b_1 \ ,
  }
 where $d_P$ is the number of real moduli of the embedding of $P$ in
$M$ and $b_1$ is the first Betti number of $P$.  The number of
right-moving bosons is of course the same since the effective string
in this case is left-right symmetric.  Note in particular that $b_1$,
not the numbers $b_1^\pm$ of self-dual and anti-self-dual one-forms in
$H^1(P)$, enter into the formulas for $N_L^B$ and $N_R^B$.  This is in
contrast to the M5-brane case considered in \cite{jaw}.  The
difference is that the field strength of the gauge field on the
D3-brane is not required to be self-dual, whereas for the M5-brane it
is.

Following the methods of \cite{jaw}, let us now compute $d_P$ and
$b_1$.  It is hoped that this simple application of the powerful
techniques of algebraic geometry will lead to a more detailed
understanding of the effective string.

To compute $d_P$ one needs to give a description of the moduli space
of embeddings of $P$ in $M$.  The holomorphic cycle $P$ cannot be
specified as the vanishing locus of a holomorphic function on $M$
because holomorphic functions on $M$ are constant.  But it is possible
to do something almost as good: one can specify $P$ as the vanishing
locus of a holomorphic section $s$ of a holomorphic line bundle $\cal
L$ over $M$.  The convenience of this description is two-fold.  First,
the Chern class of $\cal L$ is $c({\cal L}) = 1 + P$.  And second, the
moduli space is the (complex) projectivization of the Dolbeault
cohomology space $H^0(M,{\cal L})$ \cite{jaw}.  This is because the
zeroth Dolbeault space specifies precisely the different ways of
picking a section $s$ with $\bar\partial s = 0$; the projectivization
is necessary because $\lambda s$ for nonzero complex $\lambda$
specifies the same surface $P$ as does $s$.  Using an index theorem,
one obtains
  \eqn{IndTh}{\eqalign{
   \sum_{i=0}^2 (-1)^i \dim H^i(M,{\cal L}) &= 
     \int_M e^{c_1({\cal L})} \td(M)  \cr
         \fixit{\ch({\cal L}) \td{M} in general?}
     &= \int_M \left( 1 + P + \tf{1}{2} P^2 \right) 
               \left( 1 + \tf{1}{12} c_2(M) \right)  \cr
     &= \int_M \left( \tf{1}{2} P^2 + \tf{1}{12} c_2(M) \right)  \cr
     &= D + \tf{1}{12} C_2(M) \ .
  }}
 The validity of the second line depends on $c_1(M) = 0$, which is
guaranteed for K3 by $\SU(2)$ holonomy.  Since $M$ is a complex
two-manifold, $C_2(M)$ is the Euler number $\chi(M)$: for $T^4$ this
is $0$ while for K3 it is $24$.  A detailed exposition of the
properties of K3 can be found in \cite{Asp}, and \cite{egh} contains
all the information regarding characteristic classes needed for the
present discussion.  In the third line of \eno{IndTh}, $D$ is the
intersection number of $P$ with itself.  Given an integral basis
$\alpha_A$ of $H^2(M,{\bf Z})$, the intersection form is
  \eqn{IntDef}{
   D_{AB} = \tf{1}{2} \int_M \alpha_A \wedge \alpha_B \ .
  }
 Expanding the cycle $P = \sum_A p^A \alpha_A$ in the dual basis of
the homology group, one obtains $D = \tf{1}{2} D_{AB} p^A p^B$.  For
D3-brane configurations on $T^4$ T-dual to $Q_1$ D1-branes bound to
$Q_5$ D5-branes, one has $P = Q_1 \alpha_1 + Q_5 \alpha_5$ (as before,
in homology) where $\alpha_1$ and $\alpha_5$ are two tori whose cross
product is the full $T^4$: the $x^1$-$x^2$ torus and $x^3$-$x^4$ torus,
for example.  Then $D = Q_1 Q_5$.

Finally, one can use the observation \cite{jaw} that all Dolbeault
numbers but the zeroth vanish when $P$ is very ample to obtain
  \eqn{GotDp}{
   d_p = 2 D + \tf{1}{6} \chi(M) - 2 \ .
  }

The task of computing $b_1$ is conceptually more straightforward than
$d_P$, since $b_1 = 2 - \chi(P)$ where $\chi(P)$ is the Euler
characteristic of $P$.  The Euler character of $P$ is $c_1(P)$, and
using the factorization property of Chern classes one can argue that
$c_1(P) = -P$.  Heuristically, the derivation of this equality rests
on identifying fibers of ${\cal L}$ over a point of $P$ as the part of
$M$'s tangent space normal to $P$.  Roughly speaking, when we take the
tensor sum of normal and tangent bundles to $P$, we obtain the tangent
bundle of $M$ restricted to $P$, whose Chern class is simply $1$.
Chern classes multiply over tensor sums, so $c(P) c({\cal L}) = 1$ and
consequently $c(P) = 1 - P$.  The subtleties we have ignored in this
explanation are concisely dealt in \cite{jaw} via an exact sequence
argument.\comment{splitting principle works in some cases?}  Now one
can compute
  \eqn{EulerComp}{
   \chi(P) = \int_P c_1(P) = -\int_P P = -\int_M P^2 = -2 D
  }
 and obtain at last
  \eqn{GotBone}{
   b_1 = 2 D + 2 \ .
  }
 Putting everything together, 
  \eqn{ExactNLB}{
   N_L^B = \tf{2}{3} c_L = 4 D + \tf{1}{6} \chi(M) = 
    \left\{ 
     \vcenter{\openup1\jot
      \halign{\strut\ \span\TR\quad & \span\TT\cr
       4 D & for $M = T^4$  \cr
       4 D + 4 & for $M = \hbox{K3}$. \cr
     }}
    \right.
  }
 Correction terms like $\tf{1}{6} \chi{M}$, subleading for large
intersection number, were argued in \cite{jaw} to correspond to
supergravity loop corrections.

\def \hal {{1\ov 2}}

\def\D#1#2{{\partial #1 \over \partial #2}}

\def\+{^\dagger}
\def\d{d}
\def\e{e}
\def\i{i}

\def\vep{\varepsilon}
\def\ra{\rightarrow}
\def\vp{{\bf p}}
\def\al{\alpha}
\def\ab{\bar{\alpha}}
\def \bi{\bibitem}
\def \ep{\epsilon}
\def\D{\Delta}
\def \om {\omega}
\def\LL{\td \l}
\def \do {\dot}
\def\H {{\cal H}}
\def \ua {\uparrow}
\def \Q {{\hat Q}}
\def \P {{\hat P}}
\def \q {{\hat q}}
\def \bp{{\bar \psi}}

\def \k {\kappa} 
\def \F {{\cal F}}
\def \g {\gamma}
\def \del {\partial}
\def \bd {\bar \partial }
\def \na {\nabla}
\def \const {{\rm const}}
\def \na {\nabla }
\def \D {\Delta}
\def \a {\alpha}
\def \b {\beta}
\def\r {\rho}
\def \s {\sigma}
\def \p {\phi}
\def \m {\mu}
\def \n {\nu}
\def \vp {\varphi }
\def \l {\lambda}
\def \t {\tau}
\def \td {\tilde }
\def \ci {\cite}
\def \sm {$\s$-model }

\def \o {\omega}
\def \inv {^{-1}}
\def \ov {\over }
\def \four{{\textstyle{1\over 4}}}
\def \fourth{{{1\over 4}}}
\def \ha {{1\ov 2}}
\def \QQ {{\cal Q}}

\chapter{D-brane approach to fixed scalars}
\label{FixedScalars}

\section{Introduction}

Many conventional wisdoms of general relativity are being reconsidered
in the context of string theory simply because the string effective
actions for gravity coupled to matter are more general than those
considered in the past.  One of the important differences is the
presence of non-minimal scalar--gauge field couplings, leading to a
breakdown of the ``no hair'' theorem (see the discussion in
\cite{lwOne,lwTwo}).  Another new effect is the existence of certain
scalars which, in the presence of an extremal charged black hole with
regular horizon \cite{klopp,CY,CYTwo,ctt,cyy}, acquire an effective
potential \cite{gibbOne,gibbTwo} which fixes their value at the
horizon \cite{fkOne,fkTwo,fks,gkk}. These are the fixed scalars. The
absorption of fixed scalars into $D=4$ extremal black holes was
considered in \cite{kr} and found to be suppressed compared to
ordinary scalars: whereas the absorption cross-section of the latter
approaches the horizon area $A_{\rm h}$ as $\omega\rightarrow 0$
\cite{dgm}, the fixed scalar cross-section was found to vanish as
$\omega^2$.

This chapter is based on the paper \cite{cgkt}.  Our main result is
that the fixed scalar emission and absorption rates, as calculated
using the methods of semi-classical gravity, are exactly reproduced by
the effective string model of black holes based on intersecting
D-branes.  The D-brane description of the five-dimensional black holes
involves $n_1$ 1-branes and $n_5$ 5-branes with some left-moving
momentum along the intersection \cite{sv,cm}.  The low-energy dynamics
of the resulting bound state is believed to be well described by an
effective string wound $n_1 n_5$ times around the compactification
volume \cite{ms,dmw,dmOne,dmTwo,us,mast}.  This model has been
successful in matching not only the extremal \cite{sv,cm} and
near-extremal \cite{hs,hms,ms} entropies, but the rate of Hawking
radiation of ordinary scalars as well \cite{dmOne,dmTwo,us,mast}.

As part of our study, we have computed the semi-classical absorption
cross-section of fixed scalars from both extremal and
near-extremal $D=5$ black holes. In general, we find
cross-sections with a non-trivial energy dependence. In particular,
for the extremal $D=5$ black holes
with two charges equal, 
$$
\sigma_{\rm abs} = {\pi^2\over 2} R^2 r_K^3 \omega^2
    {{\omega \over 2 T_L} \over 1 - \e^{-{\omega \over 2 T_L}}}
\left (1 + {\omega^2 \over 16 \pi^2 T_L^2} \right )
$$
where $r_K$, $R \gg r_K$ and
$T_L$ are parameters related to the charges.
At low energies the cross-section vanishes as
$\omega^2$, just as in the $D=4$ case studied in
\cite{kr}. For non-extremal black holes,
however, the cross-section no longer vanishes as $\omega\rightarrow 0$.
For near-extremal $D=5$ black holes, we find (for
$\omega\sim T_H \ll T_L$)
$$ \sigma_{\rm abs} (\omega)=  {1\over 4} A_{\rm h} r_K^2
(\omega^2+ 4 \pi^2 T_H^2)
\ ,$$
where $T_H$ is the Hawking temperature.
A similar formula holds for the $D=4$ case. Thus,
even at low energies, the fixed
scalar cross-section is sensitive to 
several features of the black hole geometry.
By comparison, the limiting value of the ordinary scalar
cross-section is given by the horizon area alone.
All of the complexities of the fixed scalar emission and absorption
will be reproduced by, and find a simple explanation in,
the effective string picture.

The absorption cross-section for ordinary scalars finds its
explanation in the D-brane description in terms of the process
$scalar \to L + R$ together with its time-reversal $L + R \to scalar$, 
where $L$ and $R$ represent left-moving and right-moving modes
on the effective string \cite{cm,dmw,dmOne,dmTwo,us,mast}. 
The absorption cross-section for fixed scalars is so
interesting because, as we will show, it 
depends on the existence of eight
kinematically permitted processes: 
  \eqn{FixedDProc}{\eqalign{
   &1) \ \ scalar \to L + L + R + R \cr
   &2) \ \ scalar + L \to L + R + R \cr
   &3) \ \ scalar + R \to L + L + R \cr
   &4) \ \ scalar + L + R \to L + R \cr
  }}
 and their time-reversals.  One of the main results of this chapter is
that competition among $1$--$4$ and their time-reversals gives the
following expression for the fixed scalar
absorption cross-section,
  \eqn{SigmaDbrane}{
   \sigma_{\rm abs}(\omega) = { \pi r_1^2 r_5^2 \over 256 T^2_{\rm eff}}
   {\omega \left (\e^{\omega\over T_H} - 1 \right ) \over
   \left (\e^{\omega\over 2 T_L} - 1\right )
   \left (\e^{\omega\over 2 T_R} - 1\right ) }
   (\omega^2 + 16 \pi^2 T_L^2) (\omega^2 + 16 \pi^2 T_R^2) \ ,
  }
 where $T_L$ and $T_R$ are the left and right-moving temperatures,
$T_{\rm eff}$ is the effective string tension
\cite{lwOne,lwTwo,ctt,ms,juan,tssm,hkrs,halyo} and $r_1^2$ and $r_5^2$ are
essentially the 1-brane and 5-brane charges.  The only restriction on
the validity of \SigmaDbrane\ is that $T_L,T_R,\omega \ll 1/{r_1} \sim
1/{r_5}$ so that we stay in the dilute gas regime and keep the
wavelength of the fixed scalar much larger than the longest length
scale of the black hole.  Remarkably, the very simple effective string
result \SigmaDbrane\ is in complete agreement with the rather
complicated calculations in semi-classical gravity!  The
semi-classical calculations involve no unknown parameters, so
comparison with \SigmaDbrane\ allows us to infer $T_{\rm eff}$.  The
result is in agreement with the fractional string tension necessary to
explain the entropy of near-extremal 5-branes \cite{juan}.

To set up the semi-classical calculations, we will develop in
section~\ref{EffSGAct} an effective action technique for deriving the
equations of motion for fixed scalars.  This technique shows how the
fixed scalar equation couples with Einstein's equations when $r_1 \neq
r_5$; therefore, we restrict ourselves to the regime $r_1= r_5 =R$
where the fixed scalar equation is straightforward.  We briefly
digress to four dimensions, demonstrating how the same techniques lead
to similar equations for fixed scalars.  Clearly, comparisons
analogous to the ones made in this chapter are possible for the
four-dimensional case, where the effective string appears at the
triple intersection of M-theory 5-branes \cite{kt}.  In section~\ref{EffDBI}
we use the Dirac-Born-Infeld (DBI) action to see how various scalars
in $D=5$ couple to the effective string.  The main result of
section~\ref{EffDBI} is that the leading coupling of the fixed scalar
is to {\it four} fluctuation modes of the string.  This highlights its
difference from the moduli which couple to two fluctuation modes. In
section~\ref{MatchFix} we return to five dimensions and exhibit
approximate solutions to the fixed scalar equation, deriving the
semi-classical emission and absorption rates. In section \ref{DAgree}
we calculate the corresponding rates with D-brane methods, finding
complete agreement with semi-classical gravity. We sum up in
section~\ref{ConclusionFix}.  

\section{Field theory effective action considerations}
\label{EffSGAct}
\subsection{$D=5$ case}

First we shall concentrate on the case
of a $D=5$ black hole representing 
 the bound state of $n_1$ RR strings and  
$n_5$ RR 5-branes compactified on a 5-torus
\cite{cm}. This black hole may be viewed as a  static solution 
corresponding to the following 
 truncation of type IIB superstring effective  action  
 compactified on 5-torus
   (cf. \cite{maha,bho})
 \eqn{actit}{
S_5 = 
 {1 \ov 2 \k_5^2} \int d^5 x \sqrt{-g} \bigg[ R  - {4\ov 3}(\del_\m \p_5)^2 
 -  { 1 \ov 4} G^{pl} G^{qn}(\del_\m G_{pq} \del^\m G_{ln} 
 +  e^{{2} \p_5 } \sqrt {G} \del_\m B_{pq} \del^\m B_{ln} )
}  $$ -\ 
 \fourth e^{-{4\ov 3} \p_5   }G_{pq} F^{(K)p}_{\m\n}  F^{(K)q}_{\m\n} 
- \fourth e^{{2\ov 3}\p_5 } \sqrt {G} G^{pq}  H_{\m\n p } H_{\m\n q }
-  {1\ov 12} e^{{4\ov 3} \p_5 } \sqrt {G}  H^2_{\m\n\l}  
 \bigg]   \ ,  $$
where $\m,\n,...= 0,1,..,4; p,q,...=5,...,9$. \  
 $\p_5$ is the  5-d dilaton and $G_{pq}$ is the metric of
5-torus,  
$$\p_5 \equiv \p_{10} - \fourth \ln  G
\ , \qquad G= \det G_{pq}\  , $$ 
and $B_{pq}$ are the internal components of the  RR 2-form field.
 $F^{(K)p}_{\m\n}$ is the Kaluza-Klein
  vector field strength, while $H_{\m\n p}$ and $H_{\m\n\l}$
are  given explicitly by \cite{maha}
  \eqn{dewf}{
  H_{\m\n p} =   F_{\m\n p}  - B_{pq} F^{(K)q}_{\m\n} \ , 
  \ \ \ \  F_p = dA_p \ , \ \ \ F^{(K)p} = d A^{(K)p} \ , 
    }
  $$
  H_{\m\n\l} = \del_\m B'_{\n\l} - \hal  A^{(K)p}_\m F_{\n\l p}
  - \hal  A_{\m p} F^{(K)p}_{\n\l} + {\rm cycles} \ ,  $$
  where $A_{\m p}= B_{\m p}  + B_{pq} A^{(K)q}_\m $
  and $B'_{\m\n} = B_{\m\n}  + A^{(K)p}_{[\m} A_{\n] p} 
   -  A^{(K)p}_{\m} B_{pq} A^{(K)q}_{\n}  $
   are related to the  
 components  of the $D=10$
 RR  
 2-form field $B_{MN}$.  
 
  The shifts in the field strengths in  \dewf\ 
  will vanish for the black hole 
 background
  considered  below (for which the internal components of the 
  2-form $B_{pq}$
 will be zero and the two 
  vector fields $A^{(K)p}$ and $A_p$ will be electric), and,
   as it turns out, 
 are  also  not  relevant for the discussion of perturbations.
 
  For comparison,  a similar truncated  $D=5$  action 
 with $B_{pq}$,  $F_{\m\n p}$ and $H_{\m\n\l}$  from the NS-NS  sector has 
  the following   antisymmetric tensor terms 
  (the full action in general 
  contains both RR and NS-NS antisymmetric tensor parts)
   \cite{maha}
$$- { 1 \ov 4} G^{pl} G^{qn} \del_\m B_{pq} \del^\m B_{ln} 
- \fourth e^{-{4\ov 3}\p_5 }  G^{pq}  H_{\m\n p } H_{\m\n q }
-  {1\ov 12} e^{-{8\ov 3} \p_5 }   H^2_{\m\n\l} \ . $$

We shall assume that there are non-trivial electric charges in only 
one of the five  internal directions and  that the metric corresponding 
to  the internal 5-torus (over which the 5-brane will be wrapped) is 
\eqn{torr}{
(ds^2_{10})_{T^5} = 
e^{2\n_5} dx_5^2 +  e^{2\n} 
(dx^2_6 + dx^2_7 + dx^2_8 + dx^2_9) \ ,  } 
where $x_5$ is the string direction  and   $\n$ is the scale
of the four 5-brane directions transverse to the string.
It is useful to introduce a different basis for the scalars, defining 
the six-dimensional dilaton,  $\p$, and the scale $\l$ of the  $x_5$ 
(string) direction as measured in the $D=6$ Einstein-frame metric:
\eqn{bass}{
\p = \p_{10}-2\n= \p_5 + \ha \n_5 \ , \ \ \ \   
\l = \n_5 - \ha \p = {3\ov 4 }\n_5 - \ha \p_5 \ . }
The action  \actit\  can be expressed either in terms of
$\p_5, \n_5,\n$ or $\p,\l,\n$  (in both cases the kinetic term is
diagonal). In the latter case (we set $B_{pq}=0$) 
\eqn{acti}{
S_5 = 
 {1 \ov 2 \k_5^2} \int d^5 x \sqrt{-g}
\bigg[ R  - (\del_\m \p)^2  - {4\ov 3} (\del_\m \l)^2  
 - 4  (\del_\m \n)^2 }  $$ -\ 
 \fourth e^{{8\ov 3}\l } {F^{(K)}_{\m\n}}^2 
- \fourth e^{-{4\ov 3} \l +  4\n } F_{\m\n}^2 
-  {1\ov 12} e^{{4\ov 3} \l + 4\n } H^2_{\m\n\l}  
 \bigg]   \ . $$
Here $F^{(K)}_{\m\n}\equiv F^{(K)5}_{\m\n}$ is the 
KK vector field strength corresponding 
to the string direction, while
$F_{\m\n}\equiv F_{\m\n 5}$ and $H_{\m\n\l}$ 
correspond  the ``electric'' (D1-brane) and ``magnetic'' (D5-brane)
components of the field strength of the  RR   2-form field.
Evidently $\p$ is an ordinary ``decoupled'' scalar while $\l$ and $\n$ 
are different: they interact with the gauge charges. 
We shall see that they are examples of the so-called ``fixed scalars.''

To study spherically symmetric  
configurations corresponding to this action it is sufficient to 
 choose the five-dimensional metric in the  ``2+3'' form
 \eqn{met}{
ds^2_5 = g_{mn} dx^mdx^n + ds^2_3=
 - e^{2a}dt^2 + e^{2b} dr^2   +  e^{2c}  d\Omega_3^2  \ ,  }
where $a,b,c$ are  functions of $r$ and $t$.
Solving first the equations for   
$H_{\m\n\l}$, $F_{\m\n}$  and $F^{(K)}_{\m\n}$
and assuming that  the first two have, respectively, 
the magnetic and the electric components
(with the  charges $P$ and $Q$ 
corresponding to the D5-brane   and the D1-brane), while
the third has only the electric component with the Kaluza-Klein 
charge $Q_K$, we may  eliminate them from the action \acti. 
The result is an effective  two-dimensional theory 
with coordinates $x^m=(t,r)$ and the action given (up to the
constant prefactor) by\footnote{The full set of 
equations and constraints is derived by first keeping the 2-d metric $g_{mn}$
general and using its diagonal  gauge-fixed form only after the variation.  
In addition to choosing $g_{mn}$ diagonal as in \met, 
one  can  use the gauge freedom to  impose 
one more relation between $a $ and $b$.}
$$
S_2= \int d^2 x {\sqrt{-g}} e^{3 c } \bigg[R  +  6 (\del_m c)^2 
 - 
(\del_m \p )^2-  {4\ov 3}  (\del_m \l)^2 - 4(\del_m \n )^2 + V (c,\n,\l)  \bigg]   $$
 \eqn{ffoo}{
= \int dtdr\bigg[ -  e^{3c + b -a} 
( 6\do c \do b + 6\do c^2 - \do \p^2 -  
{4\ov 3} \do \l^2    - 4  \do \n^2  ) } $$
+ \  e^{3c + a -b} ( 6c' a' + 6c'^2  -  \p'^2 -{4\ov 3}  \l'^2 - 4 \n'^2 )  
$$
$$+ \  6 e^{a+ b + c } - 2 e^{a+b-3c} f(\n,\l) \bigg] \ .   $$
The first term in the potential 
originates from the curvature of the 
3-sphere  while  the second is 
produced by the non-trivial charges, 
 \eqn{fffo}{
 f(\n,\l) =     Q^2_K e^{-{8\ov 3}  \l}  +  e^{{4\ov 3}  \l} 
( P^2 e^{ 4\n } + Q^2 e^{-4 \n }) 
  \ .  }
 This is a special case of the more general expression following from  \actit: 
  if the electric charges corresponding to the vector 
fields in \actit\ are $Q_{Kp}$ and $Q^p$ we get 
\eqn{gene}{f(\p_5, G_{pq})
 =  e^{{4\ov 3}  \p_5 }   Q_{Kp}  Q_{Kq} G^{pq}  
+  e^{-{2\ov 3}  \p_5} 
\bigg(    P^2   {G}^{1/2}  
+    Q^p Q^q  G_{pq} {G}^{-1/2}  \bigg) \ .}
The potential $f$ in \fffo\  has the  global minimum at 
$e^{ 4\n } = QP\inv, \ e^{{4}  \l} = Q^2_KQ\inv P\inv $.
These values of $\n$ and $\l$ are thus fixed points
to which these fields are attracted on the horizon, which is
why such fields can be called fixed scalars.  By contrast, 
the decoupled scalar $\p$ can be chosen  to be equal to an 
arbitrary  constant.  

As an aside, we note that this  structure of the potential \fffo\ 
explains why one needs at least {\it three}  different charges
to get an extremal  $D=5$ black hole  with a regular horizon (i.e.{} 
with scalar fields that do not blow up): 
it is necessary to have 
  at least three exponential terms to ``confine'' 
  the  two fixed scalars. If the number of 
  non-vanishing charges is smaller than three, then
one or both scalars will blow up at the horizon.

Equivalent actions and potentials are found for theories that are
obtained from the one above by U-duality.
For example, 
in the case of the NS-NS  truncation of type II action,
which  has a $D=5$ black hole  solution representing 
a bound state of NS-NS strings and solitonic 5-branes, 
 we  can put the action in the form 
 \ffoo, where $\l$  is still the scale of the string direction
as measured by the 6-d metric, while the roles of 
$2\n$ (the scale of the 4-torus) 
and  $-\phi$ are interchanged.\footnote{There exists an equivalent  representation
of this NS-NS action where the fixed scalars are 
the 5-d dilaton and 
the scale of the string direction, 
while the scale of the 4-torus is decoupled.}

In order to find the static black hole solution to \ffoo, we define
$\r =  2c +  a, $ $d\t = - 2 e^{-3c -a + b} dr$.
 Now \ffoo\ reduces to 
a  ``particle''  action (we  choose  $\p=\const$)
\eqn{oqfw}{
S_1=  \int d\t\bigg[  {3\ov 2 } 
(\del_\t \r)^2  - {3\ov 2}(\del_\t a)^2  -{4\ov 3} 
(\del_\t \l)^2 - 4 (\del_\t \n)^2  
+ \  {3\ov 2 } e^{ 2 \r } -  {1\ov 2 } e^{ 2a} f(\n,\l) \bigg] \ ,} 
which should be supplemented by the ``zero-energy'' constraint,
$$
 {3\ov 2 } 
(\del_\t \r)^2  - {3\ov 2}(\del_\t a)^2  -{4\ov 3} 
(\del_\t \l)^2 - 4 (\del_\t \n)^2  
-  {3\ov 2 } e^{ 2 \r } +  {1\ov 2 } e^{ 2a} f(\n,\l) =0 \ . $$
The special structure of $f$ in \fffo\  makes it possible to find a
simple analytic  solution of this  Toda-type system. 
Introducing new variables 
$\a= a - {4\ov 3} \l, \ \  \b= a  + {2\ov 3} \l +2 \n  , 
\   \g= a + {2\ov 3} \l -  2\n $
and using the special form \fffo\ of $f$, we can convert \oqfw\ 
to four non-interacting Liouville-like  systems (related  only 
through  the constraint) 
\eqn{fw}{
S_1= \int d\t\bigg[  {3\ov 2 }(\del_\t \r)^2  - \ha (\del_\t \a)^2  -  
\ha (\del_\t \b)^2 -  \ha (\del_\t \g)^2 } 
$$+ \  {3\ov 2 } e^{ 2 \r } -  \ha  Q^2_K e^{2\a}  - \ha  P^2 e^{2\b}  
- \ha   Q^2 e^{2\g}  \bigg] \ .     $$

The general solution depends on the three gauge charges 
$P,Q,Q_K$ and one parameter which we will call $\mu$ which governs
the degree of non-extremality. In a convenient gauge, the solution reads 
\cite{att,cm,cyy,hms} 
\eqn{soo}{ e^{2a} = h  {\H}^{-2/3}   \ , \ 
\ \ \  e^{2b} = h\inv  {\H}^{1/3} \ ,\ \ \ 
e^{2c} = r^2  {\H}^{1/3 }\  
 ,  \ \ \ {\H}\equiv H_\P H_\Q H_{\Q_K} \ , } 
\eqn{sool}{ 
e^{2\l } = H_{\Q_K}  (H_\Q H_{\P})^{-{1\ov 2}}  \ ,
\ \  \ \ e^{4\n } =  H_{\Q} H_\P\inv  \  ,  
\ \ \  \ \ e^{2\p} =  
e^{2\p_{10,\infty}}\ ,   }  
$$ h = 1 - {2\m \ov r^2} \ , \ \ \  
H_\q = 1 + {\hat q \ov r^2} \ , \ \ \  \hat q \equiv  
\sqrt {q^2 + \m^2 } - \m \ ,  \ \ \   q=(P,Q,Q_K) \ .  $$
We have chosen the asymptotic values $\l_\infty$ and $\n_\infty$ 
to be zero. To compare with previous equations, 
we also note that $e^{2\r} = r^2(r^2-2 \m)$. 

In the  extremal limit, $\m=0$, one finds  
$$e^{-\a} = H_{Q_K},\qquad
e^{-\b} = H_P, \qquad  e^{-\g} = H_{Q}\ ,$$
where $H_q= c_q  + q\t$  and $\t=1/r^2$.
The constants  $c_{Q_K},c_P,c_Q$  must satisfy 
$ c_{Q_K} c_P c_Q =1 $ in order for the 5-d metric
to approach the Minkowski metric at  infinity.
The two remaining arbitrary constants 
correspond to the asymptotic values of $\l$ and $\n$.
As is clear from \ffoo,\fffo, shifting $\l$ and $\n$ by  constants
 is equivalent to a rescaling of $Q_K,Q,P$.
The assumption that 
$\n_{\infty} =0$ 
and 
$\l_\infty = \n_{5\infty} + \n_{\infty}  - \ha \p_{10,\infty} =0$ 
implies (setting $\a'=1$):
  \eqn{UnitChoice}{
   V_4= e^{4\n_\infty} =1\ , \qquad
   {\cal R}^2 = e^{2\n_{5\infty}} = g = e^{\p_{10,\infty}}\ ,\ \ \ \ \
   \k_5^2 = {2\pi^2 g^2 \over {\cal R} V_4} \ , 
  }
 where $(2\pi)^4 V_4$ is the volume of $T^4$
in the $(6789)$ directions, while ${\cal R}$ is the radius of the
circle in direction $5$.  Then the charges $Q_K,Q,P$ are related to
the quantized charges $n_1,n_5,n_K$ as follows:
  \eqn{IntCharges}{
   n_1 = {V_4 Q \over g} = {Q \over g} \ , \qquad 
   n_5 = {P \over g} \ , \qquad 
   n_K = {{\cal R}^2 V_4 Q_K \over g^2} = {Q_K \over g} \ .
  }
The somewhat unusual form of the last relation is due to our choice 
$\l_\infty=0$ instead of more standard  $\n_{5\infty}=0$.

In using the black hole solution \soo, \sool, we
will often find it convenient to work in terms of characteristic radii
rather than the charges, so we define
  \eqn{DefRadii}{
   r_1^2 = \hat{Q} \ , \qquad
   r_5^2 = \hat{P} \ , \qquad
   r_K^2 = \hat{Q}_K \ , \qquad
   r_0^2 = 2 \mu \ .
  }
{}From the classical GR point of view, these parameters can take on 
any values. Recent experience has shown, however, that 
when the radii satisfy \cite{mast}
\eqn{constraint}{ r_0, r_K \ll r_1, r_5 \  }
the black hole can be successfully matched to a bound state of D1-branes
and D5-branes (with no anti-branes present) carrying a dilute gas of
massless excitations propagating along the bound 
D1-branes. Evidence 
for this gas can be seen directly in the energy, entropy and temperature
of the black hole solution. Introducing a new parameter $\sigma$ through
$$ 
r_K^2 = r_0^2 \sinh^2 \sigma
$$
one finds the following expressions \cite{hms,mast}
for the ADM mass, Hawking 
temperature and the entropy in the parameter region \constraint:
\eqn{ment}{
M= { 2\pi^2\over \kappa_5^2} \left (r_1^2 + r_5^2+
\ha {r_0^2 \cosh 2\sigma} \right )\ ,
\ \ \
 \ \  T_H^{-1}={2\pi r_1 r_5\over r_0}\cosh\sigma \ , }
 $$ {\rm S}= {2\pi A_{\rm h} \over \k_5^2 } = 
 {4\pi^3 \over \kappa_5^2} r_1 r_5 r_0 \cosh\sigma
\ .$$
The entropy and energy are those of a gas of massless one-dimensional
particles with the left-movers and right-movers 
each having its own temperature \cite{mast}:
\eqn{temps}{ T_L = {r_0 \e^\sigma \over 2\pi r_1 r_5}\ ,
\qquad T_R = {r_0 \e^{-\sigma} \over 2\pi r_1 r_5}\ .
}
The Hawking temperature is related to these two temperatures by
  \eqn{THDef}{
   {2\over T_H}={1\over T_L}+{1\over T_R} \ ,  }
a fact which also has a natural thermodynamic interpretation.
These results will be heavily used in later comparisons of classical 
GR results with D-brane calculations of corresponding quantities.


Let us now turn to the discussion of the propagation of perturbations on 
this black hole background. The goal will be to calculate the classical
absorption cross-section of various scalar fields and eventually to
compare them with comparable D-brane quantities. The behavior of ``free'
scalars, like $\p$, is quite different from that of ``fixed'' scalars,
like $\l$ and $\n$. The spherically symmetric fluctuations of $\p$  
obey the standard  massless  Klein-Gordon equation in this background.
Namely, if $\delta \p = e^{i\o t} \td \p (r)$,  then 
\eqn{kgo}{
 \left[r^{-3} {d\over dr} ( h r^3 {d\over dr} ) + \omega^2    
h\inv H_\P H_\Q H_{\Q_K}   \right] \td \p  =0 \ , }
This scattering problem, and its D-brane analog, have been analyzed at
length recently and we will have no more to say about it. The spherically
symmetric fluctuations of the metric functions $a,b,c$ and the scalars  
$\l,\n$ in general obey a complicated set of coupled 
differential equations.\footnote{The spherically symmetric 
fluctuations of the gauge fields need 
not be considered explicitly: 
since the dependence on $H_{\m\n\l}$  and $F_{\m\n}$ is gaussian, they are 
automatically included when going from \acti\ to \ffoo.}
However, when the charges  $P$ and $Q$ are set 
equal, a dramatic simplification
occurs: the background value of  $\n$  in \sool\  (i.e.{} the scale of 
the transverse 4-torus) becomes  constant and its  small fluctuations 
$\delta\n$ decouple from those of the other fields.\footnote{Similar 
simplification occurs when any two of the three charges are equal.
For example, if $P=Q_K$  we may introduce  
$\l' = -\ha (  \l -3 \n) , \ \n'= -\ha (\n + \l) $ 
(in terms of  which the kinetic part 
in the action \ffoo\ preserves its diagonal form) 
to discover that $\n'$ has decoupled fluctuations. The resulting equation
for $\delta \n'$ has the  same form as the equation for $\delta \n$ in
the case of $P=Q$.}
 The gaussian effective 
action for $\delta\n$ extracted from \ffoo\ is
\eqn{ffu}{
\delta S_2= \int d^2 x {\sqrt{-g}} e^{3 c} \left[ - 4 (\del_m \delta  \n )^2
- 32 P^2 e^{-6 c +{ 4\ov 3} \l}  ( \delta  \n )^2  + ...  \right]  }
and spherically symmetric fluctuations $\delta\n=e^{i\o t}\td\n$ obey 
\eqn{kgoo}{
 \left[r^{-3} {d\over dr} ( h r^3 {d\over dr} )
 + \omega^2    h\inv  H^2_\P H_{\Q_K} 
 - 8 P^2 r^{-6} H_\P^{-2}\right] \td \n = 0 \ . }
This is the standard Klein-Gordon equation \kgo\ augmented by
a space-dependent mass term originating from the expansion of the 
effective potential $f(\n,\l)$ in \fffo. This mass term falls off as $r^{-6}$
at large $r$, and, in the extremal case, blows up like $8/r^2$ near the
horizon at $r=0$. This is the $l(l+2)/r^2$ angular momentum barrier for
an $l=2$ partial wave in $D=5$. This ``transmutation'' of angular momentum
plays an important role in the behavior of the fixed scalar cross-section. 
For later analysis, it will be convenient to rewrite this equation
using the coordinate $\t= 1/r^2$: 
\eqn{kgee}{
 \left[  { [ (1- 2\m \t){d\over d\t}]^2 } 
 + \fourth \omega^2   \t^{-3} ( 1+ \P\t)^2(1 + \Q_K\t) 
 - 2 {P^2  (1- 2\m \t) \ov  (1 + \P\t)^{2}} \right] \td \n = 0   \ . }

Note that, when all the three  charges are equal, $P=Q=Q_K$, the background
value of the other scalar, $\lambda$, is constant as well.
Then the small fluctuations of this field decouple from gravitational
perturbations and satisfy the same equation as $\nu$, \kgoo.
If only two of the charges are equal, then there is
only one fixed scalar which has a constant background value 
and decouples from gravitational perturbations.
We would also like to know the fixed scalar scattering equations 
(and solutions) for the general 
$Q_K\neq Q\neq P$ black hole. This problem 
is surprisingly complicated due to mixing with
gravitational perturbations.  The disentanglement of the equations was
achieved in \cite{kkTwo}.

To summarize, we have identified a set of scalars around the familiar
type II string $D=5$ black hole solution which merit the name of
``fixed scalars'' in that their horizon values are fixed by the
background charges.  Their fluctuations in the black hole background
satisfy the Klein-Gordon equation, augmented by a position-dependent
mass term. In section~\ref{MatchFix} we will solve the new equations
to find the absorption cross-section by the black hole for these
special scalars.

\subsection{$D=4$ case}

Previous experience  \cite{mst,myers,kt,ktI,us,gkTwo}  suggests 
that  one may be able to  extend the $D=5$ successes
in reproducing entropies and radiation rates with D-brane methods 
to $D=4$ black holes carrying 4 charges. Although we will not
pursue the $p$-brane approach to $D=4$ black hole dynamics in this
chapter, this is a natural place to discuss $D=4$ fixed scalars 
and to record their scattering equations for later use. 

A convenient representation of the $D=4$ black hole with four different 
charges \cite{CY,CYTwo,ctt} is the $D=11$ supergravity configuration 
$5\bot5\bot5$ of three 5-branes intersecting over a common string 
\cite{kt,ct}. The three magnetic charges are related to the numbers 
of 5-branes in three different hyperplanes, while the electric charge 
has Kaluza-Klein origin. The  reduction to $D=4$ of the relevant
part  of the $D=11$ supergravity (or $D=10$ type IIA) action has the form
\eqn{actip}{
S_4 = {1\ov 2 \k^2_4} 
\int d^4 x \sqrt{-g} \bigg[ R   - 2(\del_\m \n)^2 -
{3\ov 2}  (\del_\m \zeta)^2  - {4\ov 3} (\del_\m \xi )^2 - 
(\del_\m \eta )^2  }
$$
-  \fourth    e^{3 \zeta }  ( F^{(K)}_{\m\n})^2
 -  \fourth   e^{ \zeta }  \bigg( e^{-{8\ov 3} \xi }  ( F^{(1)}_{\m\n})^2 
+ e^{{4\ov 3}\xi} [ e^{2\eta }  ( F^{(2)}_{\m\n})^2
+   e^{-2\eta }  ( F^{(3)}_{\m\n})^2 ] \bigg)  \bigg] \ . $$
The scalar fields are  expressed in terms of components of the internal 
7-torus part of  the $D=11$ metric.  By the logic of the previous section,
the scale $\n$ of the 6-torus transverse to the intersection string is
a decoupled scalar, while the fields $\zeta,\xi,\eta$ 
(related to the scale of the string direction and the ratios of sizes 
of 2-tori shared by pairs of 5-branes) are fixed scalars.
If the  internal part of the $D=11$ metric   is 
$$ds^2_7=  e^{2\n_4} dx^2_{4} + 
e^{2\n_1} (dx_5^2 + dx_6^2) +e^{2\n_2} (dx_7^2 + dx_8^2) +
e^{2\n_3} (dx_9^2 + dx_{10}^2)\ ,$$
where $x_4$ is the direction of the common string, 
then 
$$\n=  \n_1 + \n_2 +\n_3 , \ \ 
\xi= \n_1 - \ha \n_2 - \ha \n_3, \  \ \eta= \n_3-\n_2,  \ \ 
\zeta= \n_4  + {2\ov 3} (\n_1 + \n_2 +\n_3) . $$

Using an ansatz for the 4-d metric similar to \met, 
solving for the vector fields, and 
substituting the result back into the action, 
we get the following effective two-dimensional action
(cf. \ffoo)
\eqn{ffe}{
S_2= \int d^2 x {\sqrt{-g}} e^{2 c} \bigg[R  + 
 2 (\del_m c)^2 - 2(\del_\m \n)^2
 -  {3\ov 2}  (\del_m \zeta)^2  - {4\ov 3} (\del_m \xi )^2 - 
(\del_m \eta )^2  }
$$ + \ 2 e^{-2c} - \ha e^{-4 c } f(\zeta,\xi,\eta) \bigg] \  , $$
where 
\eqn{ffour}{
f(\zeta,\eta,\xi)  =  
Q^2_K e^{-3 \zeta } +   e^{ \zeta } \left[ P_1^2 e^{-{8\ov 3} \xi }  
+ e^{{4\ov 3}\xi} (  P^2_2 e^{2\eta } +  P^2_3 e^{-2\eta })\right]
  \ .  }
As in the $D=5$ case, one finds that the special structure  of $f$
makes it possible to diagonalize the interaction term 
by a field redefinition and thus 
find  the static solution in a simple factorized form \cite{CY,CYTwo,ctt,ct}
(cf. \soo) 
\eqn{sooo}{
ds^2_4 = -e^{2a}dt^2 + e^{2b} dr^2 + e^{2c} d\Omega_2^3 \ ,  }
$$
e^{2a} = h  {\H}^{-1/2}   \ , \ \ \  e^{2b}= h\inv   {\H}^{1/2}  \ , 
\ \ \  e^{2c} = r^2  {\H}^{1/2 }\  
 ,  \ \ \ {\H}\equiv H_{\Q_K} H_{\P_1} H_{\P_2} H_{\P_3}\ ,  $$
\eqn{sooll}{ 
e^{2\eta } =H_{\P_3}H_{\P_2}\inv\ , 
\ \
e^{2\xi } = H_{\P_1}(H_{\P_2}  H_{\P_3})^{-1/2}\ , 
\ \ 
e^{2\zeta } = H_{\Q_K}(H_{\P_1} H_{\P_2} H_{\P_3})^{-1/3}, }
$$ h = 1 - {2\m \ov r} \ , \ \ \  
H_\q = 1 + {\hat q \ov r} \ , \ \ \  \hat q 
\equiv  \sqrt {q^2 + \m^2 } - \m \ ,  
\ \ \   q=(Q_K,P_1,P_2,P_3) \ .  $$
As in the $D=5$ case, for the
 generic values of charges the spherically symmetric 
perturbations 
of this solution obey  a complicated 
system of equations (for  discussions of perturbations
of  
single-charged dilatonic black holes see, e.g., 
\cite{gil,hw,gregOne,gregTwo}).
However, when the three magnetic charges are equal, $\eta$ and $\xi$
have constant background values, and so their 
 small spherically-symmetric fluctuations 
decouple from the metric perturbations, 
\eqn{ifr}{
\delta S_2= \int d^2 x {\sqrt{-g}} e^{2 c}
 \left[ - (\del_m \delta \eta )^2
- 2 P^2 e^{-4 c  + \zeta + {4\ov 3} \xi } 
( \delta \eta )^2  + ...  \right]  \ ,  }
leading to the  following radial 
 Klein-Gordon equation with an extra mass term:
\eqn{goo}{
 \left[r^{-2} {d\over dr} ( h r^2 {d\over dr} )
 + \omega^2    h\inv  H^3_\P H_{\Q_K} 
 - 2 P^2 r^{-4} H_\P^{-2}   \right] \td \eta = 0 \ . }
Here $\delta \eta (r,t) = e^{i\om t} \td \eta (r)$;  cf.~\kgoo. 
The same universal equation is found for $\delta \xi$.
In terms of  $\t= 1/r$ this becomes
\eqn{gee}{
 \left[  { [ (1- 2\m \t){d\over d\t}]^2 } 
 +  \omega^2   \t^{-2} ( 1+ \P\t)^3(1 + \Q_K\t) 
 - 2 {P^2  (1- 2\m \t) \ov  (1 + \P\t)^{2}} \right] \td \eta = 0   \ . }
Represented in this form 
this is  very similar to \kgee\
found in the $D=5$ case: the differential operator and mass terms are
 exactly the same,  while the  frequency terms
are related by $\omega \to 2 \omega, \   \t^{-3}( 1+ \P\t)^2 \to  
\t^{-2}( 1+ \P\t)^3$.

In the extremal case   and  with 
 all four charges  chosen to be   equal, $Q_K=P$,
\goo\  reduces to the equation 
 studied in \cite{kr}.  The characteristic 
coefficient 2 in the mass term gives the effective potential of the form
$l(l+1)/r^2$ near the horizon, with $l=1$. Away from the horizon, the
fixed scalar equation differs from that of the $l=1$ partial wave
of the ordinary scalar.

Finally, let us note that 
there exist other representations of
the  4-charge $D=4$ black hole.
For example
in the case of the  $2\bot2\bot5\bot5$ 
representation  \cite{kt}, or, equivalently,  the 
 U-dual  $D=4$ configuration in 
the NS-NS sector  with 
two  (electric and magnetic) charges coming from the  $D=10$ 
antisymmetric tensor
and two  (electric and magnetic) charges being of Kaluza-Klein  origin, 
we may parametrize the metric as
$$ds^2_{10} =  e^{2\n_4} dx^2_4 + e^{2\n_5} dx^2_5
+ e^{2\n}(dx^2_6 +  dx^2_7 + dx^2_8+ dx^2_9)\ . $$
This leads to 
the effective Lagrangian related to the above  one \ffe\
by a linear field redefinition and re-interpretation of the charges.
The  potential is (cf. \ffour)
$$f(\p,\n_4,\n_5) =e^{2\p} (  Q^2_K e^{2\n_4 } +  Q^2 e^{-2\n_4 })  
+ e^{-2\p} (  P^2_K e^{2\n_5 } +  P^2 e^{-2\n_5 }) \ ,
$$
where $\p$ is the 4-d dilaton. 
This shows that the scale  of the remaining 4-torus, $\n$, decouples.

\section{Effective string couplings}
\label{EffDBI}

We now turn to a discussion of the effective action governing the 
absorption and emission of fixed scalars by the
bound state of D1- and D5-branes. 
We use the same framework as the recent demonstrations of agreement between 
GR and D-brane treatments of the absorption of generic decoupled scalars 
\cite{cm,dmw,dmOne,dmTwo,us}. We assume that: (i) the $D=5$ black hole 
is equivalent to $n_1$ D1-branes bound to $n_5$ D5-branes, with some
left-moving momentum;
(ii) that the low-energy dynamics of this system is described by the  
DBI action for a string with an effective tension $T_{\rm eff}$, 
and 
(iii) that the relevant bosonic oscillations of this effective string are 
only in the four 5-brane directions ($i=6,7,8,9$) transverse to the 1-brane.
These assumptions serve to specify the detailed couplings of external 
closed string fields, in particular the fixed scalars, to the D-brane degrees
of freedom. This is an essential input to any calculation of absorption
and emission rates and, as we shall see, brings fairly subtle features 
of the effective action into play. 

Specifically, we assume that the low-energy excitations of our system are
described by the standard D-string action 
\eqn{nam}{
I=- T_{\rm eff} \int d^2 \s\ e^{-\p_{10}} \sqrt { - \det \g_{ab}}  + ... \ , 
\ \ \ \   \g_{ab} = G_{\m\n} (X) \del_a X^\m \del_b X^\n \ , }
where $ \p_{10}$ and $G_{\m\n}$ are the $D=10$ dilaton and string-frame 
metric. The specific dependence on $\p_{10}$ is motivated by the expected 
$1/g_{str}$ behavior of the D-string tension. The normalization constant 
of the tension, $T_{\rm eff}$, is subtle and will be discussed later. Our goal
is to
read off the couplings between excitations of the effective string and the
fluctuations of the metric and 
dilaton that correspond to the fixed scalars.

It should be noted that  the essential structure of  the effective string
 action  we are interested in
can be,  at least  qualitatively, understood using semi-classical 
effective field theory methods. A straightforward 
 generalization of  the extremal 
classical
solution \cite{att,cm} describing  a BPS bound state of a 
string and 5-brane  in which the string is localized on
the 5-brane\footnote{Instead of talking about a bound state of several
single-charged D-strings and D5-branes with coinciding centers, 
it is sufficient, at the classical level, to 
consider just a single string and a single 5-brane having
charges $Q\sim n_1$ and $P\sim n_5$.}
 has the $D=10$ string metric ($m=1,2,3,4; i=6,7,8,9$)
$$ds^2_{10}=  H^{-1/2}_1  H^{-1/2}_5 (-dt^2 + dx_5^2) 
  + H^{1/2}_1  H^{1/2}_5 dx_m dx_m  
   +  H^{1/2}_1  H^{-1/2}_5 dx_i dx_i  \ , $$
   where $ H_5=H_5 (x_m) = 1 + P/x^2_m$ and 
   $H_1=H_1(x_m,x_i)$ is a solution of 
$$ [\del^m\del_m + H_5 (x_n) \del^i\del_i] H_1 =0$$ 
   such that for $P\to 0$ it approaches  the standard string
   harmonic function, 
   $H_1 \to  1  + Q/(x^2_i + x^2_m)^3$. If one averages 
   over the $x_i$-dependence
   of $H_1$ one  returns to the  original delocalized
   case,  $H_1= 1+Q/x^2_i$,  which  corresponds to the extremal limit
   of the $D=5$ black hole \soo,\sool\ with $Q_K=0$ (here    
   we  consider the unboosted string).
    The presence of the 5-brane breaks  the $O(1,1) \times O(8)$
 symmetry of the standard  RR  string
 solution   down to   $O(1,1) \times O(4) \times O(4)$.
 Since the  localized  solution also breaks 4+4 translational 
  invariances, the string soliton has   4+4
collective coordinates: $X^m(x_5,t), X^i(x_5,t)$.
 The  corresponding  $O(4) \times O(4)$ invariant 
 effective string action thus  
should have the  following  form in the static gauge,
$$ I = \int d^2\s  [ T_0 + T_\parallel \del^a X^i \del_a X^i 
+ T_\perp  \del^a X^m \del_a X^m + ... ]\  . $$ 
 The constants 
 $T_0, T_\parallel,  T_\perp$ 
can be  determined using standard 
methods (see, e.g.,  \cite{khuri})  
by substituting the perturbed solution into the
$D=10$ effective field theory  action, etc.   
 $T_0$  is  proportional to the
ADM mass of the background, $T_0 \sim P+ Q$. The same should be 
 true  also for $T_\perp$, $T_\perp \sim P + Q$, 
  since $X^m$ describe  oscillations of
 the whole bound state in the common transverse 4-space.
At the same time,  $T_\parallel$ is the effective
 tension of the string  within the 5-brane, so that $ T_\parallel
 \sim Q$.
In the special cases $P=0$ and $Q=0$ these
  expressions  are in  obvious agreement with the standard
 results for  a  free  string  and a free  5-brane.
 In the case when $P \gg Q$, i.e.{} $n_5 \gg n_1$,
 we learn that  $T_\perp  \gg T_\parallel$,
 so that oscillations of the string in the four directions $X^m$
 transverse to the 5-brane can  be ignored.\footnote{One may
  also give the following
 argument in support of the claim that transverse oscillations of the
string
 can be ignored when  the string is light compared to the 5-brane.
The classical action for a D-string  probe moving in the above
background produced by a bound state of R-R  string  and  5-brane
  has the following form:
$I_1 = T_0 \int d^2\sigma [e^{-\phi} \sqrt {-\det (G_{\m\n} +
B^{(NS)}_{\m\n})}
+  B^{(R)} + ... ]$ (we shall set the  world-sheet gauge field
to zero and choose the  static gauge). If the string is oriented
parallel to the $x^5$ direction of the 5-brane (a BPS
  configuration),
  the non-trivial part of the potential term cancels out \cite{meprOne}.
The same is true for the dependence on the 
 $H_1$-function in the second-derivative terms which have the form
  $I_1  = T_0 \int d^2\sigma[ 1
+ \ha \del^a X^i \del_a X^i + \ha H_5 (X) \del^a X^m \del_a X^m  +
...] $.
The function $H_5=1 + {P\ov X^2_m}$  thus   determines the metric
 of the transverse part of the moduli
space (see also \cite{kabat}), i.e.{}
 it plays the role of an effective $T_\perp$
 which blows up when the string approaches the 5-brane.
 Thus the string probe  can  freely move within the 5-brane,
 but its transverse motions are suppressed.} 
If we further  assume, following \cite{ms,dmI},  that the string 
lying within the
5-brane has the effective 
length $L_{\rm eff} \sim Q P \sim n_1 n_5$,  
we finish with the following expression for the effective
string  tension  in the  relevant  directions parallel  to
the 5-brane: $T_{\rm eff}
 \sim T_\parallel/L_{\rm eff} \sim 
1/P\sim 1/n_5$.
 This picture is consistent with that suggested in \cite{ms,juan}
 and will
 pass a non-trivial test below.

  In accord with the assumption (iii), we thus drop terms involving
derivatives of $X^m= X^{1,2,3,4}$ (i.e.{} motions in the uncompactified
directions). We also eliminate two more string coordinates by choosing
the static gauge $ X^0= \s^0, \ X^5 = \s^1$ and write
  \eqn{namm}{
   \g_{ab} \equiv \eta_{ab} + \hat\gamma_{ab} 
     = G_{ab}(x) + 2G_{i(a} (x)  \del_{b)} X^i  
                 + G_{ij} (x) \del_a X^i \del_b X^j \ . 
  }
 We make the Kaluza-Klein assumption that the background fields
$\p_{10}$ and $G_{\m\n}$ depend only on the external coordinates
$x^m\equiv X^m, m=0,1,...,5$. Since we are interested in linear
absorption and emission processes, we make a weak-field expansion in
powers of $\p_{10}$ and $h_{\mu\nu} \equiv G_{\mu\nu} -\eta_{\mu\nu}$, 
splitting $h_{ij}$ into traceless and trace parts: $h_{ij} = \bar h_{ij} +
\fourth \delta_{ij} h$.  Finally, we distinguish L and R string
excitations by introducing $\del_+ = \del_0 + \del_1 \ , \del_- = -
\del_0 + \del_1$. We can then use the formula
  $$
   \sqrt { - \det (\eta_{ab} + \hat\gamma_{ab})} 
     = 1 + \ha \hat\gamma_{+-} -  {1\ov 8}\hat\gamma_{++}\hat\gamma_{--}
         + {1\ov 16 }\hat\gamma_{+-}\hat\gamma_{++}\hat\gamma_{--} + \ldots
  $$
 to expand \nam, finding the following action for $X^i$:
\eqn{expa}{
I_X = -T_{\rm eff} \int d^2 \s \bigg[1  
+  \ha  \del_+ X \del_- X   - {1\ov 8}  (\del_+ X)^2  (\del_- X)^2  + ... }
$$ 
+\  L_0 + L_1 + L_2 + L_3 + L_4 +  L'_4 + ...\bigg] \ , 
$$
\eqn{opo}{ L_0 = \ha (h_{55} - h_{00})  - \p_{10} \ , \ \  \ \ 
 L_1 =  \ha h_{5i} (\del_+  + \del_-) X^i   \ ,   }
\eqn{ttt}{
L_2 = - \ha \p   
  \del_+ X^i \del_- X^i + 
  \ha \bar h_{ij} \del_+ X^i \del_- X^j   
 - {1\ov 8} (h_{00} + h_{55}) [(\del_+ X)^2 + (\del_- X)^2] \ ,   }
\eqn{iit}{
L_3 =   - {1\ov 4}  h_{5i} [\del_- X^i    (\del_+ X)^2  +
  \del_+ X^i   (\del_- X)^2]   \ , }
\eqn{iis}{
L_4 =   {1\ov 8} (\p_{10} - \ha h)  (\del_+ X)^2  (\del_- X)^2 =
{1\ov 8} \p  (\del_+ X)^2  (\del_- X)^2
- {1\ov 4}  \nu   (\del_+ X)^2  (\del_- X)^2  \ , }
$$L'_4 =   {1\ov 16} (h_{55} +h_{00})
 \del_+ X \del_- X  [  (\del_+ X)^2  +
 (\del_- X)^2]   +  {1\ov 16} (h_{55} -  h_{00})
 (\del_+ X)^2  (\del_- X)^2
$$
\eqn{iise}{
=  {1\ov 8 } \l [ \del_+ X \del_- X  (  (\del_+ X)^2  +
 (\del_- X)^2)   +   (\del_+ X)^2  (\del_- X)^2] }
$$
    +  \  {1\ov 16 } \p [ \del_+ X \del_- X  (  (\del_+ X)^2  +
 (\del_- X)^2)   +   (\del_+ X)^2  (\del_- X)^2] + O(h_{00}) \ . 
$$
The expansion has been organized in powers of derivatives of $X^i$
and we have kept terms at most linear in the external fields (since
we don't use them in what follows, we have dropped higher-order terms involving
$\bar h_{ij}$). We have also reorganized those fields in a way appropriate 
for the compactification on a five-torus:
\eqn{defs}{
\n\equiv  {1\ov 8} h \ , \ \ \  
\p \equiv \p_{10} -  {1\ov 4} h \ , \ 
\ \ \ \l \equiv \ha h_{55}  + {1\ov 8} h  - \ha \p_{10}
 =  \ha h_{55} - \ha \p\  \ , \ \ \ 
 h\equiv h_{ii} \ ,    }
 where $\n$ is the scale of the 4-torus part of the 5-brane (if
$G_{ij} = e^{2 \nu} \delta_{ij}$ then, in the linearized
approximation, $h_{ii} = 8 \n $), $\p$ is the corresponding
six-dimensional dilaton and $\l$ is the scale of the fifth (string)
dimension measured in the six-dimensional Einstein metric. These are
the same three scalar fields that appear in the GR effective action
\acti, \ffoo.

  Since the kinetic terms in the effective action \acti, \ffoo\ are
diagonal in $\p$, $\nu$ and $\l$, we can immediately read off some
important conclusions from \ttt, \iis. The expansion in powers of
world-sheet derivatives is a low-energy expansion and, of the fields we
have kept, only the dilaton $\p$ is coupled at leading order.  It is
also easy to see that the off-diagonal components of the metric
$\bar h_{ij}$ have the same coupling as $\p$ to lowest order in
energy.  (These are the fields whose emission and absorption were
considered in \cite{dmw,dmOne,dmTwo}).  What is more interesting is that the
scalar $\n$ only couples at the next-to-leading order (fourth order in
derivatives). Note that its interaction term can be written in terms
of world-sheet energy-momentum tensors as $\n T^X_{++} T^X_{--}$.\footnote{
There is a similar coupling for $\p$ which produces a subleading
correction to its emission rate.}  The scalar $\l$ likewise does not
get emitted at the leading order and does couple at the same order as
$\p$, but with a different vertex.\footnote{The different vertices for
$\n$ and $\l$ probably reflect the different behavior of their
fluctuations (the non-decoupling of $\delta \l$ from metric
perturbations) in the case when $ Q_K$ is not equal to $Q=P$. } The
graviton components in the time and string directions $h_{00}$ and
$h_{55}$ couple to the string in a way similar to $\l$, which reflects
their mixing with $\l$ in the effective action \ffoo.  Indeed, the
vertex $(h_{00} + h_{55})(T^X_{++}+ T^X_{--})$ gives a vanishing
contribution to the amplitude of production of a closed string state:
it only couples a pair of left-movers or a pair of right-movers so
that the production is forbidden kinematically.  The important (and
non-trivial) point is that the simplest DBI action for the coupling of
the external fields to the D-brane gives the fields $\n$ and $\l$,
previously identified as the ``fixed scalars', different (and weaker)
couplings to the effective string than the fields like $\p$ previously
identified as ``decoupled scalars.'' The precise couplings will
shortly be used to make precise calculations of absorption and
emission rates.

The action \expa\ is at best the bosonic part of a supersymmetric action.
In previous discussions of D-brane emission and absorption, it has been
possible to ignore the coupling of external fields to the massless
fermionic excitations of the D-brane. For the questions that interest us,
that will no longer be possible and we make
a specific proposal for the couplings 
of world-sheet fermions. In the successful D-brane description of the entropy
of rotating black holes \cite{brekEx,brekNonEx}, five-dimensional angular momentum 
is carried by the fermions alone. There are two world-sheet fermion doublets,
one right-moving and one left-moving. 
The $SO(4)$ rotations of the 
uncompactified $(1234)$ coordinates are 
decomposed as $SU(2)_L\times SU(2)_R$
and the obvious (and correct) choice is to take the left-moving
fermions, $S^a$, to be a doublet under $SU(2)_L$ and the right-moving 
fermions, $S^{\dot a}$ to be a doublet under $SU(2)_R$. This set of fermions 
may be bosonized as two  boson fields, $\varphi^1$ and $\varphi^2$. 
As mentioned above,
the next-to-leading 
terms in \iit,\iis\ can be written in terms of the $X$-field 
stress-energy tensor,
$$ 
T^X_{++} = (\partial_+ \vec X)^2\ , \qquad T^X_{--} = (\partial_- \vec X)^2\ .
$$
The obvious guess for the supersymmetric completion of these
interaction
terms is simply to add the bosonization fields $\vec\varphi$ to the
world-sheet energy-momentum tensors:
$$ 
T^X_{++} \rightarrow T^{\rm tot}_{++}= (\partial_+ \vec X)^2+ 
(\partial_+ \vec \varphi)^2 \ ,
$$
and similarly for $T^X_{--}$. This will have a crucial effect on the 
normalization of the fixed scalar absorption rate.


\section{Semi-classical description of absorption}
\label{MatchFix}

  In this section we will mainly discuss the solution of the radial
differential equation one obtains for $s$-wave perturbations in the
fixed scalar $\nu$ related to the volume of the internal $T^4$ in
string metric \torr.  Let us start by restating some results of
section~\ref{EffSGAct}.
 {}From \met, \soo\ and \sool\ one can read off the five-dimensional
Einstein metric:
  \eqn{MetAgain}{
   d s_{5}^2 = -(H_{\hat{Q}} H_{\hat{P}} H_{\hat{Q}_K})^{-2/3} h d t^2 + 
                 (H_{\hat{Q}} H_{\hat{P}} H_{\hat{Q}_K})^{1/3} 
                  \left( {h\inv} d r^2 + r^2 d \Omega_{3}^2 \right)
 \ . }
To avoid the mixing between gravitational perturbations and the fixed
scalar, we will restrict 
ourselves to the case $P=Q$, i.e.{} $r_1 = r_5 = R$, $P^2= R^2 (R^2 +
r^2_0)$ (see \DefRadii). 
The small fluctuation equation, \kgoo, may be written as
\eqn{ExactNonExtreme}{
   \left[ \left( h r^3 \partial_r \right)^2 +
      \left (r^2+ R^2 \right )^2 (r^2 +r_K^2) \omega^2 -    
      {8 r^4 R^4 \over (r^2+ R^2)^2}
h \left (1 + {r_0^2\over R^2}\right ) \right] \tilde\nu= 0 \ ,
  }
where $h = 1 - {r_0^2 \over r^2}$. Since we work in the regime
$r_0 \ll R$, we will neglect the last factor in the last term.

 Several different radial coordinates are useful in different regions.
The ones we will use most often are $u$ and $y$ defined by the relations
  \eqn{RadVarDef}{
   1-\df{r_0^2}{r^2} = \e^{-r_0^2 / u^2} \ , \qquad
   y = \df{R^2 r_K \omega}{2 u^2} \ .
  }
 Note that $u \to 0$ and $y \to \infty$ at the horizon.

  The most efficient tool for obtaining the absorption cross-section
is the ratio of fluxes method used in \cite{mast}.  In all the cases
we will treat, the solution to \ExactNonExtreme\ whose near-horizon
limit represents purely infalling matter has the limiting forms
  \eqn{NearLim}{
   \tilde\nu \approx \e^{\i y}
  }
 near the horizon and 
  \eqn{FarLim}{
   \tilde\nu \approx \alpha \df{J_1(\omega r)}{\omega r}
    = \df{\alpha}{2} \df{H^{(1)}_1(\omega r) + H^{(2)}_1(\omega r)}{\omega r}
  }
 far from the black hole, where $J_1$ is a Bessel
function.\footnote{In fact there can be phase shifts in the arguments
of the exponential in \NearLim\ and the Bessel functions in \FarLim,
but they are immaterial for computing fluxes at leading order.}  The
term in \FarLim\ containing $H^{(2)}_1(\omega r)$ is the incoming
wave.  Once the constant $\alpha$ is known, one can compute the flux
for the incoming wave and compare it to the flux for the infalling
wave \NearLim\ to find the absorption probability.  In the present
instance, fluxes are purely radial:
  \eqn{FluxOneForm}{
   J = \df{1}{2\i} (\tilde\nu^* d \tilde\nu- {\rm c.c.}) 
     = J_r d r \ .
  }
 Observing that the number of particles passing through a sphere $S^3_r$
at radius $r$ in a time interval $[0,t]$ is 
  \eqn{NumPass}{
   \int_{S^3_r \times [0,t]} *J = 2 \pi^2 h r^3 J_r t \ ,
  }
 one concludes that the flux per unit solid angle is 
  \eqn{Flux}{
   {\cal F} = h r^3 J_r 
            = \df{1}{2\i} (\tilde\nu^* h r^3 \partial_r \tilde\nu - 
                {\rm c.c.}) \ .
  }
 The absorption probability is
  \eqn{AbsProb}{
   1 - |S_0|^2 = {{\cal F}_{\rm h} \ov  {\cal F}_\infty^{\rm incoming}} 
    = \df{2 \pi}{|\alpha|^2} R^2 r_K \omega^3 \ .
  }
 We will always be interested in cases where this probability is small.
 By the Optical Theorem, the absorption cross-section is
  \eqn{SigmaAbs}{
   \sigma_{\rm abs} = \df{4 \pi}{\omega^3} \left( 1 - |S_0|^2 \right)
     = \df{8 \pi^2}{|\alpha|^2} R^2 r_K .
  }
 Readers unfamiliar with the solution matching technology may be
helped by the analogy to tunneling through a square potential barrier
in one dimension.  If particles come from the left side of the
barrier, the wave function is to a good approximation a standing wave
on the left side of the barrier, a decreasing exponential inside the 
barrier, and a purely right moving exponential on the right side of 
the barrier.

  To obtain the familiar result $\sigma_{\rm abs} = A_{\rm h}$ for
low-energy, ordinary scalars falling into an extremal black hole, it
suffices to match the limiting value of \NearLim\ for small $y$
directly to the limiting value of \FarLim\ 
for small $r$ \cite{dgm}.  Due to
non-extremality and to the presence of the potential term in
\ExactNonExtreme, this naive matching scheme is invalid.  A more
refined approximate solution must be used, and a more physically
interesting low-energy cross-section will be obtained.  

  We will now present approximate solutions to \ExactNonExtreme\ in two
regimes most easily characterized in D-brane terms: we shall first
consider $T_R = 0$ with $\omega/T_L$ arbitrary; then we shall consider
$T_R$ much less than $T_L$ but not zero, and allow $\omega/T_R$ to
vary arbitrarily.  In \cite{kkOne} a more general case was considered:
$\omega/T_L$ and $\omega/T_R$ both arbitrary.

  When $T_R = 0$, the black hole is extremal: $r_0 = 0$ and $r = u$.
As usual,  one proceeds by joining a near horizon solution ${\bf I}$ to
a far solution ${\bf III}$ using an exact solution ${\bf II}$ to the
$\omega = 0$ equation \cite{dmOne,dmTwo,kr}.  
The dominant terms of \ExactNonExtreme\ and 
the approximate solutions in the three regions are 
  \eqn{EtaSols}{\vcenter{\openup1\jot
   \halign{\strut\span\TL & \span\TR & \span\TT & 
                 \span\TL & \span\TR\cr
   {\bf I.}\ \ &\left[ \partial_y^2 + 1 - 
      \df{2\eta}{y} - \df{2}{y^2} \right] \tilde\nu_{\bf I} = 0 &
    \qquad\qquad & \tilde\nu_{\bf I} &= G_1(y) + \i F_1(y) \cr
   {\bf II.} \ \ &\left[ (r^3 \partial_r)^2 - 
      8 \df{R^4}{H^2} \right] \tilde\nu_{\bf II} = 0 &
    \qquad\qquad & \tilde\nu_{\bf II} &= {C \over H(r)} + D H^2 (r) \cr
   {\bf III.} \ \ &\left[ (r^3 \partial_r)^2 + 
      r^6 \omega^2 \right] \tilde\nu_{\bf III} = 0 &
    \qquad\qquad &
     \tilde\nu_{\bf III} &= \alpha \df{J_1(\omega r)}{\omega r} + 
      \beta \df{N_1(\omega r)}{\omega r} \ , \cr
  }}}
 where $C,D,\a$ and $\b$ are constants, $H = 1 + R^2/r^2$, and $F_1$ and
$G_1$ are Coulomb functions \cite{hmf} whose charge parameter $\eta$ is
given by
  \eqn{EtaDef}{
   \eta = -\tf{1}{4} \left( 2 \omega r_K + \df{\omega R^2}{r_K} \right) 
        = -{1 \over 4 \pi} {\omega \over T_L} 
            \left( 1 + 2 {r_K^2 \over R^2} \right) \ .
  }
 In the last equality we have used the definition of $T_L$ in \temps.
The quantity $r_K^2 / R^2$ is small in the dilute gas approximation,
and we will neglect it when comparing the final semi-classical 
cross-section with the D-brane answer.

  By design, $\tilde\nu_{\bf I} \to \e^{\i y}$ as $y \to \infty$
up to a phase shift in $y$.  An approximate solution can be patched
together from $\tilde\nu_{\bf I,II,III}$ if one sets
  \eqn{EtaMatch}{
    C = \df{\alpha}{2} = \df{2}{3 C_1(\eta) \omega r_K} \ , \qquad 
    D = 0 \ , \qquad
    \beta = 0 \ ,
  }
where $C_1(\eta) = { 1 \over 3} e^{-\pi \eta / 2} |\Gamma(2+i\eta)| $ \cite{hmf}.
 A slightly better matching can be obtained by allowing $D$ and
$\beta$ to be nonzero, but the changes in the final solution do not
affect the fluxes ${\cal F}_{\rm h}$ and ${\cal F}_\infty$ (these
changes are however crucial in determining $S_0$ by the old methods of
\cite{unruh}, and give phase information on the scattered wave which the
flux method does not).  Having only $C \neq 0$ in region $\bf II$ is
analogous to the fact that for right-moving particles incident on a
square potential barrier, the wave function under the barrier can be
taken as a purely falling exponential with no admixture of the rising
exponential.

  {}From \EtaMatch\ and \SigmaAbs\ the cross-section is immediate:
  \eqn{EtaSigma}{
   \sigma_{\rm abs} = {\pi^2 \over 2} r_K R^2
 (\omega r_K)^2 |3 C_1(\eta)|^2
     = \tf{1}{4} A_{\rm h} (\omega r_K)^2 (1 + \eta^2) 
        \df{2 \pi \eta}{\e^{2 \pi \eta} - 1} \ ,
  }
where $A_{\rm h}$ is the area of the horizon (given in  \ment).
Note that the derivation of \EtaSigma\ does not require 
the assumption that $r_K \ll R$.

  To make the  comparison with  
  the D-brane approach, we  can 
  write \EtaSigma\ in the following 
suggestive form
  \eqn{ESAgain}{
   \sigma_{\rm abs} = {\pi^2 \over 2} r_K R^2 (\omega r_K)^2 
    {{\omega \over 2 T_L} \over 1 - \e^{-{\omega \over 2 T_L}}} 
\left (1 + {\omega^2 \over 16 \pi^2 T_L^2} \right )
     \left[ 1 + O(r_K^2/R^2) \right] \ .
  }
 In section~\ref{DAgree} we will compute the same quantity using
effective D-string method and will find agreement to the indicated
order of accuracy.  To obtain $O(r_K^2/R^2)$ corrections on the
D-brane side one would have to go beyond the dilute gas approximation.
An interesting special case where these corrections vanish is when
$T_L = 0$, corresponding to $\eta \to -\infty$.  In the brane
description, this corresponds to 1-branes and 5-branes only with no
condensate of open strings running between them: a pure quantum state
with no thermal averaging.  The limiting forms of the GR result
\EtaSigma\ and of the D-brane absorption cross-section \SigmaDbrane\
to be derived in section~\ref{DAgree} then agree exactly:
  \eqn{SigmaPure}{
   \sigma_{\rm abs} = \left( \tf{\pi}{4} \right)^3
    R^8 \omega^5 \ .
  }

  Now let us continue on to the second regime in which an approximate
solution to the radial equation \ExactNonExtreme\ is fairly
straightforward to obtain: $\omega,T_R \ll T_L$ with $\omega/T_R$ 
arbitrary.  A quantity which enters more naturally into the 
differential equations than $\omega / T_R$ is 
  \eqn{ADef}{
   B = {R^2 r_K \omega \over  r_0^2  }  = {\omega \ov \kappa } \tanh 
   \sigma\approx 
   {\omega \ov \kappa } \  , 
    \ \ \ \  \k\equiv 2\pi T_H \
   , 
  }
 where  $\k$ is the surface gravity at  the horizon, 
  and in the last step we used the fact that $\sigma \gg 1$ when
$T_R\ll T_L$.  In dropping terms from the exact equation
\ExactNonExtreme\ to obtain soluble forms in the three matching
regions, it is essential to retain $B$ as a quantity of $O(1)$;
however, $r_0/r_K$ and $\omega R^2 / r_K$ can be regarded as small
because $T_R \ll T_L$ and $\omega \ll T_L$.  In regions ${\bf II}$ and
${\bf III}$, the approximate equations turn out to be precisely the
same as in \EtaSols, but in ${\bf I}$ one obtains a more complicated
differential equation:
  \eqn{NESolI}{
\left[ \partial_y^2 + 1 - 
      { 8\ov B^2} {\e^{-2y/B} \over (1 - \e^{-2y/B})^2} \right] 
       \tilde\nu_{\bf I} = 0  \ .
  }
 This equation can be cast in the form of a supersymmetric quantum
mechanics eigenfunction problem.  Define a rescaled variable $z =
y/B$ and supercharge operators
  \eqn{SuperQ}{
   \QQ = -\partial_z + {\rm coth}\, z \ , \qquad
   \QQ^\dagger = \partial_z + {\rm coth}\, z \ .
  }
 Then \NESolI\ can be rewritten in the form
  \eqn{HOne}{
   \QQ \QQ^\dagger \tilde\nu_{\bf I}
     = \left[ -\partial_z^2 + 2 {\rm csch}^2\, z + 1 \right] 
        \tilde\nu_{\bf I}
     = \left( 1  + B^2 \right) \tilde\nu_{\bf I} \ .
  }
 The eigenfunctions of the related Hamiltonian $\QQ^\dagger \QQ =
-\partial_z^2 + 1$ are just exponentials, and from them one can read
off the solutions to \HOne: the infalling solution is
  \eqn{Infall}{
   \tilde\nu_{\bf I} 
     = {\QQ \, \e^{\i Bz} \over 1-\i B } 
     = {{\rm coth}\, z - \i B \over  1 -\i B} \,
         \e^{\i B z} 
     = { {\rm coth}\, {y\ov B} -\i B \over
         1 -\i B } \, \e^{\i y} \ .
  }
 The factor in the denominator is chosen so that $\tilde\nu_{\bf
I} \approx \e^{\i y}$ for large $y$.

  Performing the matching as usual, one obtains
  \eqn{Interp}{
   \sigma_{\rm abs} = \tf{1}{4} A_{\rm h} (\omega r_K)^2 (1 + B^{-2} )
=\tf{1}{4} A_{\rm h} (\omega r_K)^2 \left (1+ {4 \pi^2 T_H^2\over
\omega^2 }\right )\ .
  }
In section~\ref{DAgree} we will  show that the effective string
calculation gives the same result when
$\omega,T_R \ll T_L$.


\section{D-brane absorption cross-sections and emission rates}
\label{DAgree}

In this section we give a detailed calculation of the emission and 
absorption of the fixed scalar $\n$, using the interaction vertices
computed in section~\ref{EffDBI}. We recall that  $\nu$ is related (see \torr) 
to the
  volume   (measured in the string metric) 
  of the compactification 4-torus
 orthogonal to the string. To study the leading coupling
of $\nu$, it is sufficient to retain the following two terms in
the string effective action (cf. \expa):
  \eqn{Retain}{\eqalign{
    I &= \int d^2 \sigma \bigg\{ -{1\over 2} (\partial_+ \vec X\cdot
         \partial_- \vec X + \partial_+ \vec\varphi \cdot 
          \partial_- \vec\varphi) \cr
      &\qquad\ + {1\over 4 T_{\rm eff}} 
\left [(\partial_+ \vec X)^2+ (\partial_+ \vec \varphi)^2\right ]
\left [(\partial_- \vec X)^2+ (\partial_- \vec \varphi)^2\right]
 \ \nu (x)  \bigg\}
  }}
where we have absorbed $\sqrt {T_{\rm eff}}$ into the fields
to make them properly normalized.  {}From \acti\ we see that the
scalar field with the proper bulk kinetic term is $2 \nu/\kappa_5$.  
Consider the invariant amplitude for processes mediated by the quartic 
interaction in \Retain. If $p_1$ and $p_2$ are the left-moving energies, while
$q_1$ and $q_2$ are the right-moving ones, the matrix element among
properly normalized states is
  \eqn{MatrixElem}{
   {\sqrt 2\kappa_5\over T_{\rm eff}} \sqrt{ q_1 q_2 p_1 p_2\over \omega} \ .
  }

The basic assumption of the D-brane approach to black hole physics
is that the left-movers and right-movers can be treated as thermal
ensembles \cite{cm,hms}. 
Strictly speaking, they are microcanonical 
ensembles, but for our purposes the canonical ensemble is good enough
and we proceed as if we are dealing with a massless one-dimensional
gas of left-movers of temperature $T_L$ and right-movers with temperature
$T_R$. The motivation for this assumption has been explained at length
in several  recent papers \cite{cm,hms,mast}.
To compute the rate for the process $scalar \to L+L+R+R$
we have to square the normalized matrix element \MatrixElem\ and
integrate it over the possible energies of the final state particles.
Because of the presence of the thermal sea of left-movers and right-movers, 
we must insert Bose enhancement factors: for example, each left-mover in 
the final state picks up a factor of $1 + \rho_L(p_i) = -\rho_L(-p_i)$, where
  \eqn{BEDist}{
   \rho_L (p_i) = {1\over  \e^{p_i\over T_L}-1 }
  }
is the Bose-Einstein distribution.  If there were a left-mover of
energy $p_i$ in the initial state, it would pick up an enhancement
factor $\rho_L(p_i)$.  Similar factors attach to right-movers.

Conservation of energy and of momentum parallel to the effective
string introduces the factor
  \eqn{DeltaFn}{
   (2\pi)^2 \delta(p_1+ p_2 + q_1 + q_2-\omega)
    \delta(p_1+ p_2- q_1 -q_2) = {1\over 2} (2\pi)^2 \delta(p_1+ p_2-
    {\omega\over 2}) \delta(q_1 + q_2 -{\omega\over 2})
  }
into the integrals over $p_1$, $p_2$, $q_1$, and $q_2$.
Putting everything together, we find that the rate for $scalar \to
L+L+R+R$ is given by
  \eqn{firstrate}{\eqalign{
    \Gamma(1) &= \Gamma(scalar \to L+L+R+R) \cr
     &= {36\over 4} {\kappa_5^2 L_{\rm eff} \over 4 \pi^2 T^2_{\rm eff}\omega } 
         \int_0^\infty d p_1 d p_2 \, 
          \delta\left( p_1 + p_2 -{\omega\over 2} \right)
          {p_1 \over 1 - \e^{-{p_1\over T_L}}} 
          {p_2 \over 1 - \e^{-{p_2\over T_L}}} \cr
    & \qquad\qquad \times 
         \int_0^\infty d q_1 d q_2 \, 
          \delta\left( q_1 + q_2 -{\omega\over 2} \right)
          {q_1 \over 1 - \e^{-{q_1\over T_R}}}
          {q_2 \over 1 - \e^{-{q_2\over T_R}}} \ , 
  }}
where $L_{\rm eff}$ is the length of the effective string.
The factor of $36=6^2$ arises from the presence of 
six species of left-movers (four bosons and two bosonized fermions)
and six species of right-movers. We divide by $4 = 2^2$ because of
particle identity: because the two left-movers in the final state are
identical particles, the integral over $p_1,p_2$ counts every
left-moving final state twice (similarly for the right-movers).

  To write down the rates for the three other absorptions processes
(that is, processes $2$, $3$, and $4$  in   eq.~\FixedDProc), it is
convenient to define the integrals
  \eqn{LRInt}{\eqalign{
   I_L(s_1,s_2) &= \int_0^\infty d p_1 d p_2 \,
          \delta\left( s_1 p_1 + s_2 p_2 + {\omega \over 2} \right)
          s_1 p_1 \rho_L(s_1 p_1) \cdot s_2 p_2 \rho_L(s_2 p_2) \cr
   I_R(s_1,s_2) &= \int_0^\infty d q_1 d q_2 \,
          \delta\left( s_1 q_1 + s_2 q_2 + {\omega \over 2} \right)
          s_1 q_1 \rho_L(s_1 q_1) \cdot s_2 q_2 \rho_L(s_2 q_2) \ .
  }}
 The choices $s_i = 1$ and $s_i = -1$ correspond, respectively, to 
putting a particle in the initial or final state.  Then the total 
absorption rate, including all four competing processes of \FixedDProc,
is
  \eqn{AbsorbAll}{\eqalign{
   \Gamma_{\rm abs}(\omega) 
    &= \Gamma(1) + \Gamma(2) + \Gamma(3) + \Gamma(4) \cr
    &= 36 {\kappa_5^2 L_{\rm eff} \over 4 \pi^2 T_{\rm eff}^2 \omega}
        \Big[ \tf{1}{4} I_L(-1,-1) I_R(-1,-1) + 
              \tf{1}{2} I_L(-1, 1) I_R(-1,-1) \cr
    & \qquad\qquad + 
              \tf{1}{2} I_L(-1,-1) I_R(-1, 1) + 
                        I_L(-1, 1) I_R(-1, 1) \Big] \cr 
    &= {9\kappa_5^2 L_{\rm eff} \over 4\pi^2 T_{\rm eff}^2 \omega}
       \int_{-\infty}^\infty d p_1 d p_2 \, 
        \delta\left( p_1 + p_2 -{\omega\over 2} \right)
        {p_1 \over 1 - \e^{-{p_1\over T_L}}} 
        {p_2 \over 1 - \e^{-{p_2\over T_L}}} \cr
    & \qquad\qquad \times 
       \int_{-\infty}^\infty d q_1 d q_2 \, 
        \delta\left( q_1 + q_2 -{\omega\over 2} \right)
        {q_1 \over 1 - \e^{-{q_1\over T_R}}}
        {q_2 \over 1 - \e^{-{q_2\over T_R}}} \cr
    &= {\kappa_5^2 L_{\rm eff} \over (32 \pi)^2 T_{\rm eff}^2}
        {\omega \over 
         \left(1- \e^{-{\omega\over 2 T_L}} \right)
         \left(1- \e^{-{\omega\over 2 T_R}} \right) 
        }
        \left( \omega^2 + 16 \pi^2 T_L^2 \right)
        \left( \omega^2 + 16 \pi^2 T_R^2 \right) \ .
  }}
 The fractional coefficients inside the square brackets on the second
line of \AbsorbAll\ are symmetry factors for the final states (the
initial states are always simple enough so that their symmetry factors
are unity).  It is remarkable that although the individual processes
$1$--$4$ have rates which cannot be expressed in closed form, their
sum is expressible in terms of integrals which can be performed
analytically \cite{GR} because they run over all $p_1$, $p_2$, $q_1$,
and~$q_2$.

  A similar calculation may be performed for the four emission processes, 
with the result
\eqn{EmitAll}{
   \Gamma_{\rm emit}(\omega) 
    = \e^{-{\omega \over T_H}} \Gamma_{\rm abs}(\omega) 
    = -\Gamma_{\rm abs}(-\omega) \ ,
  }
where the Hawking temperature characterizing the distribution of the
emitted scalars is related to $T_R$ and $T_L$ by \THDef.
Our convention has been to compute $\Gamma_{\rm abs}(\omega)$
assuming that the flux of the incoming scalar is unity.  We have also
suppressed the phase space factor $\d^4 k / (2 \pi)^4$ for the
outgoing scalar in computing $\Gamma_{\rm emit}(\omega)$, and we have
assumed that the outgoing scalar is emitted into the vacuum state, so
that $\Gamma_{\rm emit}(\omega)$ includes no Bose enhancement factors.
These conventions were chosen because they lead to simple expressions
for $\Gamma_{\rm emit}(\omega)$ 
\EmitAll\ and $\sigma_{\rm abs}$  below,
 but they must be borne carefully in mind
when considering questions of detailed balance.  Suppose we put the
black hole in a thermal bath of scalars at temperature $T_H$.  Then
$\Gamma_{\rm abs}(\omega)$ and $\Gamma_{\rm emit}(\omega)$ pick up
Bose enhancement factors for the scalars: those factors are,
respectively, $1/(\e^{\omega/T_H}-1)$ and $1/(1-\e^{-\omega/T_H})$.
Once these factors are included, the emission and absorption rates
become equal by virtue of the first equality in \EmitAll.  The
fact that calculating $\Gamma_{\rm emit}(\omega)$ in the same way that
we calculated $\Gamma_{\rm abs}(\omega)$ leads to \EmitAll\ is a
nontrivial check on detailed balance.  This check is analogous to 
verifying that QED reproduces the Einstein $A$ and $B$ coefficients
for the decay of the first excited state of hydrogen.

  Because $\Gamma_{\rm abs}(\omega)$ was computed assuming unit flux,
one would naively guess that the absorption cross-section to be
compared with a semi-classical calculation is $\sigma_{\rm abs} =
\Gamma_{\rm abs}(\omega)$.  (Now we are back to the conventions where
$\Gamma_{\rm abs}(\omega)$ and $\Gamma_{\rm emit}(\omega)$ do not
include Bose enhancement factors for the scalars).  This is not quite
right; instead,
  \eqn{SigmaGamma}{
   \sigma_{\rm abs}(\omega) 
     = \Gamma_{\rm abs}(\omega) - \Gamma_{\rm emit}(\omega) 
     = \Gamma_{\rm abs}(\omega) + \Gamma_{\rm abs}(-\omega) \ .
  }
 To see why \SigmaGamma\ is right, we have to remember what we are
doing in a semi-classical computation.  We send in a classical wave in
the field whose quanta are the scalars of interest, and we watch to
see what fraction of it is sucked up by the black hole and what
fraction is re-emitted.  The quantum field theory analog is to send
in a coherent state of scalars with large average particle number, so
that the flux is almost fixed at its classical expectation value $\cal
{}F$.  The dominant processes are then absorption and stimulated
emission.  The Bose enhancement factors collapse to $\cal {}F$ for
both absorption and emission, up to errors which are insignificant in
the semi-classical limit.  The net rate at which particles are
absorbed is then $\Gamma_{\rm abs}(\omega) {\cal {}F} - \Gamma_{\rm
emit}(\omega) {\cal {}F}$.  But this rate is $\sigma_{\rm abs} {\cal 
{}F}$ by definition, whence \SigmaGamma.  Note that the last expression
in \SigmaGamma\ is manifestly invariant under time-reversal, which takes
$\omega \to -\omega$.

In order to obtain definite results for the absorption cross-section, we
must supply values for the effective length $L_{\rm eff}$ of the string, as
well as its effective tension $T_{\rm eff}$. It is a by-now-familiar story
that multiple D-strings bound to multiple five-branes behave like a
single D-string multiply wound about the compactification 
direction \cite{ms}.
In the case at hand it is well-understood that the effective string
length is \cite{ms,mast}
\eqn{klrel}{\kappa_5^2 L_{\rm eff} = 4 \pi^3 r_1^2 r_5^2 \ .}
With this substitution, 
the fixed scalar $\n$ absorption cross-section  becomes
\eqn{SigmaD}{
\sigma_{\rm abs}(\omega) = { \pi r_1^2 r_5^2 \over 256 T^2_{\rm eff}}
{\omega \left (\e^{\omega\over T_H} - 1 \right ) \over   
\left (\e^{\omega\over 2 T_L} - 1\right )
\left (\e^{\omega\over 2 T_R} - 1\right ) }
(\omega^2 + 16 \pi^2 T_L^2) (\omega^2 + 16 \pi^2 T_R^2)\ .
}
This is similar to, but not quite the same as, the absorption cross-section
for the ordinary ``unfixed'' scalar calculated in \cite{mast}.

The object of our exercise is to offer further evidence that the
D-brane configuration {\it is} the corresponding black hole by showing
that \SigmaD\ is identical to the corresponding quantity calculated
by standard classical GR methods. For technical reasons, the GR calculation
in a general black hole background is quite difficult and the results we
have been able to obtain (presented in section~\ref{MatchFix}) are only valid 
in certain simplifying limits. The most important simplification is to
take equal brane charges $r_1=r_5=R$. 

First we consider the extremal limit, $T_R=0$. Here \SigmaD\ 
reduces to
\eqn{simpler}{
 \sigma_{\rm abs}(\omega) = { \pi^3 r_1^2 r_5^2 T_L^3 \over 8 T^2_{\rm eff}}
\omega^2 {{\omega \over 2 T_L} \over 1 - \e^{-{\omega \over 2 T_L}}} 
\left (1 + {\omega^2 \over 16 \pi^2 T_L^2} \right )\ . }
This is to be compared with the classical fixed scalar absorption
cross-section in the extremal background (eq.~\ESAgain):
\eqn{SigmaGR}{ \sigma_{\rm abs}(\omega) = {\pi^2 \over 2} 
R^2 r_K^3 \omega^2
    {{\omega \over 2 T_L} \over 1 - \e^{-{\omega \over 2 T_L}}}
\left (1 + {\omega^2 \over 16 \pi^2 T_L^2} \right )
\ .}
Using  that in the extremal limit 
$$ 
T_L = {r_K \over \pi r_1 r_5 } \ , 
$$ 
and remembering that we were only able to do the classical calculation
for $r_1=r_5=R$, we see that the D-branes and GR match if we take the
effective string tension to be
  \eqn{tension}{
   T_{\rm eff} = {1\over 2 \pi R^2 } = {1 \over 2 \pi \alpha' g n_5} \ ,
  }
 where we have restored the dependence on $\alpha'$ that we have been
suppressing since \UnitChoice.  This value for $T_{\rm eff}$ is
precisely equal to the tension of the ``fractionated'' D-string moving
inside $n_5$ 5-branes \cite{ms,juan}. This is a highly non-trivial
independent check on the applicability of the effective string model
to fixed scalars, and also on the idea of D-string ``fractionation!''
 
Another interesting comparison to be made is for near-extremal black
holes.
For $\omega,T_R\ll T_L$ but with ratio of $\omega$ to $T_R$ otherwise
arbitrary, \SigmaD\ becomes, using \tension\ for the tension,
\eqn{neare}{
 \sigma_{\rm abs} (\omega)=  {\pi^2\over 2} R^2 r_K^3
(\omega^2+ 4 \pi^2 T_H^2) \  . } 
This is in exact agreement with the absorption cross-section 
on non-extremal black holes \Interp\ computed using general
relativity.

For the fixed scalar the coupling to $(\del X)^2$ is absent from
the D-brane action, and the cross-section we found is the leading
effect. For an ordinary decoupled scalar, such as the 6-d dilaton,
both terms are present. So, the cross-section  computed   above should be
part of the correction to the leading effect  determined  in \cite{mast}.
This is an interesting topic for future investigation.


\section{Conclusions}
\label{ConclusionFix}

Let us try to recapitulate in a few words what it has taken many
equations to state.  The main thrust of the chapter has been to
explore the behavior of the type of fixed scalar studied earlier in
\cite{gibbOne,gibbTwo,gkk,fkOne,fkTwo,fks}, and most recently in
\cite{kr} -- but now in the context of five-dimensional black holes
that can be modeled by bound states of D1-branes and D5-branes
\cite{sv,cm,ms,dmw,dmOne,dmTwo,gkOne,mast}.  For the most part we have
focused on the fixed scalar $\nu$ which corresponds to the volume of
the internal four-torus as measured by the string metric.  Through an
interesting interplay between semi-classical computations (where the
basic theory is well known but analytically intractable in general)
and D-brane computations (where the theory is less well known but very
tractable), we have arrived at a general formula \SigmaD\ for the
cross-section for low-energy fixed scalars to be absorbed into the
black hole.

The absorption cross-section \SigmaD\ has a much richer and more
interesting functional form than the simple $\omega^2$ dependence
found in \cite{kr}.  Even in the simple limit $\omega,T_R \ll T_L$ in which
comparison calculations between GR and D-branes were initially
performed \cite{dmw,dmOne,dmTwo}, the fixed scalar cross-section goes not as
$\omega^2$ but as $\omega^2 + \kappa^2$, where $\kappa = 2 \pi T_H$ is
the surface gravity at the horizon.  While we have derived the
expression \SigmaD\ in full generality only in the D-brane picture, we
have demonstrated that it agrees with semi-classical calculations of
the cross-section in the two regimes: one regime reproduces this novel
$\omega^2 + \kappa^2$ behavior, while the other deals with absorption
into extremal black holes.  Because the equations for the gravitational 
perturbations and fixed scalar perturbations 
 couple unless two of the three charges, e.g., the 1-brane charge 
 and  the
5-brane charge,  are equal to each other,
 our semi-classical computations are limited to the
equal charge case (similar equal-charge assumption 
was used  in  $D=4$  case  in \cite{kr}). 
 Modulo this limitation, we have
confidence that a full greybody factor computation along the lines of
\cite{mast} would reproduce the general result \SigmaD.

One of the reasons why the extension of the semi-classical
calculations to unequal 1-brane and 5-brane charges (but with both
still greater than the third charge, $P,Q \gg Q_K$, to remain in the
dilute gas region) would be interesting, is that the D-brane
computations involve one free parameter, the tension $T_{\rm eff}$ of
the effective string, which can be read off from a comparison with a
semi-classical calculation.  The expectation, based on the work of
\cite{ms,juan} and on the arguments given at the beginning of
section~\ref{EffDBI}, is that $T_{\rm eff} = {1 \ov 2 \pi \alpha' g
n_5}$.  Our work confirms this relation when the 1-brane and 5-brane
charges are equal.
 What a semi-classical calculation with
unequal charges should confirm  is that 
 $T_{\rm eff}$ is independent of the number of 1-branes. 

Although our ultimate goal has been to demonstrate a new agreement
between semi-classical GR and a perturbative treatment of the
effective string, we have along the way studied interesting facets of
both formalisms separately.  On the D-brane side, we have been forced
to go beyond the leading quadratic terms in the expansion of the DBI
action and examine terms quartic in the derivatives of the string 
collective coordinate fields $X^i$.  As we argued in
section~\ref{EffDBI}, the generic form of the quadratic terms is practically
inevitable given the invariances of the problem.  But the decoupling
of the fixed scalar from quadratic terms and the precise form of its
coupling to quartic terms is a signature of the DBI action.  The
agreement between the D-brane and GR cross-sections for fixed scalars
is thus a more stringent test of the DBI action than the agreements
obtained previously \cite{dmOne,dmTwo,us,mast} for ordinary scalars.

{}From the open string theory point of view,  the $(\del X)^2$ term 
in the  D-string action \nam,\expa\ originates 
 upon dimensional
reduction from the 
$F^2_{\m\n}$  term in  the $D=10$ Born-Infeld  action, while 
the $(\del X)^4$ terms correspond to  the $F^4_{\m\n}$ -terms.
It is amusing to note that the  fixed 
scalars, which are  coupled  to  the 
 Maxwell terms of the closed string  vector fields  in the 
spacetime  effective action \acti, thus 
do not couple to the Maxwell term of the open string vector field 
in the effective D-string  action,  while  
 exactly the opposite is true for the 
decoupled scalars.
It is thus the $F^4_{\m\n}$ -terms in the DBI action 
(which are important also in some other contexts) that 
are effectively responsible  for the leading contribution 
to the cross-section of fixed scalars.

At the  relevant  $(\del X)^4$ order,  
the  processes involving fermionic excitations of
the effective string contribute in the same way  as purely bosonic
processes.  Fortunately,  the  coupling of
bosonic excitations to the fixed scalar field 
  predicted by the DBI action is of a particularly
simple form, $T^X_{++} T^X_{--} \nu(x)$, which admits an obvious
generalization to include fermions: $T^{\rm tot}_{++} T^{\rm tot}_{--}
\nu(x)$.  Obtaining precise agreement with GR using this coupling and
the normalization of $T_{\rm eff}$ as in 
\cite{ms,juan} may be
viewed   as  determining a partial supersymmetrization of the 
effective string action
via D-brane spectroscopy.

On the GR side, we have to some extent systematized the study of
 spherical  black hole configurations, including  spherically symmetric
   perturbations
around the basic $D=5$ black hole with three charges,  
  by reducing the problem to an effective  two-dimensional one. 
 For time-independent configurations, this  gives a 
straightforward derivation of the basic black hole solution. 
 We were
disappointed to find, however,  that,
 despite  relative  simplicity of the
effective two-dimensional theory compared to the full supergravity
equations, it still leads  to complicated  coupled
differential equations for  time-dependent fluctuations
around the static  solution.  So far, we have been able to extract simple
equations from the intractable general case only when some pair of
charges are equal.  Then the background value of one fixed scalar
becomes constant and its fluctuations decouple from the other fields,
leading to a non-extremal five-dimensional  generalization of the
equation studied in \cite{kr}.  Similar two-dimensional effective theory
techniques with similar equal charge limitations were  applied to
the basic four-dimensional black hole with four charges. 
 In this chapter, we
have taken the four-dimensional calculations only far enough to see
that fixed scalars whose background values become constant when three
of the four charges are equal have an absorption cross-section with
the characteristic $\omega^2 + \kappa^2$ dependence.

One final comment is that we have focused almost exclusively on
absorption rather than Hawking emission.  This is not because Hawking
emission is any more difficult, but rather because agreement between
the semi-classical Hawking calculation and the D-brane result is
inevitable once a successful comparison of absorption cross-sections
is made.  To see this, one must only note that detailed balance
between emission and absorption is built into the Hawking calculation
and that it can be checked explicitly in the D-brane description.
Once detailed balance is established in both descriptions, it
obviously suffices to check that the absorption cross-section agrees
between the two in order to be sure that emission rates must agree as
well.

\chapter{Partial wave absorption by effective strings}
\label{PartialWaves}

\section{Introduction}
\label{IntroPart}

The D1-brane D5-brane bound state toroidally compactified down to five
dimensions has proven to be one of the most fruitful string theoretic
models of black holes.  Since the original paper of \cite{cm}, which
proposed the model as a way to study black hole dynamics in a
manifestly unitary string theoretic framework, and the subsequent work
in \cite{ms} clarifying the means by which a single multiply wound
effective string arises in a description of the low-energy dynamics,
there have been many exciting papers relating properties of
five-dimensional and four-dimensional black holes to the effective
string.  An explanation of the near-extremal entropy was given in
\cite{cm,brekEx,brekNonEx}, following the ideas originally laid out
in \cite{sv}.  Absorption cross-sections and the corresponding Hawking
emission rates were worked out in
\cite{dmw,dmOne,dmTwo,gkOne,mast,gkTwo,cgkt,kkOne,ja,krt} yielding
impressive agreement at low energies with the effective string model.
Suggestions that the effective string model may have some flaws or
limitations have arisen in the work of \cite{htr,dkt,kkTwo}.

In a recent paper by Strominger and Maldacena \cite{ja} it was found from a
general analysis of thermal two-point functions that the effective
string seems capable of explaining the semi-classical absorption
cross-sections for arbitrary partial waves of ordinary scalars.
Subsequent work by Mathur \cite{Mathur} exhibited more detailed agreement
between the effective string model and General Relativity for these
processes.  Ordinary scalars are scalars whose equation of motion in
five-dimensions is $\square \phi = 0$.  The canonical examples of
ordinary scalars are the off-diagonal gravitons $h_{ij}$ with both $i$
and $j$ lying within the D5-brane but perpendicular to the D1-brane.
These are the scalars for whose $s$-wave cross-section full agreement
between General Relativity and the effective string was first achieved
in \cite{dmOne}.

The results of \cite{ja,Mathur} overlap substantially with
\cite{gpartial}, on which the present chapter is based.  The
organization is as follows.  Section~\ref{semicl} covers the
semi-classical analysis of partial wave absorption and includes a
derivation of a form of the Optical Theorem for the absorption of
scalar particles which was quoted without proof in \cite{gukt}.
Section~\ref{Dbrane} presents the effective string description of the
same processes, exhibiting along the way a simple method for
performing all the phase space integrals encountered in \cite{Mathur}.
In section~\ref{llimits}, the limitation on the number of partial
waves the effective string can couple to arising from statistics and
locality is compared with the limitation imposed semi-classically by
cosmic censorship.  Section~\ref{ConclusionPart} summarizes the
results and indicates directions for further work.

\section{The semi-classical computation}
\label{semicl}

The quantity that can be conveniently computed using the matching
technique is the absorption probability.  To convert this to an
absorption cross-section, it is necessary to use properties of the
partial wave expansion in four spatial dimensions and to invoke the
Optical Theorem.  The details of this connection were worked out
independently in \cite{Mathur}, but because the derivation given
below applies for arbitrary dimensions, it seems worthwhile to present
it in full.  In all of what follows, $n = d-1$ will denote the number
of spatial dimensions.

The Optical Theorem for scattering of a scalar field off a spherically
symmetric potential states that if the scattering wave-function has
for large $r$ the asymptotic form
  \eqn{ScatteringSol}{
   \phi(\vec{r}) \sim e^{i k x} + f(\theta) {e^{i k r} \over r^{(n-1)/2}}
  }
 (here $x = r \cos \theta$), then the total
cross-section is
  \eqn{OpticalTheorem}{
   \sigma_{\rm total} = 
     -2 \left( {2 \pi \over k} \right)^{n-1 \over 2}
      \Re \left( i^{n-1 \over 2} f(0) \right) \ .
  }
 To find the partial wave expansion of \ScatteringSol, it is first
necessary to make a Neumann expansion of the exponential function,
which can be done using Gegenbauer polynomials \cite{Watson}:
  \eqn{NeumannSeries}{\eqalign{
   e^{i r \cos \theta} &= 2^{n/2-1} \Gamma(n/2-1)
    \sum_{\ell = 0}^\infty 
     i^\ell P_\ell(\cos \theta) (\ell + n/2 - 1) 
     {J_{\ell + n/2 - 1}(r) \over r^{n/2 - 1}} \cr
   P_\ell(\cos \theta) &= \sum_{m=0}^{\lfloor \ell/2 \rfloor} 
    (-1)^m 2^{\ell - 2m} 
    {\Gamma(\ell + n/2 - 1 - m) \over 
     \Gamma(n/2 - 1) m! (\ell - 2m)!} 
    \cos^{\ell - 2m} \theta \ .
  }}
 The $P_\ell(\cos\theta)$ are just the Legendre polynomials when
$n=3$.  For arbitrary $n$, they can be defined by the expansion
  \eqn{LegendreExpand}{
   (1 - 2 a \cos \theta + a^2)^{1-n/2} = 
    \sum_{\ell=0}^\infty P_\ell(\cos \theta) a^\ell \ .
  }
 An alternate normalization proves more convenient:
  \eqn{ChangeFactor}{
   \tilde{P}_\ell(\cos\theta) = \sqrt{2 \over \pi} 2^{n/2-1} 
    \Gamma(n/2-1) (\ell + n/2 - 1) P_\ell(\cos \theta) \ .
  }
 Using asymptotic properties of Bessel functions, one can now write
down the partial wave expansion of \ScatteringSol\ as
  \eqn{PartialWave}{
   e^{i k x} + f(\theta) {e^{i k r} \over r^{(n-1)/2}} 
    \sim \sum_{\ell = 0}^\infty \tf{1}{2} \tilde{P}_\ell(\cos \theta)
     {S_\ell e^{i k r} + (-1)^\ell i^{n-1} e^{-i k r} \over
      (i k r)^{(n-1)/2}} \ .
  }
 When $f(\theta) = 0$ identically, $S_\ell = 1$ for all $\ell$.

The absorption cross-section for the $\ell^{\rm th}$ partial wave can
now be computed as the difference between the total $\ell$-wave
cross-section computed via the Optical Theorem,
  \eqn{TotalScatter}{
   \sigma^\ell_{\rm total} = 
    -\left( {\sqrt{2 \pi} \over k} \right)^{n-1} 
     \tilde{P}_\ell(1) \Re (S_\ell - 1) \ ,
  }
 and the $\ell$-wave scattering cross-section,
  \eqn{SigmaScatter}{
   \sigma^\ell_{\rm scattered} = (\Vol S^{n-2}) 
    {|S_\ell - 1|^2 \over 4 k^{n-1}} 
    \int_0^\pi d\theta \, \sin^{n-2} \theta \,
     \tilde{P}_\ell(\cos \theta)^2 \ .
  }
 The final result,
  \eqn{SigmaPart}{
   \sigma^\ell_{\rm abs} = {2^{n-2} \pi^{n/2-1} \over k^{n-1}}
    \Gamma(n/2-1) (\ell + n/2 - 1) {\ell+n-3 \choose \ell} 
    \left( 1 - |S_\ell|^2 \right) \ ,
  }
 relates the absorption cross-section $\sigma^\ell_{\rm abs}$ to the
absorption probability $1 - |S_\ell|^2$.  The results for $n=3$ and
$n=4$ are
  \eqn{SigmaAbsFF}{\vcenter{\openup1\jot
   \halign{\strut\span\TL & \span\TR & \span\TT \cr 
    \sigma^\ell_{\rm abs} &= {\pi \over k^2} (2 \ell + 1) 
     \left( 1 - |S_\ell|^2 \right) & 
     \qquad in three spatial dimensions \cr
    \sigma^\ell_{\rm abs} &= {4 \pi \over k^3} (\ell + 1)^2
     \left( 1 - |S_\ell|^2 \right) &
     \qquad in four spatial dimensions. \cr
  }}}

With these results in hand, let us proceed to the semi-classical
computation of the cross-section for an ordinary scalar $\phi$ in the
$\ell^{\rm th}$ partial wave to be absorbed into a black hole.  It is
hoped that this piece of ``spectroscopic data'' will be illuminating
of the form of the effective string action.

An ordinary scalar is one whose equation of motion is just the Laplace
equation following from the black hole metric, which in five
dimensions is
  \eqn{MetricNE}{\eqalign{
   ds^2 &= -F^{-2/3} h dt^2 + 
    F^{1/3} \left( h^{-1} dr^2 + r^2 d \Omega_{S^3}^2 \right) \cr
   F &= f_1 f_5 f_K = \left( 1 + {r_1^2 \over r^2} \right)
                      \left( 1 + {r_5^2 \over r^2} \right)
                      \left( 1 + {r_K^2 \over r^2} \right) \cr
   h &= 1 - {r_0^2 \over r^2} \ .
  }}
 The mass, entropy, Hawking temperature, $U(1)$ charges, and
characteristic radii are conveniently parametrized as
  \eqn{ThermoQs}{\vcenter{\openup1\jot
    \halign{\strut\span\TC\cr
     M = {\pi \over 8} r_0^2 \sum_{i=1,5,K} \cosh 2 \sigma_i \qquad
     S = {\pi^2 \over 2} r_0^3 \prod_{i=1,5,K} \cosh \sigma_i \qquad
     \beta_H = 2 \pi r_0 \prod_{i=1,5,K} \cosh \sigma_i  \cr 
     Q_i = {r_0^2 \over 2} \sinh 2\sigma_i  \qquad
     r_i = r_0 \sinh \sigma_i  \cr
  }}}
 in five-dimensional Planck units.  Using a separation of variables, 
  $$\phi = e^{-i \omega t} P_\ell(\cos \theta) R(r) \ ,$$
 one can extract from the Laplace equation $\square \phi = 0$ the
radial equation
  \eqn{ExactNE}{
   \left[ (h r^3 \partial_r)^2 + r^6 F \omega^2 - 
    r^4 h \ell (\ell + 2) \right] R = 0 \ .
  }
 Because of the left-right symmetry of the effective string
description for five-dimen\-sional black holes, the absorption
cross-section for the near-extremal case provides essentially no more
information about the effective string than the extremal case does.
In the interest of a simple presentation, we will therefore restrict
the calculations in both this section and the next to the extremal
case.  The near-extremal generalizations of the results are summarized
at the end of each section.

In the near horizon region (denoted ${\bf I}$ for consistency with the
literature \cite{unruh,gkTwo,cgkt}), \ExactNE\ for an extremal
black hole can be approximated by a Coulomb equation \cite{hmf} in
the variable
  $y = (r_1 r_5 r_K \omega) / (2 r^2)$,
 while in the far horizon region ${\bf III}$ it can be approximated as
a Bessel equation:
  \eqn{EtaSolsTwo}{\vcenter{\openup1\jot
   \halign{\strut\span\TL & \span\TL & \span\TR & \span\TT & 
                 \span\TL & \span\TR\cr
   {\bf I.}\ \ &\left[ \partial_y^2 + 1 - 
      \df{2\eta}{y} - \df{\ell (\ell + 2)/4}{y^2} \right] R_{\bf I} &= 0 &
    \qquad\ & R_{\bf I} &= G_{\ell/2}(y) + i F_{\ell/2}(y) \cr
   {\bf III.} \ \ &\left[ (r^3 \partial_r)^2 + r^6 \omega^2 - 
     r^4 \ell (\ell + 2) \right] R_{\bf III} &= 0 &
    \qquad\ &
     R_{\bf III} &= \alpha \df{J_{\ell+1}(\omega r)}{\omega r} + 
      \beta \df{N_{\ell+1}(\omega r)}{\omega r} \ , \cr
  }}}
 where $\alpha$ and $\beta$ are constants to be determined in the
matching and 
  \eqn{EtaDefTwo}{
   \eta = -\tf{1}{4} \sum_{i=1,5,K} {r_1 r_5 r_K \omega \over r_i^2}
        \equiv -{\omega \over 4 \pi T_L}
  }
 is the charge parameter of the Coulomb functions.  The infalling
solution $R_{\bf I}$ can be matched directly onto $R_{\bf III}$
without the aid of an intermediate region ${\bf II}$, with the result
  \eqn{EtaMatchTwo}{
   \alpha = {\ell! \over C_{\ell/2}(\eta)} 
    {2^{2\ell+1} \pi^\ell \over (A_{\rm h} \omega^3)^{\ell/2}} \ , 
   \quad \beta = 0 \ .
  }
 The quantity 
  \eqn{CDef}{
   C_{\ell/2}(\eta) = {2^{\ell/2} e^{-\pi \eta / 2} 
    \Big| \Gamma\left( {\ell \over 2} + 1 + i \eta \right) \Big| \over
     \Gamma(\ell + 2)}
  }
 enters into the series expansion of Coulomb functions.  

A more accurate matching can be obtained with $\beta \neq 0$, but the
level of accuracy embodied in \EtaMatchTwo\ is sufficient for the flux
ratio method \cite{mast,gkTwo,cgkt}.  In this method, the
absorption probability is computed as the ratio of the infalling flux
at the horizon to the flux in the incoming wave at infinity.  The
result is
  \eqn{ProbAbs}{
   1-|S_\ell|^2 = {1 \over \pi} {A_{\rm h} \omega^3 \over |\alpha|^2} 
    = 4 \pi \left( A_{\rm h} \omega^3 \over 
        16 \pi^2 \right)^{\ell+1} 
       {C_{\ell/2}^2(\eta) \over \ell!^2} \ .
  }
 Now the formula \SigmaAbsFF\ comes into play to give the final
result:
  \eqn{SigmaGRPart}{\eqalign{
   \sigma_{\rm abs}^\ell &= A_{\rm h} (\ell + 1)^2 
    \left( A_{\rm h} \omega^3 \over 16 \pi^2 \right)^\ell
    {C_{\ell/2}^2(\eta) \over \ell!^2}  \cr
   &= {A_{\rm h} \over \ell!^4}
    \left( {A_{\rm h} \omega^3 \over 8 \pi^2} \right)^\ell
    e^{\omega \over 4 T_L} 
    {\textstyle \left| \Gamma\left( {\ell \over 2} + 1 - 
     i {\omega \over 4 \pi T_L} \right) \right|^2}
  }}
 The right hand side of \SigmaGRPart\ depends on $r_1$, $r_5$, and $r_K$
only through $A_{\rm h}$ and $T_L$.  Both quantities are symmetric in
the three radii, and in fact admit U-duality invariant generalizations
\cite{finn}.  The near-extremal generalization of \SigmaGRPart,
  \eqn{SigmaNonEx}{
   \sigma_{\rm abs}^\ell = A_{\rm h} 
    {(\omega r_0 / 2 )^{2\ell} \over \ell!^4}
    \left| {\Gamma\left( 1 + {\ell \over 2} - 
                {i \omega \over 4 \pi T_L} \right)
            \Gamma\left( 1 + {\ell \over 2} - 
                {i \omega \over 4 \pi T_R} \right)
     \over \Gamma\left( 1 - i {\omega \over 2 \pi T_H} \right)}
    \right|^2 \ ,
  }
 also treats the three charges on an equal footing.  The temperatures
$T_L$ and $T_R$ are given by \cite{finn}
  \eqn{TLR}{
   \beta_{L,R} = 2 \pi r_0 \left( \prod_i \cosh \sigma_i \mp
    \prod_i \sinh \sigma_i \right) 
  }
 in the general non-extremal case.

\section{The effective string analysis}
\label{Dbrane}

Despite recent progress in formulating superspace actions for branes
\cite{cederOne,cederTwo,aps} and in generalizing the DBI action to
nonabelian gauge theory (see \cite{atWhich} and references therein), a
first-principles derivation of a complete action for the effective
string, including all couplings to fields in the bulk of spacetime,
has yet to be achieved.  The goal of this section is to write down a
reasonable form for the part of the action responsible for coupling
the effective string to higher partial waves of an ordinary scalar and
see how the cross-sections it predicts compare with the semi-classical
result \SigmaGRPart.

Consider the off-diagonal graviton $h_{ij}$ with $i$ and $j$ parallel
to the D5-brane but perpendicular to the D1-brane.  The lowest-order
interaction of this field with excitations on the effective string can
be read off from the DBI action \cite{dmOne}: in static gauge where $t =
\tau$ and $x^5 = \sigma$,
  \eqn{VintSure}{
   V_{\rm int} = -t_{\rm eff} \int_0^{L_{\rm eff}} d\sigma \,
    2 h_{ij}(\tau,\sigma,\vec{x}\!=\!0) 
     \partial_+ X^i \partial_- X^j \ .
  }
 The convention in \VintSure\ and elsewhere is to sum over all $i \neq
j$.  The fields $h_{ii}$ couple somewhat differently; an exploration
of those couplings and their physical consequences was initiated in
\cite{cgkt}.
 
In \cite{juan}, an analysis of the entropy and temperature of
near-extremal 5-branes led to an effective string with $c_{\rm eff} = 6$
and $T_{\rm eff} = 1/(2 \pi r_5^2)$.  An extension of the methods used in
\cite{juan} to the case $r_1 \sim r_5$ leads to \cite{gforth} 
  \eqn{CapTeff}{
   T_{\rm eff} = {1 \over 2 \pi (r_1^2 + r_5^2)} \ .
  }
 The natural assumption is that $t_{\rm eff}$, by definition the tension
that appears in front of the DBI action, is precisely $T_{\rm eff}$.
Strangely enough, all previous scattering calculations except the
fixed scalar computation of \cite{kkTwo} (whose implications regarding the
effective string tension are unclear as yet) either do not depend on
$t_{\rm eff}$ or require $r_1 = r_5$.  Thus, purely from the point of view
of scattering computations, $t_{\rm eff}$ seems ambiguous by a factor of
the form $f(r_1/r_5)$ where $f(1) = 1$.  One of the motivations for
studying higher partial waves is to resolve this ambiguity.  The
result we will obtain is 
  \eqn{tandT}{
   t_{\rm eff} = {1 \over 2 \pi r_1 r_5} \ .
  }

Because of the evaluation of $h_{ij}$ at $\vec{x} = 0$, \VintSure\ is
a coupling to the $s$-wave of $h_{ij}$ only.  How might it be
generalized to include the dominant couplings to arbitrary partial
waves?  To begin with, a coupling to the $\ell^{\rm th}$ partial wave
should include $\ell$ derivatives of $h_{ij}$ since the wave-function
vanishes like $|\vec{x}|^\ell$.  It is the fermions on the effective
string which carry the angular momentum \cite{brekEx,brekNonEx}: the
left-moving and right-moving fermions transform in a fundamental of
$SU(2)_L$ and $SU(2)_R$ respectively, where the $SO(4)$ of rotations
in the four non-compact spatial dimensions is written as $SO(4) =
SU(2)_L \times SU(2)_R$.  Purely on group theory grounds, one thus
expects the $\ell^{\rm th}$ partial wave to couple to $\ell$
left-moving and $\ell$ right-moving fermions.  The order of the
absorption process in the string coupling can be read off from
\SigmaGRPart\ as $g^{\ell + 1}$ where $g$ is the closed string coupling.
Exactly two more open string vertex operators should be included in
the interaction to make the disk diagram come out with this power of
$g$.  The natural candidate for the interaction is
  \eqn{VintGuess}{\eqalign{
   V_{\rm int} &= -t_{\rm eff} \int_0^{L_{\rm eff}} d\sigma \,
    2 h_{ij}(\tau,\sigma,x^m\!=\!\bar\Psi \gamma^m \Psi) 
     \partial_+ X^i \partial_- X^j  \cr
    &= -t_{\rm eff} \int_0^{L_{\rm eff}} d\sigma \,
    2 \sum_{\ell=0}^\infty {1 \over \ell!} 
     \left( \prod_{k=1}^\ell \bar\Psi \gamma^{m_k} \Psi \right)
     \partial_{m_1} \cdots \partial_{m_\ell} 
     h_{ij}(\tau,\sigma,x^m\!=\!0) \,
     \partial_+ X^i \partial_- X^j \ .
  }}
 Extra derivatives on the fermion fields are possible {\it a priori},
but power counting in $\omega$ for $\omega/T_L \ll 1$ shows that they
must be absent if \SigmaGRPart\ is to be reproduced.  The same general
form of the coupling was deduced independently in \cite{ja} through a
greybody factor analysis.

The outstanding fallacy of \VintGuess\ is that the sum terminates at
$\ell = 4$ because there are only four types of left-moving fermions
and the same number of right-moving fermions.  The situation is even
worse when one factors in the restrictions from $SO(4)$ group theory.
As we shall see after \eno{VintGuessTwo}, only the $\ell = 0$ and $\ell =
1$ partial waves can be absorbed.  

In \VintGuess, the $\gamma^m$ are gamma matrices of $SO(4,1)$:
  \eqn{GammaMs}{
   \gamma^0 = \pmatrix{ -i & 0  \cr 
                         0 & i } \qquad
   \gamma^m = \pmatrix{ 0 & \tau^m_{\alpha\dot\beta}  \cr
                        \tau^{m\dot\alpha\beta} & 0 }
  }
 where 
  \eqn{TauMs}{
   \tau^m_{\alpha\dot\alpha} = 
    \left( {\bf 1},i \sigma_1,i \sigma_2,i \sigma_3 \right)  \qquad
   \tau^{m\dot\alpha\alpha} = \epsilon^{\dot\alpha\dot\beta}
    \epsilon^{\alpha\beta} \tau^m_{\beta\dot\beta} =
    \left( {\bf 1},-i \sigma_1,-i \sigma_2,-i \sigma_3 \right) \ ,
  }
 $\sigma_i$ being the usual Pauli matrices.  We follow northwest
contraction conventions for raising and lowering spinor indices, and we
set $\epsilon_{01} = \epsilon^{01} = \epsilon_{\dot0\dot1} =
\epsilon^{\dot0\dot1} = 1$.  The four-component spinor $\Psi$
decomposes into $SU(2)_L$ and $SU(2)_R$ fundamentals,
which are left-movers and right-movers on the effective string,
respectively.  These complex fermions decompose further into real
components of the 10-dimensional Majorana-Weyl spinor that one would
expect to emerge most simply from a full string theory analysis:
  \eqn{PsiDs}{\vcenter{\openup1\jot
   \halign{\strut\span\TC\cr
   \Psi = \pmatrix{ \Psi_{+\alpha} \cr \bar\Psi_-^{\dot\alpha} }  \cr
    \Psi_\pm^1 = {\psi_\pm^1 + i \psi_\pm^2 \over \sqrt{2}} \qquad
    \Psi_\pm^2 = {\psi_\pm^3 + i \psi_\pm^4 \over \sqrt{2}} \ . \cr
  }}}
 Note that complex conjugation raises or lowers a spinor index, rather
than dotting or undotting it as in the case of $SO(3,1)$.

For $\ell \ge 2$ the $\ell^{\rm th}$ term in the interaction
\VintGuess\ makes subleading contributions to the absorption of lower
partial waves because the expression $\partial_{m_1} \cdots
\partial_{m_\ell} h_{ij}$ does not pick out a pure $\ell^{\rm th}$
partial wave from a plane wave.  The cure for this is to symmetrize
$SU(2)_L$ and $SU(2)_R$ spinor indices:
  \eqn{VintGuessTwo}{\eqalign{
   V_{\rm int} &= -2 t_{\rm eff} \int_0^{L_{\rm eff}} d \sigma \,
    \sum_{\ell=0}^4 {i^\ell \over \ell!} 
    \prod_{k=0}^\ell \left( \Psi_+^{\alpha_k} \Psi_{-\dot\beta_k} + 
     \bar\Psi_+^{\alpha_k} \bar\Psi_{-\dot\beta_k} \right)
    \left( \tau^{m_1 (\dot\beta_1}_{(\alpha_1} \cdots
     \tau^{|m_\ell|\dot\beta_\ell)}_{\alpha_\ell)} \right)  \cr
   &\qquad\quad \cdot \partial_{m_1} \cdots \partial_{m_\ell} h_{ij}
    \partial_+ X^i \partial_- X^j + \ldots  
  }}
 Terms have been omitted in \VintGuessTwo\ which make subleading
contributions.  The product of fermion fields is antisymmetric in
$\alpha_1 \ldots \alpha_\ell$ and in $\dot\beta_1 \ldots
\dot\beta_\ell$.  Hence all terms in \VintGuessTwo\ vanish except
$\ell = 0$ and $\ell = 1$.  The conclusion is that only the first two
partial waves can be absorbed.  This limitation is not merely a
failing of the specific form \VintGuess; it is intrinsic to the
approach of coupling partial wave to a product of fermion fields
evaluated at a single point on the effective string without
derivatives.  To reiterate, the addition of derivatives introduces
extra powers of the energy in the final cross-section which would
cause disagreement with \SigmaGRPart.  Let us proceed with the analysis of
$\ell = 1$ and consider possible extensions to $\ell \ge 2$ later.

There are many steps involved in passing from the interaction
\VintGuessTwo\ to the cross-section for $\ell = 1$.  To avoid losing
factors it pays to be as explicit as possible.  Let us begin with mode
expansions of the fields appearing in \VintGuessTwo.  The forms of mode
expansions are dictated by the kinetic terms in the action.  In the
present case, the kinetic terms are
  \eqn{SKinetic}{\eqalign{
   S_{\rm bulk} &= {1 \over 2 \kappa_6^2} \int d^6 x \,
    \tf{1}{4} (\partial_\mu h_{ij}) (\partial^\mu h^{ij}) + \ldots \cr
   S_{\rm string} &= -2 t_{\rm eff} \int d^2 \sigma \, 
    \left[
     \partial_+ X_i \partial_- X^i + 
     \psi_+^\Delta i \partial_- \psi_+^\Delta + 
     \psi_-^{\dot\Delta} i \partial_+ \psi_-^{\dot\Delta} + 
    \ldots \right] \ ,
  }}
 resulting in the mode expansions
  \eqn{ModeExpansions}{\eqalign{
   \psi_+^\Delta(\tau + \sigma) &= 
    \sum_{k^5 \in {2 \pi \over L_{\rm eff}} ({\bf Z}^- - 1/2)}
     {1 \over \sqrt{2 L_{\rm eff} t_{\rm eff}}}
     \left( b_k^\Delta e^{i k \cdot \sigma} + {\rm h.c.} \right) \cr
   \psi_-^{\dot\Delta}(\tau - \sigma) &= 
    \sum_{k^5 \in {2 \pi \over L_{\rm eff}} ({\bf Z}^+ + 1/2)}
     {1 \over \sqrt{2 L_{\rm eff} t_{\rm eff}}}
     \left( b_k^{\dot\Delta} e^{i k \cdot \sigma} + {\rm h.c.} \right) \cr
   X^i(\sigma,\tau) &= 
    \sum_{k^5 \in {2 \pi \over L_{\rm eff}} {\bf Z}}
     {1 \over \sqrt{2 L_{\rm eff} t_{\rm eff} k^0}}
     \left( a_k^i e^{i k \cdot \sigma} + {\rm h.c.} \right) \cr
   h^{ij}(x^\mu) &= 
    \sum_{k^5,\vec{k}}
     \sqrt{2 \kappa_6^2 \over 2 V L_5 k^0}
     \left( g_k^{ij} e^{i k \cdot x} + {\rm h.c.} \right) \ .
  }}
 The sum over $k^5,\vec{k}$ has $k^5 \in {2 \pi \over L_5} {\bf Z}$
and $k^m \in {2 \pi \over \root{4}\of{V}} {\bf Z}$ for $m = 1,2,3,4$.
$V$ is the volume of a large box in which we imagine enclosing the
four uncompactified spatial dimensions.  The indices $\Delta$ and
$\dot\Delta$ run from $1$ to $4$, and since $\psi^\Delta_+$ and
$\psi^{\dot\Delta}_-$ are real, conjugation does not change the
position of the indices.  Typographical convenience will dictate the
position of $\Delta$ and $\dot\Delta$.  

It is important that $h^{ij}$ is moded differently in the $x^5$
direction from the effective string excitations: the minimal quantum
of Kaluza-Klein charge for an excitation on the effective string is
$1/(n_1 n_5)$ of the minimal quantum for a particle in the bulk
\cite{dmI,ms}.  In \ModeExpansions\ we have not been careful about
zero modes because it is the oscillator states which are important for
the absorption processes.  The factors in \ModeExpansions\ were chosen
to make the commutation relations simple:
  \eqn{ModeCommutators}{\eqalign{
   \{ b_k^\Delta,{b_q^\Gamma}\+ \} &= 
      \delta_{k^5-q^5} \delta^{\Delta\Gamma} \cr
   \{ b_k^{\dot\Delta},{b_q^{\dot\Gamma}}\+ \} &= 
      \delta_{k^5-q^5} \delta^{\dot\Delta\dot\Gamma} \cr
   [a_k^i,{a_q^j}\+] &= 
      \delta_{k^5-q^5} \delta^{ij} \cr
   [g_k^{ij},{g_q^{fh}}\+] &= 
      \delta_{\vec{k}-\vec{q}} \delta^{if} \delta^{jh} \quad
      \hbox{if $i<j$ and $f<h$} \ .
  }}

The goal now is to compute the amplitude 
  $\langle \tilde{f} | V_{\rm int} | \tilde{i} \rangle$
 for an absorption process where a scalar in the $\ell = 1$
partial wave turns into two bosons and two fermions on the
effective string.  The tildes on $| \tilde{i} \rangle$ and $|
\tilde{f} \rangle$ are meant to indicate that these state vectors are
not the real initial and final states: they include only the particles
that participate in the interaction and not the whole thermal sea of
left-movers that give the effective string its Kaluza-Klein charge.
Restoring the thermal sea is an easy exercise which will be postponed
until \eno{EnhancedGolden}.

Two other slight simplifications will be made to ease the notational
burden.  First, indices can be dropped on all the $X^i$ fields, but
then one must include an extra factor of $2$ in the rate, as shown in
\eno{ONZTEnhancedGolden}.  The $2$ accounts for the fact that $h_{ij}$ can
turn into a left-moving $X^i$ and a right-moving $X^j$ or a
left-moving $X^j$ and a right-moving $X^i$.  The second simplification
is to consider only
  \eqn{TildeIF}{\eqalign{
   |\tilde{i}\rangle &= g_k\+ |0\rangle \cr
   |\tilde{f}\rangle &= a_{p_b}\+ a_{q_b}\+ 
    {b_{p_f}^{\Delta}}\+ {b_{q_f}^{\dot\Delta}}\+ |0\rangle \ ,
  }}
 which is to say we put all the particles on the effective string into
the final state and none into the initial state.  A simple way to
account for all the crossed processes which also contribute to
absorption will be discussed after equation \eno{ILeft}.  In \TildeIF\
and below, $k$ refers to the momentum of the bulk scalar, $p$ refers
to the momentum of a left-mover on the effective string, and $q$
refers to the momentum of a right-mover.

The desired matrix element can now be read off from \VintGuessTwo\
and \ModeCommutators\ as 
  \eqn{OneNZTerm}{
   \langle\tilde{f}| V_{\rm int} |\tilde{i}\rangle =
    C_\Delta^{\dot\Delta}
     {k_1 \over 2 L_{\rm eff} t_{\rm eff}} \kappa_5
     \sqrt{p_b^0 q_b^0 \over V k^0}
     \delta_{k^5 - p_b^5 - p_f^5 - q_b^5 - q_f^5} \ ,
  }
 where $C_\Delta^{\dot\Delta}$ is the $4 \times 4$ matrix
$\diag\{1,-1,1,-1\}$.  To extract the rate is is necessary to use a
generalization of Fermi's Golden Rule that includes Bose enhancement
factors and Fermi suppression factors:
  \eqn{ONZTEnhancedGolden}{\eqalign{
   \Gamma &= 2 \sum_{|\tilde{f}\rangle} 
    (\rho_L^X(p_b^0) + 1) (\rho_R^X(q_b^0) + 1) 
    (1 - \rho_L^\psi(p_f^0)) (1 - \rho_L^\psi(q_f^0))  \cr
    &\qquad \cdot \left| \langle\tilde{f}| V_{\rm int} 
      |\tilde{i}\rangle \right|^2 
      2 \pi \delta\left( k^0 - p_b^0 - p_f^0 - 
         q_b^0 - q_f^0 \right)  \cr
    &= 2 \sum_{\Delta,\dot\Delta} 
      | C_\Delta^{\dot\Delta} |^2  
     \cdot \sum_{\rm modes}
       (\rho_L^X(p_b^0) + 1) (\rho_R^X(q_b^0) + 1) 
       (1 - \rho_L^\psi(p_f^0)) (1 - \rho_L^\psi(q_f^0))  \cr
    &\qquad \cdot 
       {k_1^2 \kappa_5^2 \over V k^0
        (2 L_{\rm eff} t_{\rm eff})^2} p_b^0 q_b^0 
        \delta_{k^5 - p_b^5 - p_f^5 - q_b^5 - q_f^5}
        2 \pi \delta\left( k^0 - p_b^0 - p_f^0 - 
         q_b^0 - q_f^0 \right)
  }}
 where
  \eqn{Rhos}{
   \rho_L^X(p_b^0) = {1 \over e^{p_b^0 / T_L} - 1} \ , \qquad
   \rho_L^\psi(p_i^0) = {1 \over e^{p_i^0 / T_L} + 1}
  }
 and similarly for the right-moving thermal occupation factors.
Again, the explicit factor of $2$ in the first line of
\ONZTEnhancedGolden\ is present to account for the two distinct 
choices, $i+$~$j-$ or $i-$~$j+$, for polarizing the bosonic fields.  

The vanishing of all cross-sections beyond $\ell = 1$ is an egregious
failing of the most naive effective string model.  The simplest fix
would be to allow an incoming scalar to couple to a product of fermion
fields evaluated at a single point on the spatial $S^1$ which the
effective string wraps, but not necessarily at a single point in the
effective string coordinates $\sigma$.  In terms of the (4,4) SCFT
from which the effective string emerges as a particular twisted
sector, this more general coupling seems very natural because it still
involves only a local operator constructed from a product of the $4
n_1 n_5$ species of fermions in the SCFT.

The present treatment extends easily to cover this more general
interaction.  Let $D_k$ and $\dot{D}_k$ be $4 n_1 n_5$-valued indices
for the left- and right-moving fermion fields, respectively.  Consider
the final state
  \eqn{MoreTildeF}{
   |\tilde{f}\rangle = a_{p_b}\+ a_{q_b}\+ 
    \left( {b_{p_1}^{D_1}}\+ {b_{q_1}^{\dot{D}_1}}\+ \cdots
           {b_{p_\ell}^{D_\ell}}\+ {b_{q_\ell}^{\dot{D}_\ell}}\+ 
    \right) |0\rangle \ .
  }
 The matrix element is now of the form 
  \eqn{NonZeroTerms}{
   \langle\tilde{f}| V_{\rm int} |\tilde{i}\rangle =
    C_{D_1 \ldots D_\ell}^{\dot{D}_1 \ldots \dot{D}_\ell} 
     {k_1^\ell \over (2 L_{\rm eff} t_{\rm eff})^\ell} \kappa_5
     \sqrt{p_b^0 q_b^0 \over V k^0}
     \delta_{k^5 - p_b^5 - \Sigma p_i^5 - q_b^5 - \Sigma q_i^5} \ .
  }
 The coefficient tensor $C_{D_1 \ldots D_\ell}^{\dot{D}_1 \ldots
\dot{D}_\ell}$ is antisymmetric in $D_1 \ldots D_\ell$ and in
$\dot{D}_1 \ldots \dot{D}_\ell$.  It encodes the $SO(4)$ group theory
factors isolating the $\ell^{\rm th}$ partial wave as well as
restrictions on the possible final states arising from D1-brane and
D5-brane Chan-Paton factors.  The rate is
  \eqn{EnhancedGolden}{\eqalign{
   \Gamma &= 2 \sum_{|\tilde{f}\rangle} 
    (\rho_L^X(p_b^0) + 1) (\rho_R^X(q_b^0) + 1) 
    \prod_{i=1}^\ell \left[ (1 - \rho_L^\psi(p_i^0)) 
           (1 - \rho_L^\psi(q_i^0)) \right]  \cr
    &\qquad \cdot \left| \langle\tilde{f}| V_{\rm int} 
      |\tilde{i}\rangle \right|^2 
      2 \pi \delta\left( k^0 - p_b^0 - {\textstyle\sum} p_i^0 - 
         q_b^0 - {\textstyle\sum} q_i^0 \right)  \cr
    &= {2 \over \ell!^2} \sum_{D_k,\dot{D}_k} 
      | C_{D_1 \ldots D_\ell}^{\dot{D}_1 \ldots 
         \dot{D}_\ell} |^2  
     \cdot \sum_{\rm modes}
       (\rho_L^X(p_b^0) + 1) (\rho_R^X(q_b^0) + 1) 
       \prod_{i=1}^\ell \left[ (1 - \rho_L^\psi(p_i^0)) 
              (1 - \rho_L^\psi(q_i^0)) \right]  \cr
    &\qquad \cdot 
       {k_1^{2 \ell} \kappa_5^2 \over V k^0
        (2 L_{\rm eff} t_{\rm eff})^{2\ell}} p_b^0 q_b^0 
        \delta_{k^5 - p_b^5 - \Sigma p_i^5 - q_b^5 - \Sigma q_i^5}
        2 \pi \delta\left( k^0 - p_b^0 - {\textstyle\sum} p_i^0 - 
         q_b^0 - {\textstyle\sum} q_i^0 \right) \ .
  }} 
 The $1/\ell!^2$ in the last expression arises because the sums over
$D_k$, $\dot{D}_k$, $p_k$, and $q_k$ are unrestricted, and there are
$\ell!^2$ different permutations of a given set of values for these
quantities which yield the same final state $|\tilde{f}\rangle$.

When the energy of the incoming scalar is much greater than the gap, a
continuum approximation can be made in \EnhancedGolden:
  \eqn{ContApprox}{
   \sum_p \to \int dp \, {L_{\rm eff} \over 2 \pi} \ , \qquad
   \delta_p \to {2 \pi \over L_{\rm eff}} \delta(p) \ .
  }
 For the sake of simplicity, only massless particle absorption will be
considered.  In that case the flux associated with the state
$g_k\+|0\rangle$ is ${\cal F} = 1/V$.  The absorption cross-section is
  \eqn{SigmaPLR}{\eqalign{
   \sigma_{\rm abs} &= 
     V \Gamma(scalar \to b_L + b_R + \ell f_L + \ell f_R) + 
     \hbox{crossed processes} \cr
    &= {\sum_{D_k,\dot{D}_k} 
      | C_{D_1 \ldots D_\ell}^{\dot{D}_1 \ldots 
         \dot{D}_\ell} |^2 \over \ell!^2 4^\ell} 
      {\kappa_5^2 L_{\rm eff} \over (2 \pi t_{\rm eff})^{2 \ell}}
       \omega^{2 \ell - 1} I_L I_R
  }}
 where
  \eqn{ILeft}{
   I_L = \int_{-\infty}^\infty d p_b^0 \prod_{i=1}^\ell d p_i^0 \,
    \delta\left( {\omega \over 2} - p_b^0 - 
     \sum_{i=1}^\ell p_i^0 \right) 
    {p_b^0 \over 1 - e^{-p_b^0 / T_L}}
    \prod_{i=1}^\ell {1 \over 1 + e^{-p_i^0 / T_L}}
  }
 and similarly for $I_R$.  
  $\Gamma(scalar \to b_L + b_R + \ell f_L + \ell f_R)$ 
 is what was computed in \EnhancedGolden.  Arbitrary crossings of the
basic process
  $scalar \to b_L + b_R + \ell f_L + \ell f_R$ 
 and their time-reversals also contribute to the net absorption rate
from which $\sigma_{\rm abs}$ is computed.  However, the simple trick
of extending the integrals in \ILeft\ over the entire real line can be
used to keep track of all of them.  A demonstration of this with
careful attention paid to symmetry factors can be found in
chapter~\ref{FixedScalars} for the special case where only two
left-moving bosons and two right-moving bosons are involved.  For
$\ell = 0$ the dependence on $t_{\rm eff}$ disappears in \SigmaPLR, as
was noted in chapter~\ref{FixedScalars}.

The integral \ILeft\ is a convolution of $\ell + 1$ simple functions
and so can be done most directly by transforming to Fourier space,
where convolutions become products.  Three integrals which are useful
for doing the Fourier transforms are
  \eqn{SechIntegrals}{\eqalign{
   \int_{-\infty}^\infty dp \, e^{i x p} {p \over 2 \sinh {p \over 2 T_L}}
    &= (\pi T_L)^2 {\rm sech}^2 \, (\pi T_L x) \cr
   \int_{-\infty}^\infty dp \, e^{i x p} {1 \over 2 \cosh {p \over 2 T_L}}
    &= (\pi T_L) {\rm sech} \, (\pi T_L x) \cr
   \int_{-\infty}^\infty dx \, e^{-i x p} (\pi T_L)^{\ell+2} 
    \sech^{\ell+2} (\pi T_L x) &=
    (2 \pi T_L)^{\ell+1} 
     {\Big| \Gamma\left( {\ell \over 2} + 1 - i {p \over 2 \pi T_L}
      \right) \Big|^2 \over (\ell+1)!} \ .
  }}
 The first two integrals are Fourier inversions of the third in the
special cases $\ell = 0$ and $-1$.  Now the computation is
straightforward: 
  \eqn{ILeftComp}{\eqalign{
   I_L &= e^{\omega \over 4 T_L} 
     \int_{-\infty}^\infty d p_b^0 \prod_{i=1}^\ell d p_i^0 \,
     \delta\left( {\omega \over 2} - p_b^0 - 
      \sum_{i=1}^\ell p_i^0 \right) 
     {p_b^0 \over 2 \sinh {p_b^0 \over 2 T_L}}
     \prod_{i=1}^\ell {1 \over 2 \cosh {p_i^0 \over 2 T_L}} \cr
    &= {e^{\omega \over 4 T_L} \over 2 \pi}
     \int_{-\infty}^\infty dx e^{-i x \omega / 2} 
      (\pi T_L)^{\ell+2} \sech^{\ell+2}(\pi T_L x) \cr
    &= {(\ell+1)! \over \pi} (\pi T_L)^{\ell+1} C_{\ell/2}^2(\eta) \ .
  }}
 The last step uses \EtaDefTwo\ and \CDef.  $I_R$ can be computed
similarly, but since $T_R = 0$ by assumption, the result is much
simpler:
  \eqn{IRight}{
   I_R = {(\omega / 2)^{\ell+1} \over (\ell+1)!} \ .
  }

Now the absorption cross-section can be given in closed form:
  \eqn{SigmaDTwo}{\eqalign{
    \sigma_{\rm abs}^\ell &= {\sum_{D_k,\dot{D}_k} 
      | C_{D_1 \ldots D_\ell}^{\dot{D}_1 \ldots 
         \dot{D}_\ell} |^2 \over \ell!^2 4^\ell} 
      {\kappa_5^2 L_{\rm eff} \over (2 \pi t_{\rm eff})^{2 \ell}}
      {\omega^{3 \ell} \over \pi}
      \left( {\pi T_L \over 2} \right)^{\ell+1} C_{\ell/2}^2(\eta) \cr
     &= A_{\rm h} \sum_{D_k,\dot{D}_k} 
      | C_{D_1 \ldots D_\ell}^{\dot{D}_1 \ldots 
         \dot{D}_\ell} |^2 
      \left( {A_{\rm h} \omega^3 \over 16 \pi^2} \right)^\ell
       {C_{\ell/2}^2(\eta) \over \ell!^2} \ .
  }}
 In the second equality two key relations have been used: 
  \eqn{TLt}{
   T_L = {r_K \over \pi r_1 r_5}  \qquad
   {\kappa_5^2 L_{\rm eff} T_L \over 2} = A_{\rm h} \ .
  }
 Both are valid when $r_K \ll r_1,r_5$.  The first can be derived by
setting the effective string entropy equal to the Bekenstein-Hawking
entropy.  The second is a limiting case of \TLR.  In addition, the
tension has at last been fixed:
  \eqn{FixedTension}{
   t_{\rm eff} = {1 \over 2 \pi r_1 r_5} \ .
  }
 Because the cross-section \SigmaGRPart\ depends on $n_1$ and $n_5$ only
through the product $n_1 n_5$ in the dilute gas regime, and because
the same is true of the quantities $L_{\rm eff}$, $T_L$, and $A_{\rm h}$
when $r_0 = 0$ and $r_K \ll r_1,r_5$, the choice $t_{\rm eff} \sim
1/\sqrt{n_1 n_5}$ seems inevitable.

Precise agreement between General Relativity and the effective string
now depends only on the relation
  \eqn{CNorm}{
   \sum_{D_k,\dot{D}_k} 
      | C_{D_1 \ldots D_\ell}^{\dot{D}_1 \ldots 
         \dot{D}_\ell} |^2 = (\ell + 1)^2 \ .
  }
 Agreement in the case $\ell = 0$ is trivial.  For $\ell = 1$, the
original treatment in terms of free fermions on the effective string
is adequate: one can easily trace through the computations and
verify that $D_1$, $\dot{D}_1$, and $C_{D_1}^{\dot{D}_1}$ can be
replaced in every equation by $\Delta$, $\dot\Delta$, and
$C_\Delta^{\dot\Delta}$.  Any numerical discrepancy could have been
fixed by introducing a multiplicative constant in the relation $x^m =
\bar\Psi \gamma^m \Psi$; however to see perfect agreement without such
artifice is pleasing and also rather suggestive of the form one
expects for a gauge-fixed kappa symmetric action.

The real test is $\ell \ge 2$.  Here it seems essential to depart from
the simplistic effective string picture and return to a more
fundamental description of the D1-D5 bound state in order to compute
the coefficients $C_{D_1 \ldots D_\ell}^{\dot{D}_1 \ldots
\dot{D}_\ell}$.

Finally, it is worth noting that if agreement can be established for
extremal absorption, agreement for the near-extremal case follows
automatically.  In the effective string computation for near-extremal
absorption one must subtract off the stimulated emission contribution
as described in chapter~\ref{FixedScalars} in order to respect
detailed balance and time reversal invariance.  Modulo this
subtraction, the result can be read off from \SigmaPLR\ and
\ILeftComp:
  \eqn{SigmaDNonEx}{\eqalign{
   \sigma_{\rm abs}^\ell &= {\sum_{D_k,\dot{D}_k} 
      | C_{D_1 \ldots D_\ell}^{\dot{D}_1 \ldots 
         \dot{D}_\ell} |^2 \over \ell!^2 (\ell+1)!^2}
    {\kappa_5^2 L_{\rm eff} T_L T_R \over T_H} 
    \left( {\omega \sqrt{T_L T_R} \over 2 t_{\rm eff}} \right)^{2 \ell}  \cr
   &\qquad \cdot \left| {\Gamma\left( 1 + {\ell \over 2} - 
                {i \omega \over 4 \pi T_L} \right)
            \Gamma\left( 1 + {\ell \over 2} - 
                {i \omega \over 4 \pi T_R} \right)
     \over \Gamma\left( 1 - i {\omega \over 2 \pi T_H} \right)}
    \right|^2  \cr
   &= {\sum_{D_k,\dot{D}_k} 
      | C_{D_1 \ldots D_\ell}^{\dot{D}_1 \ldots 
         \dot{D}_\ell} |^2 \over \ell!^2 (\ell+1)!^2}
      A_{\rm h} \left( {\omega r_0 \over 2} \right)^{2 \ell}
    \left| {\Gamma\left( 1 + {\ell \over 2} - 
                {i \omega \over 4 \pi T_L} \right)
            \Gamma\left( 1 + {\ell \over 2} - 
                {i \omega \over 4 \pi T_R} \right)
     \over \Gamma\left( 1 - i {\omega \over 2 \pi T_H} \right)}
    \right|^2 \ .
  }}
 The second line relies on a modified version of \TLt\ applicable to
the near-extremal case with $r_0,r_K \ll r_1,r_5$:
  \eqn{TLtNonEx}{
   \sqrt{T_L T_R} = {r_0 \over 2 \pi r_1 r_5} \qquad
   {\kappa_5^2 L_{\rm eff} T_L T_R \over T_H} = A_{\rm h} \ .
  }
 The same tension \FixedTension\ and the same relation \CNorm\
establish agreement between \SigmaDNonEx\ and \SigmaNonEx.

It has been suggested \cite{finn} that the effective string picture
can be used to describe in a U-duality invariant fashion black holes
with arbitrary charges, possibly even far from extremality.  The
expression on the first line of \SigmaDNonEx\ depends only only $T_L$,
$T_R$, $T_H$, and $L_{\rm eff}$---all quantities that have meaning to the
effective string considered in the abstract, independent of the
microscopic D1-D5-brane model.  It cries out to be reconciled
with \SigmaNonEx\ for arbitrary values of $r_0$, $Q_1$, $Q_5$, and
$Q_K$.  But the treatment of absorption given in this section relies
on the dilute gas approximation and thus is not general enough to be
matched in any meaningful way to General Relativity when $Q_K
\ll\!\!\!\!\!/\,\,\, Q_1,Q_5$.

\section{Limitations on partial wave absorption}
\label{llimits}

Consider the more general couplings described in the paragraph
preceding \MoreTildeF: local on spatial $S^1$ wrapped by the effective
string but not on the effective string itself.\footnote{The ideas in
this section originate largely in discussions with C.~Callan,
L.~Thorlacius, and J.~Maldacena.}  Although the details of the $SO(4)$
group theory and Chan-Paton factors have yet to be worked out fully in
the context of the (4,4) SCFT description of the D1-D5-brane bound
state, it seems clear that a coupling of the $\ell^{\rm th}$ partial
wave to an operator built out of any combination of the $4 n_1 n_5$
fermionic fields raises the maximum value of $\ell$ which the
effective string can absorb from $1$ to some number on the order $n_1
n_5$.  One might suppose that by putting derivatives on some of the
fermion fields, the problem can be avoided altogether.  But such
derivatives raise the dimension of the operator and hence suppress the
cross-section by more powers of $\omega$ than are present in the
semi-classical result.  To sum up, the assumption of locality prevents
the effective string from coupling to partial waves above a certain
maximum $\ell$ with the strength required to match General Relativity.
The situation does not seem as satisfactory as for the D3-brane, where
couplings to all partial waves exist with appropriate dimensions to
reproduce semi-classical cross-sections \cite{kleb} (normalizations
however are problematic, as we shall see in
chapter~\ref{ThreeBraneAbs}).

There is a reason, however, why one might expect not to observe
agreement between the effective string and General Relativity at high
values of $\ell$.  If the black hole absorbs some very high partial
wave, it winds up with a large angular momentum, so the geometry
before and after is appreciably different.  Back reaction is not
included in the General Relativity calculations of
section~\ref{semicl}.  In fact the only back reaction calculation to
date for the black hole under consideration \cite{kvk} is restricted
to $\ell = 0$.  But on the grounds of cosmic censorship one would
expect that absorption processes which drive the black hole past
extremality are forbidden even semi-classically.  From an adaptation
of the work of \cite{brekEx,brekNonEx} one can read off the
corresponding bound on $\ell$ as $\ell \lsim \sqrt{n_K n_1 n_5}$.

We have two different bounds on $\ell$ indicating the maximum partial
wave that the effective string should be capable of absorbing:
  \eqn{ThreeBounds}{\vcenter{\openup1\jot
    \halign{\strut\span\TL & \span\TR & \qquad\span\TT\cr
    \ell &\lsim \ell_{\rm max}^L \equiv n_1 n_5 & 
      from locality and statistics  \cr
    \ell &\lsim \ell_{\rm max}^C \equiv \sqrt{n_K n_1 n_5} & 
      from cosmic censorship.  \cr
  }}}
 Now we would like to inquire which is the more restrictive.  Using the
standard relations (see for example \cite{gkOne})
  \eqn{NRels}{
   n_1 n_5 = {4 \pi^3 r_1^2 r_5^2 \over \kappa_5^2 L_5} \qquad\quad
   n_K = {\pi L_5 r_K^2 \over \kappa_5^2}
  }
 and the formula \TLt\ for $T_L$, one can show that $\ell_{\rm
max}^C/\ell_{\rm max}^L = L_5 T_L / 2$.

The validity of any comparison between General Relativity and the
effective string model as treated in section~\ref{Dbrane} relies on being
in the dilute gas regime \cite{cm} and at low energies \cite{juanLow}:
  \eqn{StandardLims}{
   r_K \ll r_1,r_5 \ll 1/\omega \ .
  }
 These inequalities still do not determine whether $\ell_{\rm max}^L$
is larger or smaller than $\ell_{\rm max}^C$.  But if it is agreed to
examine only fat black holes \cite{ms}, which is to say if one assumes
  \eqn{FatLim}{
   L_5 \ll \sqrt{r_1 r_5} \ ,
  }
 then it is easy to obtain the inequality
  \eqn{DesiredLims}{
   \ell_{\rm max}^L \gg \ell_{\rm max}^C
  }
 by combining the dilute gas inequality in \StandardLims\ with
\FatLim.  Now, \DesiredLims\ is a hopeful state of affairs for the
effective string model, because it indicates that making couplings
local only on $S^1$ in principle enables the effective string to
absorb all the partial waves for which reasonable comparisons can be
made with General Relativity.  It is perhaps not the ideal state of
affairs: one might have hoped that the bounds $\ell_{\rm max}^L$ and
$\ell_{\rm max}^C$ would coincide, indicating that the effective
string knew about cosmic censorship.  The results of \cite{ja} suggest that
a more careful treatment of these issues using techniques of conformal
field theory would result in a translation of cosmic censorship into
unitarity of the effective string description.  Such a treatment would
need to address the problem that if one moves deep into the black
string region of parameter space by making $L_5$ large, one can obtain
$\ell_{\rm max}^L \ll \ell_{\rm max}^C$.  The perturbative D-brane
region is in fact closer to the black string region than the fat black
hole region, so it would be surprising to find such a disaster for
comparisons in the black string region when agreement seems possible
for fat black holes.

\section{Conclusion}
\label{ConclusionPart}

In this chapter we have shown that the leading order coupling of the
effective string to an ordinary scalar correctly predicts the
cross-sections for the $\ell=1$ partial wave.  Due to the Grassmannian
character of the fermionic fields which carry the angular momentum, it
is impossible for the simplest effective string model (a single long
string with $c_{\rm eff} = 6$) to couple properly to $\ell>1$ partial waves
through an operator local on the string.  Generalizing the model to
include what one might intuitively regard as multi-strand interactions
of the effective string postpones this difficulty to $\ell \gsim n_1
n_5$, a higher bound on $\ell$ for fat black holes than the one
arising from cosmic censorship.  An analysis of the unique form which
such interactions must have in order to make a leading order
contribution to the absorption of the $\ell^{\rm th}$ partial wave
demonstrates that the correct energy dependence arises from the finite
temperature kinematics.  This demonstration, together with the general
proof of the Optical Theorem for absorption of scalars given in
section~\ref{semicl}, can be viewed as a full investigation of the
kinematics involved in higher partial waves.  What is left is to
calculate the coefficients $C_{D_1 \ldots D_\ell}^{\dot{D}_1 \ldots
\dot{D}_\ell}$ and thereby verify or falsify \CNorm, on which
agreement between General Relativity and the effective string model
relies.

The means to achieve a clear description of the dynamics and hopefully
a derivation of the $C_{D_1 \ldots D_\ell}^{\dot{D}_1 \ldots
\dot{D}_\ell}$ is a more precise treatment of the low-energy SCFT
dictating the dynamics of the D1-D5-brane bound state.  The effective
string might continue to be a useful picture, perhaps supplemented by
rules governing how different strands of the effective string
interact.

In a way the finding that the effective string tension needs to scale
as $1/\sqrt{Q_1 Q_5}$ is a more serious difficulty than the vanishing
of $\ell > 1$ cross sections.  Indeed, the necessity of choosing this
peculiar value for the tension appears already at $\ell = 1$, where
the naive effective string model in other ways seems completely
adequate.  A study of T-duality in \cite{Mathur} led to the conclusion that
a tension scaling as $1/\sqrt{Q_1 Q_5}$ is more natural than $1/Q_1$
or $1/Q_5$, and it was further described how a simple modification in
the calculation of disk diagrams would lead to such a scaling.
However, in the absence of a first-principles derivation of the
tension, a $1/(Q_1+Q_5)$ scaling seems equally natural.  This scaling
is the one favored by entropy and temperature arguments.  Thus it
appears that a single energy scale does not fully characterize
effective strings in the way that $\alpha'$ does fundamental strings.

Although effective string models of black holes have recently enjoyed
a number of remarkable successes, a unifying picture has been slow in
emerging.  The General Relativity calculations for near-extremal black
holes, on which most of the evidence for effective strings is based,
are conceptually straightforward.  The difficulty of studying bound
states of solitons has caused the link between fundamental string
theory and effective strings to remain imprecise in certain
respects---most importantly in the interaction between the effective
string and fields in the bulk of spacetime.

The puzzles presented by higher partial waves may push effective
string theory in the directions it needs to go in order to become a
fully viable model of near-extremal black holes.  Discrepant results
for the tension may be a clue to nature of effective string's
interactions with bulk fields.  Rescuing the $\ell > 1$ cross-sections
must surely lead to a consideration of the multi-strand interactions
which are as yet virgin territory in the theory of effective strings.
Already, the absorption of higher partial waves into five-dimensional
black holes is the cleanest dynamical test of the role of fermions on
the effective string.  Achieving a full understanding of these
processes would constitute a major advance, not only in establishing
the viability of effective strings in black hole physics, but also in
comprehending the D1-D5-brane bound state.

\def\overleftrightarrow#1{\vbox{\ialign{##\crcr
     $\leftrightarrow$\crcr\noalign{\kern-0pt\nointerlineskip}
     $\hfil\displaystyle{#1}\hfil$\crcr}}}
\def\omicron{o}
\def\O{{\cal O}}

\chapter{Photons and fermions falling into four-dimensional black holes}
\label{PhotonAbsorption}

\section{Introduction}
\label{IntroDyad}

Microscopic models of near-extremal black holes in terms of effective
strings have recently been employed with great success to explain the
Bekenstein-Hawking entropy \cite{sv,cm,hs,hms,ms}.  These models have
also proven their value by correctly predicting certain Hawking
emission rates and absorption cross-sections at low energies
\cite{dmw,dmOne,dmTwo,gkOne,gkTwo,cgkt,dkt,htr,kvk,km,krt,mast,ja,clOne,clTwo}.

The many successful predictions have all been for scalar particles.
Most in fact have been for minimally coupled scalars.  Usually it is
said that minimally coupled scalars are those whose equation of motion
is $\square \phi = 0$.  This is a slight misnomer: a more precise way
to say it is that when the equations of motion are linearized in small
fluctuations around the black hole solution under consideration, one
of them is simply $\square \delta\phi = 0$.  Fortunately, it is
usually easy to see which scalars in the theory are minimally coupled:
for example, when one obtains a 5-dimensional black hole by toroidally
compactifying the D1-brane D5-brane bound state, the off-diagonal
gravitons with both indices lying within the D5-brane but
perpendicular to the D1-brane are clearly minimally coupled scalars in
the 5-dimensional theory.

This chapter, based on the paper \cite{gphoton}, presents a study of
particles with nonzero spin falling into effective string black holes.
As a starting point, it is easiest to consider minimally coupled
photons or fermions.  For photons, minimally coupled means that the
relevant linearized equation of motion is
  \eqn{MaxEq}{
   \nabla_\mu \delta F^{\mu\nu} = 0 \ ,
  }
 and for chiral fermions it means
  \eqn{WeylEq}{
   \sigma_\mu^{\alpha\dot\beta} \nabla^\mu \delta \psi_\alpha = 0 \ .
  }
 Minimally coupled fermions in arbitrary dimensions were studied in
\cite{dgm}.  The authors of \cite{dgm} correctly point out that the
fermions in supergravity theories do not in general obey minimally
coupled equations of motion in the presence of charged black holes.
Likewise, it is not usually the case that photons will be minimally
coupled in the presence of a supergravity black hole solution.
Indeed, it seems that the gauge fields which carry the charges of the
black hole are never minimally coupled: there is mixing between them
and the graviton.

However, for the case we shall study in this chapter, namely the equal
charge black hole \cite{klopp} of ${\cal N} = 4$ supergravity
\cite{adas,csfOne,cs,csfTwo}, two of the four Weyl fermions are in
fact minimally coupled, as are four of the six gauge fields.  The same
black hole provided the simplest framework in which to study fixed
scalars \cite{kr}.  Its metric is that of an extreme
Reissner-Nordstrom black hole.  In \cite{ja} it was shown that an
effective string model is capable of reproducing the minimally coupled
scalar cross-section of the Kerr-Newman metric, in the near-extremal
limit.  The effective string picture carries over naturally to the
equal charge extreme black hole in ${\cal N} = 4$ supergravity and its
near-extremal generalization.  The recent work \cite{clTwo} presents
evidence that the effective string can model a much broader class of
black holes which have arbitrary $U(1)$ charges and are far from
extremality.  The essential features of the effective string, however,
seem much the same in all its four-dimensional applications.  We will
show that minimally coupled fermions can be incorporated naturally
into the effective string picture through a coupling to the
supercurrent.  Minimally coupled photons fit in in a somewhat
unexpected way: the coupling of the gauge field to the string seems to
occur via the field strength rather than the gauge potential.

The organization of this chapter is as follows.  In
section~\ref{MinCoup}, the black hole solutions are exhibited and the
minimally coupled photons are identified.  In section~\ref{SepEq},
separable equations are derived for these photons.  In
section~\ref{Prob}, these equations are solved to yield absorption
cross-sections.  The parallel analysis of minimally coupled fermions
is postponed to section~\ref{OtherParticles}, in which also the axion
cross-section is computed.  The axion turns out to have the same
cross-section as the dilaton, not because it is a fixed scalar in the
usual sense of attractors \cite{fkOne,fkTwo}, but because of a
dynamical version of the Witten effect \cite{wittenDyon}.
Section~\ref{EffStr} discusses the effective string interpretation of
these cross-sections.  Although the overall normalizations of the
cross-sections are not computed, it is shown that the effective string
correctly reproduces the relative normalization of the dilaton, axion,
and minimal fermion cross-sections.  Some concluding remarks are made
in section~\ref{ConclusionDyad}.  The appendix presents some results
of the dyadic index formalism needed for the rest of the chapter.

\section{Minimally coupled photons in ${\cal N} = 4$ supergravity}
\label{MinCoup}

The fields of the $SU(4)$ version of ${\cal N} = 4$, $d=4$ supergravity
\cite{csfTwo} are the graviton $e_\mu^a$, four Majorana gravitinos
$\psi^i_\mu$, three vector fields $A^n_\mu$, three axial vectors
$B^n_\mu$, four Majorana fermions $\chi^i$, the dilaton $\phi$, and
the axion $B$.  The doubly extreme black hole of \cite{klopp} is
electrically charged under $A_3^\mu$, magnetically charged under
$B_3^\mu$, and neutral with respect to the other four gauge fields.
These four extra gauge fields are minimally coupled photons, as we
shall see shortly.

The full bosonic lagrangian of ${\cal N} = 4$ supergravity in the $SU(4)$
picture is
  \eqn{FullBL}{\eqalign{
    {\cal L} &= \sqrt{-g} \Big[ -\!R + 2 (\partial_\mu \phi)^2 + 
     2 e^{4 \phi} (\partial_\mu B)^2 - 
     e^{-2 \phi} {\textstyle \sum_n} (F_n^2 + G_n^2)  \cr
   &\qquad - \, 2 i B {\textstyle \sum_n} (F_n *F_n + G_n *G_n) \Big] 
  }}
 where $F_n = d A_n$ and $G_n = d B_n$.  The conventions used here are
those of \cite{klopp}.  In particular, Hodge duals are defined by
  \eqn{DualDef}{
   *F_{\mu\nu} = {\sqrt{-g} \over 2} \epsilon_{\mu\nu\rho\sigma}
    F^{\rho\sigma}
  }
 where $\epsilon_{0123} = \epsilon_{tr\theta\phi} = -i$.  The
equations of motion following from \FullBL\ are
  \eqn{SEOMs}{\eqalign{
   \nabla_\mu \big( e^{-2\phi} F_n^{\mu\nu} + 
     2iB *F_n^{\mu\nu} \big) &= 0  \cr 
   \nabla_\mu \big( e^{-2\phi} G_n^{\mu\nu} +
     2iB *G_n^{\mu\nu} \big) &= 0  \cr
   \square\phi - \tf{1}{2} e^{-2\phi} {\textstyle \sum_n} 
    (F_n^2 + G_n^2) - 2 e^{4\phi} (\partial_\mu B)^2 &= 0  \cr 
   \square B + 4 \partial^\mu \phi \partial_\mu B +
    \tf{i}{2} e^{-4\phi} {\textstyle \sum_n} (F_n *F_n + G_n *G_n) 
       &= 0  \cr
   R_{\mu\nu} + 2 \partial_\mu \phi \partial_\nu \phi +
    2 e^{4\phi} \partial_\mu B \partial_\nu B
    - e^{-2\phi} {\textstyle \sum_n} 
    (2 F_{n\mu\lambda} F_{n\nu}{}^\lambda - \tf{1}{2} g_{\mu\nu} F_n^2 
      \ \ \, &  \cr
     + \, 2 G_{n\mu\lambda} G_{n\nu}{}^\lambda - \tf{1}{2} g_{\mu\nu} G_n^2) 
    &= 0 \ .  
  }}
 Including the fermions introduces extra terms into these equations
involving fermion bilinears.  These terms affect neither the black
hole solution nor the linearized bosonic equations around that
solution since the fermions' background values are zero.  Duals of the
field strengths $F_n$ and $G_n$ are defined by
  \eqn{FSDuals}{
   \tilde{F} = i e^{-2 \phi} * F - 2 B F \ .
  }
 $\tilde{F}_n$ and $\tilde{G}_n$ are closed forms by the equations of
motion \SEOMs.  In the $SO(4)$ version of ${\cal N} = 4$ supergravity, one
writes $\tilde{G}_n = d \tilde{B}_n$.

The equal charge, axion free, extreme black hole solution is
  \eqn{kpSol}{\eqalign{
   ds^2 &= {1 \over (1+M/r)^2} dt^2 - 
    (1+M/r)^2 (dr^2 + r^2 d\Omega^2)  \cr
   \tilde{F}_3 &= Q \vol_{S^2} \qquad G_3 = P \vol_{S^2}  \cr
   e^{2 \phi} &= 1 \qquad B = 0 \ .
  }}
 The electric charge $Q$, the magnetic charge $P$, and the mass $M$
are related by $Q = P = M/\sqrt{2}$.  By definition, $\vol_{S^2} =
\sin\theta \, d\theta \wedge d\phi$.  Gauss' law for the electric and
magnetic charges reads
  \eqn{GaussLaw}{\vcenter{\openup2\jot
    \halign{\strut\span\TL & \span\TR \qquad & \span\TL & \span\TR\cr
   \int_{S^2} F_3 &= 0 & \int_{S^2} \tilde{F}_3 &= 4 \pi Q  \cr
   \int_{S^2} G_3 &= 4 \pi P & \int_{S^2} \tilde{G}_3 &= 0 \ .  \cr
  }}}
 Keeping the charges $Q$ and $P$ fixed but increasing the mass, one
obtains the non-extremal generalization of \kpSol:
  \eqn{NonExSol}{\eqalign{
   ds^2 &= {h \over f^2} dt^2 - 
    f^2 \left( h^{-1} dr^2 + r^2 d\Omega^2 \right)  \cr
   \tilde{F}_3 &= Q \vol_{S^2} \qquad G_3 = P \vol_{S^2}  \cr
   e^{2 \phi} &= 1 \qquad B = 0 
  }}
 where
  \eqn{hfDef}{
   h = 1 - {r_0 \over r} \qquad\qquad 
   f = 1 + {r_0 \sinh^2 \alpha \over r} \ .
  }
 The mass, charges, area, and temperature of this black hole are given
by
  \eqn{ThermQs}{\vcenter{\openup1\jot
    \halign{\strut\span\TL & \span\TR \qquad\qquad & \span\TL & \span\TR\cr
   M &= {r_0 \over 2} \cosh 2\alpha &
   Q &= P = {r_0 \over 2 \sqrt{2}} \sinh 2\alpha  \cr
   A &= 4\pi r_0^2 \cosh^4 \alpha &
   T &= {1 \over 4 \pi r_0 \cosh^4 \alpha} \ .  \cr
  }}}
 Taking $\alpha \to \infty$ with $Q$ and $P$ held fixed, one recovers
the extremal solution \kpSol.

A crucial property of \kpSol\ and \NonExSol, without which there will
be no minimally coupled photons, is the vanishing of the dilaton.
This happens only when $Q=P$.  The situation is similar to the case of
fixed scalars, where the $Q=P$ case \cite{kr} was much easier to deal
with than the $Q \ne P$ case \cite{kkTwo}.  

When the equations \SEOMs\ are linearized around the solution
\NonExSol, the variations $\delta F_3^{\mu\nu}$ and $\delta
G_3^{\mu\nu}$ appear in all the equations of motion because the
background values of $F_3$ and $G_3$ are nonzero.  
But because $F_1$, $F_2$, $G_1$, and $G_2$ do have vanishing
background values (as does the axion $B$) the variations of these
gauge fields appear in the linearized equations only as
  \eqn{OnlyAppear}{
   \nabla_\mu \big( e^{-2\phi} F^{\mu\nu} \big) = 0 \ ,
  }
 where $F = \delta F_1$, $\delta F_2$, $\delta G_1$, or $\delta G_2$.
Including $e^{-2\phi}$ in \OnlyAppear\ was unnecessary since it is
identically $1$ when the charges are equal; but \OnlyAppear\ is still
the right linearized equation of motion for these fields when the
charges are unequal (provided the background value of the axion
remains zero).  Surprisingly enough, a non-constant dilaton background
makes it much more difficult to decouple the equations for different
components of the gauge field.

\section{Separable equations for gauge fields}
\label{SepEq}

Having shown that minimally coupled gauge fields do indeed exist, let
us now show how to solve their equations of motion.  The dilaton
background will be kept arbitrary just long enough to observe why it
makes the job much harder.  The goal of this section is to convert
Maxwell's equations
  \eqn{MaxwellEqs}{
   d F = 0 \qquad d * e^{-2\phi} F = 0
  }
 into decoupled separable differential equations.  The equation of
motion for the vector potential $A$,
  \eqn{AEOM}{
   d * e^{-2\phi} d A = 0 \ ,
  }
 does not lend itself to this task.  It turns out to be easier to
dispense with $A$ altogether and analyze Maxwell's equations,
\MaxwellEqs, directly.  Even this is quite challenging if one sticks
to the traditional tools of tensor analysis.  Fortunately, several
authors \cite{teuk,teukI,Churil} in the 60's and 70's worked out an
elegant approach to this sort of problem using Penrose's dyadic index
formalism \cite{np}.  Appendix~A provides a summary of some of the
standard notation.

The field strength $F_{\mu\nu}$ (six real quantities) is replaced by a
symmetric matrix $\Phi_{\Delta\Gamma}$ (three complex quantities)
using the equation
  \eqn{PhiDef}{
   F_{\mu\nu} \sigma^\mu_{\Delta\dot\Delta} \sigma^\nu_{\Gamma\dot\Gamma} =
    \Phi_{\Delta\Gamma} \epsilon_{\dot\Delta\dot\Gamma} + 
    \bar\Phi_{\dot\Delta\dot\Gamma} \epsilon_{\Delta\Gamma} \ .
  }
 Now define $\phi_1$, $\phi_0$, and $\phi_{-1}$ as
follows:
  \eqn{MorePhiDefs}{\eqalign{
   \phi_1 &= \Phi_{00} = F_{\mu\nu} \ell^\mu m^\nu \cr
   \phi_0 &= \Phi_{10} = \Phi_{01} 
     = \tf{1}{2} F_{\mu\nu} (\ell^\mu n^\nu + \bar{m}^\mu m^\nu) \cr
   \phi_{-1} &= \Phi_{11} = F_{\mu\nu} \bar{m}^\mu n^\nu  
  }}
 where $\ell^\mu$, $n^\mu$, $m^\mu$, and $\bar{m}^\mu$ form the
complex null tetrad (see the appendix).  In the literature, it is more
common to write $\phi_0$, $\phi_1$, and $\phi_2$ instead of $\phi_1$,
$\phi_0$, and $\phi_{-1}$.  The present convention has the advantage
that the subscript is essentially the helicity.

The Bianchi identity $d F = 0$ can be rewritten as
  $D^{\Gamma\dot\Delta} \Phi^\Delta{}_\Gamma = 
   D^{\Delta\dot\Gamma} \bar\Phi^{\dot\Delta}{}_{\dot\Gamma}$.
 Using this identity one can rewrite the equation of motion 
$d * e^{-2\phi} F = 0$ as
  \eqn{EOMReWrite}{
   D^{\Gamma\dot\Delta} \Phi^\Delta{}_\Gamma = 
    \tf{1}{2} (\partial_{\Gamma\dot\Gamma} e^{-2\phi})
     (\Phi^{\Delta\Gamma} \epsilon^{\dot\Delta\dot\Gamma} + 
      \bar\Phi^{\dot\Delta\dot\Gamma} \epsilon^{\Delta\Gamma}) \ .
  }
 One immediately sees that the equations simplify greatly if the
coupling $e^{-2\phi} = 1$.  If this is not the case, then because the
right hand side involves $\bar\Phi_{\dot\Delta\dot\Gamma}$ as well as
$\Phi_{\Delta\Gamma}$, the advantage of compressing the real field
strength components into complex components of $\Phi_{\Delta\Gamma}$
is lost.  In this case, it seems difficult to decouple the equations.
It is striking that the condition for Maxwell's equations to be simple
is the same as the condition found in \cite{kr} and in
chapter~\ref{FixedScalars} for the fixed scalar equation to decouple
from Einstein's equations.
 
 Let us proceed with the case where $e^{-2\phi} = 1$, so
that Maxwell's equations can be succinctly written as
$D^{\Delta\dot\Delta} \Phi_{\Delta\Gamma} = 0$.  The spin coefficients
for the general spherically symmetric metric,
  \eqn{GenMetric}{
   ds^2 = e^{2A(r)} dt^2 - e^{2B(r)} dr^2 - 
    e^{2C(r)} \left( d\theta^2 + \sin^2 \theta d\phi^2 \right) 
  }
 are presented in \eno{DiagonalSC}.  Six of them vanish, and the remaining
six can be expressed in terms of $\gamma$, $\rho$, and $\alpha$, which
are real.  Maxwell's equations written out in components therefore
take on a particularly simple form:
  \eqn{Max}{\vcenter{\openup0\jot
    \halign{\strut\span\TL & \span\TR\cr
   (\Delta - 2 \gamma + \rho) \phi_1 &= \delta \phi_0 \cr
    (D - 2 \rho) \phi_0 &= (\bar\delta - 2 \alpha) \phi_1  
      \cr\noalign{\vskip1\jot}
   (\Delta + 2 \rho) \phi_0 &= (\delta - 2 \alpha) \phi_{-1} \cr
    (D + 2 \gamma - \rho) \phi_{-1} &= \bar\delta \phi_0  \ . \cr
  }}}
 The form of Maxwell's equations in a more general metric can be found
in \cite{np}.

A straightforward generalization of the preceding treatment can be
given for fields of arbitrary nonzero spin.  The simplest Lorentz
covariant wave equation for a massless field of spin $n/2$ is
  \eqn{SimpleEOM}{
    D^{\Delta_1 \dot\Delta} \Psi_{\Delta_1 \ldots \Delta_n} = 0
  }
 where $\Psi_{\Delta_1 \ldots \Delta_n}$ is symmetric in all its
indices.  The case $n=1$ gives the Weyl fermion equation.  The case
$n=2$ is, as we have seen, Maxwell's equations in vacuum.  The case
$n=4$ can be obtained by linearizing pure gravity around Minkowski
space, as discussed in section 5.7 of \cite{pr}.  The case $n=3$ can
be obtained in Minkowski space from the massless Rarita-Schwinger
equation, as follows.  The constraint $\gamma^\mu \psi_\mu = 0$ is
imposed on the Rarita-Schwinger field
  \eqn{RSPot}{
   \psi^\mu = 
    \pmatrix{
     \sigma^{\mu\Delta\dot\Delta} \psi_{\Delta\Gamma\dot\Delta}
       \cr\noalign{\vskip5pt}
     \sigma^{\mu\dot\Delta\Delta} 
      \bar\psi_{\dot\Delta}{}^{\dot\Gamma}{}_{\vphantom{\dot\Delta}\Delta}
    }
  }
 to project out the spin-$1/2$ components.  This constraint is
equivalent to making $\psi_{\Delta\Gamma\dot\Delta}$ symmetric in its
two undotted indices.  In the supergravity literature, the equation of
motion is usually written as $\epsilon^{\mu\nu\rho\sigma} \gamma_5
\gamma_\nu \Psi_{\rho\sigma} = 0$ where $\Psi_{\rho\sigma} =
\partial_\rho \psi_\sigma - \partial_\sigma \psi_\rho$.  The original
paper by Rarita and Schwinger \cite{rs} (see also p.~323 of
\cite{weinbergI}) proposes $\slashed\partial \psi_\mu = 0$ as the
equation of motion.  Using the constraint one can show that both are
equivalent to $\partial^{\dot\Delta\Delta}
\psi_{\Delta\Gamma\dot\Sigma} = 0$.  As a result, the field strength
  \eqn{RSField}{
   \Psi_{\Delta\Sigma\Gamma} = \partial_{\Sigma\dot\Delta}
    \psi_\Delta{}^{\dot\Delta}{}_\Gamma 
  }
 is symmetric in all its indices and obeys the $n=3$ case of
\SimpleEOM.  

Although the cases $n=3$ and $n=4$ of \SimpleEOM\ are not in general
the correct curved-space equations of motion for the gravitino and
graviton, and although for $n>2$ there are problems defining local,
gauge-invariant number currents and stress-energy tensors, still a
brief investigation of \SimpleEOM\ serves to illustrate some of the
general features one expects for fields of higher spin.  Furthermore,
the near-Minkowskian limit of the equations we will derive should be
close in form to the actual graviton and gravitino equations far from
a black hole.

Define helicity components $\psi_s$ according to
  \eqn{HComp}{
   \Psi_{\Delta_1 \ldots \Delta_n} =
    \psi_{\lower4pt\hbox{$\scriptstyle {{n \over 2} - 
     \sum\limits_i \Delta_i}$}}  \ .
  }
\vskip-5pt
 Then \SimpleEOM\ can be written out in components.  There are $2n$
equations:
  \eqn{AllEOMS}{\vcenter{\openup0\jot
    \halign{\strut\span\TL & \span\TR\cr
     \big( \Delta - n \gamma + \rho \big) \psi_{n \over 2} &=
      \big( \delta + (n-2) \alpha \big) \psi_{{n \over 2} - 1}  \cr
     \big( D - (n-2) \gamma - n \rho \big) \psi_{{n \over 2} - 1} &=
      \big( \bar\delta - n \alpha \big) \psi_{n \over 2}  \cr
     &\vdots  \cr
     \big( \Delta - 2s \gamma + (\tf{n}{2} + 1 - s) \rho \big) \psi_s &=
      \big( \delta + (2s - 2) \alpha \big) \psi_{s-1}  \cr
     \big( D - (2s - 2) \gamma - (\tf{n}{2} + s) \rho \big) \psi_{s-1} &=
      \big( \bar\delta - 2s \alpha \big) \psi_s  \cr
     &\vdots  \cr
     \big( \Delta + (n-2) \gamma + n \rho \big) \psi_{-{n \over 2} + 1} &=
      \big( \delta - n \alpha \big) \psi_{-{n \over 2}}  \cr
     \big( D + n \gamma - \rho \big) \psi_{-{n \over 2}} &= 
      \big( \bar\delta + (n-2) \alpha \big) \psi_{-{n \over 2} + 1} 
        \ .  \cr
   }}} 
 The equations \AllEOMS\ are invariant under PT, which sends $\psi_s
\to \psi_{-s}$, $\gamma \to -\gamma$, $\rho \to -\rho$, $D
\leftrightarrow \Delta$, $\delta \leftrightarrow \bar\delta$.

The commutation relations
  \eqn{KeyComs}{
   [D,\delta] = \rho \delta \ \ \quad [D,\alpha] = \rho \alpha \ \ \quad
   [\Delta,\bar\delta] = -\rho \bar\delta \ \ \quad
   [\Delta,\alpha] = -\rho \alpha
  }
 are easily established by direct computation.  They can be used to
convert the pair of equations in \AllEOMS\ relating $\psi_s$ and
$\psi_{s-1}$ into decoupled second order equations for $\psi_s$ and
$\psi_{s-1}$ separately.  In this way one obtains
  \eqn{SameEQ}{\eqalign{
   &\Big[ \big( D - (2s-2) \gamma - (\tf{n}{2}+1+s) \rho \big)
           \big( \Delta - 2s \gamma + (\tf{n}{2}+1-s) \rho \big)  \cr
   &\qquad - \big( \delta + (2s-2) \alpha \big) 
           \big( \bar\delta - 2s \alpha \big) \Big] \psi_s = 0  \cr
   &\Big[ \big( \Delta - (2s+2) \gamma + (\tf{n}{2}+1-s) \rho \big)
           \big( D - 2s \gamma - (\tf{n}{2}+1+s) \rho \big)  \cr
   &\qquad - \big( \bar\delta - (2s+2) \alpha \big)
           \big( \delta + 2s \alpha \big) \Big] \psi_s = 0 \ .
  }}
 The first of these can be derived for $s > -n/2$, while the second
can be derived for $s < n/2$.  In fact they are different forms of the
same equation, which can be written out more simply in terms of the
fields 
  \eqn{TPsi}{
   \tilde\psi_s = 
    e^{|s| A + \left( {n \over 2} - |s| + 1 \right) C} \psi_s 
  }
 as
  \eqn{TPsiEqs}{\vcenter{\openup1\jot
    \halign{\strut\span\TL & \span\TR & \span\TT\cr
     \Big[ \big( D + (2-4s) \gamma - 2s \rho \big) \Delta - 
           \big( \delta + (2s-2) \alpha \big) 
            \big( \bar\delta - 2s \alpha \big) \Big] \tilde\psi_s &= 0
      &\quad\hbox{for $s \ge 0$}  \cr
     \Big[ \big( \Delta - (2+4s) \gamma - 2s \rho \big) D - 
           \big( \bar\delta - (2s+2) \alpha \big)
            \big( \delta + 2s \alpha \big) \Big] \tilde\psi_s &= 0
      &\quad\hbox{for $s \le 0$.}  \cr
  }}}
 More explicitly, 
  \eqn{OneMaster}{\eqalign{
   &\Big[ \partial_r^2 + 
          \big( (1-2|s|) A' - B' + 2 |s| C' \big) \partial_r - 
          e^{-2A+2B} \partial_t^2 +
          2s \, e^{-A+B} (A'-C') \partial_t  \cr
   &\qquad + e^{2B-2C} \big( 
           \partial_\theta^2 + \cot\theta \partial_\theta + 
           \csc^2 \theta \partial_\phi^2 + 
           2is \cot\theta \csc\theta \partial_\phi - 
           s^2 \cot^2 \theta - |s| \big)
    \Big] \tilde\psi_s = 0 
  }}
 for all values of $s$, positive and negative.  It is interesting to
note that there is no explicit dependence on the spin $n/2$ of the
particle in \OneMaster, only on its helicity $s$.  In practice we will
mainly be interested in the equations for $\tilde\psi_{\pm {n \over
2}}$ since these are the only components that can be radiative.  The
fact that the equations for these radiative fields are identical to
equations obeyed by non-radiative components of fields of higher spin
suggests that mixing of different spins is possible.  Such mixing
between photons and gravitons was observed by Chandrasekhar in his
analysis of perturbations of the Reissner-Nordstrom black hole
\cite{Chandra}.

Equations similar to \OneMaster\ were worked out for the Kerr metric
by Teukolsky \cite{teuk,teukI}.  In that case, only the equations
for the radiative fields turned out to be separable.  But in the
present context, spherical symmetry makes the separability of
\OneMaster\ trivial: the general solution is
  \eqn{SepSol}{
   \tilde\psi_s(t,r,\theta,\phi) = 
    e^{-i \omega t} R_{s\ell}(r) Y_{s\ell m}(\theta,\phi)
  }
 where $Y_{s\ell m}$ is a spin-weighted spherical harmonic \cite{gold}
and $R_{s\ell}(r)$ satisfies the ODE
  \eqn{RadialEquation}{\eqalign{
   &\Big[ \partial_r^2 + 
          \big( (1-2|s|) A' - B' + 2|s| C' \big) \partial_r + 
          \omega^2 e^{-2 A + 2 B} - 
          2si \omega \, e^{-A + B} (A' - C')   \cr
   &\qquad - e^{2 B - 2 C} (\ell+|s|) (\ell-|s|+1) 
    \Big] R_{s\ell} = 0 \ .
  }}
 The minimal value of $\ell$ is $|s|$.  In the case of photons, $\ell
\ge 1$ indicates that the fields of lowest moment that can be radiated
are dipole fields.  For fermions, $\ell$, $s$, and $m$ are all
half-integer.

\section{Semi-classical absorption probabilities}
\label{Prob}

Returning now to the case of photons, let us investigate how an
absorption cross-section can be extracted from a solution to
\eno{RadialEquation}.  Section~\ref{Method} derives a formula for the
absorption probability.  In section~\ref{Examples} matching solutions
are exhibited and absorption probabilities calculated for the black
holes \kpSol\ and \NonExSol.

\subsection{Probabilities from energy fluxes}
\label{Method}

For the photon as for other fields of spin greater
than $1/2$, there is no gauge-invariant number current analogous to
  $J_\mu = {1 \over 2 i} \bar\phi 
    \overleftrightarrow\partial_\mu \phi$
 for spin $0$ and 
  $J_\mu = \bar\psi \gamma_\mu \psi$
 for spin $1/2$.  In order to count the photons falling into the black
hole, it is therefore necessary to examine the energy flux through the
horizon and adjust for the gravitational blueshift that the infalling
photons experience.  The stress-energy tensor can be written in terms
of $\phi_1$, $\phi_0$, and $\phi_{-1}$:
  \eqn{Tmunu}{\eqalign{
   T^{\mu\nu} &= \tf{1}{4} g^{\mu\nu} F^2 + F^{\mu\rho} F_\rho{}^\nu  
     = 2 \sigma^{\mu\Delta\dot\Delta} \sigma^{\nu\Gamma\dot\Gamma}
     \phi_{\Delta\Gamma} \bar\phi_{\dot\Delta\dot\Gamma}  \cr
    &= \Big[ |\phi_1|^2 n_\mu n_\nu + 
     2 |\phi_0|^2 (\ell_{(\mu} n_{\nu)} + m_{(\mu} \bar{m}_{\nu)}) + 
     |\phi_{-1}|^2 \ell_\mu \ell_\nu  \cr
    &\quad\ -4 \bar\phi_1 \phi_0 n_{(\mu} m_{\nu)} - 
     4 \bar\phi_0 \phi_{-1} \ell_{(\mu} m_{\nu)} + 
     2 \phi_{-1} \bar\phi_1 m_\mu m_\nu \Big] + {\rm c.c.}
  }}
 Consider a sphere $S^2$ located anywhere outside the horizon.
Taking into account the blueshift factor as described, the number of
photons passing through $S^2$ in a time interval $[0,t]$ is 
  \eqn{NumberThrough}{
   N = {1 \over \omega} \int\limits_{S^2 \times [0,t]} * (T_{tr} dr) 
     = {t \over \omega} \int\limits_{S^2} {\cal F} \vol_{S^2}
  }
 where $\vol_{S^2} = \sin\theta d\theta \wedge d\phi$, and 
  \eqn{FFlux}{
   {\cal F} = e^{A-B+2C} T_{tr} 
    = e^{2 A + 2 C} \left( |\phi_{-1}|^2 - |\phi_1|^2 \right)
    = |\tilde\phi_{-1}|^2 - |\tilde\phi_1|^2
  }
 is essentially the radial photon number flux.

The goal now is to find an approximate solution to \OneMaster\ for
photons whose wavelength is much longer than the size of the black
hole, and to extract from it an absorption probability and
cross-section.  The dominant contribution to this absorption comes
from dipole fields.  

Far from the black hole, \RadialEquation\ for dipole fields
simplifies to
  \eqn{FarPRE}{
   \left[ \partial_\rho^2 + {2 \over \rho} \partial_\rho + 1 + 
    {2 s i \over \rho} - {2 \over \rho^2} \right] R = 0
  }
 where $\rho = \omega r$.  The general solution to \FarPRE\ with $s =
1$ is 
  \eqn{FarSoln}{
   R = 2 a e^{-i \rho} \left( 1 - {i \over \rho} - 
       {1 \over 2 \rho^2} \right) + b {e^{i \rho} \over \rho^2} \ .
  }
 The general solution with $s = -1$ is just the conjugate of \FarSoln.
By using \Max\ $\phi_0$ can be calculated as well.  Let us choose the
spatial orientation by setting $m=0$ in \SepSol.  Then the final
result is
  \eqn{IIISoln}{\eqalign{
   \tilde\phi_1 &= e^{-i \omega t} \sin\theta
    \left[ 2 a e^{-i \rho} \left( 1 - {i \over \rho} - 
      {1 \over 2 \rho^2} \right) + b {e^{i \rho} \over \rho^2}
    \right]  \cr
   \tilde\phi_0 &= e^{-i \omega t} \cos\theta {\sqrt{2} i \over \omega}
    \left[ a e^{-i \rho} \left( 1 - {i \over \rho} \right) + 
           b e^{-i \rho} \left( 1 + {i \over \rho} \right) 
    \right]  \cr
   \tilde\phi_{-1} &= e^{-i \omega t} \sin\theta 
    \left[ a {e^{-i \rho} \over \rho^2} + 2 b e^{i \rho} \left( 1 +
     {i \over \rho} - {1 \over 2 \rho^2} \right) \right] \ .
  }}
 It is clear from \IIISoln\ that $\phi_{-1}$ is the radiative component
of the field for outgoing waves, while $\phi_1$ is the radiative
component for ingoing waves.  

The boundary conditions at the horizon \cite{teukI} require the radial
group velocity to point inward.  It can be shown that $\tilde\phi_1$
remains finite at the horizon while $\tilde\phi_{-1}$ vanishes.
Normalizations are fixed by requiring $|\tilde\phi_1|^2 \to \sin^2
\theta$ at the horizon.  The net flux of photons into the black hole
can be computed in two ways:
  \eqn{FluxesFN}{\vcenter{\openup1\jot
   \halign{\strut\span\TT & \span\TR\cr
    at the horizon: \quad & {\cal F}_h = -|\tilde\phi_1|^2 = 
     -\sin^2 \theta  \cr
    at infinity: \quad & {\cal F}_\infty = 
     {\cal F}^{\rm out}_\infty + {\cal F}^{\rm in}_\infty = 
     |\tilde\phi_{-1}|^2 - |\tilde\phi_1|^2 = 
      4 \left( |b|^2 - |a|^2 \right) \sin^2 \theta \ .  \cr
  }}}
 The two must agree, ${\cal F}_h = {\cal F}_\infty$, and so $|a|^2 =
|b|^2 + 1/4$.  One can thus easily perceive the equivalence of the two
common methods for computing the absorption probability.  The first
examines the deficit in the outgoing flux compared to the ingoing
flux:
  \eqn{Deficit}{
   1 - P = {{\cal F}^{\rm out}_\infty \over {\cal F}^{\rm in}_\infty} 
         = {|b|^2 \over |a|^2} \ ,
  }
 while the second simply compares the flux on the horizon to the
ingoing flux at infinity:
  \eqn{FluxRatio}{
   P = {{\cal F}_h \over {\cal F}^{\rm in}_\infty} 
     = {1 \over 4 |a|^2} \ .
  }
 For low-energy photons, the absorption probability is small and $a$
and $b$ are large and nearly equal, so from a calculational point of
view the second method is to be preferred over the first.  Indeed, it
will be standard practice in the matching calculations of later
sections to ignore the small difference between $a$ and $b$ and simply
set them equal.  This approximation suffices when \FluxRatio\ is used.
 
Finally, to obtain the absorption cross-section from the probability,
the Optical Theorem is needed.  Averaging over polarizations is
unnecessary in view of the spherical symmetry of the background.  The
result for photons in a dipole wave is 
  \eqn{OpThPh}{
   \sigma_{\rm abs} = {3 \pi \over \omega^2} P \ .
  }

\subsection{Matching solutions}
\label{Examples}

The work of previous sections can be boiled down to a simple
prescription for computing the absorption probability and
cross-section to leading order in the energy for minimally coupled
photons falling into a spherically symmetric black hole.  The
probability can be obtained by solving \RadialEquation\ with $\ell =
1$ and $s = 1$, subject to the boundary condition $R(r) \sim e^{i
f(r)}$ as $r$ approaches the horizon, $f(r)$ being some real
decreasing function of $r$.  Far from the black hole, one will find
$R(r) \sim 2 a e^{-i \omega r}$, and the probability is then given by
  \eqn{ReState}{\eqalign{
   P = {1 \over 4 |a|^2} 
     = {|R(r)|^2 \Big|_{\rm horizon} \over 
        |R(r)|^2 \Big|_\infty} \ .
  }}

First consider the extreme black hole \kpSol.  The radial equation
\RadialEquation\ with $\ell = s = 1$ is
  \eqn{ExactExtremeI}{
   \left[ \partial_r^2 + {2 \over r + M} \partial_r + 
    \omega^2 \left( 1 + {M \over r} \right)^4 + 
    {2 i \omega \over r} \left( 1 - {M^2 \over r^2} \right) - 
    {2 \over r^2} \right] R = 0 \ .
  }
 A different radial variable, $y = \omega M^2 / r$, is more natural
near the horizon.  In terms of $y$, \ExactExtremeI\ can be rewritten
as
  \eqn{ExactExtremeII}{
   \left[ (y^2 \partial_y)^2 - 
    {2 \omega M \over y + \omega M} y^3 \partial_y + 
    (y+\omega M)^4 - 2 i y (y^2 - \omega^2 M^2) - 
    2 y^2 \right] R = 0 \ .
  }
 A matching solution can be pieced together as usual from a near
region (${\bf I}$), an intermediate region (${\bf II}$), and a far
region (${\bf III}$).  In the near region, \ExactExtremeII\ is
simplified by setting to zero all terms containing explicit factors of
$\omega M$.  The intermediate region solution is obtained from
\ExactExtremeI\ with $\omega = 0$.  In the far region, we simply make
the flat space approximation, obtaining \FarPRE\ and \FarSoln.  
The solutions in the three regions are
  \eqn{OTTSoln}{\eqalign{
   R_{\bf I} &= e^{i y} \left( 1 + {i \over y} - 
    {1 \over 2 y^2} \right)  \cr
   R_{\bf II} &= {C_1 r \over 1+M/r} + {C_2 \over r^2 (1+M/r)}  \cr
   R_{\bf III} &= 2 a e^{-i \rho} \left( 1 - {i \over \rho} - 
    {1 \over 2 \rho^2} \right) + b {e^{i \rho} \over \rho^2} \ .
  }}
 A match is obtained by setting
  \eqn{MatchIt}{\eqalign{
   &C_1 = -{1 \over 2 \omega^2 M^3} \qquad C_2 = 0  \cr
   &a = b = -{3 i \over 8 (\omega M)^3} \ .
  }}
 The absorption probability and cross-section are 
  \eqn{PandSigma}{
   P = {16 \over 9} (\omega M)^6 \qquad\quad 
   \sigma_{\rm abs} = {16 \pi \over 3} \omega^4 M^6 \ .
  }

Now consider the non-extremal generalization, \NonExSol.  The radial
equation \RadialEquation\ with $\ell = 1$ and $s = 1$ is
  \eqn{ExNonEx}{
   \left[ \partial_r^2 + {2 \over fr} \partial_r + 
    \omega^2 {f^4 \over h^2} - 2 i \omega \left( \tf{1}{2} + 
    {1 \over 2h} - {2 \over f} \right) - {2 \over hr^2} \right] R = 0 \ .
  }
 As before, the far region is treated in the flat space approximation,
and a solution in the intermediate region is obtained by solving
\ExNonEx\ with $\omega = 0$.  In the near region, the useful radial
variable is $h$ itself.  Having the black hole near extremality is
useful since one can approximate $f \approx (1-h) \cosh^2 \alpha$ and
drop terms in \ExNonEx\ which are small in the limit where $\alpha \to
\infty$ and $\omega r_0 \to 0$ with
  \eqn{lambdaDef}{
   \lambda = \omega r_0 \cosh^4 \alpha = {\omega \over 4 \pi T}
  }
 held fixed.  The result is that \ExNonEx\ simplifies to
  \eqn{INonEx}{
   \left[ h (1-h) \partial_h^2 - 2h \partial_h + 
    \lambda^2 {1-h \over h} - i\lambda {1+h \over h} - 
    {2 \over 1-h} \right] R = 0 \ ,
  }
 which is representative of the general form of differential equation
which is solved by a hypergeometric function of $h$ times powers of
$h$ and $1-h$.

The solutions in the three regions are 
  \eqn{NonExSols}{\eqalign{
   R_{\bf I} &= {h^{-i \lambda} \over (1-h)^2} 
     F(-2,-1-2i\lambda,-2i\lambda;h)  \cr
    &= {h^{-i \lambda} \over (1-h)^2} \left( 1 + 
     {ih (1+2i\lambda) \over \lambda} - 
     {h^2 (1+2i\lambda) \over 1-2i\lambda} \right)  \cr
   R_{\bf II} &= C_1 {h \over f (1-h)} + 
    C_2 {h \over f} \left( 1 + {1 \over h} + 
    {2 \log h \over 1-h} \right)  \cr
   R_{\bf III} &= 2 a e^{-i \rho} \left( 1 - {i \over \rho} - 
    {1 \over 2 \rho^2} \right) + b {e^{i \rho} \over \rho^2} \ ,
  }}
 and a match is obtained by setting
  \eqn{MatchNonEx}{\eqalign{
   &C_1 = {i \cosh^2 \alpha \over \lambda (1-2i\lambda)} \qquad 
    C_2 = 0  \cr
   &a = b = -{3 \cosh^2 \alpha \over 4 \omega r_0 \lambda
    (1-2i\lambda)} \ .
  }}
 The absorption probability and cross-section are 
  \eqn{NonExProb}{
   P = \tf{4}{9} (\omega r_0)^3 \lambda (1+4\lambda^2) \qquad\quad 
   \sigma_{\rm abs} = {4 \pi \over 3} 
    \omega r_0^3 \lambda (1+4\lambda^2) \ .
  }

\section{The axion and minimally coupled fermions}
\label{OtherParticles}

Because the solution \kpSol\ preserves a quarter of the supersymmetry,
it is clearly of interest to compare cross-sections of particles with
different spins related by the unbroken supersymmetry.  The other
particles in ${\cal N} = 4$ supergravity whose cross-sections are
straightforward to compute are the dilaton, the axion, and those
fermions which obey the Weyl equation.  The dilaton has been dealt
with at length in the fixed scalar literature \cite{kr,kkOne,kkTwo}.
The axion in fact is also a fixed scalar, as section
\ref{Axion} will show.  Of the four massless fermions, two
are minimally coupled.  The purpose of section
\ref{Fermions} is to demonstrate this fact and to compute
the minimal fermion cross-section.  For comparison with \NonExProb\
let us quote here the final results:
  \eqn{AFProbs}{\vcenter{\openup1\jot
    \halign{\strut\span\TT\ \ & \span\TL & \span\TR \qquad & 
            \span\TL & \span\TR\cr
     axion, dilaton: &    
      P &= (\omega r_0)^2 (1+4\lambda^2) &
      \sigma_{\rm abs} &= \pi r_0^2 (1+4\lambda^2)  \cr
     minimal fermions: &
      P &= {(\omega r_0)^2 \over 4} (1 + 16 \lambda^2) &
      \sigma_{\rm abs} &= {\pi r_0^2 \over 2} (1 + 16 \lambda^2)  \cr
  }}}
 where $\lambda = \omega / (4 \pi T)$, as in \NonExProb.  Note that in
the extremal limit the bosonic and fermionic absorption probabilities
quoted in \AFProbs\ coincide.\footnote{Similar results on the agreement of
absorption probabilities due to residual supersymmetry have appeared
elsewhere in the literature \cite{OkaOne,OkaTwo} for the case of $N=2$
supergravity.  See also \cite{krw}.  Thanks to G.~Horowitz and A.~Peet
for bringing these papers to my attention.}

\subsection{The axion}
\label{Axion}

The strategy for deriving the linearized equation of motion for the
axion is the same as the one used in \cite{kr} for the dilaton: only
spherical perturbations of the solution \NonExSol\ are considered, and
a gauge is chosen where the only components of the metric that
fluctuate are $g_{rr}$ and $g_{tt}$.  The minimally coupled gauge
fields $F_1$, $F_2$, $G_1$, and $G_2$ do not affect the linearized
axion equation because they have no background value and enter into
the axion equation quadratically.  Spherical symmetry dictates that
only $tr$ and $\theta\phi$ components of these field strengths can
fluctuate, corresponding respectively to radial electric and radial
magnetic field fluctuations.  These fluctuations are constrained
further by Gauss' law \GaussLaw:
  \eqn{GLConsequence}{\vcenter{\openup1\jot
    \halign{\strut\span\TL & \span\TR \qquad & \span\TL & \span\TR\cr
   F^3_{\theta\phi} &= 0 & 
   F_3^{tr} &= {Q \sin\theta \over \sqrt{-g}} e^{2\phi}  \cr
   G^3_{\theta\phi} &= P \sin\theta &
   G_3^{tr} &= {2 B e^{2\phi} \over \sqrt{-g}} G^3_{\theta\phi} \ .  \cr
  }}}
 The asymmetry between $F_3$ and $G_3$ arises because the black hole
is electrically charged under $F_3$ and magnetically charged under
$G_3$.  The axion $B$ is a dynamical theta-angle for the gauge fields,
so the last relation in \GLConsequence\ should be viewed as a
dynamical version of the Witten effect \cite{wittenDyon}: when the
axion fluctuates, an object that was magnetically charged picks up
what seems like an electric charge in that there are radial electric
fields.

Using \GLConsequence\ and ignoring the dilaton terms in the axion
equation of \SEOMs\ (which is valid for the purpose of deriving the
linearized axion equation because the dilaton has zero background
value), one obtains
  \eqn{SimpleAxion}{
   \square B + \tf{i}{2} (F_3 * F_3 + G_3 * G_3) = 
   \square B + i (G_3^{tr} * G^3_{tr} + G_3^{\theta\phi} * G^3_{\theta\phi}) =
   \left[ \square + {4 P^2 \over f^4 r^4} \right] B = 0 \ .
  }
 This is indeed identical to the linearized equation for the dilaton,
although the ``mass'' term for the dilaton receives equal
contributions $2 Q^2 + 2 P^2$ from the electric and magnetic charges
in place of the $4 P^2$ we see in \SimpleAxion.

The radial equation for $B$ and $\phi$ is 
  \eqn{DilRad}{
   \left[ (hr^2 \partial_r)^2 + \omega^2 r^4 f^4 - 
    {h r_0^2 \sinh^2 2\alpha \over 2 f^2} \right] R = 0  \ .
  }
 By an analysis sufficiently analogous to the treatments in
\cite{kr,kkOne} that it seems superfluous to present the details, one
obtains the result already quoted in \AFProbs:
  \eqn{DilProb}{
   P = (\omega r_0)^2 (1+4\lambda^2) \qquad\quad 
   \sigma_{\rm abs} = \pi r_0^2 (1+4\lambda^2) \ .
  }

\subsection{Minimally coupled fermions}
\label{Fermions}

It was shown in \cite{csfTwo} that the complete fermionic equations of
motion for simple ${\cal N} = 4$, $d=4$ supergravity take on a simple form when
written in terms of the supercovariant derivatives introduced in
\cite{cs}.  The relevant one of these equations for spin-$1/2$
fermions is
  \eqn{FEOM}{
   i \hat{\slashed{D}} \Lambda_I - \tf{3}{2} e^{2 \phi} 
    (\hat{\slashed{D}} B) \Lambda_I = 0 \ , 
  }
 where $\hat{D}_\mu$ denotes a supercovariant derivative.\footnote{The
spinor and gamma matrix conventions conventions used here are those
described in Appendix~A of \cite{klopp}.  $\Lambda_I$ is a chiral
spinor with $\gamma_5 \Lambda_I = \Lambda_I$ which replaces the
Majorana spinor $\chi^i$ of \cite{csfTwo}.  The gravitinos are also
written in terms of chiral spinors $\Psi_\mu^I$ with $\gamma_5
\Psi_\mu^I = \Psi_\mu^I$.  $I$ runs from $1$ to $4$.  Conversion to
these conventions from those of \cite{csfTwo} is discussed in
\cite{klopp}.  It is worth adding that in the current conventions,
each field is identified with $K$ times its counterpart in
\cite{csfTwo}, and for notational simplicity $K$ is then set equal to
$1/2$.  With this choice of the gravitational constant, $\phi$ and
$\Lambda_I$ are not canonically normalized; rather, they are twice the
canonically normalized fields.  So for example the kinetic term of
$\phi$ in \FullBL\ is $2 (\partial_\mu \phi)^2$ rather than the
canonical ${1 \over 2} (\partial_\mu \phi)^2$.}

Supercovariant derivatives in general can be read off from the
supersymmetry variations of a field: if $\delta f = F_I \epsilon^I$,
then $\hat{D}_\mu f = D_\mu f - \tf{1}{4} F_I \Psi_\mu^I$.  Thus the
supercovariant derivative of the axion is $\hat{D}_\mu B =
\partial_\mu B + \hbox{(two fermion terms)}$.  Again, terms in the
equations of motion which are quadratic in fields with zero background
value do not contribute to the linearized first-varied equations of
motion.  Because the background values of the axion as well as all
fermions are zero for the solution \NonExSol, the second term in
\FEOM\ can be discarded.

A further simplification of \FEOM\ can be made by dropping terms from
$\hat{\slashed{D}} \Lambda_I$ which are quadratic in fields with zero
background value.  The supersymmetry variation of $\Lambda_I$ is
  \eqn{SUSYvs}{
   \delta \Lambda_I = \sqrt{2} \sigma^{\rho\sigma} 
    (F^3_{\rho\sigma} \alpha^3_{IJ} - \tilde{G}^3_{\rho\sigma}
     \beta^3_{IJ}) \epsilon^J 
  }
 plus terms which vanish for the solution \NonExSol.  So
  \eqn{LambdaD}{
   \hat{D}_\mu \Lambda_I = D_\mu \Lambda_I - 
    {1 \over 2 \sqrt{2}} \sigma^{\rho\sigma} 
    (F^3_{\rho\sigma} \alpha^3_{IJ} - \tilde{G}^3_{\rho\sigma}
     \beta^3_{IJ}) \Psi_\mu^J 
  }
 plus terms quadratic in fields with zero background value.  In
\SUSYvs\ and \LambdaD, we have introduced 
  $\sigma^{\rho\sigma} = \tf{1}{4} [\gamma^\rho,\gamma^\sigma]$ 
 and the matrices
  \eqn{ABDef}{
   \alpha^3_{IJ} = \pmatrix{ 0 & 1 & 0 & 0 \cr
                            -1 & 0 & 0 & 0 \cr
                             0 & 0 & 0 & 1 \cr
                             0 & 0 &-1 & 0 } \qquad\quad
    \beta^3_{IJ} = \pmatrix{ 0 &-1 & 0 & 0 \cr
                             1 & 0 & 0 & 0 \cr
                             0 & 0 & 0 & 1 \cr
                             0 & 0 &-1 & 0 } \ .
  }
 The matrices $\alpha^n_{IJ}$ and $\beta^n_{IJ}$ were introduced in
\cite{gso} and used in \cite{csfTwo} to establish the $SU(4)$
invariance of ${\cal N} = 4$ supergravity.

Since $F^3_{\rho\sigma} = \tilde{G}^3_{\rho\sigma}$, the first
variation of the equation of motion \FEOM\ is
  \eqn{FirstFEOM}{
   i \hat{\slashed{D}} \delta \Lambda_I = 
    i \slashed{D} \delta \Lambda_I - {i \over 2 \sqrt{2}} \sigma^{\rho\sigma}
    F^3_{\rho\sigma} (\alpha^3_{IJ} - \beta^3_{IJ}) 
     \delta \Psi_\mu^J = 0 \ .
  }
 For $I=3$ and $4$, the gravitino part gets killed and \FEOM\ is
nothing but the Weyl equation \WeylEq.  For $I=1$ and $2$, the
gravitino mixes in.\footnote{The gravitino equation of motion has the form
of the Rarita-Schwinger equation plus interactions.  This equation
also has an $SO(4)$ index structure which is block diagonal, and the
spin-$1/2$ particles decouple from the $I = 3,4$ equations.  The
resulting gravitino equation,
  $\epsilon^{\mu\nu\rho\sigma} \gamma_\nu \hat{\Psi}^I_{\rho\sigma} = 0$,
 is less simple than the equations studied previously because the
supercovariant field strength $\hat{\Psi}^I_{\rho\sigma}$ involves a
non-vanishing combination of the field strengths $F_3$ and
$\tilde{G}_3$.  The problem of extracting from it a separable PDE like
\OneMaster\ is under investigation.}

The key point in this analysis is that the supersymmetry
transformation \SUSYvs\ leaves two of the $\Lambda_I$ invariant no
matter what $\epsilon^J$ is chosen to be.  In view of the form of
\FEOM\ and the prescription for reading off supercovariant derivatives
from the supersymmetry transformation laws, this makes it inevitable
that two of the $\Lambda_I$ are minimally coupled fermions.  The
situation is related to unbroken supersymmetries, but only loosely:
the non-extremal black hole has the same minimally coupled fermions
that the extremal one does, and the extremal black hole preserves only
one supersymmetry but admits two minimally coupled fermions.  The
condition for an unbroken supersymmetry is that the supersymmetry
variations of all the fermions fields must vanish.  So one would
expect that there are at least as many minimally coupled fermions as
unbroken supersymmetries.  More precisely, suppose that there are $n$
{\em broken} supersymmetries and $m$ fermions in the theory whose
equation of motion is of the form $i \hat{\slashed{D}} \Lambda = 0$,
up to terms quadratic in fields with zero background value.  Then
there must be at least $m-n$ minimally coupled fermions.  For
axion-free solutions to pure ${\cal N} = 4$ supergravity, $m = 4$.

The existence of minimally coupled fermions having been established,
the computation of their absorption cross-section now proceeds in
parallel to the case of minimally coupled photons.  If $R(r)$ is a
solution of the radial equation \RadialEquation\ with $\ell = s =
1/2$, then the absorption probability is once again
  \eqn{ProbAgain}{
   P = {|R(r)|^2 \Big|_{\rm horizon} \over 
        |R(r)|^2 \Big|_\infty} \ .
  }
 The justification for \ProbAgain\ is a little different than for its
analog \ReState\ for photons because for fermions there is a conserved
number current, 
  \eqn{JDef}{
   J_\mu = -\sqrt{2} \sigma_\mu^{\dot\Delta\Delta} \bar\Psi_{\dot\Delta} 
    \Psi_{\vphantom{\dot\Delta}\Delta} \ .
  }
 Here $\Psi_\Delta$ is the dyadic version of $\delta \Lambda_3$ or
$\delta \Lambda_4$.  The number of fermions passing through a sphere in
a time $t$ is
  \eqn{NPass}{
   N = \int\limits_{S^2 \times [0,t]} * (J_r dr) 
     = t \int\limits_{S^2} {\cal F} \vol_{S^2}
  }
 where
  \eqn{FPass}{
   {\cal F} = e^{A-B+2C} J_r 
    = e^{A+2C}
       \left( |\psi_{-1/2}|^2 - |\psi_{1/2}|^2 \right)
    = |\tilde\psi_{-1/2}|^2 - |\tilde\psi_{1/2}|^2 \ .
  }
 The component $\psi_{1/2}$, like $\phi_1$ in the case of photons, is
both the radiative component at infinity for infalling solutions and
the nonzero component at the black hole horizon.  The formula
\eno{ProbAgain} thus follows from \FPass\ by the same analysis that gave
\ReState\ from \eno{FFlux}.

The radial equation \RadialEquation\ with $\ell = s = 1/2$ is 
  \eqn{FermionRE}{
   \left[ \partial_r^2 + {1 \over r} 
    \left( \tf{1}{2} + {1 \over 2 h} \right) + 
    \omega^2 {f^4 \over h^2} - i \omega {f^2 \over h} {1 \over r}
     \left( \tf{1}{2} + {1 \over 2 h} - {2 \over f} \right) - 
     {1 \over h r^2} \right] R = 0 \ .
  }
 A matching solution can be obtained in the usual fashion:
  \eqn{FMSoln}{\eqalign{
   R_{\bf I} &= {h^{-i \lambda} \over 1-h} 
     F(-1,-\tf{1}{2} - 2 i \lambda,\tf{1}{2} - 2 i \lambda; h)  \cr
    &= {h^{-i \lambda} \over 1-h} \left( 1 + 
     {1 + 4 i \lambda \over 1 - 4 i \lambda} h \right)  \cr
   R_{\bf II} &= C_1 {1+h \over 1-h} + C_2 {\sqrt{h} \over 1-h}  \cr
   R_{\bf III} &= 2a \, e^{-i \rho} \left( 1 - {i \over 2\rho} \right) +
     ib {e^{i \rho} \over \rho}
  }}
 with 
  \eqn{MatchF}{\eqalign{
   &C_1 = {1 \over 1 - 4 i \lambda} \qquad C_2 = 0  \cr
   &a = b = {i \over \omega r_0} {1 \over 1 - 4 i \lambda}
  }}
 where as before $\lambda = \omega / (4 \pi T)$.  The absorption
probability and cross-section are
  \eqn{FPandSigma}{
   P = {(\omega r_0)^2 \over 4} (1 + 16 \lambda^2) \qquad\quad 
   \sigma_{\rm abs} = {2 \pi \over \omega^2} P = 
    {\pi r_0^2 \over 2} (1 + 16 \lambda^2) \ .
  }
 The $\lambda \to 0$ limit of \FPandSigma\ agrees with the general
result of \cite{dgm}.

\section{The effective string model}
\label{EffStr}

The usual approach to effective string calculations (see for example
\cite{dmOne,cgkt,gkOne,krt}) has been to derive couplings between bulk
fields and the effective string by expanding the Dirac-Born-Infeld
(DBI) action to some appropriate order.  The leading terms in the
expansion specify a conformal field theory (CFT) which would describe
the effective string in the absence of interactions with bulk fields.
The interactions are dictated at leading order by terms in the
expansion which are linear in the bulk fields: for a scalar field
$\phi$, a typical coupling would be
  \eqn{CoupTyp}{
   S_{\rm int} = \int d^2 x \, \phi(t,x,\vec{x}\!=\!0) \O(t,x)
  }
 where $\O(t,x)$ is some local conformal operator in the CFT.  The
integration is over the effective string world-volume, and $\vec{x}$
is set to zero because this is the location of the effective string in
transverse space.  One can then consider tree-level processes mediated
by $S_{\rm int}$ where a bulk particle is converted into excitations
on the effective string.  Although the applicability of the DBI action
to D-brane bound states can be called into question, the prescription
described here for computing absorption or emission rates appears to
be very robust.  In the spirit of \cite{ja}, where effective string
calculations were used to account for properties of Reissner-Nordstrom
black holes without reference to any underlying microscopic picture,
let us examine the consequences of couplings of the general form
\CoupTyp.

Consider the absorption of a quanta of $\phi$ with energy $\omega$,
momentum $p$ along the effective string, and transverse momentum
$\vec{p}$.  We shall continue to use ${+}{-}{-}{-}{-}$ signature, so
for example $p \cdot x = \omega t - p x - \vec{p} \cdot \vec{x}$.  The
absorption cross-section can be calculated by setting
$\phi(t,x,\vec{x}\!=\!0) = e^{-i p \cdot x}$ and then treating
\CoupTyp\ as a time-dependent perturbation to the CFT which describes
the effective string in isolation.  Stimulated emission would be
calculated by choosing $e^{i p \cdot x}$ rather than $e^{-i p \cdot
x}$.  The $t$ and $x$ dependence of $\O(t,x)$ is fixed by the free
theory:
  \eqn{FreeEv}{
   \O(t,x) = e^{i \hat{p} \cdot x} \O(0,0) 
    e^{-i \hat{p} \cdot x}
  }
 where $\hat{p} \cdot x = Ht - Px$, $H$ and $P$ being the Hamiltonian
and momentum operators of the CFT.  If one considers the perturbation
\CoupTyp\ to act for a time $t$, then Fermi's Golden Rule gives the
thermally averaged transition probability as
  \eqn{FermiGR}{
   {\cal P} = \sum_{i,f} {e^{-\beta \cdot p_i} \over Z} P_{i \to f} = 
    Lt \sum_{i,f} {e^{-\beta \cdot p_i} \over Z} 
     (2 \pi)^2 \delta^2(p + p_i - p_f) 
      \left| \langle f | \O(0,0) | i \rangle \right|^2 \ .
  }
 This formula is valid for when the length $L$ of the effective string
is much larger than the Compton wavelength of the incoming scalar.
In \FermiGR, $\beta$ has two components, $\beta^+ = \beta_L$ and
$\beta^- = \beta_R$.  The partition function splits into left and
right sectors:
  \eqn{PartDef}{
   Z = \tr e^{-\beta \cdot \hat{p}} = 
    (\tr_L e^{-\beta_L \hat{p}_+}) (\tr_R e^{-\beta_R \hat{p}_-}) \ .
  }
 For simplicity we take the momentum $p = (\omega,0,\vec{p})$ of the
incoming particle perpendicular to the brane, but clearly \FermiGR\
remains valid for the case of particles with Kaluza-Klein charge.

The summation over final states can become tedious when the scalar
turns into more than two excitations on the effective string.  Already
in the case of fixed scalars (chapter~\ref{FixedScalars}), which split
into two right-movers and two left-movers, the evaluation of this
summation was a nontrivial exercise.  It therefore seems worthwhile to
develop further a method employed in \cite{ja} in which the absorption
probability is read off from the two-point function of the operator
$\O$ in the effective string CFT.

Allow $t$ to take on complex values, defining $\O(t,x)$ by \FreeEv\
for arbitrary complex $t$.  According to usual notational conventions
\cite{Fetter}, $\O\+(t,x)$ is no longer the adjoint of $\O(t,x)$
except when $t$ is real; instead, $\O\+(t,x)$ is evolved from
$\O\+(0,0)$ using \FreeEv.  The conventional thermal Green's function
takes $t = -i \tau$ where $\tau$ is the Euclidean time:
  \eqn{TGreen}{
   {\cal G}(-i \tau,x) = \langle \O\+(-i \tau,x) \O(0,0) \rangle = 
    \tr \left( \rho \,
     T_\tau \left\{ \O\+(-i \tau,x) \O(0,0) \right\} \right)
  }
 where $\rho = e^{-\beta \cdot \hat{p}} / Z$.  One can continue to
arbitrary complex $t$, defining $T_\tau$ to time-order with respect to
$-\Im(t)$.  The convenience of doing this is that the integral
  \eqn{GreenGotcha}{
   \int d^2 x \, e^{i p \cdot x} {\cal G}(t - i \epsilon,x) = 
    \sum_{i,f} {e^{-\beta \cdot p_i} \over Z} 
     (2 \pi)^2 \delta^2(p + p_i - p_f) 
      \left| \langle f | \O(0,0) | i \rangle \right|^2
  }
 reproduces the right-hand side of \FermiGR.  The proof of
\GreenGotcha\ proceeds by inserting $\sum_i |i\rangle \langle i|$ and
$\sum_f |f\rangle \langle f|$ into \TGreen\ before $\O\+$ and $\O$
respectively.

Now let us turn to the evaluation of ${\cal G}(t,x)$.  Assume that
$\O(t,x)$ has the form
  \eqn{OForm}{
   \O(t,x) = \O_+(x^+) \O_-(x^-)
  }
 where $x^\pm = t \pm x$ and $\O_+$ and $\O_-$ are primary fields of
dimensions $h_L$ and $h_R$, respectively.  Set $z = i x^-$ so that, for
$x$ real and $t = -i \tau$ imaginary, $\bar{z} = i x^+$.  The
singularities in ${\cal G}(t,x)$ are determined by the OPE's of $\O_+$
and $\O_-$ with themselves:
  \eqn{RLOPEs}{\eqalign{
   \O_+(\bar{z}) \O_+\+(\bar{w}) &= 
    {C_{\O_+} \over (\bar{z}-\bar{w})^{2 h_L}} + 
     \hbox{less singular}  \cr
   \O_-(z) \O_-\+(w) &= 
    {C_{\O_-} \over (z-w)^{2 h_R}} + \hbox{less singular.}
  }} 
 ${\cal G}(t,x)$ factors into a left-moving and right-moving piece.
The imaginary time periodicity properties of each piece, together with
their singularities, suffice to fix the form of ${\cal G}(t,x)$
completely:\footnote{Actually, there is a subtlety here: the information
from periodicity and singularities must be supplemented by a sum rule
\cite{Fetter} on the spectral density to squeeze out an ambiguity in
the analytic continuation.}
  \eqn{TPForm}{
   {\cal G}(t,x) = {C_\O \over i^{2 h_L + 2 h_R}}
    \left( {\pi T_L \over \sinh \pi T_L x^+} \right)^{2 h_L}
    \left( {\pi T_R \over \sinh \pi T_R x^-} \right)^{2 h_R} 
  }
 where $C_\O = C_{\O_+} C_{\O_-}$.

At nonzero temperature, the absorption cross-section cannot be
calculated straight from \FermiGR: for bosons, the stimulated emission
probability must be subtracted off in order to obtain a result
consistent with detailed balance, as described in
chapter~\ref{FixedScalars}.  The net result is to set $\sigma_{\rm
abs} {\cal F} t = {\cal P} (1 - e^{-\beta \cdot p})$ where ${\cal F}$
is the flux and ${\cal P}$ is read off from \FermiGR.  For fermions,
the presence of an incoming wave inhibits by the Exclusion Principle
emission processes leading to another fermion in the same state as the
incoming wave.  The absorption cross-section must therefore be
calculated using $\sigma_{\rm abs} {\cal F} t = {\cal P} (1 +
e^{-\beta \cdot p})$.  Again, this result is in accord with detailed
balance.

The considerations of the previous paragraph can be restated compactly
in terms of the Green's function: 
  \eqn{SigmaForm}{\eqalign{
   \sigma_{\rm abs} &= {L \over {\cal F}} 
    \int d^2 x \, e^{i p \cdot x} \big( {\cal G}(t-i\epsilon,x) - 
      {\cal G}(t+i\epsilon,x) \big)  \cr
    &= {L C_\O \over {\cal F}} 
       {(2 \pi T_L)^{2h_L-1} (2 \pi T_R)^{2h_R-1} \over 
        \Gamma(2h_L) \Gamma(2h_R)}
       {e^{\beta \cdot p / 2} - 
        (-1)^{2 h_L + 2 h_R} e^{-\beta \cdot p / 2} \over 2}  \cr
    &\qquad\cdot
       \left| \Gamma \left( h_L + i {p_+ \over 2 \pi T_L} \right)
              \Gamma \left( h_R + i {p_- \over 2 \pi T_R} \right)
       \right|^2 \ .
  }}
 One way to perform the integral in the first line is first to
separate into $x^+$ and $x^-$ factors and then to deform the contours
in the separate factors by setting $\epsilon = \beta_L/2$ or
$\beta_R/2$.  Assuming that bosons and fermions couple, respectively,
to conformal fields with $h_L + h_R$ an integer or half an odd
integer, one indeed obtains the factor $1 \mp e^{-\beta \cdot p}$
required by detailed balance.

The formula \SigmaForm\ represents almost the most general functional
form for an absorption cross-section that the effective string model
is capable of predicting.  One possible generalization is for the bulk
field to couple to a sum of different operators $\O(t,x)$, in which
case a sum of terms like \SigmaForm\ would be expected.  Another
generalization can arise from a coupling of a bulk field $\phi$ to the
effective string not through its value $\phi(t,x,\vec{x}\!=\!0)$ on
the string, as shown in \CoupTyp, but rather through its derivatives:
for instance $\partial_i \phi(t,x,\vec{x}\!=\!0)$ where $i$ labels a
transverse dimension.  In case of fields without Kaluza-Klein charge,
the effect of $n$ such derivatives is simply to introduce an extra
factor $\omega^{2n}$ on the right hand side of \SigmaForm.  Since the
flux is ${\cal F} = \omega$ for a canonically normalized scalar, the
$\omega$ dependence of the cross-section is 
  \eqn{SigmaZE}{ 
   \sigma_{\rm abs} \sim \omega^{2n-1} 
    \sinh \left( \omega \over 2 T_H \right) 
   \left| \Gamma \left( h_L + i {\omega \over 4 \pi T_L} \right)
          \Gamma \left( h_R + i {\omega \over 4 \pi T_R} \right) 
    \right|^2 \ .
  }
 As we shall see in a specific example below, the flux factor for
massless fermions cancels out a similar factor in $C_\O$.  So for a
fermionic field which couples to the effective string through a term
in the lagrangian of the form $\partial^n \psi \O$, the energy
dependence of the cross-section is
  \eqn{FermionGrey}{
   \sigma_{\rm abs} \sim \omega^{2n} 
      \cosh \left( \omega \over 2 T_H \right) 
    \left| \Gamma\left( h_L + i {\omega \over 4 \pi T_L} \right)
           \Gamma\left( h_R + i {\omega \over 4 \pi T_R} \right)
    \right|^2 \ .
  }

The remarkable fact is that numerous classical absorption calculations
that have appeared in the literature
\cite{dmw,dmOne,dmTwo,gkOne,gkTwo,cgkt,dkt,htr,kvk,km,krt,mast,ja,%
clOne,clTwo,dgm} all give results consistent with \SigmaForm\ or
\SigmaZE\ in the near-extremal limit.  As an example, consider
massless minimally coupled scalars falling into the four-dimensional
black hole considered in \cite{gkTwo}, whose effective string model is
derived from the picture of three intersecting sets of M5-branes.  The
absorption cross-section for the $\ell^{\rm th}$ partial wave is
  \eqn{ScalarEll}{
   \sigma_{\rm abs}^\ell = {2 \over \omega^2}
    {(\omega T_H A_{\rm h})^{2 \ell + 1} \over (2 \ell)!^2
     (2 \ell + 1)!!} \sinh \left( \omega \over 2 T_H \right) 
     \left| \Gamma \left( \ell + 1 + i {\omega \over 4 \pi T_L} \right)
            \Gamma \left( \ell + 1 + i {\omega \over 4 \pi T_R} \right)
     \right|^2 \ ,
  }
 consistent with a coupling to the effective string of the form
$(\partial^\ell \phi) \O$ where $\O$ has dimensions $h_L = h_R = \ell
+ 1$.  Another interesting example is the fixed scalar
\cite{cgkt,kkOne}.  The cross-section for the $s$-wave in the
four-dimensional case \cite{kkOne}, with three charges equal and much
greater than the fourth ($R = r_1 = r_2 = r_3 \gg r_K \sim r_0$) is
  \eqn{FixedS}{
   \sigma_{\rm abs} = {r_0^3 \over 2 \omega R^2} 
    \sinh \left( \omega \over 2 T_H \right) 
     \left| \Gamma \left( 2 + i {\omega \over 4 \pi T_L} \right)
            \Gamma \left( 2 + i {\omega \over 4 \pi T_R} \right)
     \right|^2 \ ,
  }
 consistent with a coupling of the form $\phi T_{++} T_{--}$.  It is
not hard to convince oneself that the analysis of the two-point
function works the same when $\O(t,x) = T_{++}(x^+) T_{--}(x^-)$ as it
did when $\O(t,x)$ was the product of left and right moving primary
fields.

One naturally expects that an equal charge black hole whose metric is
extreme Reissner-Nordstrom, like the ${\cal N} = 4$ example we focused
on in sections \ref{MinCoup}, \ref{Examples}, and
\ref{OtherParticles}, can be obtained in the effective string picture
by taking $T_L \gg T_R,\omega$.  Indeed, it was found in \cite{ja}
that ordinary scalar cross-sections in this metric, and even in the
Kerr-Newman metric, have precisely the form one would expect from an
effective string with $T_L \gg T_R,\omega$.  The cross-sections have
no dependence on $T_L$ in this limit, and the authors of \cite{ja}
suggested a model which made no reference to the left-moving sector.
Note however that left-movers seem the natural explanation for the
finite entropy of extremal black holes---a subject not addressed in
\cite{ja}.  In \cite{hlm} it was argued in the context of ${\cal N} =
8$ compactifications that the CFT on the effective string is a $(0,4)$
theory with central charges $c_L = c_R = 6$.  The (local) $SU(2)$
$R$-symmetry of the right-moving sector was identified with the group
$SO(3)$ of spatial rotations of the black hole.  The same
identification of a local $SU(2)$ on the effective string with $SO(3)$
was used in \cite{ja}.  We will assume that an effective string
description with $4$ supersymmetries and $c=6$ in the right-moving
sector also applies to the equal charge black hole of ${\cal N} = 4$
supergravity.

Without committing to specific assumptions about the nature or
existence of left-movers, one can conclude that the general form for
an effective string absorption cross-section of massless particles is
  \eqn{OneSided}{\vcenter{\openup1\jot
    \halign{\strut\span\TT \quad& \span\TL & \span\TR & \quad\span\TT\cr
    bosons: & \sigma_{\rm abs} &\sim \omega^{2n} 
      \left| \Gamma(h_R + 2 i \lambda) \over 
             \Gamma(1 + 2 i \lambda) \right|^2 = 
      \omega^{2n} \prod_{r=1}^{h_R-1} (r^2 + 4 \lambda^2)
     & if $h_R \in {\bf Z}$  \cr
    fermions: & \sigma_{\rm abs} &\sim \omega^{2n} 
      \left| \Gamma(h_R + 2 i \lambda) \over 
             \Gamma({1\over 2} + 2 i \lambda) \right|^2 = 
      \omega^{2n} \prod_{r=1/2}^{h_R-1/2} (r^2 + 4 \lambda^2)
     & if $h_R \in {\bf Z}+\tf{1}{2}$  \cr
  }}}
 where $r$ runs over integers in the case of bosons and integers plus
$1/2$ in the case of fermions, and 
  \eqn{LambdaAgain}{
   \lambda = {\omega \over 4 \pi T} = {\omega \over 8 \pi T_R} \ .
  }
 The cross-section \NonExProb\ for minimally coupled photons fits the
form \OneSided\ with $h_R=2$ and $n=1$.  The form of the coupling of
the minimal photon to the effective string is further constrained by
rotation invariance.  If we represent the right-moving sector using
$4$ free chiral bosons (which are neutral under $SU(2)$) and an
$SU(2)$ doublet of fermions, then simplest coupling to the minimal
photon with the right group theoretic properties is
  \eqn{PhotonCoupling}{
   {\cal L}_{\rm int} = 
    \phi_{\alpha\beta} \Psi_-^\alpha \partial_- \Psi_-^\beta F_+ + {\rm h.c.} 
  }
 where $F_+$ is a left-moving field which, as noted previously, does
not affect the form of the absorption cross-section.  Recall that
$\phi_{\alpha\beta}$ is a field strength: one derivative is hidden
inside it, so indeed $n=1$.  The indices on $\Psi_-^\alpha$ are for
the group $SO(3)$ of spatial rotations, while the indices on
$\phi_{\alpha\beta}$ are for the $SU(2)_L$ half of the Lorentz group
$SO(3,1)$.  But they can be contracted as shown in a static gauge
description since the generators of $SO(3)$ are just sums of the
generators of $SU(2)_L$ and $SU(2)_R$.  With the current spinor index
conventions, an upper dotted index is equivalent to a lower undotted
index if only spatial $SO(3)$ rotations are considered.

The dilaton and axion cross-sections \AFProbs\ fit the form \OneSided\
with $h_R = 2$, and so does the minimal fermion cross-section with
$h_R = 3/2$.  The natural guess is a coupling of these fields to the
stress-energy tensor and the supercurrents: to linear order in all the
fields,
  \eqn{SuperCoupling}{
   {\cal L}_{\rm int} = \big[ (\phi + i B) T_{--} +
     \Lambda_{3\alpha} T_{F-}^\alpha -
     i \Lambda_{4\alpha} T_{F-}^{\dagger\alpha} \big] F_+ + {\rm h.c.}
  }
 The form of \SuperCoupling\ is dictated by the quarter of the ${\cal N} = 4$
supersymmetry which is preserved by the extreme black hole.  The terms
in the supersymmetry variation of ${\cal L}_{\rm int}$ with no
derivatives can shown to cancel using
  \eqn{SuperVars}{\eqalign{
   \delta (\phi + i B) &= \epsilon^\alpha \Lambda_{3\alpha} +
    \bar\epsilon_{\dot\alpha} i \sqrt{2} \sigma^{0\dot\alpha\beta}
     \Lambda_{4\beta}  \cr
   \delta T_{F-}^\alpha &= \epsilon^\alpha T_{--} \ .
  }}
 Here $\epsilon_\alpha$ parametrizes what was referred to in
\cite{klopp} as the $\epsilon^{34}_+$ supersymmetry.  Note that in the
conventions outlined in the appendix, both $\sqrt{2}
\sigma^{0\dot\alpha\beta}$ and $\sqrt{2} \sigma^0_{\alpha\dot\beta}$
are numerically the identity matrix.  The left-moving sector of the
CFT is assumed to be neutral under supersymmetry.

The equivalence (in static gauge) of upper undotted and lower dotted
indices has been used to simplify \SuperCoupling.  The matrices
$\sqrt{2} \sigma^{0\dot\alpha\beta}$ and $\sqrt{2}
\sigma^0_{\alpha\dot\beta}$ can be used to convert between them.  The
index on $T_{F-}\+$ has been raised in \SuperCoupling\ using
$\epsilon^{\alpha\beta}$.  A consequence of the second line in
\SuperVars\ is thus $\delta T_{F-}^{\dagger\alpha} =
-\bar\epsilon_{\dot\alpha} T_{--}$.

 The normalization of $T_{F-}^\alpha$ used here differs from \cite{ss}
by a factor of $\sqrt{2}$: if $G_n^\alpha$ and $L_n$ are the
supercurrent and Virasoro generators of \cite{ss}, then the present
conventions are to set $T_{F-}^\alpha(z) = {1 \over \sqrt{2}} \sum_n
z^{-n-3/2} G_n^\alpha$ and $T(z) = \sum_n z^{-n-2} L_n$.  On the
complex plane, the nonzero two-point functions are 
  \eqn{HolTwoPt}{\eqalign{
   \langle T_{F-}^\alpha(z) T_{F-\beta}\+(w) \rangle &= 
    {c \over 3} {\delta^\alpha_\beta \over (z-w)^3}  \cr
   \langle T_{--}(z) T_{--}(w) \rangle &= {c/2 \over (z-w)^4} \ .
  }}
 Let us also assume that the only nonzero two-point function of $F_+$
and $F_+\+$ is
  \eqn{RightTwoPt}{
   \langle F(\bar{z}) F\+(\bar{w}) \rangle = 
    {C_F \over (\bar{z} - \bar{w})^{2 h_R}} \ .
  }

Now we are ready to compute effective string cross-sections.  For the
dilaton, the operator $\O(t,x)$ entering into the analysis of
\CoupTyp-\SigmaForm\ is 
  \eqn{WhichO}{
   \O(t,x) = T_{--}(x^-) \big(F_+(x^+) + F_+\+(x^+) \big) \ .
  }
 The fact that $T_{--}$ is not primary does not
alter the periodicity properties of its two-point function.  Thus the
arguments leading from \OForm\ to \SigmaForm\ still apply, and $C_\O =
c C_F = 6 C_F$.  Because of the non-canonical normalization of the
dilaton field in the action \FullBL, the particle flux in a wave $\phi
= e^{-i p \cdot x}$ is ${\cal F} = 4 \omega$, four times the usual
value.  Plugging these numbers into \SigmaForm\ and taking the large
$T_L$ limit, one obtains
  \eqn{DilCross}{
   \sigma_{\rm abs} = {L C_F \over 8 T} (2 \pi T_L)^{2 h_L - 1}
    (2 \pi T_R)^3 {\Gamma(h_L)^2 \over \Gamma(2 h_L)}
    (1 + 4 \lambda^2) \ .
  }
 The axion of course yields the same result.

The minimal fermions clearly have the same cross-section, so let us
consider only $\Lambda_3$.  Let the incoming wave be
$\Lambda_{3\alpha} = u_\alpha e^{-i p \cdot x}$.  With $\O(t,x) =
u_\alpha T_{F-}^\alpha(x^-) F_+(x^+)$, the analysis leading to
\SigmaForm\ goes through as usual, yielding
  \eqn{CEval}{
   C_\O = \bar{u}_{\dot\alpha} \delta_\alpha^\beta u_\beta 
      {c C_F \over 3}  
    = \bar{u}_{\dot\alpha} \sqrt{2} \sigma^{0\dot\alpha\beta} u_\beta \,
       2 C_F \ .
  }
 In the second equality the equivalence of lower undotted and upper
dotted indices has again been used.  Recall that $\sqrt{2}
\sigma^{0\dot\alpha\beta}$ is indeed the identity matrix.  The flux is
${\cal F} = 4 \bar{u}_{\dot\alpha} \sqrt{2} \sigma^{0\dot\alpha\beta}
u_\beta$.  (As for the dilaton, the $4$ here is due to the
non-canonical normalization of the fermion field: see the footnote at
the beginning of section~\ref{Fermions}.)  Now \SigmaForm\
can be used again to give
  \eqn{FermCross}{
   \sigma_{\rm abs} = {\pi L C_F \over 16} (2 \pi T_L)^{2 h_L - 1}
    (2 \pi T_R)^2 {\Gamma(h_L)^2 \over \Gamma(2 h_L)}
    (1 + 16 \lambda^2) \ .
  }
 The effective string cross-sections \DilCross\ and \FermCross\ stand
in the same ratio as the semi-classical cross-sections for fixed
scalars and minimal fermions quoted in \AFProbs.

Of course, it would be highly desirable to carry out calculations
similar to the ones presented here for the black holes in five
dimensions that can be modeled using the D1-brane D5-brane bound
state.  There one can hope that an understanding of the soliton
picture can fix overall normalizations; but again one expects the
residual supersymmetry to fix relative normalizations between
fermionic and bosonic cross-sections.

\section{Conclusion}
\label{ConclusionDyad}

One of the main technical results of this chapter has been to show that
when the equations of motion for photons or fermions are simple enough
to be analyzed by the dyadic index methods of
\cite{teuk,teukI,Churil}, they lead to ordinary differential equations
whose near-horizon form is hypergeometric.  That fact alone gives
their low-energy absorption cross-sections a form which is capable of
explanation in the effective string description.

Dyadic index methods are not essential to the analysis of minimally
coupled fermions; indeed, the Weyl equation for fermions has been
analyzed recently in \cite{dgm} in arbitrary dimensions using more
conventional techniques.  However, the dyadic index method provides an
efficient, unified treatment of minimal fermions and minimal photons.
Indeed, the photon radial equations \ExactExtremeI\ and \ExNonEx\ seem
difficult to derive by other means.  The existence of minimally
coupled photons seems to depend essentially on the equal charge
condition, which makes the dilaton background constant.  Minimally
coupled fermions, as suggested in section~\ref{Fermions}, may be more
common because their existence depends on the vanishing of their
supersymmetry variations in the black hole background.

The greybody factors computed in \NonExProb\ and \AFProbs\ are
polynomials in the energy $\omega$ rather than quotients of gamma
functions as found in \cite{mast}.  This is characteristic of an
effective string whose left-moving temperature $T_L$ is much greater
than $T_R$ and $\omega$.  These polynomial greybody factors are
sufficient to determine the conformal dimension of the right-moving
factor in the operator through which a field couples to the effective
string, and the number of derivatives in that coupling.  For example,
the minimal photon couples through its field strength times a $h_R=2$
operator.  But \NonExProb\ and \AFProbs\ do not yield any information
regarding the left-movers.  To see the effects of left-movers, one
might try to generalize the present treatment to black holes far from
extremality, as was done recently in \cite{clTwo} for minimally
coupled scalars.

The absence of a string soliton description of the equal charge black
hole in pure $d=4$, ${\cal N} = 4$ supergravity precludes a precise comparison
of cross-sections between the effective string and semi-classical
descriptions.  However, by assuming that the effective string
world-sheet theory is a $(0,4)$ super-conformal field theory whose
right-moving $R$-symmetry group, $SU(2)$, is identified with the group
of spatial rotations, it has been possible to show that the relative
normalizations of the dilaton, axion, and minimally coupled fermion
cross-sections are correctly predicted by the effective string.  The
proposed couplings of these fields to the effective string are
simple: the scalars couple to the stress-energy tensor while the
fermions couple to the supercurrent.  One would expect that it
possible to extend this picture to a manifestly supersymmetric
specification of how all the massless bulk fields couple to the
effective string at linear order.

There is a simple point which nevertheless is worth emphasizing: the
cross-sections of the dilaton, axion, and minimal fermions are related
by supersymmetry despite their different energy dependence.  The
energy dependence (also known as the greybody factor) arises from
finite-temperature kinematics.  Unsurprisingly, the kinematic factors
are different for particles of different spin; but their form turns
out to be fixed by the conformal dimension of the field by which a
field couples to the effective string.  Supersymmetry acts on the
$S$-matrix, relating the coefficients we have called $C_\O$ in
section~\ref{EffStr}.  The predictions of supersymmetry regarding the
absorption cross-sections of different particles in the same multiplet
thus have more to do with the relative normalization than the energy
dependence.

\newpage


\section*{Appendix: Dyadic index conventions}

This appendix presents in a pedestrian fashion the aspects of the
Newman-Penrose formalism relevant to the rest of the chapter.  A
readable introduction can be found in \cite{Wald}; for an authoritative
treatment the reader is referred to \cite{pr}.

Sign conventions vary by author, and the ones used here are as close
as possible to those of the original paper by Newman and Penrose
\cite{np} and to those of Teukolsky \cite{teuk,teukI}.  First 
consider flat Minkowski spacetime with
mostly minus metric, $\eta_{ab} =
\diag(1,-1,-1,-1)$.  The conventions on raising and lowering spinor
indices are those of ``northwest contraction:''
  \eqn{NW}{\vcenter{\openup1\jot
    \halign{\strut\span\TL & \span\TR & \span\TT & \span\TL & \span\TR\cr
   \psi^\alpha &= \epsilon^{\alpha\beta} \psi_\beta &\qquad&
    \psi_\alpha &= \psi^\beta \epsilon_{\beta\alpha} \cr
   \bar\psi^{\dot\alpha} &= 
     \epsilon^{\dot\alpha\dot\beta} \bar\psi_{\dot\beta} &\qquad&
    \bar\psi_{\dot\alpha} &= 
     \bar\psi^{\dot\beta} \epsilon_{\dot\beta\dot\alpha} \cr
  }}}
 where the sign of the antisymmetric tensors is fixed by
  $\epsilon_{01} = \epsilon^{01} = 
   \epsilon_{\dot{0}\dot{1}} = \epsilon^{\dot{0}\dot{1}} = 1$.
 In flat space, the conventional choice of the matrices 
$\sigma^a_{\alpha\dot\beta}$ which map bispinors to vectors is 
  \eqn{sigmas}{
   \sigma^a = {1 \over \sqrt{2}} 
    \left( 1,\tau_3,\tau_1,-\tau_2 \right) 
  }
 where the matrices $\tau_i$ are the standard Pauli matrices.  
Vector indices are interchanged with pairs of spinor indices using 
the formulae
  \eqn{VectorSpinor}{
   v^a = \sigma^a_{\alpha\dot\alpha} v^{\alpha\dot\alpha} \qquad
   v^{\alpha\dot\alpha} = \sigma_a^{\alpha\dot\alpha} v_a 
  } 
 where 
  $\sigma_a^{\alpha\dot\alpha} = \eta_{ab} \epsilon^{\alpha\beta}
   \epsilon^{\dot\alpha\dot\beta} \sigma^b_{\beta\dot\beta}$,
 consistent with our northwest contraction rules.  
 There is no need to define matrices $\bar\sigma^{a\dot\alpha\beta}$. 
The metric has a simple form when written with spinor indices:
  \eqn{MetricForm}{
   \eta_{ab} \sigma^a_{\alpha\dot\alpha} \sigma^b_{\beta\dot\beta} =
    \epsilon_{\alpha\beta} \epsilon_{\dot\alpha\dot\beta} \ .
  }

In curved spacetime, the metric $g_{\mu\nu}$ is again chosen with
${+}{-}{-}{-}$ signature.  In this chapter, the metric is always of the
form
  \eqn{DiagonalMetric}{
   ds^2 = e^{2A(r)} dt^2 - e^{2B(r)} dr^2 - 
    e^{2C(r)} \left( d\theta^2 + \sin^2 \theta d\phi^2 \right) \ .
  }
 It will turn out to be useful to define not only the standard
diagonal vierbein,
  $$e_\mu^a = \diag(\sqrt{g_{tt}},\sqrt{-g_{rr}},\sqrt{-g_{\theta\theta}},
   \sqrt{-g_{\phi\phi}}) \ ,$$
 but also a complex null tetrad\footnote{In the literature it is common to
see factors of $g_{tt}$ included in the definitions of $\ell^\mu$ and
$n^\mu$ so that seven rather than six of the spin coefficients vanish.
This however complicates the time-reversal properties of the
solutions.}
  \eqn{ComplexNullTetrad}{\vcenter{\openup1\jot
    \halign{\strut\span\TL & \span\TR & \span\TT & \span\TL &
     \span\TR\cr
   \ell^\mu &= {e^\mu_t + e^\mu_r \over \sqrt{2}} &\qquad&
    n^\mu &= {e^\mu_t - e^\mu_r \over \sqrt{2}} \cr
   m^\mu &= {e^\mu_\theta + i e^\mu_\phi \over \sqrt{2}} &\qquad&
    \bar{m}^\mu &= {e^\mu_\theta - i e^\mu_\phi \over \sqrt{2}} \ . \cr
  }}}
 One of the conveniences of working with spinors is that a spinor is a 
sort of square root of a null vector: for any spinor $\psi^\alpha$, 
  $v^\mu = e^\mu_a \sigma^a_{\alpha\dot\alpha} 
   \psi^\alpha \bar\psi^{\dot\alpha}$ 
is a null vector, and any null vector can be written in this form.  It is
important to note that in the context of the Newman-Penrose formalism,
spinor components are ordinary commuting numbers, not Grassmann numbers.
It is possible to introduce a basis $(\omicron^\alpha,\iota^\alpha)$ 
for spinor space with the properties
  \eqn{SpinorBasis}{\vcenter{\openup1\jot
    \halign{\strut\span\TC\cr
   \omicron_\alpha \iota^\alpha = 1 \cr
   \ell^{\alpha\dot\alpha} = \omicron^\alpha \bar\omicron^{\dot\alpha}
    \quad
   n^{\alpha\dot\alpha} = \iota^\alpha \bar\iota^{\dot\alpha} \quad
   m^{\alpha\dot\alpha} = \omicron^\alpha \bar\iota^{\dot\alpha} \quad
   \bar{m}^{\alpha\dot\alpha} = \iota^\alpha \bar\omicron^{\dot\alpha}
    \ . \cr
  }}}
 A particular choice of $(\omicron^\alpha,\iota^\alpha)$ is
  \eqn{OmicronIota}{
   \omicron^\alpha = \pmatrix{1 \cr 0 } \qquad
   \iota^\alpha = \pmatrix{0 \cr 1 } \ .
  }
 Dyadic indices are introduced by defining 
  $\xi_0^\alpha = \omicron^\alpha$,
  $\xi_1^\alpha = \iota^\alpha$   
 and writing $\psi_\Gamma$ for the components of the spinor
$\psi_\alpha$ with respect to the basis $\xi_\Gamma^\alpha$:
  \eqn{DyadSum}{
   \psi_\Gamma = \xi_\Gamma^\alpha \psi_\alpha \qquad
   \psi_\alpha = -\xi^\Gamma_\alpha \psi_\Gamma \ .
  }
 The minus sign in the second equation is the result of insisting on 
the same raising and lowering conventions for dyadic indices as for 
spinor indices: 
  $\xi^\Gamma_\alpha = \epsilon^{\Gamma\Delta} \xi_\Delta^\beta 
   \epsilon_{\beta\alpha}$.

It is a familiar story \cite{bd} how the minimal $SO(3,1)$
connection $\omega_\mu{}^a{}_b$ on the local Lorentz bundle is induced 
from the Christoffel connection: one defines
  \eqn{SpinConnection}{
   \omega_\mu{}^a{}_b = e^a_\nu \partial_\mu e^b_\nu + 
    e^a_\nu \Gamma^\nu_{\mu\rho} e_b^\rho 
  }
 so that 
  \eqn{Consist}{
   \nabla_\mu v^a = \partial_\mu v^a + \omega_\mu{}^a{}_b v^b
     = e^a_\nu \nabla_\mu v^\nu 
     = e^a_\nu (\partial_\mu v^\nu + \Gamma^\nu_{\mu\rho} v^\rho) \ .
  }
 The Newman-Penrose spin coefficients are defined in an exactly
analogous way.  In fact, they are merely special (complex) linear
combinations of the $\omega_\mu{}^a{}_b$.  It is conventional in the
literature to make dyadic indices ``neutral'' under the covariant
derivative $\nabla_\mu$: $\nabla_\mu v_\Gamma = \partial_\mu v_\Gamma$.  The
covariant derivatives of spinors, by contrast, are defined using the
connection induced from $\omega_\mu{}^a{}_b$.  It is convenient to
define a ``completely covariant'' derivative $D_\mu$ and a connection
$\gamma_\mu{}^\Sigma{}_\Gamma$ with the defining properties
  \eqn{DerDef}{
   D_\mu \psi_\Gamma = 
    \partial_\mu \psi_\Gamma - \psi_\Sigma \gamma_\mu{}^\Sigma{}_\Gamma = 
    \xi_\Gamma^\alpha \nabla_\mu \psi_\alpha \ . 
  }
 A brief way of characterizing the covariant derivative is to say that
under $\nabla_\mu$, the quantities $g_{\mu\nu}$, $\eta_{ab}$,
$\epsilon_{\alpha\beta}$, $e^a_\mu$, and $\sigma^a_{\alpha\dot\alpha}$
(together with their alternative incarnations $\eta^{ab}$,
$\epsilon^{\dot\alpha\dot\beta}$, etc.) are covariantly constant.
Under $D_\mu$, the quantities $\xi_\Gamma^\alpha$ are covariantly
constant as well.  From \DerDef\ it is immediate that\footnote{Note
that the choice of sign for $\gamma_{\Delta\dot\Delta\Gamma\Sigma}$
follows the convention of \cite{teuk} and \cite{np} rather than of
\cite{Wald}.}
  \eqn{SpinCoefDef}{
   \gamma_{m\Sigma\Gamma} = 
    -\xi_{\Gamma\alpha} \nabla_\mu \xi_\Sigma^\alpha \ . 
  }
 The quantities 
  $\gamma_{\Delta\dot\Delta\Gamma\Sigma} = 
   \sigma^\mu_{\Delta\dot\Delta} \gamma_{m\Gamma\Sigma}$
 are the spin coefficients.  They have the symmetry 
  $\gamma_{\Delta\dot\Delta\Gamma\Sigma} = 
   \gamma_{\Delta\dot\Delta\Sigma\Gamma}$.
 With twelve independent complex components they represent the same
information as the forty real components of the Christoffel connection
$\Gamma^\mu_{\nu\rho}$.  It is useful to note that
  \eqn{SigmaDyad}{
   \sigma^\mu_{\Delta\dot\Delta} = 
    e^\mu_a \xi_\Delta^\alpha \bar{\xi}_{\dot\Delta}^{\dot\alpha}
     \sigma^a_{\alpha\dot\alpha} = 
    \pmatrix{ \ell^\mu & m^\mu \cr
              \bar{m}^\mu & n^\mu } \ .
  }
 A useful formula for calculating the spin coefficients can be given 
in terms of $\sigma^\mu_{\Delta\dot\Delta}$:
  \eqn{EasyDef}{
   \gamma_{\Delta\dot\Delta\Gamma\Sigma} = 
    -\tf{1}{2} \xi_\Gamma^\beta \bar\xi_{\dot\Gamma}^{\dot\beta} 
     \sigma^\nu_{\Delta\dot\Delta} \nabla_\nu 
     ( \xi_{\Sigma\beta} \bar\xi^{\dot\Gamma}_{\dot\beta} ) = 
    -\tf{1}{2} \sigma^\mu_{\Gamma\dot\Gamma}
     \sigma^\nu_{\Delta\dot\Delta} \nabla_\nu
     \sigma_{\mu\Sigma}{}^{\dot\Gamma} \ .
  }
 Some further notational definitions are conventional in dyadic index
papers:
  \eqn{DyadNotation}{\vcenter{\openup1\jot
    \halign{\strut\span\TL & \span\TR & \span\TT & \span\TL & \span\TR\cr
   \gamma_{0\dot{0}\Gamma\Sigma} &= 
    \pmatrix{ \kappa & \epsilon \cr \epsilon & \pi } &\qquad&
   \gamma_{0\dot{1}\Gamma\Sigma} &=
    \pmatrix{ \sigma & \beta \cr \beta & \mu } \cr
   \gamma_{1\dot{0}\Gamma\Sigma} &=
    \pmatrix{ \rho & \alpha \cr \alpha & \lambda } &\qquad&
   \gamma_{1\dot{1}\Gamma\Sigma} &=
    \pmatrix{ \tau & \gamma \cr \gamma & \nu } \cr
  }}}
  \eqn{DerNotation}{
   D = \ell^\mu \nabla_\mu \qquad \Delta = n^\mu \nabla_\mu \qquad
   \delta = m^\mu \nabla_\mu \qquad \bar\delta = \bar{m}^\mu \nabla_\mu \ .
  }
 For the metric \DiagonalMetric, one finds
  \eqn{DiagonalSC}{\eqalign{
    \kappa &= \pi = \sigma = \lambda = \tau = \nu = 0 \cr
    \epsilon &= \gamma = {e^{-B} A' \over 2 \sqrt{2}} \qquad
      \beta = -\alpha = {e^{-C} \cot \theta \over 2 \sqrt{2}} \qquad
      \mu = \rho = -{e^{-B} C' \over \sqrt{2}} 
  }}
 where primes denote derivatives with respect to $r$.  \DiagonalSC\
represents a remarkably economical way of describing the connection of
an arbitrary spherically symmetric spacetime: there are only three
independent nonzero spin coefficients, and they are real.

\def\vep{\varepsilon}
\def\ra{\rightarrow}
\def\vp{{\bf p}}
\def\al{\alpha}
\def\ab{\bar{\alpha}}
\def \bi{\bibitem}
\def \ep{\epsilon}
\def\D{\Delta}
\def \om {\omega}
\def\LL{\td \l}
\def \do {\dot}
\def\H {{\cal H}}
\def \B {{\cal B}}
\def \ua {\uparrow}
\def \Q {{\hat Q}}
\def \P {{\hat P}}
\def \q {{\hat q}}
\def \bp{{\bar \psi}}

\def \k {\kappa} 
\def \F {{\cal F}}
\def \g {\gamma}
\def \del {\partial}
\def \bd {\bar \partial }
\def \na {\nabla}
\def \const {{\rm const}}
\def \na {\nabla }
\def \D {\Delta}
\def \a {\alpha}
\def \b {\beta}
\def\r {\rho}
\def \s {\sigma}
\def \p {\phi}
\def \m {\mu}
\def \n {\nu}
\def \vp {\varphi }
\def \l {\lambda}
\def \t {\tau}
\def \td {\tilde }
\def \ci {\cite}
\def \sm {$\s$-model }

\def \o {\omega}
\def \inv {^{-1}}
\def \ov {\over }
\def \four{{\textstyle{1\over 4}}}
\def \fourth{{{1\over 4}}}
\def \ha {{1\ov 2}}
\def \QQ {{\cal Q}}

\chapter{Absorption by non-dilatonic branes}
\label{ThreeBraneAbs}

\section{Introduction}
\label{IntroNon}

Dirichlet branes provide an elegant embedding of 
Ramond-Ramond charged objects into string theory \cite{JP}.
The D-brane
description of the dynamics of these solitons may be compared 
with corresponding results in the semi-classical low-energy
supergravity. In particular, a counting of degeneracies for
certain intersecting D-branes reproduces, in the limit of large
charges, the Bekenstein-Hawking entropy of the corresponding
geometries \cite{sv,cm}.
Furthermore, calculations of emission and absorption rates for
scalar particles agree
with a simple ``effective string'' model for the dynamics of the
intersection \cite{cm,dmw,dmOne,dmTwo,gkOne,mast,gkTwo,cgkt,kkOne}.
Questions have been raised, however, whether the simplest model
is capable of incorporating all the complexities of black holes physics 
\cite{km,htr,dkt}. Indeed, it is possible that 
the dynamics of intersecting D-branes, which is not yet
fully understood,
cannot be captured by one simple model.

This chapter, based on the paper \cite{gukt}, is concerned with the
simpler configurations which involve parallel D-branes only.  Their
string theoretic description is well understood in terms of
supersymmetric $U(N)$ gauge theory on the world-volume \cite{Witten}.
Perturbative string calculations of scattering \cite{us,shkm,gmm} and
absorption \cite{HK} are fairly straightforward for the parallel
D-branes, and their low-energy dynamics is summarized in the DBI
action.

To leading order in the string coupling, $N$ coincident
D$p$-branes are described by $O(N^2)$ free fields in $p+1$ dimensions.
This result may be compared with the Bekenstein-Hawking entropy
of the near-extremal $p$-brane solutions in supergravity.
In \cite{gkp,ktPrime} it was found the the scaling of the 
Bekenstein-Hawking entropy with the temperature agrees
with that for a massless gas in $p$ dimensions only for the
non-dilatonic $p$-branes.  The only representative of this class
which is described by parallel D-branes is the self-dual three-brane.
Other representatives include the dyonic string in $D=6$, and
the two-branes and five-branes of M-theory.
In \cite{hp} a way of reconciling the differing scalings 
for the dilatonic branes \cite{Duff} was proposed.\footnote{
Other ideas on how to find
agreement for the dilatonic branes 
were put forward in \cite{pope}.}
Nevertheless, in \cite{kleb} it was shown that the non-dilatonic
branes (and especially the self-dual three-brane) have a number of
special properties that allow for more detailed comparisons between
semi-classical gravity and the microscopic theory.
For example, the string theoretic calculation of the absorption 
cross-section by three-branes
for low-energy dilatons was found to agree {\it exactly}
with the classical calculation in the background of the extremal
classical geometry \cite{kleb}.

What are the features that make the three-brane so special? 
In classical supergravity, the extremal three-brane is the only
RR-charged solution that is perfectly non-singular \cite{dlu,ght,dgt}.
For $N$ three-branes, the curvature of the classical solution is bounded
by a quantity of order
$$ {1\over \sqrt{N\kappa_{10}}} \sim {1\over \alpha'\sqrt{N g_{\rm str}}}
\ .$$ 
Thus, to suppress the string scale corrections to the classical metric,
we need to take the limit $N g_{\rm str}\rightarrow\infty $.
This fact seems to lead to a strongly coupled theory on the world-volume
and raises questions about the applicability of string perturbation
theory to macroscopic three-branes. However, in \cite{kleb} it was shown
that the dimensionless expansion parameter that enters the
string theoretic calculation of the
absorption cross-section is actually
\eqn{param}{ N\kappa_{10} \omega^4 \sim N g_{\rm str} \alpha'^2 \omega^4
\ ,}
where $\omega$ is the incident energy. Thus, we may consider
a ``double scaling limit,''
\eqn{dsl}{ N g_{\rm str}\rightarrow\infty\ ,
\qquad \omega^2 \alpha'\rightarrow 0\ ,
}
where the expansion parameter \param\ is kept small.
Moreover, the classical absorption cross-section is naturally
expanded in powers of $\omega^4\times {\rm (curvature)}^{-2}$, which
is the same expansion parameter \param\ as the one governing the
string theoretic description of the three-branes. The two expansions
of the cross-section thus may indeed be compared, and 
the leading term agrees exactly \cite{kleb}. This 
provides strong evidence in favor of
absorption by extremal three-branes being a unitary process.
While in the classical calculation the information carried by the
dilaton seems to disappear down the infinite throat of the classical
solution, the stringy approach indicates that the information is not
lost: it is stored in the quantum state of the back-to-back massless
gauge bosons on the world-volume which are
produced by the dilaton. Subsequent decay of the three-brane
back to the ground state proceeds via annihilation of the gauge
bosons into an outgoing massless state, and there seems to be no
space for information loss.

The scenario mentioned above certainly deserves a more careful
scrutiny. In this chapter, we carry out further comparisons between
string theory and classical gravity of the three-brane.  In string
theory there is a variety of fields that act as minimally coupled
massless scalars with respect to the transverse dimensions.  The
dilaton, which was investigated in \cite{kleb}, is perhaps the
simplest one to study. In this chapter we turn to other such fields:
the RR scalar and the gravitons polarized parallel to the
world-volume.  We show that in classical gravity all these fields
satisfy the same equation and, therefore, are absorbed with the same
rate as the dilaton. However, their cubic couplings to the massless
world-volume modes dictated by string theory \cite{HK} are completely
different. In particular, the longitudinally polarized gravitons can
turn into pairs of gauge bosons, scalars, or fermions.  Adding up
these rates we find that the total cross-section has the same
universal value, in agreement with classical gravity.

Another interesting check concerns the absorption of scalars in 
partial waves higher than $\ell=0$. In \cite{kleb} it was shown that the
scaling of the relevant cross-sections with $\omega$ and $N$ agrees
for all $\ell$. For $\ell=1$ we show that the coefficient agrees as well.
For $\ell>1$ the simplest assumption about the
effective action does not yield a coefficient
which agrees with the classical calculation. Normalization of the
effective action is a subtle matter, however, 
and we suspect that its
direct determination from string amplitudes
will yield agreement with the classical cross-sections.

While a microscopic description of RR-charged branes is
by now well known in string theory, the situation is not as simple
for the $p$-brane solutions of the 11-dimensional supergravity.
There is good evidence that a yet unknown M-theory underlies
their fundamental description. Comparisons with semi-classical gravity
provide consistency checks on this description. 
In \cite{kleb} the scalings of the classical absorption cross-sections
were found to agree with the two-brane and five-brane effective action
considerations. 
Here we calculate the absorption cross-section
of a longitudinally polarized graviton by a single five-brane.
Using the world-volume  effective action, we find that the 
absorption cross-section is $1/4$ of the rate formally predicted
by classical gravity. In fact, one could hardly expect perfect
agreement for a single five-brane -- the classical description is expected
to be valid only for a large number of coincident branes.
It is interesting, nevertheless, how close the two calculations come
to agreeing with each other. We also carry out a similar comparison
for a single two-brane, but find the discrepancy in
the coefficient to be far greater than in the five-brane case.

The structure of this chapter is as follows. In section~\ref{STEff} we
exhibit the spacetime effective actions and derive the classical
equations satisfied by various fields. In section~\ref{SelfThree} we
present the new three-brane calculations. In section~\ref{CompAb} we
carry out the comparisons of classical 11-dimensional supergravity
with the predictions of the M-brane world-volume actions.  We conclude
in section~\ref{Conc}.


\section{Spacetime effective actions and perturbations around
non-dilatonic $p$-brane solutions}
\label{STEff}
\subsection{Perturbations of three-branes }
The bosonic part of the field equations 
of $d=10$ type IIB supergravity, which is the 
low-energy limit of the type IIB superstring \cite{schw},   can be 
derived from the following    action:
  \cite{bho,bbb}
\eqn{efec}{  S_{10} 
= {1\ov 2\k_{10}^2}  \int d^{10} x \bigg[ \sqrt{-G} \big( e^{-2\p} [ R + 4 (\del \p)^2
- { \textstyle{1\ov 12}} (\del B_2)^2 ] 
} 
$$ -  \  { \textstyle{1\ov 2}} (\del C)^2 
 - { \textstyle{1\ov 12}} (\del C_2  - C  \del B_2) ^2
- { \textstyle{1 \ov 4\cdot 5!}}  F^2_5 \big)
- { \textstyle{1\ov 2\cdot 4! \cdot (3!)^2    }}
 {\ep_{10}} C_4 \del C_2 \del B_2 + ... \bigg] \ ,  $$
where\footnote{We use the following 
notation.  The signature is $(-+...+)$. 
$M,N,...$ label the  coordinate 
indices of $d=10$ or $d=11$ theory.
 The indices $\a,\b,...$ ($a,b,..$)
  will label the  spacetime  (spatial) coordinates parallel
to the $p$-brane  world-volume, i.e.{} $\a =(0,a), \ a=1,...,p$;  
  the indices $i,j,...,$   
will label the coordinates transverse
to  the $p$-brane, $i=p+1,...,9 (10)$. We  shall  also use
$\m,\n,...$ for the indices of the dimensionally reduced
theory obtained by compactifying the internal directions
parallel to the  $p$-brane, 
i.e.{} $\m= (0,i)$. $\ep_d$ will  stand for the totally antisymmetric
 symbol (density). Contractions over repeated 
lower-case indices are always performed with the flat  metric, 
and $A_{[ab]} \equiv  \ha (A_{ab} - A_{ba})$.}
$$(\del B_2)_{MNK} \equiv 3 \del_{[M} B_{NK]}\ , \ \ \ 
\  \ (\del C_4)_{MNKLP} \equiv 5 \del_{[M} C_{NKLP]} \ , $$ 
$$ F_5= \del C_4 + {5} (B_2 \del C_2 - C_2 \del B_2) \  . $$
Following \cite{bbb} we assume   that the self-duality constraint 
 \ $F_5= \tilde{F}_5$
may be  added  at the level of the equations of motion.
This action is a useful tool for deriving the dimensionally reduced 
forms of type IIB supergravity action  \cite{bbb} and also 
for discussing perturbations near the solitonic $p$-brane 
solutions given below.
Written in the Einstein frame ($g_{MN} = e^{-\p/2} G_{MN}$)
it takes the following $SL(2,R)$ covariant form: 
\eqn{efc}{  S_{ 10} 
= {1\ov 2\k_{10}^2}  \int d^{10} x \bigg( \sqrt{-g_{10}} \big[ \ R 
 - { \textstyle{1\ov 2}} (\del \p)^2
- { \textstyle{1\ov 12}} e^{-\p}   (\del B_2)^2 
} 
$$ -  \  { \textstyle{1\ov 2}}  e^{2 \p} (\del C)^2 
 - { \textstyle{1\ov 12}} e^{ \p}  (\del C_2  - C  \del B_2) ^2
- { \textstyle{1 \ov 4\cdot 5!}}  F^2_5\ \big]
- { \textstyle{1\ov 2\cdot 4! \cdot (3!)^2    }}
 {\ep_{10}} C_4 \del C_2 \del B_2 + ... \bigg) \ . $$
The extremal three-brane of type IIB theory is represented
by the background \cite{hsTwo,dlu}
\eqn{tree}{
ds^2_{10 } = H^{-1/2}(-dt^2 + dx_a dx_a) + H^{1/2} dx_i dx_i  \ , }
\eqn{tre}{ 
(\del C_4)_{abcdi} = \ep_{abcd} \del_i H\inv \ , \ \ \ \ 
(\del C_4)_{ijklm} = \ep_{ijklmn}\del_n H \ , \ \ \ \  
H= 1 + {R^4\ov r^4} \ , \ \ r^2=x_ix_i\ , }
with all other  components  and  fields  vanishing (i.e.{} $\p=C=B_2=C_2=0$, \ 
$F_5= \del C_4$).
The scalar curvature $\sim (\del C_4)^2  \sim R^{-2}$ near $r=0$,
In fact, this metric may be extended to a geodesically complete
 non-singular geometry \cite{ght,dgt}. This is the only RR-charged
$p$-brane for which this is possible.

Let us consider small perturbations near this background
which depend only on $x_\m$, i.e.{}  on the 
time $t$  and the transverse coordinates $x_i$.
Our aim is to identify the modes which have simple 
Klein-Gordon type equations.  A  guiding principle
is to look  for  (components of) the fields
that have trivial background values. 

The obvious examples are the  
dilaton $\p$ 
and  the Ramond-Ramond scalar $C$ perturbations which  are  decoupled
from the 
$C_4$-background and thus  have the action ($\vp=(\p,C)$) 
\eqn{scal}{ S_{scal.} =- {1\ov 4\k_{10}^2} 
 \int d^{10} x  \sqrt{-g_{10}}\  g^{\m\n} \del_\m \vp \del_\n \vp 
= - {1\ov 4\k_{10}^2} 
 \int d^{10} x  \big[
   \del_i \vp \del_i \vp  -  H(r)  \del_0 \vp \del_0 \vp  \big]  , }
where we have used the fact that for the three-brane background 
the Einstein and the string frame
metrics are identical, $G_{MN}=g_{MN}$,
which implies 
  $$ g_{\m\n} =
\diag (-H^{-1/2}, H^{1/2}\delta_{ij} ) \ , \qquad\qquad
  \sqrt{-g_{10}}= H^{1/2} \ .$$

Perturbations of the metric are, in general, mixed with
perturbations of components of  $C_4$.  An
important  exception is 
the traceless part of the longitudinal (polarized along the
three-brane) graviton  perturbations 
$h_{ab}$. These perturbations give rise to scalars upon
dimensional reduction to $d=7$. 
Let us  split the metric in the  ``7+3''  fashion
$$ ds^2_{10E}= g_{\m\n} dx^\m dx^\n + g_{ab} dx^a dx^b \ ,$$
and assume that all  the fields depend only on $x^\m$
(this is equivalent to  reduction to 7 dimensions).
Then the relevant part of \efc\ becomes 
\eqn{efic}{  S_{10} 
= {1\ov 2\k_{10}^2}  \int d^{10} x  \sqrt{-g_7} \sqrt {g_3}  
\big( R_7 
 - { \textstyle{1\ov 4}} g^{ab} g^{cd} 
 \del_\m  g_{ac}\del^\m  g_{bd}
- { \textstyle{1 \ov 48}}  F_{\m \n abc} F^{\m \n}_{ \ \  \  def} 
g^{ad} g^{be}  g^{cf}  + ... \big) , }
where $g_7 \equiv \det g_{\m\n}, 
\ g_3\equiv \det {g_{ab}}$ 
and we have written down explicitly the only 
$F^2_5$ term that could potentially couple
$h_{ab}$ to the gauge field strength
background.  Introducing 
the ``normalized'' metric $\g_{ab} =  g_3^{-1/3} g_{ab}$
which has unit determinant, we find
\eqn{ecf}{  S_{scal.grav.} 
= {1\ov 2\k_{10}^2}  \int d^{10} x  \sqrt{-g_7} \sqrt{ g_3 } 
\big( R_7 
 - { \textstyle{1\ov 4}} \g^{ab} \g^{cd} 
 \del_\m  \g_{ac}\del^\m  \g_{bd}   }
$$ 
- \  { \textstyle{1\ov 12}} g_3^{-2}  \del_\m g_3 \del^\m g_3 
- { \textstyle{1 \ov 8}}  g_3^{-1}
 F_{\m \n} F^{\m \n} + ... \big) \ , $$
where $F_{\m\n} = F_{\m\n 123}$ is the $d=7$ vector field strength
which, according to \tre, describes   an  electrically
charged extremal black hole. Thus, only
the determinant of $g_{ab}$, which is related to
the 7-dimensional dilaton, 
couples to the gauge field 
background. 
This  is consistent with  
the fact that $\g_{ab}$ has a trivial background value,
$\g_{ab} =\delta_{ab}$,  while the value of $\det g_{ab}$ in \tree\ is 
\  $g_3= H^{-3/2}$. 

The fluctuations  $h_{ab}  = \g_{ab}- \delta_{ab}$ 
(which are traceless, $h_{aa}= 0$, to keep $\det \g_{ab}=1$) 
 thus have the 
same  quadratic part of the action as the scalar fields in \scal\
\eqn{ligr}{ S_{scal.grav.} =- {1\ov 8\k_{10}^2} 
 \int d^{10} x  \sqrt{-g_{10}}\  \del_\m h_{ab} \del^\m h_{ab} } $$
=  - {1\ov 8 \k_{10}^2} 
 \int d^{10} x  \big[
   \del_i h_{ab} \del_i h_{ab}  -  H(r)  \del_0 h_{ab} \del_0 h_{ab}  \big]  \ . $$
Similar conclusions hold for other $p$-brane solutions because
the ``normalized'' internal space
metric  $\g_{ab}$ has a flat background value.

\subsection{Perturbations of two-branes and five-branes}
The  discussion of perturbations around the two-brane and five-brane solutions
 \cite{ds,guv}
of the $d=11$ supergravity  is very similar to the analysis in the
preceding subsection. 
The starting point is the bosonic  part of the  
$d=11$ supergravity action \cite{cjs} 
\eqn{eleva}{  S_{11} 
= {1\ov 2\k_{11}^2}  \int d^{11} x \bigg( \sqrt{-g_{11}} \big[ R 
 - { \textstyle{1\ov 2\cdot 4!}} (\del C_3)^2 \big] 
+  { \textstyle{1\ov (12)^4 }}  {\ep_{11}} C_3 \del C_3 \del C_3 + ... \bigg) \ , }
where 
$
(\del C_3)_{MNKL} = 4 \del_{[M} C_{NKL]}$. 
The two-brane and five-brane
backgrounds  are respectively ($a=1,...,p, \ i=p+1,...,10$, \ $p=2,5$)
\eqn{twe}{
ds^2_{11}  = H^{-2/3}(-dt^2 + dx_a dx_a) + H^{1/3} dx_i dx_i  \ , }
$$ 
(\del C_3)_{0abn} = \ep_{ab}\del_n H\inv  \ , \ \ \ \  
H= 1 + {R^6\ov r^6} \ ,$$
\eqn{five}{
ds^2_{11}  = H^{-1/3}(-dt^2 + dx_a dx_a) + H^{2/3} dx_i dx_i  \ , }
$$
(\del C_3)_{ijkl} = \ep_{ijkln}\del_n H \ , \ \ \ \  
H= 1 + {R^3\ov r^3} \ .$$
Both for the two-brane and for the five-brane, the
scalar curvature is $\sim (\del C_3)^2  \sim R^{-2}$ near $r=0$.
There is a subtle difference, however, in that the five-brane
metric may be extended to a geodesically complete non-singular
geometry, while the two-brane metric  cannot \cite{ght,dgt}.

We consider small perturbations near these backgrounds
which depend only on $x_\m=(t,x_i)$.
There are no scalars like $\p$ or $C$ in the $d=11$ theory, but 
it is easy to check that,  as  in \efic,\ecf,  the
``normalized'' internal part of the metric 
 $\g_{ab} = g_p^{-1/p} g_{ab}, \   g_p\equiv \det g_{ab}$
($p=2,5$),   which has a flat background value, 
 is decoupled from the $(\del C_3)^2$
term. As a result, the gravitons polarized parallel to the brane,
$h_{ab} = \g_{ab} - \delta_{ab}, \ h_{aa}=0,$
have the minimal action  as in \ligr, 
\eqn{lgr}{ S_{scal.grav.} =- {1\ov 8 \k_{11}^2} 
 \int d^{11} x  \big[
  \del_i h_{ab} \del_i h_{ab}  -  H(r)  \del_0 h_{ab} 
\del_0 h_{ab}  \big] \ . }

\section{The self-dual three-brane}
\label{SelfThree}

\subsection{Classical absorption}

This section focuses on minimally coupled scalars, such as
those that satisfy the Klein-Gordon equation following from \scal, \ligr\
with $H(r)$ given in  \tre. 
Because the classical analysis of absorption 
for all partial waves of such a
scalar was sketched in \cite{kleb} and is similar to the matching
calculations that have appeared in many other places in the literature
\cite{dmw,dmOne,dmTwo,gkOne,mast,gkTwo,cgkt,kkOne,unruh,ageTwo,dgm,kr},
details will not be presented here.
In order to employ analytical rather than numerical techniques, it is
necessary as usual to take the Compton wavelength much larger than
the typical radii of the black hole.  Fortunately, it is precisely in
this region where agreement with string theoretic models is
expected~\cite{juanLow}.

Recall from chapter~\ref{PartialWaves} that in $d$ spacetime
dimensions, the absorption cross-section $\sigma^\ell_{\rm abs}$ of a
massless scalar with energy $\omega$ in the $\ell$-th partial wave is
related to the absorption probability $1 - |S_\ell|^2$ by
  \eqn{SigmaAbsGen}{
   \sigma^\ell_{\rm abs} = {2^{d-3} \pi^{(d-3)/2} \over \omega^{d-2}}
    \Gamma((d-3)/2) (\ell + (d-3)/2) {\ell+d-4 \choose \ell} 
    \left( 1 - |S_\ell|^2 \right) \ .
  }
 For the three-brane one takes $d=7$ since the other $3$~dimensions are
compactified on $T^3$.  
A matching calculation outlined for all partial waves in
\cite{kleb} gives the absorption probability
$$ 1 - |S_\ell|^2= {\pi^2 (\omega R)^{8+4\ell}\over
[(\ell+1)!]^4 (\ell+2)^2 4^{2\ell+2}}
\ .$$
Therefore, the classical
result for the cross-section to absorb the $\ell$-th partial
wave of a minimally coupled scalar 
is, to leading order in $(\omega R)^4$,
  \eqn{SigmaGRTwo}{
   \sigma_{3{\rm \ class.}}^\ell = {\pi^4 \over 24} 
    {(\ell+3) (\ell+1) \over [(\ell+1)!]^4} 
    \left( {\omega R \over 2} \right)^{4\ell} \omega^3 R^8 \ . 
  }
The  scale  parameter $R$  of the classical three-brane solution \tre\ 
is related \cite{gkp,kleb} to the number $N$ of coinciding microscopic 
three-branes
by the equation 
\eqn{chaq}{ R^4 = {\kappa_{10}\over 2\pi^{5/2}} N \ , 
}
which follows from the quantization of the three-brane charge.

There is a number of minimally coupled scalars in the theory: the
dilaton, the RR scalar, and off-diagonal gravitons polarized with both
indices parallel to the three-brane world-volume.  The goal of the next
section will be to demonstrate that the universality of leading order
cross-section for these scalars, which is so obvious in the
semi-classical framework, also follows from the D-brane description.

\subsection{Universality of the absorption cross-section
for minimally coupled scalars}

The three-brane is the case where we know the world-volume theory the
best: at low energies (and in flat space) 
 it is ${\cal N}=4$ supersymmetric $U(N)$ gauge theory 
where $N$ is the number of parallel three-branes 
\cite{Witten}.  Thus, the massless fields on the
world-volume are the gauge field, 6 scalars, and 4 Majorana fermions,
all in the adjoint representation of $U(N)$. 
As we will see below, the universality of the cross-section is
not trivial in the world-volume description:
while for dilatons and RR scalars leading absorption proceeds by
conversion into a pair of gauge bosons only, for the
gravitons polarized along the brane
it involves a summation over conversions into
world-volume scalars, fermions, and gauge bosons.

The world-volume action, excluding all couplings to external fields, is
  \eqn{NFourS}{
   S_3 = T_3 \int d^4 x \, \tr \left[ -\tf{1}{4} F_{\alpha\beta}^2 + 
    \tf{i}{2}
 \bar\psi^I \gamma^\alpha \partial_\alpha \psi_I - 
    \tf{1}{2}
 (\partial_\alpha X^i)^2 + {\rm interactions} \right] \ .
  }
 The indices $I=1,...,4$ and $i=4,...,9$ label representations of the
R-symmetry group, $SO(6)$.  It is clear from the $d=10$ origin of this
theory that the R-symmetry group is the group of spatial rotations in
the uncompactified dimensions.  Under this group, the gauge fields are
neutral and the scalars $X^i$ form a ${\bf 6}$.  When the fermions are
written in chiral components,
  $
   \psi_I = {\lambda_I \choose \bar\lambda^I}
  $,
 the fields $\lambda_I$ form a ${\bf 4}$ of $SO(6) = SU(4)$.

In order to properly normalize the amplitudes, one needs the kinetic
part of the action for the bulk fields, which is given by
\scal\ and \ligr.

We also need to know how the three-brane
world-volume fields couple to the bulk fields of type~IIB
supergravity.  For a single three-brane,  a $\k$-symmetric version of the
DBI action in a non-trivial type IIB background 
was constructed in \cite{cederOne,cederTwo} (see also \cite{jhs,bt}),
which, by use of superfields, captures 
all such couplings to leading order in derivatives
of external fields. 
For the terms necessary to us, the generalization from $U(1)$ to
$U(N)$ gauge group is straightforward. One can, in principle, 
obtain  detailed  information  about the structure of the non-abelian 
action by directly computing
string amplitudes as done in \cite{HK}.  The three-brane is
well suited to this line of attack because the string theory
description is known exactly and is not
complicated by the difficulties one encounters 
in bound states of intersecting branes, such as
the D1-D5-brane system \cite{aki}.

$SO(6)$ invariance and power counting in the string coupling greatly
restrict the possible couplings between the bulk and world-volume
fields in the interaction part of the action $S_{\rm int}$.  In both
the semi-classical and world-volume computations, the dimensionless
expansion parameter is not $N g_{\rm str}$ but rather $(\omega R)^4$.
A cross-section which involves $n$ powers of this parameter arises
from a coupling of the bulk field to a local operator ${\cal O}$ of
dimension $n+4$.

The leading   bosonic terms in the action for a  single three-brane 
in a type IIB supergravity background  are \cite{dbiOne,dbiTwo,dbiThree,tse}
\eqn{teea}{
 S_3=-T_3 \int d^4 x \bigg(
    \sqrt {-\det (\hat  g  +  e^{-\p/2} \F) }
    +{\textstyle  {1\ov 4!} } \ep^{\a\b\s\r} \hat C_{\a\b\s\r} 
+   \tf{1}{2}    \hat C_{\a\b} \td \F^{\a\b}
+  \four  C   \F_{\a\b}\td  \F^{\a\b} \bigg) 
 \ , }
where 
$$ \F_{\a\b}= F _{\a\b}+ \hat B_{\a\b}\ , \qquad
\td F^{\a\b} = {{\textstyle{1\over 2}}} \ep^{\a\b\s\r}F_{\s\r}\ , \qquad
\hat g_{\a\b} = g_{MN} \del_\a X^M \del_\b X^N \ , {\rm etc.}, 
$$
and the background fields are functions of $X^M$.
In the static gauge ($X^\a = x^\a$, $\a=0,\ldots, 3$) one has 
$$\hat g_{\a\b} = g_{\a\b} + 2g_{i(\a } \del_{\b )} X^i +
 g_{ij } \del_{\a} X^i\del_{\b} X^j 
\ .$$
In flat space the  fermionic terms 
may be found  by replacing  
$\hat  g_{\a\b}  +  e^{-\p/2} \F_{\a\b}  $ by \cite{cederOne,cederTwo,jhs,bt} 
$$\hat  g_{\a\b }  +  e^{-\p/2}[ \F_{\a\b}  -  
2 \bar \Psi (\Gamma_\a + \Gamma_i 
\del_\a X^i)\del_\b  \Psi  + \bar \Psi \Gamma^i \del_\a  \Psi 
\bar \Psi \Gamma_i \del_\b  \Psi]\ , $$
where $\Psi$ 
is  the $d=10$ Majorana-Weyl spinor (which can be split into 
four $d=4$ Majorana spinors $\psi_I$ in \NFourS) 
and $\Gamma_M$ are the $d=10$ Dirac matrices. 

The  leading-order 
interaction of the dilaton with world-volume fields 
implied by \teea\   was discussed 
in \cite{kleb}.  The coupling of the RR scalar $C$  is  
similar, being related by $SL(2,R)$ 
duality.
The coupling of  the gravitons polarized parallel to the brane, 
$h_{\alpha\beta}=g_{\a\b}-\eta_{\a\b}$,
 to the bosonic world-volume fields can be
deduced  by expanding the action \teea.
 At  the leading order  it is  given by 
  $\tf{1}{2} h^{\alpha\beta} T^{\rm bosons}_{\alpha\beta}$
 where $T^{\rm bosons}_{\alpha\beta}$ is  the energy-momentum
tensor  of $A_\a$ and $X^i$. The 
  only possible supersymmetric extension
is for $h_{\alpha\beta}$ to couple to the complete energy-momentum
tensor corresponding to \NFourS.

Generalizing to $U(N)$, we find that
the part of $S_{\rm int}$ that is relevant to the leading-order 
absorption processes we wish to consider is\footnote{We  ignore 
the  fermionic couplings like $ \phi  \bar\psi^I \gamma^\alpha
 \partial_{\a} \psi_I $ and  similar ones for $C$ and $h_{\a\b}$ 
 which are proportional to the 
fermionic equations of motion and thus  
give vanishing contribution to the $S$-matrix elements.}
\eqn{RelSint}{
   S_{\rm int} = T_3 \int d^4 x \, \left[ \tr \left(
 \tf{1}{4}   {\phi} F_{\alpha\beta}^2  -
  \tf{1}{4}   {C} F_{\alpha\beta} \td {F}^{\alpha\beta} \right) +  
     \tf{1}{2} h^{\alpha\beta} T_{\alpha\beta} \right] \ , 
  }
 where 
  \eqn{Tab}{
   T_{\alpha\beta} = \tr \big[ F_{\alpha}^{\ \gamma} F_{\beta\gamma} - 
    \tf{1}{4} \eta_{\alpha\beta} F_{\gamma\delta}^2 -
   \tf{i}{2} 
\bar\psi^I \gamma_{(\alpha} \partial_{\beta)} \psi_I 
  +
\partial_\alpha X^i \partial_\beta X^i  - 
     \tf{1}{2} \eta_{\alpha\beta} (\partial_\gamma X^i)^2
\big] \ .
  }

Let us first consider an off-diagonal graviton
polarized along the brane, say
$h_{xy}$, which is an  example of a traceless
perturbation $h_{ab} $
 whose quadratic action is given in \ligr). {}From \RelSint\ one
 can read off the invariant amplitudes
for absorption into two scalars, two fermions, or two gauge bosons:
  \eqn{InvAmpsOne}{\vcenter{\openup1\jot
   \halign{\strut\span\TT & \span\TL & \span\TR\cr
    scalars:\ \  \ & {\cal M} &= -\sqrt{2 \kappa_{10}^2} 
     (p_{1x} p_{2y} + p_{1y} p_{2x})  \cr
    fermions:\ \ \ & {\cal M} &= -\tf{1}{2} \sqrt{2 \kappa_{10}^2}
     \bar{v}(-p_1) (\gamma_x p_{2y} + \gamma_y p_{2x}) u(p_2) \cr
    gauge bosons:\ \ \ & {\cal M} &= -\sqrt{2 \kappa_{10}^2} 
      (f^{(1)}_x{}^\beta f^{(2)}_{y\beta} + 
      f^{(1)}_y{}^\beta f^{(2)}_{x\beta})  \cr
  }}}
 where $p_1$ and $p_2$ are the momenta of the outgoing particles, and 
the field strength polarization tensors $f^{(s)}_{\alpha\beta}$ are
given by 
  \eqn{fDef}{
   f^{(s)}_{\alpha\beta} = 
    i p_{s\alpha} \epsilon^{(s)}_\beta - 
    i p_{s\beta} \epsilon^{(s)}_\alpha \ .
  }
 Summation over the spins of the outgoing particles can be performed
using
\eqn{SpinSums}{
  \sum_s u_{(s)}(p) \bar{u}_{(s)}(p) = p\!\!\!/ \ , \ \ \ \  \ 
  \sum_s v_{(s)}(p) \bar{v}_{(s)}(p) = -p\!\!\!/ \ , }
$$
  \sum_s \epsilon_\alpha^{(s)}  \epsilon_{\beta}^{(s)*} = 
    \eta_{\alpha\beta} 
\ .$$
 Summing as well over different species of particles available (six
different $X^i$, for example), one obtains
  \eqn{AvAmpSq}{\vcenter{\openup1\jot
   \halign{\strut\span\TT & \span\TL & \span\TR\cr
    scalars:\ \ \ & \overline{|{\cal M}|}^2 &= 
     3 \kappa_{10}^2 \omega^4 n_x^2 n_y^2  \cr
    fermions:\ \ \ & \overline{|{\cal M}|}^2 &= 
     \kappa_{10}^2 \omega^4 (n_x^2 + n_y^2 - 4 n_x^2 n_y^2)  \cr
    gauge bosons:\ \ \ & \overline{|{\cal M}|}^2 &= 
     \kappa_{10}^2 \omega^4 (1 - n_x^2 - n_y^2 + n_x^2 n_y^2) \cr
  }}}
 where $\vec{n}$ is the direction of one of the outgoing particles.
In \AvAmpSq\ we have anticipated conservation of energy and momentum by
setting $\vec{p}_1 + \vec{p}_2 = 0$ and $\omega_1 + \omega_2 =
\omega$. 
 It is remarkable that the sum of these three quantities is
independent of $\vec{n}$.  Thus, if one performs
the spin sums not just over all polarizations and species of particles
of a given spin, but rather over all the states in the ${\cal N}=4$
super-multiplet, the result is isotropic:
  \eqn{TotAmpSq}{
   \overline{|{\cal M}|}^2 = \kappa_{10}^2 \omega^4 \ .
  }
 The absorption cross-section is evaluated from $\overline{|{\cal
M}|}^2$ in precisely the same way that decay rates of massive
particles are calculated in conventional 4-dimensional field theories:
  \eqn{SigmaFormTwo}{
   \sigma_{3{\rm \ abs}} = {N^2 \over 2} {1 \over 2 \omega} 
    \int {d^3 p_1 \over (2 \pi)^3 2 \omega_1}
         {d^3 p_2 \over (2 \pi)^3 2 \omega_2} \,
    (2 \pi)^4 \delta^4 \big( q - {\textstyle \sum\limits_i} p_i \big) \ 
     \overline{|{\cal M}|}^2 \ .
  }
 The leading factor of $N^2$ accounts for the multiple branes; the
$1/2$ is present because the outgoing particles are identical (this is
true also of the fermions because we are working with
Majorana spinors).  The cross-section following from \TotAmpSq\ agrees
with the semi-classical $\ell = 0$ result \SigmaGRTwo:
  \eqn{SigmaGrav}{
   \sigma_{3{\rm\ abs}} = {\kappa_{10}^2  \omega^3 N^2\over 32 \pi} 
     = \sigma_{3{\rm\ class.}} \ ,
  }
where we have used the relation \chaq\ 
between $N$ and $R$.
 
Compared to the off-diagonal graviton, the calculation of the
cross-section for the RR scalar is relatively simple.  Inspection of
the leading order amplitudes for absorption of a dilaton and a RR
scalar makes it  obvious that they have the same cross-section:
  \eqn{InvAmpsTwo}{\vcenter{\openup1\jot
   \halign{\strut\span\TT & \span\TL & \span\TR\cr
    dilaton:\ \ \ & {\cal M} &= \tf{1}{2} \sqrt{2 \kappa_{10}^2} 
     f^{(1)}_{\alpha\beta} f^{(2)\alpha\beta}  \cr
    RR scalar:\ \ \ & {\cal M} &= -\tf{1}{2} \sqrt{2 \kappa_{10}^2}
     f^{(1)}_{\alpha\beta} \tilde{f}^{(2)\alpha\beta} \ .  \cr
  }}}
 To show that $\overline{|{\cal M}|}^2 = \kappa_{10}^2 \omega^2$ in both
cases, it suffices to prove the relation
  \eqn{EpsRel}{
   \sum_{\rm spins}
 f^{(1)}_{\alpha\beta} f^{(1)*}_{\gamma\delta}  
 f^{(2)\alpha\beta} f^{(2)\gamma\delta*} = 
    \sum_{\rm spins} 
 f^{(1)}_{\alpha\beta} f^{(1)\gamma\delta *}  
\tilde{f}^{(2)\alpha\beta} 
     \tilde{f}^{(2)\gamma\delta*} = 8 (p_1 \cdot p_2)^2 \ .
  }
 The verification is straightforward algebra.  The formula \SigmaFormTwo\
applies as written to the RR scalar as well, so the agreement with the 
semi-classical calculation is clear.

\subsection{Higher partial waves}

A more difficult comparison is the absorption cross-section for higher
partial waves.  It is possible to argue, based on power counting and
group theory, that the interaction term suggested in \cite{kleb}, 
  \eqn{LthPartial}{
   S_{\rm int}^\ell = - T_3 \int d^4 x \, \ \tf{1}{4 \cdot \ell!}
    {\partial_{i_1} \cdots \partial_{i_\ell} \phi}
    \tr \left( X^{i_1} \cdots X^{i_\ell} 
      F_{\alpha\beta} F^{\alpha\beta}\right) 
  }
 is the only one that could possibly contribute at leading order to a
given partial wave. It was shown in \cite{kleb} that this term predicts a
cross-section for the $\ell$-th partial wave whose scaling with
$\omega$ and $N$ is in agreement with classical gravity.  Here we show
that, if we use the specific normalization given in \LthPartial, we
obtain precise agreement for $\ell = 0,1$ but disagreement
for higher $\ell$.  Later on we will argue that the disagreement for
$\ell>1$ may be due simply to an incorrect normalization of the
necessary effective action terms.

Let us restrict our attention to the dilaton.  It will be obvious that
all our arguments apply equally well to the RR scalar, and perhaps
with a bit more attention to details to the off-diagonal gravitons.
The absorption cross-section \SigmaGRTwo\ is of order $\kappa_{10}^{\ell
+ 2}$ for the $\ell$-th partial wave.  Processes which can contribute
to this absorption at leading order must be mediated by an operator
${\cal O}$ which involves $\ell + 2$ fields.  All interaction terms
which involve the dilaton must include at least two powers of
$F_{\alpha\beta}$ because of the restricted way in which the dilaton
enters the DBI action.  Since the gauge bosons are neutral under
$SO(6)$, the other $\ell$ fields must be responsible for balancing the
$SO(6)$ transformation properties of the $\ell$-th partial wave.  We
want to argue that the only way this can be done is to use $\ell$
powers of the scalar fields $X^i$.  To do this it is necessary to know
something about the addition rules for representations of $SO(6) =
SU(4)$.  The $SU(4)$ Young tableaux for the fields in question are
  \eqn{YoungT}{
   \hbox{
    \vtop{\hbox{$\lambda_I$:}}\enspace
    \vtop{\null\nointerlineskip\vskip-10pt
          \hbox{\psfig{figure=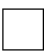}}}\qquad
   \vtop{\hbox{$\bar\lambda^I$:}}\enspace
     \vtop{\null\nointerlineskip\vskip-10pt
          \hbox{\psfig{figure=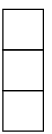}}}\qquad
   \vtop{\hbox{$X^i$:}}\enspace
    \vtop{\null\nointerlineskip\vskip-10pt
          \hbox{\psfig{figure=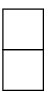}}}\qquad
   \vtop{\hbox{$\phi_\ell$:}}\enspace
    \vtop{
      \hbox{
       \vtop{\null\nointerlineskip\vskip-10pt
             \hbox{\psfig{figure=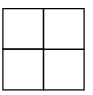}}}
       \vtop{\hbox{$\ldots$}}\enspace
      \vtop{\null\nointerlineskip\vskip-10pt
             \hbox{\psfig{figure=figC.eps}}}
     }\nointerlineskip\vskip-1pt
     \hbox{$\ \displaystyle
       \underbrace{\qquad\qquad\qquad}_{\ell\rm\;columns}$}
     }
   }    
  }
 where $\phi_\ell$ stands for the $\ell$-th partial wave.  There
is a complicated general procedure known as the Littlewood-Richardson
rule for taking tensor products of representations of $SU(N)$.  It
becomes much simpler when one of the factors is a fundamental
representation ($k$ boxes in a single column) \cite{Wass}.  In that case, one
adds these $k$ boxes to the other factor's Young tableaux in all
possible ways, modulo the restriction that not more than one box can
be added to any given row.  Then one eliminates all columns
containing $N$ boxes.  Using this rule it is easy to show that with
$\ell$ fundamental representations as the factors in a tensor product,
it is impossible to obtain the tableau for $\phi_\ell$ unless all the
factors are ${\bf 6}$ representations.

A shorter argument can be made based on purely dimensional grounds.
The dimension of ${\cal O}$ must be $\ell+4$ or the cross-section
would be suppressed by extra powers of $\omega$.  Since
$F_{\alpha\beta}^2$ has dimension $4$, the other $\ell$ fields must
add $\ell$ to the total dimension of ${\cal O}$.  The only possibility
is $\ell$ scalars.

The upshot is that the only possible term in the action which can
contribute at leading order to the absorption of a dilaton in the
$\ell$-th partial wave is, up to normalization, given by
\LthPartial.
 The normalization given there is the one arising from Taylor expanding
$\phi(X)$.\footnote{As we argue below, it is possible that this
local expansion is not consistent with the string amplitude
calculations. This may remove the discrepancy for
$\ell>1$.}  Actually, this action is not
quite the one we want for $\ell > 1$: in addition to mediating the
process which contributes at leading order to the absorption of the
$\ell$-th partial wave, it makes subleading contributions to
lower partial waves.  In order to isolate the $\ell$-th partial
wave, the product $X^{i_1} \cdots X^{i_\ell}$ should be replaced by an
expression which transforms irreducibly under $SO(6)$.  For $\ell = 2$
and $\ell = 3$, the appropriate replacements are
  \eqn{LTwoThree}{\vcenter{\openup1\jot
   \halign{\strut\span\TT & \span\TL & \span\TR\cr
    $\ell = 2$: & X^{i_1} X^{i_2} &\to 
     X^{i_1} X^{i_2} - \tf{1}{6} \delta^{i_1 i_2} X^2  \cr
    $\ell = 3$: & \ X^{i_1} X^{i_2} X^{i_3} &\to
     X^{i_1} X^{i_2} X^{i_3} - 
      \tf{3}{8} \delta^{(i_1 i_2} X^{i_3)} X^2  \ . \cr
  }}}
 One thus subtracts an $\ell = 0$ contribution from the putative $\ell
= 2$ term in \LthPartial, and an $\ell = 1$ contribution from $\ell =
3$.  The whole Taylor series can be thus reshuffled.

Because the replacements in \LTwoThree\ project out a final state with
definite $\ell$, it is permissible to let the initial dilaton
wave-function be $e^{-i \omega (t-x^1)}$ as usual.  The derivatives
$\partial_i$ in \LthPartial\ can then be replaced by $i \omega
\delta^1_i$.  The spin-summed amplitudes squared for absorption of the
first few partial waves are 
  \eqn{FewAmps}{\vcenter{\openup1\jot
   \halign{\strut\span\TT & \span\TL & \span\TR\cr
    $\ell = 0$:\ \ \ & \overline{|{\cal M}|}^2 &= 
     4 \kappa_{10}^2 (p_1 \cdot p_2)^2  \cr
    $\ell = 1$:\ \ \ & \overline{|{\cal M}|}^2 &= 
     {4 \over \sqrt\pi} \kappa_{10}^3 \omega^2 (p_1 \cdot p_2)^2  \cr
    $\ell = 2$:\ \ \ & \overline{|{\cal M}|}^2 &= 
     {10 \over 3 \pi} \kappa_{10}^4 \omega^4 (p_1 \cdot p_2)^2  \cr
    $\ell = 3$:\ \ \ & \overline{|{\cal M}|}^2 &= 
     {5 \over 2 \pi^{3/2}} \kappa_{10}^5 \omega^6 (p_1 \cdot p_2)^2 \ .  \cr
  }}}
 As before, $p_1$ and $p_2$ are the momenta of the outgoing gauge
bosons.  The gauge bosons are identical particles, as are the $\ell$
outgoing scalars, whose momenta will be labeled $p_3$, $\ldots$,
$p_{\ell+2}$.  To avoid over-counting in the integral over phase
space, we must include a factor of $1/(2 \cdot \ell!)$ in the
expression for the cross-section.  The cross-section also includes an
explicit factor of $1/(2 \omega)$ from the normalization of the
incoming dilaton.  Altogether,
  \eqn{CrossF}{
   \sigma_{\rm D}^\ell = {N^{\ell+2} \over 2 \cdot \ell!} {1 \over 2 \omega}
    \int {d^3 p_1 \over (2 \pi)^3 2 \omega_1} \cdots
     {d^3 p_{\ell+2} \over (2 \pi)^3 2 \omega_{\ell+2}} \,
     (2 \pi)^4 \delta^4 \big( q - {\textstyle \sum\limits_i} p_i \big) 
    \  \overline{|{\cal M}|}^2 \ .
  }
 The momentum integrations take the following form:
  \eqn{IlDef}{
   I_\ell = \int {d^3 p_1 \over 2 \omega_1} \cdots
       {d^3 p_{\ell+2} \over 2 \omega_{\ell+2}} \,
     \  \delta^4 \big( q - {\textstyle \sum\limits_i} p_i \big) \ 
       (p_1 \cdot p_2)^2 
    = {3 \pi^{\ell+1} \over 2^{\ell+1}}
       {\omega^{2 \ell + 4} \over (\ell+2)! (\ell+3)!} \ .
  }
 The quickest way to establish \IlDef\ is to Fourier transform to
position space.  

The final results for the first few $\ell$ are
  \eqn{FewSigmas}{\eqalign{
   \sigma^{\ell=0}_{\rm D-brane} &= \sigma^{\ell=0}_{\rm \ class.}  \cr
   \sigma^{\ell=1}_{\rm D-brane} &= \sigma^{\ell=1}_{\rm \ class.}  \cr
   \sigma^{\ell=2}_{\rm D-brane} &= 
    \tf{9}{5} \sigma^{\ell=2}_{\rm \ class.}  \cr
   \sigma^{\ell=3}_{\rm D-brane} &= 
    \tf{24}{5} \sigma^{\ell=3}_{\rm \ class.} \ .  \cr
  }}
 If the $\ell=2$ and $\ell=3$ D-brane cross-sections had been smaller
than the corresponding 
classical results, one might have wondered if the
argument around \YoungT\ might possibly be invalidated by some
peculiar interaction.  But any such additional interaction could only
increase the D-brane cross-section, making the disagreement even
worse.

It is worth noting that nowhere in the literature have higher partial
waves been compared with complete success between D-brane models and
General Relativity.  Agreement up to numerical factors was obtained in
\cite{ja} for effective string models of four- and five-dimensional
black holes, and independently in \cite{gunp} for the 
five-dimensional case; but in
the absence of a well articulated prescription for coupling the
effective string to the bulk fields it is difficult to tell much about
numerical coefficients.  

The agreement of the $\ell=1$ cross-section certainly encourages us to
believe that the D-brane description is capable of handling partial
waves correctly.  On the level of effective field theory this may seem
peculiar because, in this low-energy description, the D-brane has no
thickness.  How can an object with no extent in transverse dimensions
absorb particles with angular momentum?  Power counting and group
theory alone seem to dictate the answer \LthPartial, up to
normalization.  

Fortunately, with three-branes the string theory prescription for the
couplings to external fields is more directly accessible than for
other solitonic models of black holes.  While it does indeed appear
that $\ell>1$ partial waves present a test which the DBI action,
supplemented by the prescription of \cite{kleb} to obtain the normalization
of \LthPartial\ via a Taylor expansion, fails to pass, we expect that
a proper string theoretic treatment will once again yield agreement.
For $\ell=1$, a full-fledged disk amplitude computation verifies both
the form of \LthPartial\ and the normalization shown.  The $\ell=1$
disk amplitude, which involves one bulk insertion and three boundary
operators, is easy to deal with because the insertion of one scalar
vertex operator on the boundary simply generates a spacetime
translation.  When more than one such operators are present, one
encounters singularities in their mutual collisions that need a
careful treatment. The necessary calculations look quite complicated:
for example, computing the full $\ell=2$ amplitude would be equivalent
via the prescription of \cite{gmm} to computing a six point type~I disk
amplitude. They nevertheless seem
highly worthwhile as a means to refine our understanding of the 
world-volume action.

\section{Absorption cross-sections for M-branes}
\label{CompAb}
A remarkable aspect of the agreement between 
the string theoretic and 
the classical
results for three-branes is that it holds exactly 
for any value of  $N$, including
$N=1$. As explained in the introduction, the classical geometry should
be trusted only in the limit $g_{\rm str} N \rightarrow \infty$.
Thus, the agreement of the absorption cross-sections for $N=1$
suggests that our calculations are valid even in the limit
$g_{\rm str} \rightarrow \infty$, provided that $g_{\rm str}
\alpha'^2 \omega^4$ is kept small. The supersymmetric 
non-renormalization theorems are probably at work here, insuring that 
there are no string loop corrections.

In this section we would like to ask whether the exact agreement
between the absorption cross-sections is also found for the two-branes
and five-branes of M-theory. These branes have much in common with the
self-dual three-brane of type IIB theory: their entropies scale with
the temperature in agreement with the scaling for a gas of massless
fields on the world-volume \cite{ktPrime}, while the scaling of their absorption
cross-sections with $\omega$ agrees with estimates from the
world-volume effective theory \cite{kleb}. While the effective actions for one
two-brane \cite{bst} and one five-brane \cite{ffffOne,ffffTwo,ffffThree} are now known in some detail,
their generalizations to $N>1$ remain somewhat obscure.\footnote{It is
believed, for instance, that $N>1$ coincident five-branes are described
by non-trivial conformally invariant theories in $5+1$
dimensions. Comparisons with classical gravity of the kind made in
\cite{ktPrime,kleb} and here are among the ways of learning more about this
theory.}  In this section we compare the cross-sections for $N=1$ and
find that, in contrast to the three-branes, there is no exact agreement
in the normalizations. This is probably due to the fact that M-theory
has no parameter like $g_{\rm str}$ that can be dialed to make the
classical solution reliable.

First we discuss absorption of longitudinally polarized
gravitons by an two-brane. The massless fields in the effective action 
are 8 scalars and 8 Majorana fermions.
The longitudinal
graviton couples to the energy momentum tensor on the
world-volume, $T_{\alpha\beta}$.
The terms in the effective action necessary to describe the absorption of
$h_{xy}$ are ($i= 3, \ldots, 10$;\ $I=1, \ldots, 8$)
\eqn{action}{\eqalign{& S_2= T_2
\int d^3 x\  
\bigg [ -\tf{1}{2}  \partial_{\alpha} X^i
\partial^{\alpha} X^i  
+ \tf{i}{2} \bar \psi^I \gamma^\alpha\partial_\alpha
\psi^I\cr & +  \sqrt 2 \kappa_{11} 
h_{xy} \big (\partial_x X^i \partial_y X^i 
- \tf{i}{4}  \bar \psi^I
(\gamma_x \partial_y + \gamma_y \partial_x) \psi^I \big ) 
\bigg ]
\ , }
}
where $h_{xy}$ is the canonically normalized field which enters the
$D=11$ spacetime action as (cf. \lgr)
$$ - {1\over 2} \int d^{11} x\ \partial_M h_{xy} \partial^M h_{xy} 
\ .$$

The absorption cross-section is found using the Feynman rules in
a manner analogous to the three-brane calculation of section~\ref{SelfThree}.
For the 8 scalars, we find that the matrix element squared 
(with all the relevant factors included) is
$$ {\kappa_{11}^2 \omega^4 \over 2 } 4 n_x^2 n_y^2
\ ,$$
where $\vec n$ is the unit vector in the direction of one of the
outgoing particles.
For the 8 Majorana fermions, the corresponding object 
summed over the final polarizations is
$$ {\kappa_{11}^2 \omega^4 \over 2 } (n_x^2- n_y^2 )^2 \ . 
$$
Adding them up, we find that the dependence on direction cancels out,
just as in the three-brane case. The sum must be multiplied by
the phase space factor 
$$
{1\over 2\omega} {1\over 2\pi} \int {d^2 p_1\over 2 \omega_1} 
\int {d^2 p_2\over 2 \omega_2} 
\delta^2 (\vec p_1 + \vec p_2) \delta (\omega_1+ \omega_2- \omega)=
{1\over 8 \omega^2}
\ ,$$ 
so that the total cross-section is
\eqn{total}{ \sigma_{2{\rm \ abs}} = {\kappa_{11}^2 \omega^2 \over 16 } 
\ .}
This does not agree with the classical result for $N$ set to 1
\cite{kleb,emparan},
$$ \sigma_{2{\rm \ class.}}= {\pi^4\over 12} \omega^2 R^9= {1\over
6\sqrt 2\pi}
\kappa_{11}^2 \omega^2 N^{3/2}
\ ,$$
where we have used the two-brane charge quantization to express $R^9$
in terms of $N$.
Notice that even the power of $\pi$ does not match.
This situation is reminiscent of the
discrepancy in the near-extremal entropy where the relative factor was 
a transcendental number involving $\zeta(3)$\  \cite{ktPrime}. 

Now we show that, just as for the three-brane, the transversely
polarized gravitons have the same absorption cross-section as
the longitudinally polarized gravitons.
The coupling of $h_{67}$ to scalars is given by
\eqn{couu}{- T_2\int d^3 x\ \sqrt 2 \kappa_{11} h_{67} 
\partial_{\alpha} X^6 \partial^{\alpha} X^7  
\ ,}
while pairs of fermions are not produced because the coupling 
to them vanishes on shell.
The matrix element squared is $\kappa_{11}^2 \omega^4/2$.
Multiplying this by the phase space factor, we again find the 
cross-section \total.

Now we turn to the five-brane.
The massless fields on the five-brane form a tensor
multiplet consisting of 5 scalars, 2 Weyl fermions and the
  antisymmetric tensor $\B_{\alpha\beta}$  with anti-self-dual 
strength \cite{modesOne,modesTwo}.
The transverse gravitons are again the easier case because they produce
pairs of scalars only (the coupling to 2 fermions vanishes on shell).
The necessary coupling of $h_{67}$ is  similar to \couu:
$$-T_5 \int d^6 x\ \sqrt 2 \kappa_{11} h_{67} 
\partial_{\alpha} X^6 \partial^{\alpha} X^7 \ . 
$$
The matrix element squared is $\kappa_{11}^2 \omega^4/2$,
while the phase space factor is now
\eqn{phase}{ {1\over 2\omega}
{1\over (2\pi)^4} \int {d^5 p_1\over 2 \omega_1} 
\int {d^5 p_2\over 2 \omega_2} 
\delta^5 (\vec p_1 + \vec p_2) \delta (\omega_1+ \omega_2- \omega)=
{\omega\over 2^7 \cdot 3 \pi^2}
\ ,}
so that the absorption cross-section is
\eqn{fivecross}{\sigma_{5{\rm \ 
class.}}= {\kappa_{11}^2 \omega^5\over 2^8 \cdot 3 \pi^2}
\ .}

To discuss the absorption of longitudinally polarized 
gravitons, $h_{xy}$, we need the action ($i=6, \ldots, 10$; \ 
$I=1, 2$)
\eqn{fiveaction}{\eqalign{& S= T_5
\int d^6 x\  
\bigg [ - \ \tf{1}{2} \partial_{\alpha} X^i
\partial^{\alpha} X^i - \tf{1}{12} \H_{\alpha\beta\gamma}^2 
+i \bar \psi^I \gamma^\alpha\partial_\alpha
\psi^I\cr &   + \sqrt 2 \kappa_{11} h_{xy} 
\big (\partial_x X^i \partial_y X^i 
+  \tf{1}{2} \H^-_{x\beta\gamma} \H_y^{- \beta\gamma}
-  \tf{i}{2} \bar \psi^I
(\gamma_x \partial_y + \gamma_y \partial_x) \psi^I \big)
\bigg ]
\ . }
}
To describe interaction of the anti-self-dual antisymmetric tensor
with external field we  follow the covariant approach of 
\cite{alvw} using the standard unconstrained propagator for $\B_{\a\b}$
and replacing $\H$ by  its anti-self-dual part 
$\H^-=\ha(\H- \td \H)$ in the vertices. 

For the 5 scalars, we find that the matrix element squared 
(with all the relevant factors included) is
\eqn{scalar}{
{\kappa_{11}^2 \omega^4 \over 2 } {5\over 2} n_x^2 n_y^2
\ .}
For the 2 Weyl fermions, the corresponding object 
summed over the final polarizations is\footnote{
It is interesting to observe that, if we consider an 
${\cal N}=1$ multiplet
consisting of 1 Weyl fermion and 4 scalars, then the ``tensor term''
$n_x^2 n_y^2$ cancels out in the direction dependence. 
The same cancelation occurs for the three-brane 
(both for the ${\cal N}=1$ vector
multiplet and for the ${\cal N}=1$ hypermultiplet).}
\eqn{weyl}{
{\kappa_{11}^2 \omega^4 \over 2 } (n_x^2+ n_y^2 - 4 n_x^2 n_y^2 )
\ .}
Finally,
the contribution  of the anti-self-dual gauge field
turns out to be equal to that  of the
usual, unconstrained 
$\H_{\a\b\g}$  divided by 2.
The matrix element squared is, therefore,\footnote{Since the propagator of 
$\B_{\a\b}$  is taken to be non-chiral, it is sufficient to do
the  replacement $\H\H \to \H^-\H^-$ in only 
one of the two  stress tensor factors  in $|{\cal M}|^2$.
The relevant part of 
$\H^-\H^-$ is  $\four (\H \H + \td \H \td \H)$ which is equal to
$ \ha \H \H$ for off-diagonal components of the stress tensor.}
\eqn{anti}{|{\cal M}|^2= {1\over 2} 2 \kappa_{11}^2 
{1\over 4} \sum \ [\H^{(1)}_{x\beta\gamma} \H_y^{(2)\beta\gamma}
+ \H^{(2)}_{x\beta\gamma} \H_y^{(1)\beta\gamma}]
\H^{(1)*}_{x\alpha\delta} \H_y^{(2)\alpha\delta *}
\ , }
where we 
have also included $1/2$ because the outgoing particles are identical.
Sums over polarizations are to be performed with
$$ \sum \epsilon_{\alpha\beta}  \epsilon^*_{\gamma\delta}=
\eta_{\alpha\gamma} \eta_{\beta\delta}- \eta_{\alpha\delta}
\eta_{\beta\gamma}
\ .$$
The entire calculation is lengthy, but the end result
is simple,\footnote{
This answer passes also a number of heuristic checks. 
For instance, for the ${\cal N}=1$ multiplet
including the gauge field, 1 scalar and 1 Weyl fermion, 
the $n_x^2 n_y^2$ terms again cancels out.} 
\eqn{endi}{ \overline{ |{\cal M}|}^2= {\kappa_{11}^2 \omega^4 \over 2 } 
\left (1- n_x^2- n_y^2 + {3\over 2} n_x^2 n_y^2\right ) 
\ .}

Adding up the contributions of the entire tensor multiplet, we find
that all the direction-dependent terms cancel out, just as they did
for the three-brane and the two-brane.
Multiplying by the phase space factor \phase, we find that 
the total cross-section for the longitudinally
polarized gravitons is again given
by \fivecross. This turns out to be
a factor of 4 smaller than the classical result,
\eqn{ende}{
 \sigma_{5{\rm \ class.}}= {2\pi^3\over 3} \omega^5 R^9= 
{\kappa_{11}^2 N^3 \omega^5\over 2^6 \cdot 3 \pi^2}
\ ,  } 
evaluated for $N=1$, i.e.{} 
$\sigma_{5{\rm \ abs}} = \four \sigma^{(N=1)}_{5{\rm \ class.}}$.
This discrepancy is relatively minor and is of a kind that could easily
be produced by a calculational error. However, having checked
our calculations a number of times, we believe that the 
factor of 4 discrepancy is real.

A conclusion that we may draw from this section is that, although the
single M-brane cross-sections scale with the energy in the same way
as the classical cross-sections, the normalizations do not agree.
The five-brane comes much closer to agreement than the two-brane,
which may be connected to
the fact that the five-brane supergravity solution is
completely non-singular. For $N=1$, however, the curvature of the 
solution is of order of the 11-dimensional Planck scale.
Obviously, the 11-dimensional supergravity is at best a low-energy
approximation to M-theory. The M-theory effective action should contain
higher-derivative terms weighted by powers of $\kappa_{11}$, by
analogy with the $\alpha'$ and $g_{\rm str}$  
expansions of the string effective action.
Thus, for $N=1$, the classical solution may undergo corrections of order
one which we believe to be the source of the discrepancy.
For large $N$, however, we expect the M-theory cross-section to agree
exactly with the classical cross-section.
We hope that these considerations 
will serve as a useful guide in constructing the world
volume theory of $N$ coincident five-branes.

\section{Conclusions}
\label{Conc}
In this chapter we have provided new evidence, furthering the earlier
results of \cite{kleb}, that there exists exact agreement between the 
classical and the
D-brane descriptions of the self-dual three-brane of type
IIB theory. The specific comparisons that we have carried out involve
probing an extremal three-brane with low-energy massless quanta incident
from the outside. As argued in \cite{kleb} and here, the 
great advantage of the
three-brane is that both the perturbative
string theory and the classical supergravity
calculations are under control and yield expansions in the same
dimensionless expansion parameter 
$(\omega R)^4 \sim N g_{\rm str} \alpha'^2 \omega^4$,
which may be kept small.

While the low-energy physics of the three-brane is described by a ${\cal N}=4$
SYM theory, probing it from the extra dimensions provides a new point
of view and allows for a variety of interesting calculations.
One example is the absorption of a graviton polarized parallel
to the brane which, as discussed in section~\ref{SelfThree},
couples to the energy-momentum tensor in
3+1 dimensions. Because of the existence
of the transverse momentum, the kinematics for this process is that of
a decay of a massive spin-two particle into a pair of
massless world-volume modes. Summing over all the states
in the ${\cal N}=4$ multiplet we find that the rate is completely
isotropic, which is undoubtedly related to the conformal invariance
of the theory. We should emphasize, however, that we are exploring
the properties of this theory away from the BPS limit; 
therefore, they are not determined by supersymmetry alone.
For this reason, we find it remarkable 
that the net absorption cross-section
has the same value as that of the dilaton and the RR scalar, 
and which agrees with the cross-section found in 
classical supergravity.

We have also carried out similar explorations of the physics
of M-branes. These studies are hampered, to a large extent, by 
insufficient understanding of the multiple coincident branes of
M-theory. For a single two-brane, a formal application of classical 
gravity gives an answer which is off by a 
factor of $4\sqrt 2/(3\pi)$,
while for a single five-brane -- off only by a factor of 4.  
For the five-brane the discrepancy is relatively minor which, we are
tempted to speculate, is due to the fact that its classical geometry
is completely non-singular, just like that of the three-brane.
We should keep in mind, however, that for singly charged branes
the quantum effects of M-theory are expected to be
important, and the classical reasoning should not be trusted.

While in this chapter we have probed three-branes  with massless 
particles,  we may contemplate another interesting use of the three-brane:
it may  be used as a probe by itself \cite{mdlate}.
For example, $N$ coincident three-branes may be probed by
a three-brane parallel to them. This situation is described by
a $3+1$ dimensional ${\cal N}=4$ supersymmetric 
$U(N+1)$ gauge theory, with gauge symmetry broken
to $U(N)\times U(1)$. A more complicated theory will arise if
a three-brane is used to probe a stringy
$d=4$ black hole.
As was  shown in \cite{kt,bl}, 
the $d=4$
extremal black holes with regular horizons 
(which are parametrized by 4 charges \cite{CYTwo})  
can be represented by 1/8 supersymmetric configurations of four intersecting 
three-branes wrapped over a  6-torus.
 Following
\cite{meprOne,meprTwo}  one can find an action for a classical three-brane 
probe moving in this geometry. If the probe is oriented parallel
to one of the four source three-branes, the resulting moduli space 
metric is 
$$ds^2_6=  H_1 H_2 H_3  (dr^2 + r^2 d\Omega_2^2)  + 
H_1 dy^2_1  +  H_2 dy^2_2 +  H_3 dy^2_3
\ , \ \ \ \ \ H_i =1 + {R_i\ov r}
\  , $$
where $y_i$ are toroidal coordinates transverse to the probe.
In contrast to the case of the 5-brane-plus-string-plus-momentum
configuration   describing  $d=5$  regular  extremal 
black holes with 3 charges \cite{att} 
where one  finds   \cite{dpl} that the non-compact part of the moduli space 
 metric is multiplied by the product of two  harmonic functions, here 
we obtain  the product of  {\it three}
 harmonic functions. This is, however,
exactly what is needed to get the same near-horizon 
($r\to 0$) behaviour as found in \cite{dpl}, 
i.e.{} that the 3-dimensional non-compact
part of the moduli space metric becomes flat: 
$ds^2_3 \to  R_1R_2R_3 (d\r^2 +  \r^2   d\Omega_2^2) $, \
 $\r\equiv {r}^{-1/2} \to \infty$.
This close similarity 
between a $d=5$ black hole probed by a string  
and a $d=4$ black hole probed by a three-brane
strongly suggests that 
one can obtain important information 
about these black holes  by studying the corresponding 
three-brane world-volume theory.
This theory in the presence of intersecting  branes is necessarily 
more intricate than the simpler, parallel brane case
discussed in the main part of this chapter.

Clearly, there is much work to be done before we claim a complete
understanding of the remarkable duality between the D-brane and the
classical descriptions of the three-brane.  The complexities that
affect the higher-spin classical equations remain to be disentangled.
The numerical discrepancies for $\ell>1$ partial waves suggest an
incomplete understanding of the low-energy effective action.  As we
have argued in section~\ref{SelfThree}, a direct string calculation is
necessary to check normalizations.  And finally, we need to push our
methods away from extremality where the physics is necessarily more
complicated, involving thermal field theory in 3+1 dimensions. We hope
that detailed insight into such a theory will help explain the
specific factors appearing in the near-extremal entropy \cite{gkp}.

We feel that further efforts in the directions mentioned above are
worthwhile because they offer a promise of building a theory
of certain black holes
in terms of a manifestly unitary theory -- perturbative string
theory.

\newcommand{\beq}{\begin{equation}}
\newcommand{\eeq}{\end{equation}}
\newcommand{\beqs}{\begin{eqnarray}}
\newcommand{\eeqs}{\end{eqnarray}}
\newcommand{\laa}{\lambda_{IIA}}
\newcommand{\lop}{\lambda_{I'}}
\newcommand{\lhet}{\lambda_{E_8}}
\newcommand{\rop}{R_{I'}}
\newcommand{\rhet}{R_{E_8}}
\newcommand{\da}{{\dot a}}
\newcommand{\db}{{\dot b}}

\renewcommand{\epsilon}{\varepsilon}
\def\equno#1{(\ref{#1})}
\def\equnos#1{(#1)}
\def\sectno#1{section~\ref{#1}}
\def\figno#1{Fig.~(\ref{#1})}
\def\D#1#2{{\partial #1 \over \partial #2}}
\def\df#1#2{{\displaystyle{#1 \over #2}}}
\def\tf#1#2{{\textstyle{#1 \over #2}}}
\def\d{d}
\def\e{e}
\def\i{i}
\def\Leff{L_{\rm eff}}
\def \ci {\cite}
\def \sm {$\s$-model }

\def \o {\omega}
\def \inv {^{-1}}
\def \ov {\over }
\def \four{{\textstyle{1\over 4}}}
\def \fourth{{{1\over 4}}}
\def \ha {{1\ov 2}}
\def \QQ {{\cal Q}}

\def \t {\tau}
\def \ci {\cite}
\def \sm {$\s$-model }

\def \o {\omega}
\def \inv {^{-1}}
\def \ov {\over }
\def \four{{\textstyle{1\over 4}}}
\def \fourth{{{1\over 4}}}
\def \ha {{1\ov 2}}
\def \QQ {{\cal Q}}

\def \lr { \lref}
\def\np {{  Nucl. Phys. }}
\def \pl {{  Phys. Lett. }}
\def \mpl {{ Mod. Phys. Lett. }}
\def \prl {{  Phys. Rev. Lett. }}
\def \pr  {{ Phys. Rev. }}
\def \ap  {{ Ann. Phys. }}
\def \cmp {{ Commun.Math.Phys. }}
\def \ijmp {{ Int. J. Mod. Phys. }}
\def \jmp {{ J. Math. Phys.}}
\def \cqg {{ Class. Quant. Grav. }}

\chapter{Cross-sections and Schwinger terms in the world-volume}
\label{SchwingerTerms}

\section{Introduction}
\label{IntroCent}

Extremal black holes with non-vanishing horizon area may be embedded
into string theory or M-theory using intersecting $p$-branes
\cite{cttPrime,ctt,sv,cm,mst,myers,at,kt,blTwo}.  These configurations
are useful for a microscopic interpretation of the Bekenstein-Hawking
entropy.  The dependence of the entropy on the charges, the
non-extremality parameter, and the angular momentum suggests a
connection with $1+1$ dimensional conformal field theory
\cite{sv,cm,cyFour,mst,myers,kt}.  This ``effective string'' is
essentially the intersection of the $p$-branes.  Calculations of
emission and absorption rates
\cite{cm,dmw,dmOne,gkOne,mast,gkTwo,cgkt,kkOne,km,htr,dkt,kkTwo,krt,Mathur,gpartial,clOne,clTwo}
provide further tests of the ``effective string'' models of $D=5$
black holes with three charges and of $D=4$ black holes with four
charges.  For minimally coupled scalars the functional dependence of
the greybody factors on the frequency agrees exactly with
semi-classical gravity, providing a highly non-trivial verification of
the effective string idea \cite{mast,gkTwo}. Similar successes have
been achieved for certain non-minimally coupled scalars, which were
shown to couple to higher dimension operators on the effective string.
For instance, the fixed scalars \cite{fkOne,fkTwo,kr} were shown in
chapter~\ref{FixedScalars} to
couple to operators of dimension $(2,2)$ \cite{cgkt} while the
``intermediate'' scalars \cite{krt} -- to operators of dimension
$(2,1)$ and $(1,2)$. Unfortunately, there is little understanding of
the ``effective string'' from first principles, and some of the more
sensitive tests reveal this deficiency. For instance, semi-classical
gravity calculations of the fixed scalar absorption rates for
general black hole charges reveal a gap in our understanding of higher
dimension operators \cite{kkTwo}.  A similar problem occurs when one
attempts a detailed effective string interpretation of the higher
partial waves of a minimally coupled scalar, as we saw in
chapter~\ref{PartialWaves}.  Even
the $s$-wave absorption by black holes with general charges is complex
enough that it is not reproduced by the simplest effective string
model \cite{km,clOne,clTwo}. These difficulties by no means invalidate the
general qualitative picture, but they do pose some interesting
challenges.  In order to gain insight into the relation between
supergravity and D-branes it is useful to study, in addition to the
intersecting branes, the simpler configurations which involve parallel
branes only.  This chapter, based on the paper \cite{gkThree},
shows how absorption calculations in linearized supergravity are
related to certain Schwinger terms in the operator algebra of the
D-brane world-volume gauge theory.

A microscopic interpretation of the entropy of near-extremal
$p$-branes was first studied in \cite{gkp,ktPrime}. It was found that
the scaling of the Bekenstein-Hawking entropy with the temperature
agrees with that for a massless gas in $p$ dimensions only for the
``non-dilatonic $p$-branes'': namely, the self-dual three-brane of the
type IIB theory, and the 2- and 5-branes of M-theory.  Their further
study was undertaken in \cite{kleb,gukt}, where low-energy absorption
cross-sections for certain incident massless particles were compared
between semi-classical supergravity and string or M-theory. For
three-branes, exact agreement was found for the leading low-energy
behavior of the absorption cross-sections for dilatons \cite{kleb} as
well as R-R scalars and for gravitons polarized along the brane
\cite{gukt}.  The string-theoretic description of macroscopic three-branes
can be given in terms of many coincident D3-branes \cite{dlp,JP}.
Indeed, $N$ parallel D3-branes are known to be described by a $U(N)$
gauge theory in 3+1 dimensions with ${\cal N}=4$ supersymmetry
\cite{Witten}. This theory has a number of remarkable properties,
including exact S-duality, and we will be able to draw on a known
non-renormalization theorem in explaining the absorption by the
three-branes.

 From the point of view of supergravity, the three-brane is also
special because its extremal geometry,
\beq
d s^2 =\left (1+{R^4\over r^4}\right )^{-1/2}
(-dt^2 + dx_1^2 + dx_2^2+ dx_3^2 ) +
\left (1+ {R^4\over r^4}\right )^{1/2} (dr^2 + r^2 d\Omega_5^2)\ ,
\eeq
is non-singular \cite{ght},
while the dilaton background is constant.  Instead of a singularity 
at $r=0$ we
find an infinitely long throat whose radius is determined by the
charge (the vanishing of the horizon area is due to the longitudinal
contraction).  Thus, for a large number $N$ of coincident branes, the
curvature may be made arbitrarily small in Planck units. For instance,
for $N$ D3-branes, the curvature is bounded by a quantity of order
 \beq
{1\over \sqrt{N\kappa}} \sim {1\over \alpha'\sqrt{N g_{\rm str}}}
 \ .
\eeq
In order to suppress the string scale corrections to the classical
metric, we need to take the limit $N g_{\rm str}\rightarrow\infty $.

The tension of a D3-brane depends on $g_{\rm str}$ and $\alpha'$ only
through the ten-dimensional gravitational constant $\kappa = 8
\pi^{7/2} g \alpha'^2$:
 \beq
T_{(3)}={\sqrt \pi \over\kappa}\ . 
\eeq
 This suggests that we can compare the expansions of various
quantities in powers of $\kappa$ between the microscopic and the
semi-classical descriptions. Indeed, the dominant term in the
absorption cross-section at low energy is (see
chapter~\ref{ThreeBraneAbs}) 
 \beq
\label{three}
\sigma= {\pi^4\over 8}\omega^3 R^8
={\kappa^2  \omega^3 N^2\over 32 \pi} \ , 
\eeq
which agrees between semi-classical supergravity and string theory.

It is important to examine the structure of higher power in $g_{\rm
str}$ corrections to the cross-section \cite{Das}.  In semi-classical
supergravity the only quantity present is $\kappa$, and corrections to
(\ref{three}) can only be of the form
 \beq\label{ACor}
  a_1 \kappa^3 \omega^7+ a_2 \kappa^4 \omega^{11}+ \ldots
 \eeq
 However, in string theory we could in principle find corrections even
to the leading term $\sim \omega^3$, so that
 \beq\label{BCor}
\sigma_{\rm string}= {\kappa^2  \omega^3 N^2\over 32 \pi} 
\left (1+ b_1 g_{\rm str} N+ b_2 (g_{\rm str} N)^2+ \ldots \right )
+ {\cal O} (\kappa^3 \omega^7)\ .\eeq
 Presence of such corrections would spell a manifest disagreement with
supergravity because, as we have explained, the comparison has to be
carried out in the limit $N g_{\rm str}\rightarrow\infty $.  The main
purpose of this chapter is to show that, in fact, $b_i=0$ due to a
non-renormalization theorem in $D=4$ ${\cal N}=4$ SYM theory.

In section~\ref{DBApp} we present a detailed argument for the
absence of such corrections in the absorption cross-section of
gravitons polarized along the brane. These gravitons couple to the
stress-energy tensor on the world volume and we will show that the
absorption cross-section is, up to normalization, the central term in
the two-point function of the stress-energy tensor. The fact that
$b_i=0$ follows from the fact that the one-loop calculation of the
central charge is exact in $D=4$ ${\cal N}=4$ SYM theory.

The connection between absorption of gravitons polarized along the
brane and Schwinger terms in the stress-energy correlators of the
world volume theory is a general phenomenon that holds for all branes.
In section~\ref{ExtOther} we explore this connection to deduce some
properties of the stress-energy tensor OPE's for multiple 2-branes and
5-branes of M-theory, as well as for multiple 5-branes of string
theory.

\section{D-brane approach to absorption}
\label{DBApp}

It was probably Callan who first realized that, in terms of D-brane
models, absorption cross-sections correspond up to a simple overall
factor to discontinuities of two-point functions of certain operators
on the D-brane world volume \cite{cp}.  This realization was exploited
in \cite{ja} and in chapter~\ref{PhotonAbsorption}.  Consider massless
scalar particles in ten dimensions normally incident upon D3-branes.
If the coupling to the brane is given by
  \begin{equation}\label{GenCoup}
   S_{\rm int} = \int d^4 x \, \phi(x,0) {\cal O}(x) \ ,
  \end{equation}
 where $\phi(x,0)$ is a canonically normalized scalar field evaluated
on the brane, and ${\cal O}$ is a local operator on the brane, then
the precise correspondence is
  \begin{equation}\label{SigmaDisc}
   \sigma = {1 \over 2 i \omega} \Disc \Pi(p) 
    \bigg|_{p^0 = \omega \atop \vec{p} = 0} \ .
  \end{equation}
 Here $\omega$ is the energy of the incident particle, and
  \begin{equation}\label{PiP}
   \Pi(p) = \int d^4 x \, e^{i p \cdot x} 
    \langle {\cal O}(x) {\cal O}(0) \rangle \ .
  \end{equation}
 When ${\cal O}$ is a scalar in the world volume theory, $\Pi(p)$
depends only on $s = p^2$, and $\Disc \Pi(p)$ is computed as the
difference of $\Pi$ evaluated for $s = \omega^2 + i \epsilon$ and $s =
\omega^2 - i \epsilon$.  In the case of the graviton, we shall see
that $\Pi(p)$ is a polynomial in $p$ times a function of $s$, so the
evaluation of $\Disc \Pi(p)$ is equally straightforward.

The validity of (\ref{SigmaDisc}) depends on $\phi$ being a
canonically normalized field.  In form it is almost identical to the
standard expression for the decay rate of an unstable particle of mass
$\omega$.  The dimensions are different, however: $\sigma$ is the
cross-section of the three-brane per unit world volume, and so has
dimensions of $(\hbox{length})^5$.  Similar formulas can be worked out
for branes of other dimensions and for near-extremal branes, although
away from extremality (\ref{PiP}) would become a thermal Green's
function (see chapter~\ref{PhotonAbsorption}).

Now let us work through the example of the graviton.  The three-brane
world volume theory is ${\cal N}=4$ supersymmetric $U(N)$ gauge
theory, where $N$ is the number of parallel three-branes \cite{Witten}.
Thus, the massless fields on the world volume are the gauge field, six
scalars, and four Majorana fermions, all in the adjoint representation
of $U(N)$.  To lowest order in $\kappa$, and ignoring the couplings to
the bulk fields, the world volume-action is ($I=1,...,4$; $i=4,...,9$)
  \beq
   S_3 = \int d^4 x \, \tr \left[ -\tf{1}{4} F_{\alpha\beta}^2 + 
    \tf{i}{2} \bar\psi^I \gamma^\alpha \partial_\alpha \psi_I + 
    \tf{1}{2} (\partial_\alpha X^i)^2 + 
     \hbox{interactions} \right] \ .  \label{wvAction}
  \eeq
 The interactions referred to here are the standard renormalizable
ones of ${\cal N} = 4$ super-Yang-Mills (see for example
\cite{Sohnius} for the complete flat-space action).  In
equation~\NFourS, a factor of the three-brane tension $T_{(3)}$ appeared
in front of the action.  It is convenient to work with canonically
normalized fields, and so, relative to the conventions of
chapter~\ref{ThreeBraneAbs} we have absorbed a factor of
$\sqrt{T_{(3)}}$ into $A_\mu$, $\psi^I$, and $X^i$.  Another
difference between chapter~\ref{ThreeBraneAbs} and our present
conventions is that we work here with a mostly minus metric and the
spinor conventions of \cite{Sohnius}.

The full (and as yet unknown) action for multiple three-branes is
non-polynomial, and (\ref{wvAction}) includes only the dimension $4$
terms.  The higher dimension terms will appear with powers of
$1/T_{(3)}$, which is to say positive powers of $\kappa$.  They could
give rise to corrections of the form (\ref{ACor}), but not
(\ref{BCor}).

Despite our ignorance of the full action for multiple coincident
three-branes, one can be fairly confident in asserting that the
external gravitons polarized parallel to the brane couple via 
 \beq \label{sint}
    S_{\rm int} = \int d^4 x \, \tf{1}{2} h^{\alpha\beta} 
     T_{\alpha\beta}  \ ,
 \eeq 
 where $h_{\alpha\beta} = g_{\alpha\beta} - \eta_{\alpha\beta}$ is the
perturbation in the metric, and $T_{\alpha\beta}$ is the stress-energy
tensor:
  \beq\label{ConfT}
   \vcenter{\openup1\jot
   \halign{\strut\span\TL & \span\TR\cr
    T_{\alpha\beta} &= \tr \big[ -F_{\alpha}^{\ \gamma} F_{\beta\gamma} + 
      \tf{1}{4} \eta_{\alpha\beta} F_{\gamma\delta}^2 +
     \tf{i}{2} \bar\psi^I \gamma_{(\alpha} \partial_{\beta)} \psi_I  
      \cr
     &\quad + 
      \tf{2}{3} \partial_\alpha X^i \partial_\beta X^i  - 
       \tf{1}{6} \eta_{\alpha\beta} (\partial_\gamma X^i)^2 -
       \tf{1}{3} X^i \partial_\alpha \partial_\beta X^i + 
       \tf{1}{12} \eta_{\alpha\beta} X^i \square X^i  \cr
     &\quad +
      \hbox{interactions} \, \big] \ .  \cr
   }}
  \eeq 
 This ``new improved'' form
of the stress-energy tensor \cite{ccj} is chosen so that $\partial_\mu
T^{\mu\nu} = 0$ and $T^\mu_\mu = 0$ on shell.  The scalar terms differ
from the canonical form
  \beq
   T^{\rm (can)}_{\alpha\beta} = 
    \tr \left[ \partial_\alpha X^i \partial_\beta X^i - 
     \tf{1}{2} \eta_{\alpha\beta} (\partial_\gamma X^i)^2 + 
     \ldots \, \right] \ .
  \eeq
 The difference arises from adding a term $-\tf{1}{12} \sqrt{-g} R \tr
X^2$ to the lagrangian so that the scalars are conformally coupled.

Choosing the scalars to be minimally or conformally coupled does not
affect the one-loop result for the cross-section.  But it is the
traceless form of $T_{\alpha\beta}$ presented in (\ref{ConfT}) which
has a non-renormalized two-point function.  The reason is that the
conformal $T_{\alpha\beta}$ is in the same supersymmetry multiplet as
the supercurrents and the $SU(4)$ R-currents.  ${\cal N} = 4$
super-Yang-Mills theory is finite to all orders and anomaly free in
flat space.  The Adler-Bardeen theorem guarantees that any anomalies
of the $SU(4)$ R-currents can be computed exactly at one loop.
Because $\partial_\alpha R^\alpha$ and $T^\alpha_\alpha$ are in the
same supermultiplet (the so-called ``multiplet of anomalies''), the
one-loop result for the trace anomaly must also be exact.\footnote{We
thank D. Anselmi for pointing out the relevance of the Adler-Bardeen
theorem.}  Thus the trace anomaly in a curved background is the same
as in the free theory:
  \beq\label{TraceAnom}
   \langle T^\mu_\mu \rangle = 
    -{1 \over 16 \pi^2} \left( c F - 
      {2 c \over 3} \square R - b G \right) \ ,
  \eeq
 where $F = C_{\alpha\beta\gamma\delta}^2$ is the square of the Weyl
tensor and $G = R_{\alpha\beta\gamma\delta}^2 - 4 R_{\alpha\beta}^2 +
R^2$ is the topological Euler density.  Thus the second and third
terms in (\ref{TraceAnom}) are total derivatives.  The coefficients
$c$ and $b$ are given by \cite{bd}
  \beq\label{CentralC}
   \vcenter{\openup1\jot
   \halign{\strut\span\TL & \span\TR\cr
    c &= {12 N_1 + 3 N_{1/2} + N_0 \over 120} = {N^2 \over 4}  \cr
    b &= {124 N_1 + 11 N_{1/2} + 2 N_0 \over 720} = {N^2 \over 4} \ . \cr
   }}
  \eeq
 $N_1 = N^2$, $N_{1/2} = 4 N^2$, and $N_0 = 6 N^2$ are the numbers of
spin-one, Majorana spin-half, and real spin-zero fields in the
super-Yang-Mills theory.  Note that spin-one, spin-half, and spin-zero
particles make contributions to $c$ in the ratio $2:2:1$.  The
different spins contribute to the cross-section in precisely the same
ratio, as was demonstrated in chapter~\ref{ThreeBraneAbs}.

It remains to make the connection between $\langle T^\alpha_\alpha
\rangle$ and $\langle T_{\alpha\beta}(x) T_{\gamma\delta}(0) \rangle$.
Suppressing numerical factors and Lorentz structure, the OPE of
$T_{\alpha\beta}$ with $T_{\gamma\delta}$ is \cite{AnsOne}
  \beq\label{TOPE}
   T(x) T(0) = {c \over x^8} + \ldots + {T(0) \over x^4} + \ldots \ .
  \eeq
 We have omitted terms involving the Konishi current as well as terms
less singular than $1/x^4$, and we have anticipated the conclusion
that the coefficient on the Schwinger term is precisely the central
charge $c$ appearing in (\ref{TraceAnom}).  A clean argument to this
effect is presented in \cite{erd} and summarized briefly below.  The
Schwinger term is nothing but the two-point function: with all factors
and indices written out explicitly\footnote{The field normalization
conventions in \cite{AnsTwo} differ from those used here.}
\cite{AnsTwo},
  \beq\label{TPF}
   \langle T_{\alpha\beta}(x) T_{\gamma\delta}(0) \rangle = 
    {c \over 48 \pi^4} X_{\alpha\beta\gamma\delta} 
    \left( 1 \over x^4 \right)
  \eeq
 where
  \beq\label{XDef}
   \vcenter{\openup1\jot
   \halign{\strut\span\TL & \span\TR\cr
     X_{\alpha\beta\gamma\delta} &= 
      2 \square^2 \eta_{\alpha\beta} \eta_{\gamma\delta} - 
       3 \square^2 (\eta_{\alpha\gamma} \eta_{\beta\delta} + 
        \eta_{\alpha\delta} \eta_{\beta\gamma}) -
      4 \partial_\alpha \partial_\beta 
        \partial_\gamma \partial_\delta  \cr
      &\quad - 2 \square 
        (\partial_\alpha \partial_\beta \eta_{\gamma\delta} +
         \partial_\alpha \partial_\gamma \eta_{\beta\delta} +
         \partial_\alpha \partial_\delta \eta_{\beta\gamma} +
         \partial_\beta \partial_\gamma \eta_{\alpha\delta} +
         \partial_\beta \partial_\delta \eta_{\alpha\gamma} +
         \partial_\gamma \partial_\delta \eta_{\alpha\beta}) \ .  \cr
   }}
  \eeq
 The argument of \cite{erd} starts by obtaining an expression for the
flat space three-point function $\langle T^\alpha_\alpha(x)
T_{\beta\gamma}(y) T_{\rho\sigma}(z) \rangle$ by expanding
(\ref{TraceAnom}) around flat space to second order.  Using Ward
identities for the conservation of $T_{\mu\nu}$, one can then derive a
relation on two-point functions which can only be satisfied if the
coefficients $c$ in (\ref{TPF}) and (\ref{TraceAnom}) are identical.

We are finally ready to compute the cross-section.  Fourier
transforming the two-point function (\ref{TPF}), one obtains
  \beq\label{PiPT}
   \Pi_{\alpha\beta\gamma\delta}(p) = \int d^4 x \, 
    e^{i p \cdot x} 
    \langle T_{\alpha\beta}(x) T_{\gamma\delta}(0) \rangle = 
    {c \over 48 \pi^4} \hat{X}_{\alpha\beta\gamma\delta} 
     \int d^4 x \, {e^{i p \cdot x} \over x^4}
  \eeq
 where $\hat{X}_{\alpha\beta\gamma\delta}$ is just the
$X_{\alpha\beta\gamma\delta}$ of (\ref{XDef}) with $\partial \to -i p$.
The integral is evaluated formally as 
  \beq\label{FormalInt}
   \Pi(s) = \int d^4 x \, {e^{i p \cdot x} \over x^4} = 
    \pi^2 \log(-s) + \hbox{(analytic in s)}
  \eeq
 where, as before, $s = p^2$.  One easily reads off the discontinuity
across the positive real axis in the $s$-plane: 
  \beq
   \Disc \Pi(s) = \Pi(s + i \epsilon) - \Pi(s - i \epsilon) = 
    -2 \pi^3 i \ ,
  \eeq
 and so
  \beq
   \Disc \Pi_{\alpha\beta\gamma\delta}(p) = -{i c \over 24 \pi} 
    \hat{X}_{\alpha\beta\gamma\delta} \ .
  \eeq
 For the sake of definiteness, let us consider a graviton polarized in
the $x^1$--$x^2$ direction.  $\hat{X}_{1212} = -3 \omega^4$ for
normal incidence, so
  \beq\label{SigmaGravTwo}
   \sigma = {2 \kappa^2 \over 2 i \omega} \Disc \Pi_{1212}(p)
     \bigg|_{p^0 = \omega \atop \vec{p} = 0}
    = {c \over 8 \pi} \kappa^2 \omega^3 \ ,
  \eeq
 in agreement with the classical result when $c = N^2 / 4$.  The extra
factor of $2 \kappa^2$ in the second expression in (\ref{SigmaGravTwo})
comes from the fact that $h_{\alpha\beta}$ as defined in the text
following (\ref{sint}) is not a canonically normalized scalar field;
instead, $h_{\alpha\beta} / \sqrt{2 \kappa^2}$ is.

In summary, the non-renormalization argument is as follows: the
graviton cross-section is read off at leading order in $\kappa$, but
correct to all orders in $g_{\rm str}$, from the two-point function of
the stress-energy tensor.  The two-point function is not renormalized
beyond one loop because the Schwinger term in the OPE (\ref{TOPE}) is
similarly non-renormalized.  That in turn is due to the fact that the
central charge appearing in the Schwinger term is precisely the
coefficient of the Weyl tensor squared in the trace anomaly
(\ref{TraceAnom}).  The trace anomaly is not renormalized past one
loop because $T^\alpha_\alpha$ is related by supersymmetry to the
divergence of the $SU(4)$ $R$-current, which is protected by the
Adler-Bardeen theorem against anomalies beyond one loop.  Another,
more heuristic, reason why the central charge should not be
renormalized is that there is a critical line extending from $g_{\rm
str} N=0$ to $g_{\rm str}N= \infty$.  The central charge is expected
to be constant along a critical line.  In four dimensions, this
expectation is supported by the work of \cite{AnsTwo}.\footnote{A more
definitive argument has been given in four dimensions for the
constancy of flavor central charges along fixed lines
\cite{AnsThree}.}  Therefore, $c$ can be calculated in the $g_{\rm
str} N\rightarrow 0$ limit where it is given by one-loop diagrams.

We could adopt a different strategy and invert our arguments.
Requiring that the world volume theory of $N$ coincident three-branes
agrees with semi-classical supergravity tells us that in the $g_{\rm
str} N\rightarrow \infty$ limit its central charge approaches
$N^2/4$. In the next section we will similarly deduce the Schwinger
terms in the two-point functions of the stress-energy tensor for other
branes. Furthermore, we can study two-point functions of other
operators by calculating the semi-classical absorption cross-section
for particles that couple to them. For instance, the dilaton couples
to $\tr F^2$. The dilaton absorption cross-section was calculated in
\cite{kleb}. The semi-classical result implies that the Schwinger term
here is again $\sim N^2$ in the $g_{\rm str} N\rightarrow \infty$
limit.  The comparisons with the gauge theory calculation in
chapter~\ref{ThreeBraneAbs} suggest that the one-loop result is again
exact. It will be interesting to extract more results from these
connections between gravity and gauge theory.

\section{Extension to other branes}
\label{ExtOther}

In this section we further explore the connection between the
absorption cross-sections in semi-classical supergravity and central
terms in two-point functions calculated in corresponding world volume
theories. We proceed in analogy to the three-brane discussion presented in
the previous section, and consider absorption of gravitons polarized
along the branes. In the world volume theories such gravitons couple
to components of stress-energy tensor, $T_{\alpha\beta}$. On the other
hand, in supergravity such gravitons satisfy the minimally coupled
scalar equation with respect to the coordinates transverse to the
brane \cite{gukt}, as we have seen in chapter~\ref{ThreeBraneAbs}.
This establishes a general connection between the absorption
cross-section of a minimally coupled scalar and the central term in
the algebra of stress-energy tensors.  For $N$ coincident three-branes the
world volume theory is known, and we have shown that the conformal
anomaly is in exact agreement with this principle. There are cases,
however, where little is known about the world volume theory of
multiple branes. In such cases we can use our method to find the
Schwinger term without knowing any details of the world volume theory.

Consider, for instance, the 2-branes and the 5-branes of M-theory.
While the world volume theory of multiple coincident branes
is not known in detail, the extreme
supergravity solutions are well-known. The absorption
cross-sections for low-energy
gravitons polarized along the brane were calculated
in \cite{kleb,gukt,emparan}, with the results
\beq
\sigma_2= {1\over 6\sqrt 2 \pi} \kappa_{11}^2 \omega^2 N^{3/2}\ ,
\qquad\qquad
\sigma_5={1\over 3\cdot 2^6 \pi^2} N^3\omega^5\ .
\eeq
For $N$ coincident M2-branes we may deduce that the schematic
structure of the stress-energy tensor OPE is
\beq
T(x) T(0) = {c_2\over x^6} + \ldots
\eeq
where the central charge behaves as $c_2\sim N^{3/2}$
in the large $N$ limit.
For $N$ coincident M5-branes we instead have
\beq
T(x) T(0) = {c_5\over x^{12}} + \ldots \ .
\eeq
Now the central charge behaves as $c_5\sim N^3$
in the large $N$ limit.
These results have an obvious connection with properties of
the near-extremal entropy found in \cite{ktPrime}.
Indeed, the near-extremal entropy of a large number $N$
of coincident M2-branes is formally
reproduced by ${\cal O}(N^{3/2})$
massless free fields in 2+1 dimensions, while that of
$N$ coincident M5-branes is reproduced by ${\cal O}(N^3)$
massless free fields in 5+1 dimensions. 

As a final example we consider the 5-branes of string theory.  In
\cite{juan} it was shown that their near-extremal entropy is
reproduced by a novel kind of string theory, rather than by massless
fields in 5+1 dimensions. For $N$ coincident D5-branes the string
tension turns out to be that of a D-string divided by $N$. This
suggests that the degrees of freedom responsible for the near-extremal
entropy are those of ``fractionated'' D-strings bound to the
D5-branes.  An S-dual of this picture suggests that the entropy of
multiple NS-NS 5-branes comes from fractionated fundamental strings
bound to them.\footnote{New insights into the world volume theory of
NS-NS 5-branes were recently obtained in
\cite{dvvTwo,ns,abkss,wittenHiggs}.} We would like to learn more about
these theories by probing them with longitudinally polarized gravitons
incident transversely to the brane.

The extreme Einstein metric of both the NS-NS and the R-R 5-branes
is
\beq
d s_E^2 = \left (1 + {R^2\over r^2}\right )^{-1/4}
(-d t^2+ dx_1^2+ \ldots + dx_5^2)+
\left (1 + {R^2\over r^2}\right )^{3/4} (dr^2 + r^2 d\Omega_3^2)\ .
\eeq
The $s$-wave Laplace equation in the background of this metric is
\beq
\left [ \rho^{-3} {d\over d\rho}
 \rho^3 {d\over d\rho} + 1 + {(\omega R)^2\over \rho^2}
\right ] \phi (\rho)=0\ ,
\eeq
where $\rho =\omega r$.
Remarkably, this equation is exactly solvable in terms of 
Bessel functions. The two possible solutions are
\beq \rho^{-1} J_{\pm \sqrt{1-(\omega R)^2}} (\rho)
\ .\eeq 
Clearly, there are two physically different regimes.
For $\omega R > 1$ the label of the Bessel function is
imaginary, and the requirement that the wave is incoming for
$\rho\rightarrow 0$ selects the solution
\beq \rho^{-1} J_{- i \sqrt{(\omega R)^2- 1}} (\rho)
\ .\eeq 
{}From the large $\rho$ asymptotics we find that the absorption
probability is
\beq
{\cal P} = 1-
e^{- 2\pi \sqrt{(\omega R)^2- 1}} \ .
\eeq
Hence, for $\omega R\geq 1$ the absorption cross-section is 
\beq \label{cs}
\sigma = {4\pi\over \omega^3} \left (1-
e^{- 2\pi \sqrt{(\omega R)^2- 1}} \right )\ .
\eeq
For $\omega R < 1$ the question of how to choose the solution
is somewhat more subtle. It is clear that
$\rho^{-1} J_{\sqrt{1-(\omega R)^2}} (\rho)$ is better behaved near
$\rho =0$ than
$\rho^{-1} J_{-\sqrt{1-(\omega R)^2}} (\rho)$.
If we approach the extreme 5-brane as a limit of a near-extreme
5-brane, we indeed find that
$\rho^{-1} J_{\sqrt{1-(\omega R)^2}} (\rho)$ is the solution
that is selected. Since this solution is real, there is no
absorption for $\omega R <1$.\footnote{The fact that an
extremal 5-brane in 10 dimensions 
does not absorb minimally coupled scalars
below a certain threshold was noted in \cite{blTwo}.}
This result agrees with the extremal
limit of the 5-brane absorption cross-section calculated in
\cite{km}.

It is not hard to generalize our calculation to higher partial waves.
For the $\ell$-th partial wave we find that the absorption
cross-section vanishes for $\omega R \leq \ell+1$. Above this threshold
it is given by
\beq 
\sigma_\ell = {4\pi (\ell+1)^2\over \omega^3} \left (1-
e^{- 2\pi \sqrt{(\omega R)^2- (\ell+ 1)^2}} \right )\ .
\eeq

{}From (\ref{cs}) we reach the surprising conclusion that 
  \beq
   \langle T(\omega, \vec 0) T(-\omega, \vec 0) \rangle
  \eeq
 vanishes identically for $\omega <1/R$, which implies that gravity
does not couple to the massless modes of the world volume theory!  The
threshold energy $1/R$ is precisely $1/\sqrt{\alpha'_{\rm eff}}$,
where $\alpha'_{\rm eff} = 1 / (2 \pi T_{\rm eff})$ and
  \beq
   T_{\rm eff}= {1\over 2\pi R^2}
  \eeq
 is the tension of the fractionated strings.  The ordinary superstring
has its first massive excited state at mass $m^2 = 2/ \alpha'$.  The
threshold energy squared is half this value with $\alpha'$ replaced
by $\alpha'_{\rm eff}$.  If one imagines producing a single massive
string at $\omega = 1/R$, then its mass is $m = 1/\sqrt{\alpha'_{\rm
eff}}$.  Perhaps this is the first excited level of the non-critical
string living on the 5-brane. Similarly, the higher partial wave
thresholds might correspond to higher excited levels of mass
$(\ell+1)/\sqrt{\alpha'_{\rm eff}}$. If instead $\omega \geq 1/R$
corresponds to the pair production threshold of the first massive
state of fractionated strings, then the mass would be $m = 1 / (2
\sqrt{\alpha'_{\rm eff}})$.  Neither picture yields any obvious
explanation of the behavior
  \beq\label{scalingE}
   \langle T(\omega, \vec 0) T(-\omega, \vec 0) \rangle \sim
   \left[ (\omega R)^2- 1 \right]^{1/2}
  \eeq
 just above threshold.  Pair production in a weakly interacting theory
would predict a $7/2$ power in (\ref{scalingE}).  It would be
interesting to find an explanation of the observed square-root scaling
in (\ref{scalingE}).

We believe that our discussion applies to a large number of coincident
D5-branes, as well as to solitonic 5-branes of type IIA and IIB
theories. This is because all these solutions have the same Einstein
metric. Perhaps in the large $N$ limit some properties of the world
volume theories of these different branes become
identical.\footnote{We thank J. Maldacena for suggesting this
possibility to us.}
For $N$ coincident NS-NS 5-branes,
\beq
R^2\sim N\alpha'\ .
\eeq
Thus, the absorption cross-section (\ref{cs}) is formally independent
of $g_{\rm str}$. Therefore, our formula should
be applicable in the $g_{\rm str}\rightarrow 0$
limit proposed in \cite{ns}. There it was argued that in this limit
the NS-NS 5-branes decouple from the bulk modes. We indeed find that
incident gravitons are not absorbed for sufficiently low energies.
However, above a critical energy of order $1/\sqrt{N\alpha'}$
the 5-branes do appear to couple to the bulk modes. As we have 
commented, this is probably related to the fact that the scale
of the string theory living on the 5-brane is
\beq
\alpha'_{\rm eff} = N\alpha'\ ,
\eeq
i.e.{} the fundamental strings become fractionated \cite{juan}.

In summary, we note that probing branes with low-energy particles
incident from transverse directions is a useful tool for extracting
correlation functions in their world volume theory.  Here we have
given an application of this technique to M2-branes and to 5-branes,
but a more general investigation would be worthwhile.

\chapter{Green's functions from supergravity}
\label{GreensFunctions}

\section{Introduction}
\label{IntroGreen}

The nature of the relation between gauge fields and strings is an old,
fascinating, and largely unanswered question.  Its full answer is of
great importance for theoretical physics.  On one hand, it should
provide us with a theory of quark confinement by explaining the
dynamics of color-electric fluxes.  On the other hand, it will perhaps
uncover the true ``gauge'' degrees of freedom of the fundamental
string theories, and therefore of gravity.  There is still a long way
to go before we can hope to give a satisfactory string theory account
of the gauge theories observed in the real world.  In this chapter,
based on the paper \cite{gkPol}, we attempt nevertheless to make some
forward progress by considering the relation of maximally
supersymmetric gauge theory to supergravity via D3-branes.

The Wilson loops of gauge theories satisfy loop equations which
translate the Schwinger-Dyson equations into variational equations on
the loop space \cite{AP,mm}.  These equations should have a solution
in the form of the sum over random surfaces bounded by the loop.
These are the world surfaces of the color-electric fluxes.  For the
$SU(N)$ Yang-Mills theory they are expected to carry the 't~Hooft
factor \cite{GT}, $N^\chi$, where $\chi$ is the Euler
character. Hence, in the large $N$ limit where $g_{YM}^2 N$ is kept
fixed only the simplest topologies are relevant.

Until recently, the action for the ``confining string''
had not been known. In \cite{Sasha} it was suggested that it
must have a rather unusual structure. Let us describe it briefly.
First of all, the world surface of the electric flux propagates
in at least 5 dimensions. This is because the non-critical strings
are described by the fields
\eqn{fields}{X^\mu (\sigma)\ ,\qquad g_{ij} (\sigma) =
e^{\varphi (\sigma)} \delta_{ij}\ ,
}
where $X^\mu$ belong to 4-dimensional (Euclidean) space and
$g_{ij} (\sigma)$ is the world sheet metric in the conformal gauge.
The general form of the world sheet lagrangian compatible with
the 4-dimensional symmetries is
\eqn{ws}{
{\cal L}= {1\over 2} (\partial_i \varphi)^2 + a^2 (\varphi)
(\partial_i X^\mu)^2 + \Phi (\varphi) ^{(2)}R + {\rm Ramond-Ramond
\ backgrounds}\ ,
}
where $^{(2)}R$ is the world sheet curvature, $\Phi(\varphi)$
is the dilaton \cite{ft}, while the field 
$$ \Sigma (\varphi) = a^2 (\varphi)
$$
defines a variable string tension. In order to reproduce the zig-zag
symmetry of the Wilson loop, the gauge fields must be
located at a certain value $\varphi=\varphi_*$ such 
that $a(\varphi_*)=0$.
We will call this point ``the horizon.''

The background fields $\Phi(\varphi)$, $a(\varphi)$ and others must be
chosen to satisfy the conditions of conformal invariance on the
world sheet \cite{cmpf}. 
After this is done, the relation between gauge fields and 
strings can be described as an isomorphism between the general Yang-Mills
operators of the type
\eqn{gaugeop}{
\int d^4 x e^{ip\cdot x}\tr \left (\nabla_{\alpha_1}\ldots F_{\mu_1 \nu_1}
\ldots \nabla_{\alpha_n}\ldots F_{\mu_n \nu_n} (x)\right )
}
and the algebra of vertex operators of string theory, which have the form
\eqn{stringop}{ V^{\alpha_1 \ldots \alpha_n} (p)= \int d^2 \sigma 
\Psi_p^{i_1\ldots i_n j_1\ldots j_m} \big (\varphi (\sigma) \big )
e^{ip\cdot X(\sigma)} \partial_{i_1} X^{\alpha_1}
\ldots \partial_{i_n} X^{\alpha_n}\partial_{j_1}\varphi\ldots
 \partial_{j_m} \varphi  
\ ,
}
where the wave functions $\Psi_p^{i_1\ldots i_n j_1\ldots j_m} (\varphi)$
are again determined by the conformal invariance on the world sheet.
The isomorphism mentioned above implies the coincidence of the correlation 
functions of these two sets of vertex operators.

Another, seemingly unrelated, development is connected with the
Dirichlet brane \cite{JP} description of black three-branes
in \cite{gkp,kleb,gukt,gkThree}. 
The essential observation is that, on the one hand, the black
branes
are solitons which curve space \cite{hs} 
and, on the other hand, the world volume of $N$
parallel D-branes is described by supersymmetric $U(N)$ gauge theory with
16 supercharges \cite{Witten}.
A particularly interesting system is 
provided by the limit of
a large number $N$ of coincident D3-branes \cite{gkp,kleb,gukt,gkThree}, 
whose world volume is described by ${\cal N}=4$ supersymmetric $U(N)$ 
gauge theory in $3+1$ dimensions. For large $g_{YM}^2 N$ the
curvature of the classical geometry becomes small compared to the
string scale \cite{kleb}, 
which allows for comparison
of certain correlation functions between the supergravity and the gauge
theory, with perfect agreement \cite{kleb,gukt,gkThree}. 
Corrections in powers of $\alpha'$ times the curvature
on the string theory side correspond to corrections in powers of
$(g_{YM}^2 N)^{-1/2}$ on the gauge theory side.
The string loop corrections are suppressed by powers 
of $1/N^2$.

The vertex operators introduced in \cite{kleb,gukt,gkThree} describe
the coupling of massless closed string fields to the world volume.
For example, the vertex operator for the dilaton is
\eqn{dilvert}{ \int d^4 x e^{ip\cdot x} \tr F_{\mu\nu} F^{\mu\nu} (x) }
while that for the graviton polarized along the three-branes is
\eqn{gravvert}{ \int d^4 x e^{ip\cdot x} T^{\mu\nu}(x)\ ,
}
 where $T^{\mu\nu}$ is the stress tensor.  The low energy absorption
cross-sections are related to the two-point functions of the vertex
operators, and turn out to be in complete agreement with conformal
invariance and supersymmetric non-renormalization theorems, as we saw
in chapter~\ref{SchwingerTerms}.  An earlier calculation \cite{gkp} of
the entropy as a function of temperature for $N$ coincident D3-branes
exhibits a dependence expected of a field theory with $O(N^2)$
massless fields.  As we saw in chapter~\ref{EntUnp}, the entropy turns
out to be $3/4$ of the free field answer.  This is not a discrepancy
since the free field result is valid for small $g_{YM}^2 N$, while the
result of \cite{gkp} is applicable as $g_{YM}^2 N\rightarrow
\infty$. We now regard this result as a non-trivial prediction of
supergravity concerning the strong coupling behavior of ${\cal N} = 4$
supersymmetric gauge theory at large $N$ and finite temperature.

The non-critical string and the D-brane approaches to 3+1 dimensional
gauge theory have been synthesized in \cite{jthroat} by rescaling the
three-brane metric and taking the limit in which it has conformal
symmetry, being the direct product $AdS_5\times S^5$.\footnote{The
papers that put an early emphasis on the anti-de~Sitter nature of the
near-horizon region of certain brane configurations, and its relation
with string and M-theory, are \cite{modesOne,ght}. Other ideas on the
relation between branes and AdS supergravity were recently pursued in
\cite{Hyun,Boonstra,Sfetsos}.}  This is exactly the confining string
ansatz \cite{Sasha} with \eqn{confan}{ a(\varphi) = e^{\varphi/R}\ , }
corresponding to the case of constant negative curvature of order
$1/R^2$. The horizon is located at $\varphi_*=-\infty$.  The Liouville
field is thus related to the radial coordinate of the space transverse
to the three-brane. The extra $S^5$ part of the metric is associated with
the 6 scalars and the $O(6)$ R-symmetry present in the ${\cal N}=4$
supersymmetric gauge theory.

The purpose of this chapter is to 
make the next step and show how the excited states of the ``confining
string'' are related to the anomalous dimensions of the SYM theory.
Hopefully this analysis will help future explorations of asymptotically
free gauge theories needed for quark confinement. 

We will suggest a potentially very rich and detailed means of
analyzing the throat-brane correspondence: we propose an
identification of the generating function of the Green's functions of
the superconformal world-volume theory and the supergravity action in
the near horizon background geometry.
We will find it necessary to introduce a boundary of the $AdS_5$
space near the place where the throat turns into the asymptotically
flat space. Thus, the anti-de~Sitter coordinate $\varphi$ is 
defined on a half-line $(-\infty, 0]$,
similarly to the Liouville coordinate of the 2-dimensional string
theory \cite{Polch,DJ}. 
The correlation functions are specified by the boundary terms
in the action, again in analogy with the $c=1$ case. 
One new prediction that we will be able to extract this way is for the 
anomalous dimensions of the gauge
theory operators that correspond to massive string
states. For a state at level $n$ we find that, for large 
$g_{YM}^2 N$, the anomalous dimension grows as
$2 \sqrt n (2 g_{YM}^2 N)^{1/4}$.

\section{Green's functions from the supergravity action}
\label{Green}

The geometry of a large number $N$ of coincident D3-branes is
  \eqn{DThreeGeom}{
   ds^2 = \left( 1 + {R^4 \over r^4} \right)^{-1/2}
    \left( -dt^2 + d\vec{x}^2 \right) + 
    \left( 1 + {R^4 \over r^4} \right)^{1/2} 
    \left( dr^2 + r^2 d\Omega_5^2 \right) \ .
  }
 The parameter $R$, where
  \eqn{DThreeVars}{
   R^4 = {N \over 2\pi^2 T_3}\ , \qquad\qquad 
   T_3 = {\sqrt{\pi} \over \kappa}
  }
 is the only length scale for the geometry.  $T_3$
is the tension of a single D3-brane, and $\kappa$ is the
ten-dimensional gravitational coupling.  The near-horizon geometry of
$N$ D3-branes is $AdS_5 \times S^5$, as one can see most easily by
defining the radial coordinate $z = R^2/r$.  Then 
  \eqn{NearGeom}{
   ds^2 = {R^2 \over z^2} \left( -dt^2 + d\vec{x}^2 + dz^2 \right) + 
    R^2 d\Omega_5^2 \ .
  }
The relation to the coordinate $\varphi$ used in the previous section
is
\eqn{zphi}{ z= R e^{-\varphi/R}\ .
}
 Note that the limit $z \to 0$ is far from the brane.  Of course, for
$z \lsim R$ the $AdS$ form \NearGeom\ gets modified, and for $z \ll R$
one obtains flat ten-dimensional Minkowski space.  We will continue to
use the phrase ``far from the brane'' to emphasize that the
all-important boundary conditions at small $z$ are not imposed far
down the throat, but rather at the border region where the throat
merges into flat space.  The geometry is geodesically complete and
nonsingular, so there is no sense in which any one point on $AdS_5$ is
the location of the brane.

The basic idea is to identify the generating functional of connected
Green's functions in the gauge theory with the minimum of the
supergravity action, subject to some boundary conditions at 
$z=R$ and $z=\infty$:
  \eqn{Ident}
{
   W[g_{\mu\nu}(x^\lambda)] = K[g_{\mu\nu}(x^\lambda)] = 
     S[g_{\mu\nu}(x^\lambda,z)] \ .
  }
 $W$ generates the connected Green's functions of the gauge theory; $S$ is 
the supergravity action on the $AdS$ space; while $K$ is the minimum of $S$ 
subject to the boundary conditions.
 We have kept only the metric $g_{\mu\nu}(x^\lambda)$ of the
world-volume as an explicit argument of $W$.  The boundary conditions
subject to which the supergravity action $S$ is minimized are
  \eqn{BndCond}{
   ds^2 = {R^2 \over z^2} 
    \left( g_{\mu\nu} dx^\mu dx^\nu + dz^2 \right) + O(1) \qquad
    \hbox{as $z \to R$} \ .
  }
All fluctuations have to vanish as $z\rightarrow \infty$.

 A few refinements of the identification \Ident\ are worth commenting
on.  First, it is the generator of connected Green's functions which
appears on the left hand side because the supergravity action on the
right hand side is expected to follow the cluster decomposition
principle.  Second, because classical supergravity is reliable only
for a large number $N$ of coincident branes, \Ident\ can only be
expected to capture the leading large $N$ behavior.  Corrections in
$1/N$ should be obtained as loop effects when one replaces the
classical action $S$ with an effective action $\Gamma$.  This is
sensible since the dimensionless expansion parameter $\kappa^2/R^8
\sim 1/N^2$. We also note that, since $(\alpha')^2/R^4 \sim
(g^2_{YM} N)^{-1}$, 
the string theoretic $\alpha'$ corrections
to the supergravity action translate into gauge theory
corrections proportional to inverse powers of $g^2_{YM} N$. 
Finally, the fact that there is
no covariant action for type IIB supergravity does not especially
concern us: to obtain $n$-point Green's functions one is actually
considering the $n^{\rm th}$ variation of the action, which for $n>0$
can be regarded as the $(n-1)^{\rm th}$ variation of the covariant
equations of motion.

In section~\ref{TwoPt} we will compute a two-point function 
of massless vertex operators from \Ident, compare it
with the absorption calculations in \cite{kleb,gukt,gkThree}
and find exact agreement.
However it is instructive first to examine
boundary conditions.

\subsection{Preliminary: symmetries and boundary conditions}
\label{SymBC}

As a preliminary it is useful to examine the appropriate boundary
conditions and how they relate to the conformal symmetry.  In this
discussion we follow the work of Brown and Henneaux \cite{hen}.  In
the consideration of geometries which are asymptotically
anti-de~Sitter, one would like to have a realization of the conformal
group on the asymptotic form of the metric.  Restrictive or less
restrictive boundary conditions at small $z$ (far from the brane)
corresponds, as Brown and Henneaux point out in the case of $AdS_3$,
to smaller or larger asymptotic symmetry groups.  On an $AdS_{d+1}$
space,
  \eqn{AnyADS}{
   ds^2 = G_{mn} dx^m dx^n  
     = {R^2 \over z^2} \left( -dt^2 + d\vec{x}^2 + dz^2 \right)
  }
 where now $\vec{x}$ is $d-1$ dimensional, the boundary conditions
which give the conformal group as the group of asymptotic symmetries
are 
  \eqn{AllowedSize}{
   \delta G_{\mu\nu} = O(1)\ , \qquad
   \delta G_{z\mu} = O(z)\ , \qquad
   \delta G_{zz} = O(1) \ .
  }
 Our convention is to let indices $m,n$ run from $0$ to $d$ (that is,
over the full $AdS_{d+1}$ space) while $\mu,\nu$ run only from $0$ to
$d-1$ (i.e.{} excluding $z = x^d$).  Diffeomorphisms which preserve
\AllowedSize\ are specified by a vector $\zeta^m$ which for small $z$
must have the form
  \eqn{AllowedZeta}{\eqalign{
   \zeta^\mu &= \xi^\mu - {z^2 \over d} \eta^{\mu\nu} 
     \partial_\nu (\xi^\kappa{}_{,\kappa}) + O(z^4)  \cr
   \zeta^z &= {z \over d} \xi^\kappa{}_{,\kappa} + O(z^3) \ .
  }}
 Here $\xi^\mu$ is allowed to depend on $t$ and $\vec{x}$ but not $r$.
\AllowedZeta\ specifies only the asymptotic form of $\zeta^m$ at large
$r$, in terms of this new vector $\xi^\mu$.

Now the condition that the variation
  \eqn{LieVar}{
   \delta G_{mn} = {\cal L}_\zeta G_{mn} = 
     \zeta^k \partial_k G_{mn} + G_{kn} \partial_m \zeta^k + 
      G_{mk} \partial_n \zeta^k
  }
 be of the allowed size specified in \AllowedSize\ is equivalent to 
  \eqn{CKE}{
   \xi_{\mu,\nu} + \xi_{\nu,\mu} = {2 \over d} \xi^\kappa_{,\kappa}
  }
 where now we are lowering indices on $\xi^\mu$ with the flat space
Minkowski metric $\eta_{\mu\nu}$.  Since \CKE\ is the conformal
Killing equation in $d$ dimensions ($d=4$ for the three-brane), we see
that we indeed recover precisely the conformal group from the set of
permissible $\xi^\mu$.

The spirit of \cite{hen} is to determine the central charge of an
$AdS_3$ configuration by considering the commutator of deformations
corresponding to Virasoro generators $L_m$ and $L_{-m}$.  This method
is not applicable to higher dimensional cases because the conformal
group becomes finite, and there is apparently no way to read off a
Schwinger term from commutators of conformal transformations.
Nevertheless, the notion of central charge can be given meaning in
higher dimensional conformal field theories, either via a curved space
conformal anomaly (also called the gravitational anomaly) or as the
normalization of the two-point function of the stress energy tensor
\cite{erd}.  We shall see in section~\ref{TwoPt} that a calculation
reminiscent of absorption probabilities allows us to read off the
two-point function of stress-energy tensors in ${\cal N} = 4$
super-Yang-Mills, and with it the central charge.

\subsection{The central charge from the stress tensor two-point function}
\label{TwoPt}

As a preliminary, it is instructive to consider the case of a 
minimally coupled massless
scalar propagating in the anti-de~Sitter near-horizon geometry 
(one example of such a scalar is the dilaton $\phi$ \cite{kleb}).
As a further simplification
we assume for now that $\phi$ is in the $s$-wave
(that is, there is no variation over $S^5$).  Then the action becomes
  \eqn{MinAction}{\eqalign{
   S &= {1 \over 2 \kappa^2} \int d^{10} x \sqrt{G} 
    \left[ \tf{1}{2} G^{MN} \partial_M \phi \partial_N \phi \right]  \cr
     &= {\pi^3 R^8 \over 4 \kappa^2} \int d^4 x \,
      \int_R^\infty {dz \over z^3} \left[ (\partial_z \phi)^2 +  
       \eta^{\mu\nu} \partial_\mu \phi \partial_\nu \phi \right] \ .
  }}
 Note that in \MinAction, as well as in all the following equations,
we take $\kappa$ to be the ten-dimensional gravitational constant.
The equations of motion resulting from the variation of $S$ are
  \eqn{MinEOM}{
   \left[ z^3 \partial_z {1 \over z^3} \partial_z + 
    \eta^{\mu\nu} \partial_\mu \partial_\nu \right] \phi = 0 \ .
  }
 A complete set of normalizable solutions is
  \eqn{CompleteSet}{
   \phi_k(x^\ell) = \lambda_k e^{i k \cdot x} \tilde\phi_k(z) \quad 
    \hbox{where} \quad \tilde\phi_k(z) = 
{z^2 K_2(kz)\over R^2 K_2 (k R)} \ ,
   }
$$ k^2 = \vec k^2 - \omega^2
\ .$$
We have chosen the modified Bessel function $K_2(kz)$
rather than $I_2(kz)$ because the functions $K_\nu (kz)$ fall off
exponentially for large $z$, 
while the functions $I_\nu (kz)$ grow exponentially.
In other words, the requirement of regularity at the horizon (far
down the throat) tells us which solution to keep.
A connection of this choice with the absorption calculations of
\cite{kleb} is provided by the fact that,
for time-like momenta, this is the incoming wave which
corresponds to absorption from the small $z$ region. 
$\lambda_k$ is a coupling constant, and
the normalization factor has been chosen so that $\tilde\phi_k
=1$ for $z=R$.

Let us consider a coupling 
  \eqn{WVCoup}{
   S_{\rm int} = \int d^4 x \, \phi(x^\lambda) {\cal O}(x^\lambda)
  }
 in the world-volume theory.  If $\phi$ is the dilaton then according
to \cite{kleb} one would have ${\cal O} = \tf{1}{4} \tr F^2$.  Then
the analogue of \Ident\ is the claim that 
  \eqn{SimpIdent}{
   W[\phi(x^\lambda)] = K[\phi(x^\lambda)] = S[\phi(x^\lambda,z)]
  }
 where $\phi(x^\lambda,z)$ is the unique solution of the equations of
motion with $\phi(x^\lambda,z) \to \phi(x^\lambda)$ as $z \to R$.
Note that the existence and uniqueness of $\phi$ are guaranteed
because the equation of motion is just the Laplace equation on the
curved space.  (One could in fact compactify $x^\lambda$ on 
very large $T^4$ and
impose the boundary condition $\phi(x^\lambda,z) = \phi(x^\lambda)$ at
$z = R$.  Then the determination of $\phi(x^\lambda,z)$ is just the
Dirichlet problem for the laplacian on a compact manifold with
boundary).  

Analogously to the work of \cite{Polch} on the $c=1$ matrix model, we
can obtain the quadratic part of
$K[\phi(x^\lambda)]$ as a pure boundary term through
integration by parts,
  \eqn{KBound}{\eqalign{
   K[\phi(x^\lambda)] &= {\pi^3 R^8 \over 4 \kappa^2} 
     \int d^4 x \, \int_R^\infty {dz \over z^3} 
     \left[ -\phi \left( z^3 \partial_z {1 \over z^3} \partial_z + 
      \eta^{\mu\nu} \partial_\mu \partial_\nu \right) \phi  + 
      z^3 \partial_z \left( \phi {1 \over z^3} \partial_z \phi \right)
     \right]  \cr
    &={1\over 2}\int d^4 k d^4 q \lambda_k \lambda_q
(2\pi)^4 \delta^4(k+q) {N^2 \over 16 \pi^2} \int d^4 x \, {\cal F}
\ ,  }}
where we have expanded
$$\phi (x^\lambda) = \int d^4 k \lambda_k e^{i k\cdot x}\ .
$$ 
The ``flux factor'' ${\cal F}$ (so named because of its
resemblance to the particle number flux in a scattering calculation)
is
  \eqn{FluxDef}{
   {\cal F} = 
\left[ \tilde \phi_k {1 \over z^3} \partial_z 
\tilde \phi \right]_R^\infty \ .
  }
 In \KBound\ we have suppressed the boundary terms in the $x^\lambda$
directions---again, one can consider these compactified on 
very large $T^4$ so
that there is no boundary.  We have also used \DThreeVars\ to simplify
the prefactor.  Finally, we have cut off the integral at $R$ as a
regulator of the small $z$ divergence.  This is in fact appropriate
since the D3-brane geometry is anti-de~Sitter only for $z \gg R$.
Since there is exponential falloff in $\tilde\phi_k$ as $z\to\infty$,
only the behavior at $R$ matters.

To calculate the two-point function of ${\cal
O}$ in the world-volume theory, we differentiate $K$ twice
with respect to the parameters $\lambda_i$:\footnote{The 
appearance of the logarithm 
here is analogous to the logarithmic scaling violation in the
$c=1$ matrix model.}
  \eqn{TwoPtO}{\eqalign{
   \langle {\cal O}(k) {\cal O}(q) \rangle &= 
    \int d^4 x d^4 y \, e^{i k \cdot x + i q \cdot y} 
     \langle {\cal O}(x) {\cal O}(y) \rangle  \cr
    &= {\partial^2 K\over \partial \lambda_k \partial \lambda_q }=
(2\pi)^4 \delta^4(k+q) {N^2 \over 16 \pi^2} {\cal F}  \cr
    &= -(2\pi)^4 \delta^4(k+q) {N^2 \over 64 \pi^2} 
        k^4 \ln (k^2 R^2) + \hbox{(analytic in $k^2$)}
  }}
 where now the flux factor has been evaluated as
  \eqn{EvalFlux}{
   {\cal F} = \left[ \tilde\phi_k
{1 \over z^3} \partial_z \tilde\phi_k 
    \right]_{z=R} =
\left[ {1 \over z^3} \partial_z  \ln (\tilde\phi_k) \right]_{z=R}
     = -\tf{1}{4} k^4 \ln (k^2 R^2) + 
      \hbox{(analytic in $k^2$)} \ .
  }
Fourier transforming back to position space, we find 
\eqn{poscorr}{
\langle \tr F^2 (x) \tr F^2 (y) \rangle \sim {N^2\over
\vert x - y \vert ^8 }\ .
}
This is consistent with the free field result for small $g_{YM}^2 N$.
Remarkably, supergravity tells us that this formula continues to
hold as $g_{YM}^2 N\rightarrow \infty$.

Another interesting application of this analysis is to the two-point
function of the stress tensor, which with the normalization
conventions of chapter~\ref{SchwingerTerms} is
  \eqn{TwoPtFct}{
   \langle T_{\alpha\beta}(x) T_{\gamma\delta}(0) \rangle
     = {c \over 48 \pi^4} X_{\alpha\beta\gamma\delta}
       \left( {1 \over x^4} \right) \ ,
  }
 where the central charge $c = N^2/4$ and 
  \eqn{XFourDef}{
   \vcenter{\openup1\jot
   \halign{\strut\span\TL & \span\TR\cr
     X_{\alpha\beta\gamma\delta} &= 
      2 \square^2 \eta_{\alpha\beta} \eta_{\gamma\delta} - 
       3 \square^2 (\eta_{\alpha\gamma} \eta_{\beta\delta} + 
        \eta_{\alpha\delta} \eta_{\beta\gamma}) -
      4 \partial_\alpha \partial_\beta 
        \partial_\gamma \partial_\delta  \cr
      &\quad - 2 \square 
        (\partial_\alpha \partial_\beta \eta_{\gamma\delta} +
         \partial_\alpha \partial_\gamma \eta_{\beta\delta} +
         \partial_\alpha \partial_\delta \eta_{\beta\gamma} +
         \partial_\beta \partial_\gamma \eta_{\alpha\delta} +
         \partial_\beta \partial_\delta \eta_{\alpha\gamma} +
         \partial_\gamma \partial_\delta \eta_{\alpha\beta}) \ .  \cr
   }}}
 For metric perturbations $g_{\mu\nu} = \eta_{\mu\nu} + h_{\mu\nu}$
around flat space, the coupling of $h_{\mu\nu}$ at linear order is
  \eqn{SInt}{
   S_{\rm int} = \int d^4 x \, \tf{1}{2} h^{\mu\nu} T_{\mu\nu} \ .
  }
 Furthermore, at quadratic order the supergravity
action for a graviton polarized along the brane,
$h_{xy}(k)$, is exactly the minimal scalar action, 
provided the momentum $k$ is orthogonal to the $xy$ plane.
We can therefore carry over the result \TwoPtO\ to obtain
  \eqn{TwoTT}{
   \langle T_{xy}(k) T_{xy}(q) \rangle 
    = - (2\pi)^4 \delta^4(k+q) {N^2 \over 64 \pi^2} 
        k^4 \ln (k^2 R^2) + 
         \hbox{(analytic in $k^2$)} \ ,
  }
 which upon Fourier transform can be compared with \TwoPtFct\ to
give $c = N^2/4$.  In view of the conformal symmetry of both the
supergravity and the gauge theory, the evaluation of this one
component is a sufficient test.

The conspiracy of overall factors to give the correct normalization of
\TwoTT\ clearly has the same origin as the successful prediction of
the minimal scalar $s$-wave absorption cross-section examined in
\cite{kleb} and chapters~\ref{ThreeBraneAbs} and~\ref{SchwingerTerms}.
The absorption cross-section is, up to a constant of proportionality,
the imaginary part of \TwoTT.  In \cite{kleb} the absorption
cross-section was calculated in supergravity using propagation of a
scalar field in the entire three-brane metric, including the asymptotic
region far from the brane.  Here we have, in effect, replaced
communication of the throat region with the asymptotic region by a
boundary condition at one end of the throat. The physics of this is
clear: signals coming from the asymptotic region excite the part of
the throat near $z=R$. Propagation of these excitations into the
throat can then be treated just in the anti-de~Sitter approximation.
Thus, to extract physics from anti-de~Sitter space we must introduce a
boundary at $z=R$ and take careful account of the boundary terms that
contain the dynamical information.

It now seems clear how to proceed to three-point functions: on
the supergravity side one must expand to third order in the perturbing
fields, including in particular the three point vertices.  
At higher
orders the calculation is still simple in concept (the classical
action is minimized subject to boundary conditions), but the
complications of the ${\cal N}=8$ supergravity theory seem likely to
make the computation of, for instance, the four-point function,
rather tedious. We leave the details of such calculations for
the future. It may be very useful to compute at least the three point
functions in order to have a consistency check on the normalization of
fields.  

One extension of the present work is to consider what fields couple to
the other operators in the ${\cal N} = 4$ supercurrent multiplet.  The
structure of the multiplet (which includes the supercurrents, the
$SU(4)$ $R$-currents, and four spin $1/2$ and one scalar field)
suggests a coupling to the fields of gauged ${\cal N} = 4$ supergravity.  The
question then becomes how these fields are embedded in ${\cal N} = 8$
supergravity.  We leave these technical issues for the future, but
with the expectation that they are ``bound to work'' based on
supersymmetry. 

The main lesson we have extracted so far is that, for
certain operators that couple to the massless string states, the
anomalous dimensions vanish. We expect this to hold for all
massless vertex operators. This may be the complete set of operators
that are protected by supersymmetry. As we will
see in the next section, other operators acquire anomalous dimensions
that grow for large 't~Hooft coupling.

\section{Massive string states and anomalous dimensions}
\label{MassiveString}

Before we proceed to the massive string states, 
a useful preliminary is to discuss the higher partial waves of a
minimally coupled massless scalar.
The action in five
dimensions (with Lorentzian signature), the equations of motion, and
the solutions are
  \eqn{PartAction}{
   S = {\pi^3 R^8 \over 4 \kappa^2} \int d^4 x \int_R^\infty
        {dz \over z^3} \left[ (\partial_z \phi)^2 +
         (\partial_\mu \phi)^2 + {\ell (\ell + 4) \over z^2} \phi^2
        \right]
  }
  \eqn{PartEOM}{
   \left[ z^3 \partial_z {1 \over z^3} \partial_z +
    \eta^{\mu\nu} \partial_\mu \partial_\nu -
    {\ell (\ell + 4) \over z^2} \right] \phi = 0
  }
  \eqn{PartSols}{
   \phi_k(x^\ell) = e^{i k \cdot x} \tilde\phi_k(z) \quad
    \hbox{where} \quad
   \tilde\phi_k(z) = {z^2 K_{\ell+2}(kz) \over
    R^2 K_{\ell+2}(kR)} \ .
  }
We have chosen the normalization such that
$\tilde\phi_k(z) =1$ at $z=R$.
The flux factor is evaluated by expanding
  \eqn{PhiBehave}{\eqalign{
   K_{\ell+2}(kz) &= 2^{\ell+1}\Gamma(\ell+2) (k z)^{-(\ell+2)}  \cr
    &\qquad \cdot \left (1 + \ldots +
     {(-1)^\ell \over 2^{2\ell+3} (\ell+1)! (\ell+2)!}
     (kz)^{2(\ell+ 2)} \ln kz + \ldots \right )\ ,
  }}
where in parenthesis we exhibit the leading non-analytic term.
We find
  \eqn{GotFlux}{
   {\cal F} =
\left[ {1 \over z^3} \partial_z  \ln (\tilde\phi_k) \right]_{z=R}=
 {(-1)^\ell \over 2^{2 \ell + 2}
    [(\ell+1)!]^2 } k^{4+ 2l} R^{2\ell} \ln kR \ .
  }
 As before, we have neglected terms containing analytic powers of
$k$ and focused on the leading non-analytic term.
This formula indicates that the operator that couples to $\ell$-th
partial wave has dimension $4+\ell$.
In \cite{kleb} it was shown that such operators with the $SO(6)$
quantum numbers of the $\ell$-th partial wave have the form
\eqn{higher}{
\int d^4 x e^{ik\cdot x} \tr \left [ \left (X^{(i_i}\ldots X^{i_\ell )}
+ \ldots \right )
F_{\mu\nu} F^{\mu\nu} (x) \right ]\ , }
where in parenthesis we have a traceless symmetric tensor of $SO(6)$.
Thus, supergravity predicts that
their non-perturbative dimensions equal their bare dimensions.

Now let us consider massive string states.  
Our goal is to use supergravity to calculate the anomalous dimensions of
the gauge theory operators that couple to them.
To simplify
the discussion, let us focus on excited string states which are
spacetime scalars of mass $m$.  The propagation equation for such a
field in the background of the three-brane geometry is
\eqn{Coulone}{
\left [{d^2\over d
r^2} + {5\over r} {d\over dr} - k^2 \left (1 + {R^4\over r^4}\right )
-m^2 \left (1 + {R^4\over r^4}\right )^{1/2} \right ]
\tilde \phi_k =0\ . } 
For the state at excitation level $n$,
$$ m^2 = {4 n\over
\alpha'} \ . $$ 
In the throat region, $z\gg R$, \Coulone\ simplifies to 
\eqn{Coultwo}{ \left [{d^2\over d
z^2} - {3\over z} {d\over dz} -k^2 
-{m^2 R^2\over z^2} \right ]\tilde \phi_k =0\ . } 
Note that a massive particle with small energy $\omega \ll m$, which would
be far off shell in the asymptotic region $z\ll R$, can nevertheless
propagate in the throat region (i.e.{} it is described by an oscillatory
wave function).
 
Equation \Coultwo\
is identical to the equation encountered in the analysis of higher
partial waves, except
the effective angular momentum is not in general an integer:
in the centrifugal barrier term $\ell (\ell+4)$ is replaced by $(m R)^2$.
Analysis of the choice of wave function goes through as before,
with $\ell+2$ replaced by $\nu$, where
\eqn{anom}{\nu = \sqrt{ 4 + (m R)^2 } \ .}
In other words, the wave function falling off exponentially for
large $z$ and normalized to $1$ at $z=R$ is\footnote{
These solutions are reminiscent of the loop correlators
calculated for $c\leq 1$ matrix models in \cite{mss}.}
\eqn{massivewave}{
   \tilde\phi_k(z) = {z^2 K_\nu(kz) \over
    R^2 K_\nu (k R)} \ .}
Now we recall that
$$ K_\nu = {\pi\over 2\sin (\pi\nu)} \left (I_{-\nu} - I_\nu \right )\ ,
$$
$$ I_\nu (z) = \left ({z\over 2}\right )^\nu
\sum_{k=0}^\infty {(z/2)^{2k} \over k! \Gamma (k+\nu+ 1)}\ .
$$
Thus, 
\eqn{genexpand}{ K_\nu (kz) = 2^{\nu -1}\Gamma (\nu)
(k z)^{-\nu} \left (1+ \ldots -  
\left ({z k\over 2 }\right )^{2\nu } 
{\Gamma (1-\nu)\over \Gamma (1+\nu )}+\ldots \right )\ ,
} 
where in parenthesis we have exhibited the leading non-analytic term.
Calculating the flux factor as before, 
we find that the leading non-analytic term of the
two-point function is
  \eqn{OOFct}{
   \langle {\cal O}(k) {\cal O}(q) \rangle =-(2\pi)^4 \delta^4(k+q)
{N^2 \over 8\pi^2} {\Gamma (1-\nu)\over \Gamma (\nu )}
    \left ({kR\over 2}\right )^{2\nu} R^{-4}\ .
  }
This implies that the dimension of the corresponding SYM
operator is equal to $2+\nu$.

Now, let us note that
$$ R^4 = 2 N g_{YM}^2 (\alpha')^2\ ,
$$
which implies
$$ (m R)^2 = 4 n g_{YM} \sqrt {2 N} 
\ .
$$
Using \anom\ we find that the spectrum of dimensions
for operators that couple to massive string states is,
for large $g_{YM} \sqrt N$,
\eqn{anomform}{ 
h_n \approx 2\left (n g_{YM} \sqrt {2 N} \right )^{1/2}
\ .}
Equation \anomform\ is a new 
non-trivial prediction of the string theoretic
approach to large $N$ gauge theory.\footnote{We do not expect this equation
to be valid for arbitrarily large $n$, because application of linearized
local effective actions to arbitrarily excited string states 
is questionable.
However, we should be able to trust our approach for moderately excited
states.}

We conclude that, for large 't~Hooft coupling, the anomalous
dimensions of the vertex operators corresponding to massive string
states grow without bound.  By contrast, the vertex operators that
couple to the massless string states do not acquire any anomalous
dimensions. This has been checked explicitly for gravitons, dilatons
and RR scalars \cite{kleb,gukt,gkThree}, and we believe this to be a
general statement.\footnote{A more general set of such massless fields
is contained in the supermultiplet of AdS gauge fields, whose
reduction to the boundary was recently studied in \cite{ff}.} Thus,
there is an infinite class of operators that do not acquire anomalous
dimensions (this is probably due to the fact that they are protected
by SUSY). The rest of the operators are not protected and can receive
arbitrarily large anomalous dimensions.  Such large dimensions may be
expected to ``freeze out'' the operators from the local operator
algebra, just as the corresponding massive string states are frozen
out of the supergravity description.  While we cannot yet write down
the explicit form of these operators in the gauge theory, it seems
likely that they are conventional local operators, such as \gaugeop.
Indeed, the coupling of a highly excited string state to the world
volume may be guessed on physical grounds.  Since a string in a
D3-brane is a path of electric flux, it is natural to assume that a
string state couples to a Wilson loop ${\cal O} = \exp \left( i
\oint_\gamma A \right)$. Expansion of a small loop in powers of $F$
yields the local polynomial operators.

There is one potential problem with our treatment of massive
states. The minimal linear equation \Coulone\ where higher
derivative terms are absent may be true only for a particular field
definition (otherwise corrections in positive
powers of $\alpha' \nabla^2$ will be present in the equation).
Therefore, it is possible that there are energy dependent leg
factors relating the operators $\cal O$ in 
\OOFct\ and the gauge theory operators of the form \gaugeop.
We hope that these leg factors do not change our conclusion about the
anomalous dimensions. However, to completely settle this issue we
need to either find an exact sigma-model which incorporates all
$\alpha'$ corrections or to calculate the three-point functions
of massive vertex operators.

\section{Conclusions}
\label{ConclusionGreen}

There are many unanswered questions that we have left for the future.
So far, we have considered the limit of large 't~Hooft coupling,
since we used the one-loop sigma-model calculations for
all operators involved. If this coupling is not large, then we have
to treat the world sheet theory as an exact conformal field theory
(we stress, once again, that the string loop corrections are
$\sim 1/N^2$ and, therefore, vanish in the large $N$ limit).
This conformal field theory is a sigma-model on a hyperboloid.
It is plausible that, in addition to the global $O(2,4)$ symmetry,
this theory possesses the $O(2,4)$ Kac-Moody algebra.
If this is the case, then the sigma model is tractable
with standard methods of conformal field theory.

Throughout this work we detected many formal similarities of our
approach with that used in $c\leq 1$ matrix models. These models
may be viewed as early examples of gauge theory -- non-critical string
correspondence, with the large $N$ matrix models playing the role
of gauge theories. Clearly, a deeper understanding of
the connection between the present work and the $c\leq 1$ matrix models
is desirable.

\chapter{Concluding Remarks}
\label{FurtherDirections}

Because this dissertation represents the author's contribution to a
rapidly evolving field, it is perhaps more worthwhile to look ahead to
where the field may be moving rather than to simply recapitulate the
contents of chapters~\ref{EntUnp}-\ref{GreensFunctions}.  There are
certainly a number of avenues for further research, but perhaps the
most attractive revolve around the idea of focusing on the
near-horizon geometry of $p$-branes and the relation to gauge
theory.

The near-horizon geometry of simple D-brane configurations is in
general a homogeneous space: for example, D3-branes have as their
near-horizon geometry $AdS_5 \times S^5 = SO(4,2)/SO(4,1) \times
SO(6)/SO(5)$.  Writing the anti-de~Sitter factor as a quotient of
groups makes the conformal symmetry manifest.  There is certainly
great interest in investigating further how the properties of
conformal field theories can be reflected by supergravity on
homogeneous spaces; but from a mainstream physics point of view the
most significant extensions of the work in
chapter~\ref{GreensFunctions} would involve non-conformal gauge
theories, such as pure ${\cal N} = 2$, $1$, or $0$ gauge theory in
four dimensions.  These theories have non-vanishing beta-functions,
and for ${\cal N} < 2$ there is confinement.  Already in
\cite{WitHolTwo} a way of studying non-conformal theories and
confinement ``holographically'' has been suggested, using brane
configurations at finite temperature.  A very clean way to proceed
would be to perturb the D3-brane solution by what in the gauge theory
are relevant operators (for instance, scalar or fermion masses).  The
resulting geometry should reflect softly broken ${\cal N} = 4$ gauge
theory, which in general does confine in the infrared.  To put it
another way, ${\cal N} = 4$ would be the ultraviolet regulator of a
confining theory.  Several ideas for studying soft breakings in
supergravity have been proposed, but the main difficulty lies in
finding solutions of the full non-linear supergravity equations with
less symmetry than the usual three-brane solution.

The greatest conceptual limitation of the supergravity--gauge theory
correspondence is the requirement of large $N$ and strong 't~Hooft
coupling $g_{YM} \sqrt{N}$.  Let us review the reasons for these
requirements.  The formulas will be specific to the case of the
three-brane, but similar observations apply to other brane
configurations like the D1-D5 bound state.  Supergravity has a scale
(the Planck scale) because the gravitational coupling $\kappa$ is
dimensionful.  The near-horizon geometry has length scale $R$ and is
an exact solution to the tree-level supergravity equations.  Loop
corrections in supergravity are suppressed only if $\kappa/R^4$ is
small.  But $\kappa/R^4 \sim 1/N$, so we see the origin of the large
$N$ requirement.  Also, supergravity is the $\alpha' \to 0$ limit of
string theory, and the $\alpha'$ corrections can be neglected only if
$\alpha'/R^2 \sim 1/(g_{YM} \sqrt{N})$ is small---hence the
requirement of large $g_{YM} \sqrt{N}$.

Gauge theory at large $N$ is dominated by planar graphs, so one way to
regard the relation
  \eqn{FavoriteAgain}{
   Z_{YM}[\phi_0] \sim 
       \maximum_{\phi=\phi_0\ {\rm on}\ \partial AdS} e^{-S[\phi]} 
    \quad \hbox{as $g_{YM}^2 N \to \infty$} 
  }
 is to say that classical supergravity captures the large $g_{YM}
\sqrt{N}$ asymptotics of planar graphs.  

One would hope to improve \FavoriteAgain\ to an exact equality by
replacing the right hand side with a full quantum theory of gravity:
namely, type IIB string theory in the D3-brane background.  As
mentioned in chapter~\ref{Introduction}, there are formidable
difficulties standing in the way of formulating string theory in a
background involving Ramond-Ramond fields.  Ramond-Ramond fields enter
into the world-sheet description of strings as bilinears of spin
fields with additional ``superghost'' dressings which maintain
two-dimensional superconformal invariance.  At the time of writing we
are unaware of any proposals for formulating a ghost-free non-linear
sigma-model action.

Notwithstanding the technical difficulties, we can ask what stringy
corrections to \FavoriteAgain\ might be like.  By including the full
tower of massive excited string states, a string theory description
would reintroduce the non-chiral gauge theory operators that were
argued in chapter~\ref{GreensFunctions} to freeze out in the large
coupling limit on account of large anomalous dimensions.  The ability
to make $\alpha'/R^2$ arbitrary is precisely what we need in order to
get down to weak coupling where perturbative gauge theory is reliable.
Special quantities, like Schwinger terms in the OPE's of conserved
currents, can be expected to be protected against radiative
corrections, as we saw in the case of the central charge in
chapter~\ref{SchwingerTerms}.  Tests of the gravity--gauge theory
correspondence that rely in no way on symmetries, anomalies, or
non-renormalizations are much harder to come by.  This is perhaps
typical of duality conjectures: expansion parameters get shuffled in
such a way that what is simple on one side of the duality is very
difficult to verify on the other.

One can nevertheless try to see what improvements to \FavoriteAgain\
might be the most accessible to calculation.  One idea is to compute
the leading $1/(g_{YM} \sqrt{N})$ correction to the entropy
from the first $\alpha'$ correction to supergravity (which comes at
order $\alpha'^3$ for type IIB).  Another possibility is to consider
$1/N$ corrections via low order loop effects in quantized supergravity
which are not troubled by divergences.  For example, as explained in
\cite{WitHolOne}, the $U(1)$ part of the world-volume gauge theory is
not visible in the $AdS$ geometry, because the low-dimension conformal
fields in the $U(1)$ sector would correspond to tachyons violating the
Breitenlohner-Freedman bound.  Consequently, supergravity in the
throat region should reflect a central charge $c = (N^2 - 1)/4$ rather
than the $c = N^2/4$ calculated in chapter~\ref{SchwingerTerms}.  This
slight difference may be one of the simplest $1/N^2$ effects.  To see
it in supergravity, one would want to evaluate the two-point function
of the stress-tensor via graviton propagation in the $AdS$ geometry,
but including one-loop corrections to the propagator.  The
identification of parameters discussed in the previous paragraph
guarantees that this will lead to some $1/N^2$ effect; whether it
comes out with the correct coefficient is a check on whether
\FavoriteAgain\ can be extended beyond the level of planar graphs.

Another direction, building upon the results in
chapters~\ref{ThreeBraneAbs} and~\ref{GreensFunctions}, is to consider
what effects in the world-volume theory become visible when one
considers the full three-brane metric rather than just its
near-horizon limit.  Crudely speaking, the $AdS$ space is ``cut off''
at a radius $R$ by the asymptotically flat portion of the three-brane
metric.  Modes of the supergravity fields with $\omega R \ll 1$ do not
see the effects of this cut-off because they approach a pure power law
dependence at radii much larger than $\omega R^2$.  Supergravity
effects at these low energies seem to be all that is relevant to the
renormalizable gauge theory on the world-volume.  The full
world-volume theory (presumably some non-abelian version of the DBI
action) contains non-renormalizable terms which contribute to Green's
functions with inverse powers of $R$.  By considering the double
scaling limit \dsl\ with finite $\omega R$, one should be able to
probe the complicated dynamics of the full DBI theory.  The $\omega R
\to 0$ limit is all that is explored by the analysis of the
near-horizon geometry.  In \cite{ghkk} the leading corrections in
$\omega R$ to $s$-wave minimal scalar absorption were shown to
correspond to the effects of breaking conformal invariance in the
world-volume theory by retaining the leading non-renormalizable term
in the DBI action.  Further work in this direction has begun to
suggest the appearance of anomalous dimensions at finite $\omega R$
for what in the conformal theory are protected operators whose
dimensions are fixed by the superconformal algebra.

A better understanding of the double scaling limit may also be
interesting from the point of view of black hole physics, for the
following reason.  In dynamical processes like absorption or emission,
one usually restricts to low energy.  The physics picked out by this
limit is the same as by the decoupling limit (where gravity is taken
arbitrarily weak), the throat limit (where attention is focused on the
near-horizon geometry), and the renormalizable gauge theory lagrangian
on the world-volume.  To put it another way, the ``universal''
low-energy behavior \TwoCut\ of the cross sections has to do with a
renormalizable theory on the brane decoupled from the bulk.  If we
instead we explore the double scaling limit with $(\omega R)^4 \sim N
\kappa \omega^4$ held finite, then the cross-sections retain $\kappa$
dependence.  We may speculate that couplings between the brane and the
bulk are being studied in a controlled way.  The world-volume gauge
theory is now non-renormalizable, and its time evolution cannot be
regarded as strictly unitary because of the ultraviolet divergences.
The energy scale of the non-renormalizable effects is $1/R$.  Is it
possible that this divergence-related non-unitarity is related to the
apparent lack of unitarity in the evolution of black holes?  If so,
then string theory resolves information loss by providing a perfectly
unitary description of the system at the (much higher) energy scale
$1/\sqrt{\alpha'}$.  This discussion has focused on the D3-brane, but
in principle such considerations should extend to the effective string
model.  As mentioned before, the string theory description becomes
less simple because of the complicated structure of the D1-D5 bound
state.  However, non-unitarity issues may be more straightforward to
analyze in $1+1$ dimensions.

\addcontentsline{toc}{chapter}{Bibliography}
\bibliography{phd}
\bibliographystyle{ssg}

\end{document}